\documentclass[preprint,3p,times]{elsarticle}

\usepackage{amssymb}
\usepackage{amsthm}
\usepackage{bm}
\usepackage[utf8]{inputenc}
\usepackage{microtype}                
\usepackage{mathtools}
\usepackage{amsmath}
\usepackage{yhmath}
\usepackage{amssymb}
\usepackage{amsthm}
\usepackage[german,english]{babel}         
                                    
\usepackage{12many}                           
\usepackage[suftesi]{frontespizio}  
\usepackage[binding=5mm]{layaureo}  
\usepackage{booktabs}                       

\usepackage{tabularx}                       
\usepackage{graphicx}                       
\usepackage{subfig}                         
\usepackage{caption}                        
\usepackage{listings}                       
\usepackage[font=small]{quoting}            
\usepackage{epigraph}   
\usepackage{amsmath, amssymb, amsthm, braket, empheq}         

\usepackage[dvipsnames]{xcolor}             
\usepackage{lipsum}                         
\usepackage{eurosym}                        
\usepackage{hyperref}                       
\usepackage{bookmark}                       
                              
\usepackage{calc}
\usepackage{float}                          
\usepackage[normalem]{ulem}                 
\usepackage{pdfpages}
\usepackage{dsfont}
\usepackage{bbold}
\usepackage{empheq}                         
\usepackage{mparhack,fixltx2e,relsize}      
\usepackage{enumitem}                       
\usepackage{esint}                          

\usepackage{epigraph}                       
\hypersetup{
    colorlinks,
    linkcolor={red!50!black},
    citecolor={blue!50!black},
    urlcolor={blue!80!black}
}

\biboptions{numbers,sort&compress}

\usepackage[figuresright]{rotating}

\graphicspath{{Figures/}} 

\DeclarePairedDelimiter{\abs}{\lvert}{\rvert}

\newcommand{\ir}{\mathrm{i}}
\newcommand{\cl}{\text{cl}}
\newcommand{\Teh}{T_{\text{Ehr}}}
\newcommand{\beq}{\begin{equation}}
\newcommand{\eeq}{\end{equation}}
\newcommand{\up}{\uparrow}
\newcommand{\down}{\downarrow}

\newcommand{\mpar}[1]{\marginpar{\footnotesize \it #1}}

\def\bxi{\boldsymbol {\xi}}

\def\heff{\hbar_{\text{eff}}}
\def \H{\hat{H}}
\def\ro{\hat{\rho}}

\def\a{\alpha}

\def\a{\alpha}

\def\Tr{\text{Tr}}
\DeclarePairedDelimiter\floor{\lfloor}{\rfloor}

\def\summary#1{\bigskip \framebox{\parbox{0.89\textwidth}{\small {\bf{Summary}}: { #1}}}\\}

\begin{document}

\begin{frontmatter}




\title{Out-of-equilibrium dynamics of quantum many-body systems with long-range interactions}


\author[ND]{Nicol\`o Defenu}
\address[ND]{Institute for Theoretical Physics, ETH Zürich, Wolfgang-Pauli-Str.\,27, 8093 Zürich, Switzerland}
\ead{ndefenu@phys.ethz.ch}
\author[AL]{Alessio Lerose}
\address[AL]{Department of Theoretical Physics, University of Geneva, Quai Ernest-Ansermet 30, 1205 Geneva, Switzerland}
\ead{alessio.lerose@unige.ch}
\author[SP]{Silvia Pappalardi}
\address[SP]{Institut f\"ur Theoretische Physik, Universit\"at zu K\"oln, Z\"ulpicher Straße 77, 50937 K\"oln, Germany}
\ead{pappalardi@thp.uni-koeln.de}

\begin{abstract}
Experimental progress in atomic, molecular, and optical platforms in the last decade has stimulated strong and broad interest in the quantum coherent dynamics of many \emph{long-range interacting} particles. 
The prominent collective character of these systems enables novel non-equilibrium phenomena with no counterpart in conventional quantum systems with local interactions. 
Much of the theory work in this area either focussed on the impact of variable-range interaction tails on the physics of local interactions or relied on mean-field-like descriptions based on the opposite limit of all-to-all infinite-range interactions.
In this Report, we present a systematic and organic review of recent advances in the field.  
Working with prototypical interacting quantum spin lattices without disorder, our presentation hinges upon a versatile theoretical formalism that interpolates between the few-body mean-field physics and the many-body physics of quasi-local interactions. Such a formalism allows us to connect these two regimes, providing both a formal quantitative tool and basic physical intuition. 
We leverage this unifying framework to review several findings of the last decade, including the peculiar non-ballistic spreading of quantum correlations, counter-intuitive slowdown of entanglement dynamics, suppression of thermalization and equilibration, anomalous scaling of defects upon traversing criticality, dynamical phase transitions, and genuinely non-equilibrium phases stabilized by periodic driving.
The style of this Report is on the pedagogical side, which makes it accessible to readers without previous experience in the subject matter.
\end{abstract}


\end{frontmatter}
\tableofcontents

\section{Introduction}
\label{introduction}

An increasing interest in quantum many-body physics with long-range interactions is being driven by growing experimental capabilities in controlling and manipulating atomic, molecular, and optical systems (AMO). Currently, various platforms such as Rydberg atoms, dipolar quantum gases, polar molecules, quantum gases in optical cavities, and trapped ions, have native two-body long-range interactions which can be modeled as algebraically decaying $J/(\Delta r)^\alpha$ with the distance $\Delta r$\,\cite{blatt2012quantum,britton2012engineered,richerme2014nonlocal,baumann2010dicke,barredo2018synthetic,yan2013observation}. The exponent $\alpha$ can in some cases be experimentally tuned --- e.g. through off-resonant coupling of internal levels of trapped ions to motional degrees of freedom\,\cite{blatt2012quantum,monroe2021programmable}, or trapping neutral atoms in photonic modes of a cavity\,\cite{ritsch2013cold,mivehvar2021cavity}. Additionally, the effective interaction range can be efficiently tuned in systems of Rydberg atoms in one- and two-dimensional arrays\,\cite{endres2016atom, labuhn2016tunable} or by Rydberg dressing \cite{zeiher2017coherent,hollerith2022realizing}.

The versatility of the aforementioned AMO platforms spurred intense theoretical and experimental explorations. These studies established that long-range interactions provide clear routes to circumventing the constraints imposed by either conventional thermalization~\cite{polkovnikov2011colloquium} or conventional bounds on information spreading~\cite{lieb1972finite}. 
Accordingly, the prominent collective character of systems with long-range interactions can lead to 
a kaleidoscope of novel phenomena which cannot be observed in systems with local interactions.
Major examples include: the observation of ``super-luminal'' correlation and entanglement spreading\,\cite{mottl2012roton,jurcevic2014quasiparticle,richerme2014nonlocal} (to be contrasted with the conventional light-cone behavior in presence of local interactions\,\cite{cheneau2012no}); dynamical phase transitions in low dimensions\,\cite{baumann2011exploring, brennecke2013real,klinder2015observation,landig2015quantum, leonard2017monitoring,zhang2017observationdpt,muniz2020exploring}; exotic defect scaling\,\cite{keesling2019quantum,klinder2015dynamical}; self-organized criticality\,\cite{helmrich2020signatures}; time-translation symmetry breaking\,\cite{zhang2017observationdtc,choi2017observation,rovny2018observation}; {quantum many-body chaos}\,\cite{garttner2017measuring,wei2018exploring,joshi2020quantum}.  As such, control of long-range interacting assemblies stands out as a promising ingredient for future quantum-technological applications, including quantum metrology and quantum computation.

While this great diversity of platforms and research directions largely contributes to generate widespread excitement about long-range interactions, it has at the same time certain drawbacks. The backgrounds and interests of the numerous research groups active in this area span a very wide spectrum.
 On one hand, experimental interpretations are often based on a few-body, mean-field-like way of thinking\,\cite{ritsch2013cold,mivehvar2021cavity,leroux2010implementation,bilitewski2021dynamical}. 
 Albeit remarkably simple and powerful, this perspective may fail to fully capture the complexity of non-equilibrium phenomena with long-range interactions. 
 On the other hand, theoretical investigations have often prioritized mathematically rigorous efforts aimed at characterizing the departure from known properties of locally-interacting systems\,\cite{else2018improved,tran2019localityheating,tran2020locality,guo2019signaling,chen2019finite,storch2015interplay,kastner2011diverging,hernandez2017correlation,kuwahara2020area}. 
 Albeit sometimes in synergy with experiments\,\cite{monroe2021programmable}, this perspective may obscure the construction of an intuitive physical picture applicable to the broad range of out-of-equilibrium phenomena mentioned above. 
 Despite recent attempts to recompose the corresponding mosaic in equilibrium\,\cite{defenu2021longrange}, 
this complementarity of perspectives on similar phenomena still struggle to come together and cement a unified research field and community. 
As a consequence, the current understanding of the out-of-equilibrium dynamics of long-range interacting quantum many-body systems still seems to lack a systematic organization comparable to that of quantum locally interacting~\cite{polkovnikov2011colloquium} or classical long-range interacting\,\cite{campa2009statistical} systems.
 
The purpose of this Report is to provide a systematic and intuitive theoretical approach to non-equilibrium phenomena arising from non-random long-range interactions in quantum many-body systems. Our effort aims at bridging the various complementary views in this wide research area and creating a unifying framework. We will review a selection of significant findings in the field, emphasizing how they can be encompassed within a common basic theoretical language and formalism. 
The approach reviewed in this Report is suited to bridge the simple mean-field description --- which applies to infinite-range interactions, i.e. $\alpha=0$ --- to the description of systems with quasi-local interactions, i.e. $\alpha\gg d$ (where $d$ is the dimension of the system), which allow a well-defined notion of locality in spite of non-local interaction tails. The \emph{strong long-range regime} in between, i.e. $0<\alpha<d$, will be the focus of this Report; we will frequently emphasize the \textit{leitmotiv} that \emph{the physics in this regime interpolates between conventional few-body and conventional many-body physics}.
The reach of this unifying framework will be illustrated using prototypical models of interacting quantum spin lattices. This choice does not only serve the purpose of directly relating our results to paradigmatic locally-interacting systems\,\cite{sachdev1999quantum,tauber2009critical}, but it is also allows us to make direct connections with the major AMO experimental platforms recalled above. 

\medskip

This Report is organized as follows:
\begin{itemize}
\item Our journey will start in Sec.\,\ref{sec_eq} with a review of \emph{equilibrium properties} of ferromagnetic quantum spin systems exemplified by a variable-range quantum XY model Sec.~\ref{sec_21}, including a discussion of the equilibrium phase diagram upon varying parameters and interaction range (via $\alpha$) Sec.~\ref{sec_22} and a critical examination of the mean-field limit Sec.~\ref{sec_alpha0eq}. Hence, in Sec.\,\ref{low_en_theory} we will review the low-energy description in terms of bosonic excitations (spin waves) across the phase diagram, with emphasis on spectral properties arising from a long interaction range. Finally in Sec.~\ref{sec_beyondlsw} we will discuss spectral properties beyond linear spin-wave theory. 
\item {The low-energy description} reviewed in Sec.\,\ref{sec_eq} can be used to investigate near-equilibrium dynamics. This setup allows to study the peculiar properties of spatial propagation of quantum correlation in presence of long-range interactions\,\cite{ cevolani2016spreading, cevolani2017universal} as well as their unusual equilibration dynamics\,\cite{defenu2021metastability}, reviewed in Sec.\,\ref{sec_31} and\,\ref{sec_metastability} respectively. Both these phenomena can be studied in quantum quenches lying within the supercritical phase and, therefore, only represent a small departure from equilibrium. More surprisingly, we are going to show that the low-energy description is also capable of addressing dynamical scaling phenomena arising after quenches across the critical point. This is the case of the universal defect formation following a quasi-static sweep across the quantum critical point\,\cite{defenu2018dynamical, defenu2021quantum}, see Sec.\,\ref{kzm_gen} and the rise of dynamical quantum phase transitions\,\cite{heyl2018dynamical}, which we will treat in Sec.\,\ref{sec_34} (the latter, however, will require to modify the simple low-energy description employed before).
\item Section\,\ref{dyn_high_exc} is devoted to the study of \emph{dynamics far away from equilibrium}, induced by \emph{quantum quenches}. We will first consider  in Sec.\,\ref{sec_classLim} the fully-connected model with all-to-all uniform interactions, and examine the simplest instances of dynamical phenomena in this limit, which can be understood in terms of few-body semiclassical dynamics. For finite-range interactions, however, the motion of semiclassical collective degrees of freedom is coupled to many quantum-fluctuation modes with various wavelengths. In Sec.\,\ref{sec_42} we review a systematic approach to the resulting complex many-body problem, originally developed in Refs.~\cite{lerose2018chaotic,lerose2019impact}. As a first implication stemming from this approach we will review lower bounds on thermalization time scales associated with long-range interactions, establishing the genuinely non-equilibrium nature of dynamical phenomena in these systems. Hence, we will examine the impact of many-body quantum fluctuations on dynamical criticality and quantum information spreading far away from equilibrium upon tuning the interaction range. Throughout, we will highlight the role of long-range interactions in generating novel phenomena.

\item In Sec.\,\ref{sec_5} we will employ the methodology of Sec.~\ref{dyn_high_exc} to describe coherent dynamics subject to \emph{periodic driving}.
Here we will review how long-range interactions allow to stabilize genuinely non-equilibrium phases, without an equilibrium counterpart, in low-dimensional quantum systems of the kind routinely realized in AMO experiments. This will include phases that may be viewed as quantum many-body realizations of the celebrated Kapitza pendulum (Sec.~\ref{sec_kapitza}) as well as discrete-time crystals, which spontaneously break time-translation symmetry (Sec.~\ref{sec_tc}). 
\item Finally, in the conclusive Section, we spell out the topics which are \emph{not} covered in this Report: from effects of inhomogeneities, to frustrated, random, or noisy interactions, to dissipative and monitored dynamics.
\end{itemize}

\medskip
Throughout the presentation, our goal is to provide both physical intuition and systematic theoretical understanding of experimentally relevant phenomena. 
We kept the style of the Report on the pedagogical side, as we hope this work will also be useful to readers who are interested in taking their first dive into the realm of quantum dynamics in presence of long-range interactions.

\section{Equilibrium properties of long-range interacting quantum spin systems}
\label{sec_eq}

In this Section we summarize and discuss basic \emph{equilibrium} properties of quantum spin lattices with variable interaction range, which will prove to be useful in the rest of this Report.
For definiteness we will focus on a class of ferromagnetic XY quantum spin models, introduced in Sec.~\ref{sec_21}, and review its equilibrium phase diagram in Sec.~\ref{sec_22}.
We will then work out its low-energy description in terms of bosonic excitations (``spin waves'') both in the fully-connected limit (Sec.~\ref{sec_alpha0eq}) and with finite-range interactions (Sec.~\ref{low_en_theory}), with emphasis on the peculiar features of the quasiparticle spectrum such as discreteness \cite{lerose2020origin, defenu2021metastability}, divergent group velocity \cite{hauke2013spread, cevolani2015protected, cevolani2016spreading, lepori2015effective}, and dressing effects.
Finally, in Sec.~\ref{sec_beyondlsw} we will discuss finer low-energy properties beyond spin-wave theory, including domain-wall (de)confinement \cite{liu2019confined, lerose2019quasilocalized}. 

In this Section we will keep the model parameters fully general. In the rest of the Report we will frequently restrict the model for simplicity, but all the results can always be straightforwardly extended, drawing on the general setup introduced here and in Sec.~\ref{sec_421} below.

\subsection{Variable-range quantum XY model}
\label{sec_21}

Throughout this work we will consider a prototypical model implemented in AMO platforms, a quantum XY spin model with tunable interaction range.
We take a $d$-dimensional square lattice of $N=L^d$ quantum spins-$s$, i.e. with $(\hat s^x)^2+(\hat s^y)^2+(\hat s^z)^2=s(s+1)\hat{\mathbb{1}}$, described by a Hamiltonian of the form
\beq
\hat H_{\alpha} = -   \sum_{\mathbf{r},\mathbf{r'}} J_{\mathbf{r},\mathbf{r'}}(\alpha) \bigg( \frac{1+\gamma} 2 \hat\sigma^x_{\mathbf{r}} \hat\sigma^x_{\mathbf{r'}}
+ \frac{1-\gamma} 2 \hat\sigma^y_{\mathbf{r}} \hat\sigma^y_{\mathbf{r'}} \bigg)
 - h  \sum_{\mathbf{r}}  \hat\sigma^z_{\mathbf{r}} 
 \label{eq_H}
\eeq
In this equation $\hat \sigma^\mu_\mathbf{r} = \hat s^\mu_\mathbf{r} /s$ are operators corresponding to the normalized spin components in the $\mu=x,y,z$ direction, acting on site $\mathbf{r}=(r_1,\dots,r_d)$ denote the sites of the square lattice $r_1,\dots,r_d=1,\dots,L$.
This represents a generalization of the standard spin-$1/2$ case, where $\hat \sigma^\alpha_\mathbf{r}$ reduce to the standard Pauli matrices. Such a normalization allows us to keep track of the role of quantum fluctuations, which are suppressed in the classical limit $s\to\infty$. 
 The quantity $\gamma$  parametrizes the XY anisotropy.  In this Report we will consider anisotropic spin systems, i.e. $\gamma\neq0$; for definiteness we assume $\gamma>0$ (negative values equivalent upon rotating the spins around the $z$-axis by $\pi/2$), and we will frequently set $\gamma=1$ (quantum Ising model). We will occasionally comment on the isotropic limit $\gamma\to0$ when relevant.
The quantity $h$ represents the transverse magnetic field strength, which we assume $h\ge 0$ (negative values are equivalent upon rotating the spins around the $x$-axis by $\pi$). Throughout this report we will always use units such that Planck's constant is $\hbar\equiv 1$.

The ferromagnetic couplings $J_{\mathbf{r},\mathbf{r'}}(\alpha) \equiv J_{\Delta r}(\alpha)$ depend on the  distance $\Delta r=\lvert\lvert\mathbf{r}-\mathbf{r'}\rvert\rvert$ between the two involved sites, and we will be interested in tuning their spatial range through the parameter $\alpha$. 
Specifically, we consider interactions algebraically decaying with the distance,\footnote{Note that for spins-$1/2$ the terms $\mathbf{r}=\mathbf{r'}$ produce an inconsequential additive constant $E=\sum_{\mathbf{r}} J_{\mathbf{r},\mathbf{r}}  /2$, as Pauli matrices square to $1$. 
Diagonal terms may be important for higher-spin Hamiltonians. As a rule we will set $J_{\mathbf{r},\mathbf{r}}=0$. {We will occasionally comment on interesting phenomena associated with spin self-interactions further below.}}
\begin{equation}
\label{lr_coup}
J_{\mathbf{r},\mathbf{r'}}(\alpha)  = \frac{J}{\lvert \lvert \mathbf{r} - \mathbf{r'} \rvert \rvert^\alpha}\qquad \mathrm{for}\,\,r\neq r'.
\end{equation}
To impose periodic boundary conditions, various equivalent choices of distance function are possible; we take $\lvert \lvert \mathbf{r} - \mathbf{r'} \rvert \rvert \equiv \sqrt{\sum_{\mu=1}^d [\min(|r_\mu-r'_\mu|, L-|r_\mu-r'_\mu|)]^2}$.
The overall constant $J>0$ is chosen in such a way to fairly compare the models with different $\alpha$, i.e. to make the mean-field interaction strength $J_0$ independent of $\alpha$:
\begin{equation}
\label{kac_norm}
J = \frac{J_0}{2\mathcal{N}_{\alpha,L}}, \qquad \mathcal{N}_{\alpha,L} \equiv \frac 1 2 \sum_\mathbf{r}  \frac{1}{\lvert \lvert \mathbf{r} - \mathbf{r'} \rvert \rvert^\alpha}  \, .
\end{equation}
This prescription --- known as Kac normalization~\cite{kac1963van} --- 
 is necessary to make the thermodynamic limit well defined for $\alpha \le d$, where the divergence $\mathcal{N}_{\alpha,L} \sim L^{d-\alpha} $ with the system size ensures that energy scales extensively. For $\alpha>d$ the Kac rescaling factor saturates to a finite value in the thermodynamic limit  $\mathcal{N}_{\alpha,L} \to \mathcal{N}_{\alpha} $.
Note that for ferromagnetic interactions, the specific choice of the lattice does not alter the qualitative properties discussed in this review, as we will further comment below.

\summary{We consider a $d$-dimensional quantum XY spin model with non-random ferromagnetic interactions that decay algebraically with exponent $\alpha$. The interaction strength is rescaled to make the ground-state energy density independent of $\alpha$.}

\subsection{Equilibrium phase diagram in a nutshell}
\label{sec_22}


For $\gamma\neq 0$ \mpar{$T=0$ phase diagram} and for ferromagnetic interactions $J_{\mathbf r, \mathbf r'} \ge 0$ the system has an equilibrium zero-temperature phase transition for small enough $\abs{h}$, associated with the spontaneous breaking of its $\mathbb{Z}_2$ spin-inversion symmetry of the $x$-component.
The longitudinal magnetization $\braket{\hat \sigma_{\bold r}^x}$ undergoes an abrupt change from $\braket{\hat \sigma_{\bold r}^x}=0$ in the unique paramagnetic ground state for $\abs{h} > h_{\text{cr}}$ to $\braket{\hat \sigma_{\bold r}^x}_{\pm}=\pm m(h)\ne0$ in the two degenerate ferromagnetic ground states  for $\abs{h}<h_{\text{cr}}$. 
When interactions are local, the universality class of this quantum phase transition is the same as that of the $(d+1)$-dimensional classical Ising model\,\cite{sachdev1999quantum}. 

The possible emergence \mpar{$T>0$ phase diagram} of an ordered phase at \emph{finite temperature} $T>0$, i.e. of long-range ordered excited states at finite energy density, depends on the dimensionality $d$ and on the decay exponent $\alpha$ of the interactions. 
For strictly finite range, finite-temperature order is stable only for $d\ge 2$; 
in this case, the universal properties of the thermal phase transition are the same as that of the corresponding $d$-dimensional classical Ising model\,\cite{sachdev1999quantum,lazo2021finite}. 
Increasing the interaction range, however, enhances the effective lattice connectivity, somewhat similarly to the effect of increasing the lattice dimensionality\,\cite{joyce1966spherical,sak1973recursion,defenu2015fixed}. While frustration prevents antiferromagnetic interactions from creating collective ordering, ferromagnetic interactions do cooperate to suppress the effect of spatial fluctuations, generally resulting in a qualitative enhancement of the system's ability to order as the interaction range increases\,\cite{dyson1969existence}.

\begin{figure}[t]
\centering
\subfloat[ Ising Model]{\label{Fig_SecII_1a}
\includegraphics[width=.55\linewidth]{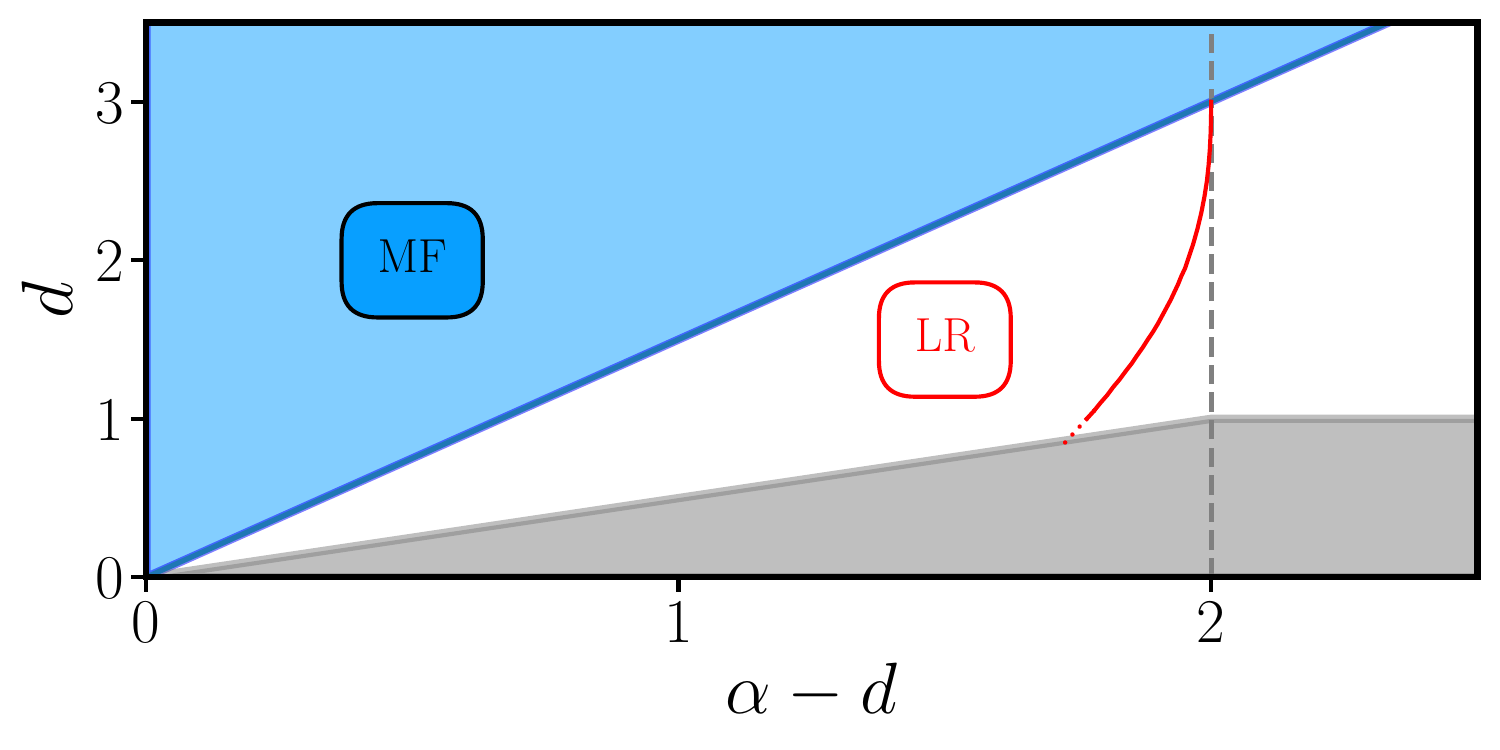}
}
\subfloat[ Finite temperature ]{\label{Fig_SecII_1b}
\includegraphics[width=.45\linewidth]{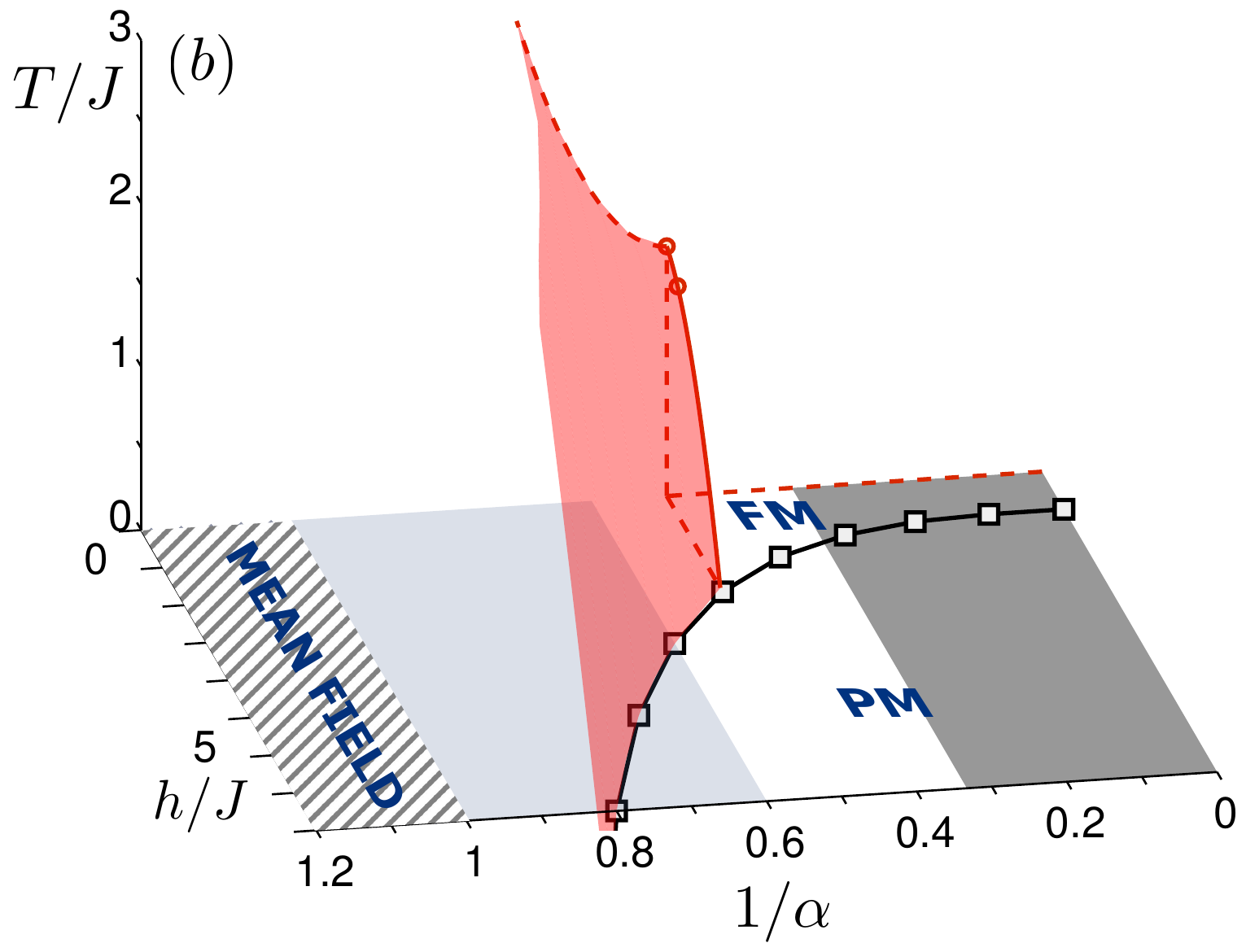}}
\caption{\textbf{Phase diagram of long-range interacting quantum Ising model.} Panel (a) draws the phase diagram of the Hamiltonian in Eq.\,\eqref{eq_H} with $\gamma\neq 0$. There, mean-field theory properly describes the universal scaling behavior in the cyan shaded region. The white background with the red $\rm LR$ label represents the region in the phase diagram where the universal behavior is correlated and influenced by the presence of relevant long-range couplings. The boundary  of the LR region is simply $\alpha=d+2$ in mean-field theory (vertical shaded line), but gets displaced to $\alpha_{*}=d+2-\eta_{\rm sr}$ by two-loop corrections (red line). Finally, the white area on the right of the red line  signals the region of irrelevant long-range couplings where the universal behavior is controlled by the local part of the interactions. Figure reproduced from Ref.\,\cite{defenu2017criticality}. (b) Finite temperature phase-diagram of the one-dimensional Ising model, in the transverse field $h$, interaction exponent $\alpha$, and temperature $T$ space, in the units of $J$ (without Kac normalization). For $\alpha<1$, the system is in the mean-field regime (striped region), for $\alpha<5/3$ the mean-field universality is exact, while for $\alpha>3$ the model is in the same universality class of the short-range Ising model (dark grey).  Figure adapted from Ref.\cite{knap2013probing}. 
}
\label{Fig_SecII_1}
\end{figure} 

The analogy \mpar{Raise of effective dimension by $\alpha$} between integer-dimension long-range systems and local systems in lower fractional dimensions has been quantitatively tested in multiple studies in recent years, both in classical\,\cite{angelini2014relations,defenu2015fixed, horita2016upper,behan2017long} and quantum\,\cite{dutta2001phase,defenu2017criticality} long-range systems. Leading-order perturbation theory results support the exact correspondence between the universal behavior of long-range interacting systems with dimension $d$ and decay exponent $\alpha$ and locally-interacting system with dimension $d_{\rm eff}=2(d+z)/(\alpha-d)$, where $z$ is the dynamical critical scaling exponent
\,\cite{dutta2001phase, monthus2015dyson, monthus2016real}. Advanced renormalization group studies highlighted deviations from this correspondence, which only occur beyond the leading order and hence remain small\,\cite{defenu2017criticality}. Therefore, it is possible to employ the effective-dimension relation above to get the qualitative shape of the phase diagram: For $\alpha<\frac{5}{3}d$ the universal scaling behavior is captured by mean-field theory, while for $\frac{5}{3}d<\alpha<\alpha_{*}$ the system displays correlated critical behavior influenced by the presence of long-range interactions. Finally, for $\alpha>\alpha_{*}$ the interaction tails become irrelevant and the critical exponents coincide with the ones of the model with local interactions\,\cite{defenu2017criticality}.

The location of $\alpha_{*}$ was subject to multiple controversies, but the result $\alpha_{*}=d+2-\eta_{\rm sr}$\,\cite{angelini2014relations, defenu2015fixed}, with $\eta_{\rm sr}$ the anomalous dimension of the model with local interactions, appears now to be established\,\cite{defenu2017criticality} in agreement with extensive numerical simulations in classical models\,\cite{horita2016upper}. Due to the dependence of $\alpha_{*}$ on the universal equilibrium properties of the local model, this boundary \emph{does} depend on the particular model. For the model considered in this Report, the phase diagram of the equilibrium critical problem is displayed in Fig.\,\ref{Fig_SecII_1}: in Fig.\,\ref{Fig_SecII_1a} we report the universality properties for different dimension \cite{defenu2017criticality}, while in Fig.\,\ref{Fig_SecII_1b} we show the phase diagram for finite temperature of the one-dimensional model \cite{knap2013probing}. Decreasing the decay exponent below the dimension of the system, i.e.  $\alpha<d$, does not produce any major implications in the equilibrium critical scaling, but it does modify the thermodynamic properties. In the regime $\alpha<d$ the boundary contribution to thermodynamic quantities cannot be neglected (a property referred to as non-additive), leading to the violation of several established equilibrium properties including the equivalence of thermodynamic ensembles\,\cite{barre2001inequivalence}.  

Given this scenario, long-range interacting systems can be classified in the following way \cite{defenu2021longrange}: 
for $\alpha < d$ they are in the so-called \emph{strong long-range regime}, for $d< \alpha< \alpha  ^* $ they are in the \emph{weak long-range regime}, while for $\alpha>\alpha^*$ one retrieves short-range properties.\footnote{We warn the readers that the nomenclature we adopt here is far from being universally established in the vast literature on long-range interactions. }
In this Report we will mainly focus on quantum spin systems around the strong long-range regime, i.e. $0\le \alpha \lesssim d$, and we aim at providing a cohesive picture for their distinctive dynamics.

In the light of the above discussion, the role of the spatial dimension is diminished in systems with variable-range interactions, as the relevant parameter in equilibrium is the effective dimension $d_{\rm eff}$. Yet, the one-dimensional case is particularly interesting: In presence of local interactions, $d=1$ systems cannot exhibit ordering at finite temperature, because isolated topological defects of a ferromagnetically ordered pattern (domain-wall-like excitations) cost a finite energy\,\cite{dyson1969existence}; a longer range of ferromagnetic interactions induces \emph{binding} between domain walls, and hence a tendency to stabilize ferromagnetic order.
The effect of long-range interactions is thus most dramatic for $d=1$:
The algebraically decaying interactions in Eq.~\eqref{eq_H} allow to stabilize ferromagnetic order  in the thermodynamic limit upon decreasing $\alpha$ below $2$.
This happens as the interaction potential between two domain walls becomes \emph{confining} at large distances, such that free isolated domain walls cost an infinite energy.

\summary{In equilibrium, the universal critical properties with $J/(\Delta r)^\alpha$-interactions are close to those of the locally interacting version of the system ($\alpha=\infty$) in a higher effective dimension $d_{\rm eff}=2(d+z)/(\alpha-d)$. For $d=1$ and $\gamma\ne0$, finite-temperature ordering becomes possible for $\alpha\le2$.}

\subsection{Low-energy spectrum with infinite-range interactions ($\alpha=0$)}
\label{sec_alpha0eq}

Let us start \mpar{$\alpha=0$} by discussing the exactly solvable infinite-range limit $\alpha\to0$. This will be the starting point to analyze the behavior for $\alpha>0$.


\subsubsection{Mean-field theory as an exact classical limit}
\label{sec_231}

Increasing \mpar{Collective spin} the range of interactions, by decreasing the exponent $\alpha\to0$, weakens spatial fluctuations, leading the system toward its mean-field limit --- similarly to the effect of increasing the system dimensionality $d\to\infty$.
This can be seen explicitly by rewriting the Hamiltonian~\eqref{eq_H} in terms of the collective spin components 
\begin{equation}
\label{Stot}
{\hat{S}}^{\mu}= \sum_{i=1}^N {\hat{s}}^\mu_i \, , \quad \mu=x,y,z \, ,
\end{equation}
which gives the expression 
\beq
\label{eq_Halpha0}
{\hat H}_{\alpha=0} =   
- \frac{J_0}{N\, s^2}
\left ( \frac{1+\gamma}{2} (\hat S^x)^2 +   \frac{1-\gamma}{2} (\hat S^y)^2\right) -  \frac h s \, \hat S^z \, \ ,
\eeq
where we used $J=J_0/(N-1)\approx J_0/N$. 
This expression highlights that the $\alpha=0$ Hamiltonian is a function of a single degree of freedom: the collective spin. All other non-collective spin modes are frozen and do not participate in dynamics.
The collective spin magnitude ${ {\hat S}}^2=(\hat S^x)^2+(\hat S^y)^2+(\hat S^z)^2={S(S+1)}$ with $S=Ns, Ns-1,Ns-2,\dots,0$ or $1/2$ is conserved,
\beq
\big [  {\hat S}^2,\hat H_{\alpha=0}\Big] = 0 \ .
\eeq
The Hilbert space sector $\mathcal{H}_{\hat S^2=S(S+1)}$ associated with the quantum number $S$ contains $g_{N,S}$ copies of a spin-$S$ representation of SU(2), where $g_{N,S}=\rm{dim} \, \mathcal{H}_{\hat S^z=S}-\rm{dim} \, \mathcal{H}_{\hat S^z=S+1}$. This combinatorial number depends implicitly on $s$; in the simplest case $s=1/2$ we have 
\beq
\label{eq_sectorsdim}
g_{N,S} =  \binom{N}{N/2-S} - \binom{N}{N/2-S-1} = \frac{2S+1}{N+1}\binom{N+1}{N/2-S }.
\eeq
In each such $(2S+1)$-dimensional space, the Hamiltonian acts as Eq.~\eqref{eq_Halpha0} and it effectively describes a single spin of size $S$.

For all states \mpar{Classical limit} with large $S$ growing with $N$, the thermodynamic limit $N\to\infty$ is equivalent to a semiclassical limit for the collective spin: The rescaled spin satisfies commutation relations of the form
\beq
\left [ \frac{\hat S^{\mu}}{S} , \frac{\hat S^{\nu}}{S} \right ] = \frac i {S} \; \epsilon_{\mu\nu\rho} \;\frac{ \hat S^{\rho}}{S}  \ ;
\eeq
and the Hamiltonian can be rewritten in terms of the rescaled spin as
\beq
\label{eq_Halpha0rescaled}
{\hat H}_{\alpha=0} = (S/s) \bigg\{   -  J_0 \rho\left [ \frac{1+\gamma}{2} \bigg(\frac{\hat S^x}{S}\bigg)^2 +   \frac{1-\gamma}{2} \bigg(\frac{\hat S^y}{S}\bigg)^2\right] -  h \, \frac{\hat S^z}{S}  \bigg\}\, ,
\eeq
where $\rho\equiv S/(Ns)$  is a constant depending on the collective spin sector, with $0\le \rho\le 1$. 
Thus, the system manifestly has an effective Planck constant $\hbar_{\text{eff}} \equiv 1 /S$.
Keeping in mind that a meaningful thermodynamic limit requires to take $J_0\rho$ as a constant independent of $N$, 
we conclude that the limit $N\to\infty$ realizes a classical limit with a continuous spin 
\beq
 \frac{\langle {\hat {\vec{S}}} \rangle} {S} \leadsto \vec{\mathcal S}
\eeq
of (conserved) length $1$ governed by the classical Hamiltonian $\hat H_{\alpha=0}/(S/s) \leadsto \mathcal{H}_{\text{cl}}$,
\beq
\label{eq_Hcl}
\mathcal{H}_{\text{cl}}(\vec{\mathcal{S}}) =    -  {\rho J_0} \left ( \frac{1+\gamma}2 (\mathcal S^x)^2 +  \frac{1-\gamma}2 (\mathcal S^y)^2\right ) - h \mathcal S^z \, ,
\eeq
Canonical variables can be taken as, e.g., $\mathcal{S}^z=\cos\theta$ and $\arctan_2 (\mathcal{S}^x,\mathcal{S}^y)=\phi$.

The absolute ground state \mpar{Ground state} minimizes energy across all sectors; for ferromagnetic interactions the ground state is realized for maximal collective spin polarization, $S = Ns$, i.e. for $\rho=1$.
A rigorous implication of the classical limit~\cite{aizenman2020dimerization} is that, as ${N\to\infty}$, the ground state expectation values $\braket{ \vec{\hat{S}}}_{\text{GS}}/S$ of the collective spin components converge to the minimum point $\vec{\mathcal{S}}^*$ of the classical Hamiltonian $\mathcal{H}_{\text{cl}}$ on the unit sphere. For later purpose it is convenient to define a rotated reference frame $(\mathbf{X},\mathbf{Y},\mathbf{Z})$ adapted to the ground state polarization, i.e., such that $\mathbf{Z}\equiv \vec{\mathcal{S}}^*$. Using spherical coordinates we can parametrize 
\beq
\label{eq_rotatedframe}
\bold{X} \equiv 
\left( \begin{matrix}
\cos\theta \cos\phi \\
\cos\theta\sin\phi \\
-\sin\theta
\end{matrix} \right) ,  \quad
\bold{Y} \equiv 
\left( \begin{matrix}
-\sin\phi \\
\cos\phi \\
0 
\end{matrix} \right)
 , \quad
\bold{Z} \equiv 
\left( \begin{matrix}
\sin\theta \cos\phi \\
\sin\theta\sin\phi \\
\cos\theta
\end{matrix} \right) .
\eeq
Crucially, the quantum uncertainty $ \sqrt{\big\langle\big(\hat{S}^X\big)^2\big\rangle  \big\langle\big(\hat{S}^Y\big)^2\big\rangle}$ associated with spin fluctuations in the transverse directions $\mathbf{X}$ and $\mathbf{Y}$ spans a phase-space area of order $h_{\rm eff}=2\pi/S$, which is vanishingly small as $N\to\infty$.\footnote{This point will be further discussed at length in Sec.~\ref{sec_413}.}

The discussion above is valid for generic infinite-range Hamiltonians. For our model in Eq.~\eqref{eq_Halpha0}, minimization of $\mathcal{H}_{\rm cl}$ on the unit sphere gives the ground-state polarization  $\vec{\mathcal{S}}^* = (\pm\sin\theta^*,0,\cos\theta^*)$, with
\beq
\label{eq_thetastar}
\theta^* = \left\{
\begin{split}
0 & \quad \text{ for } h>h_{\rm cr}\equiv\rho J_0(1+\gamma)\\
\arccos\left(\frac h {\rho J_0(1+\gamma)}\right) &\quad    \text{ for } 0\le h \le h_{\rm cr}\equiv \rho J_0 (1+\gamma)
\end{split}
\right.
\eeq
The ferromagnetic phase transition at $h=h_{\rm cr} $ is associated with the bifurcation of the minimum.
The ferromagnetic energy $\mathcal{E}\equiv \mathcal{H}_{\rm cl}(\vec{\mathcal{S}}^*) =-\rho J_0(1+\gamma)/2-h^2/[\rho J_0(1+\gamma)]$ is extremal for $\rho=1$, in agreement with the claim anticipated above and with intuition. See plots in Figs.~\ref{fig2a} and~\ref{fig2b}.

\summary{The fully-connected Hamiltonian with $\alpha=0$ is a function of collective spin variables only. The thermodynamic limit realizes a semiclassical limit for the collective spin with an effective Planck constant scaling as $\hbar_{\text{eff}} \propto 1/N$. 
}

\subsubsection{Collective quantum fluctuations and excitations}
\label{sec_232}

It is important \mpar{Zero-point collective fluctuations} to stress that, in spite of the exact classical limit, the ground-state wavefunction is \emph{not} a product state of $N$ spins pointing in the direction $\vec{\mathcal{S}}^*$: Collective interactions generate global (\emph{multipartite}) quantum entanglement among all spins. Such quantum correlations stem from quantum fluctuations of the collective spin around the average direction $\vec{\mathcal{S}}^*$. Such effects can be understood via semiclassical analysis to leading order in $\hbar_{\rm eff}$.


Let us first compute the low-energy spectrum of the infinite-range Hamiltonian~\eqref{eq_Halpha0} thought as a single-spin Hamiltonian, with $S$ growing with $N$.
The collective spin moves in an energy landscape whose depth grows with $N$. The ground state wavefunction is localized around its global minimum (or minima).
Expansion of $\mathcal{H}_{\rm cl}$ around the minimum gives access to the ground-state fluctuations and low-lying harmonic excitations. This can be conveniently done via a Holstein-Primakoff transformation \cite{wannier1987statistical}: Recalling $\hat S^\pm = \hat S^x \pm i \hat S^y$, 
\beq
\label{eq_hpbasic}
\left\{
\begin{split}
 \hat S^- &= \hat b^\dagger \sqrt{2S-\hat b^\dagger \hat b}  \, , \\
 \hat S^+ &=  \sqrt{2S-\hat b^\dagger \hat b} \;\; \hat b \, , \\
  \hat S^z &= S - \hat b^\dagger \hat b \, .
 \end{split}
 \right.
\eeq
These equations represent an exact embedding of a quantum spin into a bosonic mode.

This procedure is simplest in the paramagnetic phase.\mpar{Single-spin spectrum in the symmetric phase}
For large $h \gg h_{\rm cr}$ the ground state approaches the uncorrelated state fully polarized along $z$, and the elementary excitations  approach the tower of spin lowering excitations. For finite $h>h_{\rm cr}$ the collective spin fluctuates along the transverse directions --- more prominently along the ``soft'' direction $x$ and more weakly along the ``stiff'' direction $y$.\footnote{Recall that we assumed $\gamma>0$ for definiteness.} Such fluctuations can be described by mapping $\hat S^x$ and $\hat S^y$ to canonical bosonic operators via Eq.~\eqref{eq_hpbasic}, 
\beq
\label{eq_hpcollective}
\left\{
\begin{split}
 \hat S^x &\approx \sqrt{S} \; \hat q \, , \\
 \hat S^y &\approx \sqrt{S} \; \hat p \, , \\
  \hat S^z &= S -  \hat n_0 = S - \frac{\hat q^2+\hat p^2-1}{2} \, .
 \end{split}
 \right.
\eeq
Using $[ \hat q, \hat p]=i$ one can check that for large $S$ the spin commutation relations are satisfied by the right-hand sides of Eqs.~\eqref{eq_hpcollective} to leading order.
In a classical phase-space description, the approximation given by the above truncated Holstein-Primakoff transformation corresponds to replacing the surface of the sphere by its tangent plane at the North pole.

Using Eq.~\eqref{eq_hpcollective}, the Hamiltonian~\eqref{eq_Halpha0} can be approximated 
by neglecting terms of order $1/S$, and hence easily diagonalized. 
We find: 
\begin{subequations}\label{eq_expansionH>}
\begin{align}
	\label{eq_expansionH>1}
		 \hat H_{\alpha=0} &\approx -  N \rho h + \frac{h}{s} \frac{ \hat q^2+ \hat p^2-1}{2} -  \frac{\rho J_0}{s}  \bigg(\frac{1+\gamma} 2 \hat q^2 + \frac{1-\gamma} 2 \hat p^2 \bigg)  \\
	\label{eq_expansionH>2}
   &=  - N \rho h  + \frac{1}{s} \bigg( \frac{\omega_>-\omega_{>}^{(0)}}{2}\bigg) +  \frac{1}{s} \omega_> \;  \hat n \  ,
\end{align}
\end{subequations}
where
\beq
\label{eq_MFsmallosc>}
\omega_{>}=\sqrt{[h-\rho J_0 (1-\gamma)][h-\rho J_0 (1+\gamma)]}, \qquad \omega_{>}^{(0)}=h \ .
\eeq
The first term in the last line of Eq.~\eqref{eq_expansionH>2} represents the classical energy, and the second one is the variation of the zero-point energy due to quantum fluctuations around the classical configuration. In the last term, $\hat n$  is the harmonic excitation quanta of energy $\omega_> $ 
(not to be confused with the ``bare'' spin-lowering excitation quanta $\hat n_0$).

\begin{figure}[t]
\centering
\subfloat[ Order parameter]{\includegraphics[width=.3\linewidth]{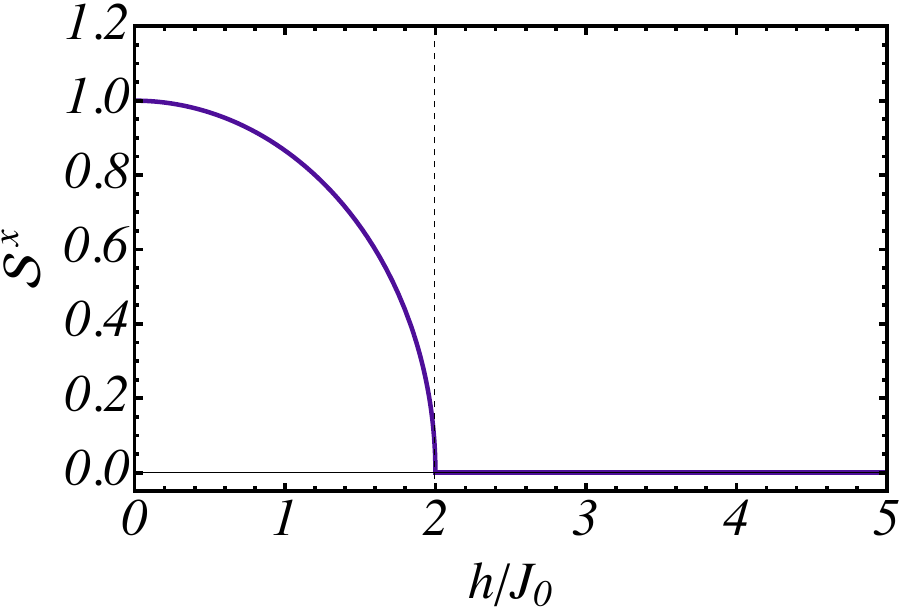}\label{fig2a}} \hspace{0.2cm}
\subfloat[Ground-state energy density]{\includegraphics[width=.3\linewidth]{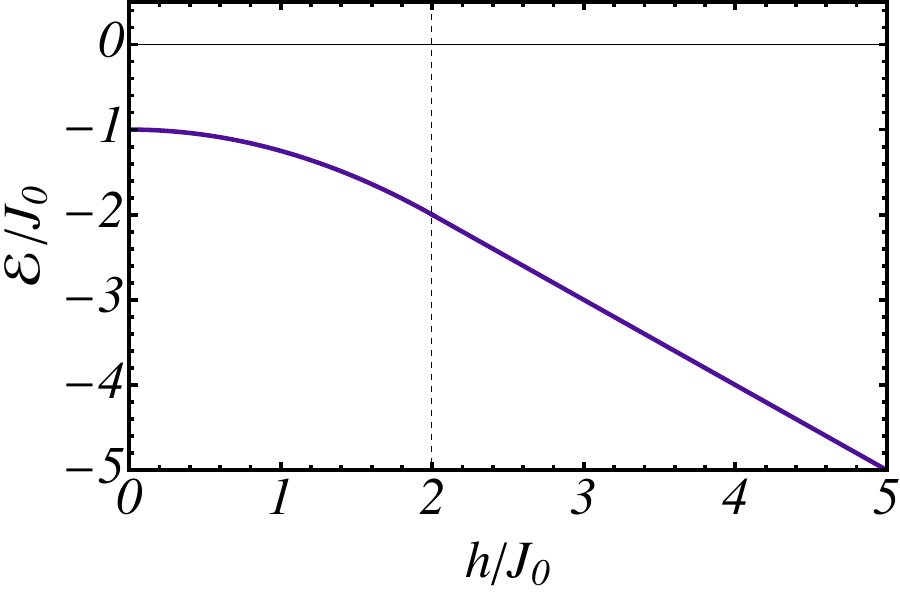}\label{fig2b}} \\ 
\subfloat[ Excitation gap ]{\includegraphics[width=.3\linewidth]{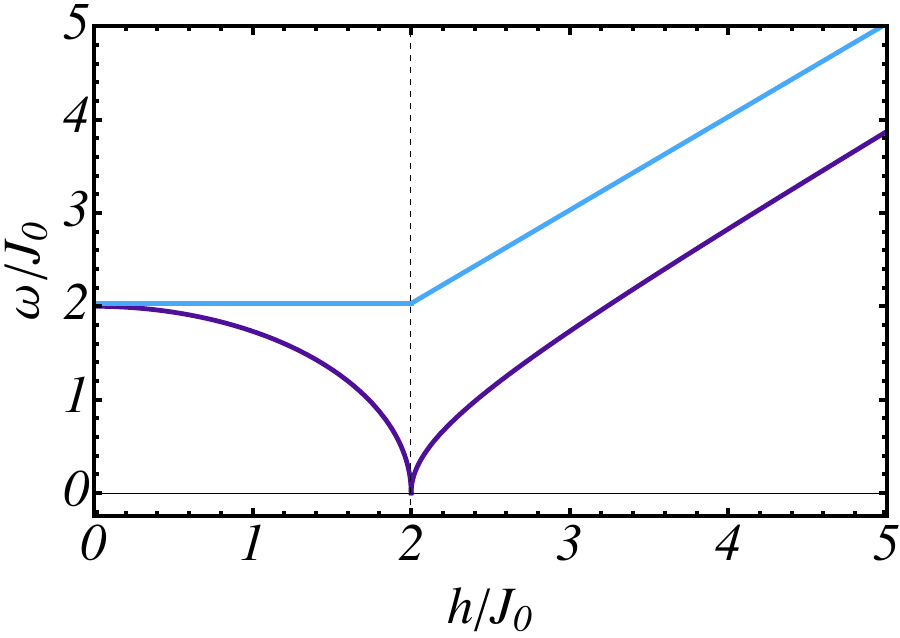}\label{fig2c}}   \hspace{0.2cm}
\subfloat[Polarization depletion]{\includegraphics[width=.3\linewidth]{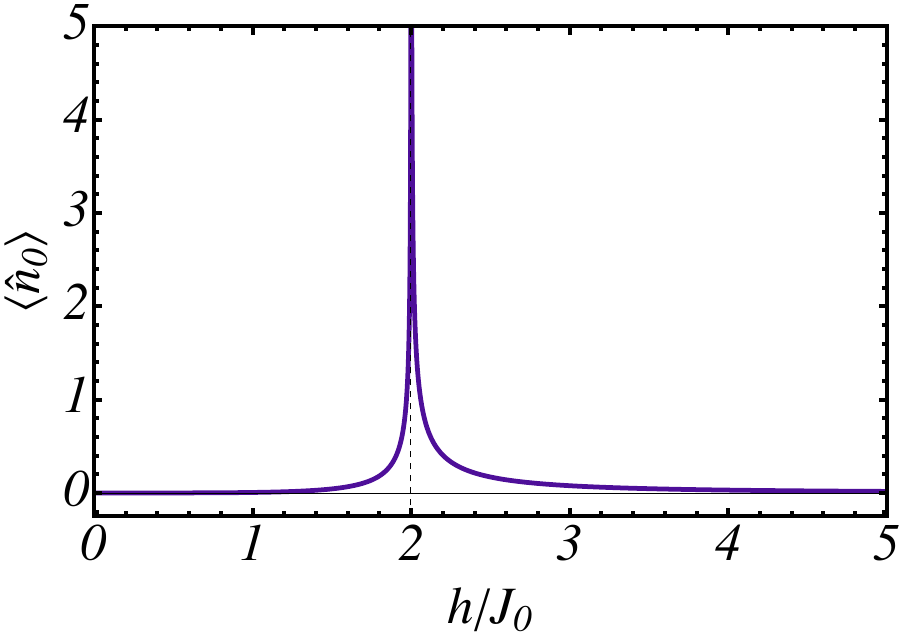}\label{fig2d}}  \hspace{0.2cm}
\subfloat[ Entanglement entropy]{\includegraphics[width=.3\linewidth]{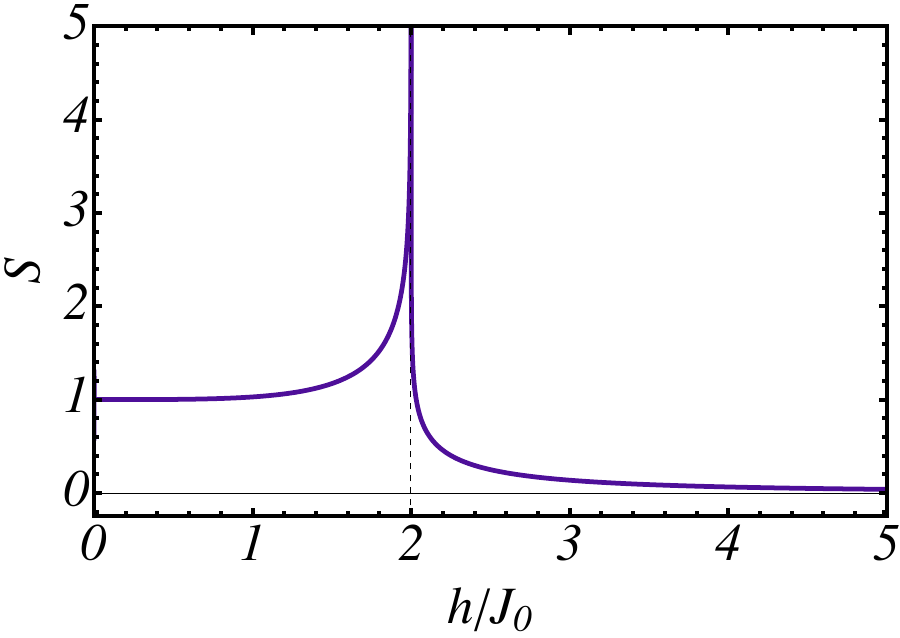}\label{fig2e}}
\caption{\textbf{Equilibrium properties of the fully-connected quantum Ising model across the phase diagram.}  
All quantities are computed for the Hamiltonian~\eqref{eq_Halpha0} with $\gamma=1$, and display singularities at the phase transition $h=2J_0$ with mean-field critical exponents. (The behavior of the corresponding quantities for a general anisotropic XY model $\gamma\neq0$ is analogous.) 
Panel (a): The order parameter is determined by Eq.~\eqref{eq_thetastar}.
Panel (b): The ground-state energy density is the minimum of the classical energy~\eqref{eq_Hcl} on the unit sphere.
Panel (c): The dark-blue curve corresponds to $\omega_<$ in Eq.~\eqref{eq_MFsmallosc<} and $\omega_>$ in Eq.~\eqref{eq_MFsmallosc>}, respectively below and above the phase transition; the  light-blue curve corresponds to $\omega_{\rm sw}$, cf. Eq.~\eqref{eq_spectrum}.
Panel (d): The number of collective spin excitations is computed by diagonalizing Eqs.~\eqref{eq_expansionH>1} and~\eqref{eq_Hferrocollectiveboson}, respectively below and above the phase transition.
Panel (e): The bipartite (half-system) entanglement entropy in fully-connected spin models is a function of $\langle \hat n_0 \rangle$ only, see Sec.~\ref{sec_415} below; in the quantum ferromagnetic phase, the finite-size ground state approaches a symmetric superposition of two symmetry-breaking ground states for large $N$, which yields an extra bit of entropy; the divergence at criticality is logarithmic in system size, see Refs.~\cite{latorre2005entanglement, vidal2004entanglement}.
 }
\label{fig_alpha0eq}
\end{figure}

For \mpar{Phase transition}
$h>h_{\text{cr}}$, the number $\braket{\hat n_0} = \braket{\hat q^2 + \hat p^2 - 1} / 2$ of bare collective spin excitations in the ground state is finite, and it diverges as $h \searrow h_{\text{cr}}$,  signaling a critical phenomenon (see Fig.~\ref{fig2d}).
Indeed the energy gap $\omega_{>}/s$ closes at $h=h_{\text{cr}}$, with a mean-field critical exponent $1/2$ (see Fig.~\ref{fig2c}). 
For $h<h_{\text{cr}}$ the frequency $\omega_>$ becomes imaginary, which signals instability of the paramagnetic state. 

In \mpar{Single-spin spectrum in the broken-symmetry phase}
order to determine the ground state and the elementary excitations in the broken-symmetry phase, let us start from some general considerations.
For $h<h_{\text{cr}}$ the classical landscape presents two symmetric minima, as discussed above. 
Below the energy $E_{\text{dyn}} \equiv \mathcal{H}_{\text{cl}}(\theta=0)$ of the classical phase-space separatrix, two symmetric families of classical trajectories fill the two energy wells. 
In the thermodynamic limit, this corresponds to two towers of  pairwise degenerate energy levels, associated with wavefunctions localized in the two wells. At finite size $N$, however, the energy eigenstates below the critical energy are nondegenerate and alternately even and odd with respect to the $\mathbb{Z}_2$ symmetry of the Hamiltonian. For large $N$, they approach even and odd superpositions of the localized wavefunctions. The energy splitting between each pair of quasidegenerate eigenstates is proportional to the quantum tunneling amplitude across the energy barrier, which is exponentially small in the height of the barrier\,\cite{landau1965quantum}, and hence exponentially small in $N$. Accordingly, tunneling between the two broken-symmetry sectors is practically suppressed even for moderate system sizes, and it is extremely fragile to tiny symmetry-breaking perturbations. For these reasons it makes sense to consider the two towers of symmetry-breaking states independently of each other.

To compute the spectrum explicitly, it is convenient to introduce a procedure which will lend itself to powerful generalizations in the rest of this Report. 
We rewrite the components of the collective spin in a rotated frame $(\bold{X},\bold{Y},\bold{Z})$, cf. Eq.~\eqref{eq_rotatedframe}, 
by angle $\theta$ in the $xz$-plane, i.e.,
\beq
\label{eq_rotatedspins}
\hat S^x = \cos \theta \, \hat S^X + \sin \theta\,  \hat S^Z, \qquad \hat S^y =  \hat S^Y, \qquad \hat S^z = -\sin \theta \, \hat S^X + \cos \theta \, \hat S^Z \, .
\eeq
Performing a Holstein-Primakoff transformation with rotated quantization axis $\bold{Z}$ and neglecting terms of order $1/S$,
\beq
\label{eq_hpcollectiveferro}
\left\{
\begin{split}
 \hat S^X &\approx \sqrt{S} \; \hat q, \\
 \hat S^Y &\approx \sqrt{S} \; \hat p, \\
  \hat S^Z &= S -  \hat n_0 = S - \frac{\hat q^2+\hat p^2-1}{2}, 
 \end{split}
 \right.
\eeq
 we get
\beq
\label{eq_Hferrocollectiveboson}
\begin{split}
\hat H_{\alpha=0}  \approx & 
- N \left( h \rho \cos\theta + J_0 \frac{1+\gamma}2 \rho^2 \sin^2\theta\right)  \\
       &
- \sqrt{\frac{\rho N}{s}} \sin\theta \Big(
 h  -  \rho J_0 (1+\gamma)  \cos\theta
 \Big) \hat q
 \\
     & 
     + \frac{1}{s} \bigg[ \left( \rho J_0 (1+\gamma) \sin^2 \theta + h \cos\theta \right)\frac{\hat q^2 + \hat p^2 -1}{2} - \rho J_0 (1+\gamma) \cos^2 \theta \, \frac {\hat q^2} 2
     - \rho J_0 (1-\gamma)  \, \frac {\hat p^2} 2
      \,  \bigg] .
\end{split}
\eeq
In order for the bosonic variables to describe quantum fluctuations it is necessary to align the frame with the classical configuration, in such a way that linear terms in the second line vanish. This condition leads to $\theta^*$ as in Eq.~\eqref{eq_thetastar}. The resulting quadratic Hamiltonian can then be readily diagonalized:
\beq
\label{eq_expansionH<}
 \hat H_{\alpha=0}  \approx 
  - 
  \frac N 2 \bigg(\frac{h^2}{J_0(1+\gamma)} + \rho^2 J_0 (1+\gamma)  \bigg) 
   + \frac{1}{s} \bigg( \frac{\omega_<-\omega_{<}^{(0)}}{2}\bigg)
   + \frac{1}{s}   \omega_< \; \hat n ,
\eeq
where
\beq
\label{eq_MFsmallosc<}
\omega_{<}=\sqrt{\left[ \rho^2 J_0^2 (1+\gamma)^2  - h^2 \right]   \frac{2\gamma}{1+\gamma}}, \qquad 
\omega_{<}^{(0)} =  \rho J_0 (1+\gamma) \,  .
\eeq
Analogously to Eq.~\eqref{eq_expansionH>}, the first term on the right-hand side of Eq.~\eqref{eq_expansionH<} represents the classical energy, the second one expresses the shift in the zero-point energy due to quantum fluctuations around the classical minimum configuration, while the last one [arising from diagonalization of Eq.~\eqref{eq_Hferrocollectiveboson}] is the energy of the harmonic excitations, with $n=0,1,2,\dots$.

 {In Fig.~\ref{fig_alpha0eq} \mpar{Validity of semiclassical analysis} we plotted the exact ground state energy density $\mathcal{E}$ [panel (b)], the energy gap of collective spin excitations $\omega_{>,<}$ (dark-blue curve) [panel (c)], and the number of ``bare'' collective spin excitations $\langle \hat n_0 \rangle$ [panel (d)], of the infinite-range quantum Ising model ($\gamma=1$) in the thermodynamic limit $N\to\infty$, as a function of the ratio $h/J_0$.}
The results \eqref{eq_expansionH>} and \eqref{eq_expansionH<} are asymptotically exact for $n/N\to 0$, and \emph{fully nonperturbative} in the Hamiltonian parameters $h,J_0,\gamma$. Systematic improvements in powers of $n/N$ can be worked out with a more refined analysis\,\cite{dusuel2005continuous}. This is particularly relevant to understand  the finite-size scaling $\omega \sim N^{-1/3}$ of the energy gap at criticality $h=h_{\text{cr}}$: see~\ref{app_seiclassicalspectrum} for an elementary semiclassical derivation.

(For completeness, Fig.~\ref{fig2e} also reports the ground-state bipartite entanglement entropy across the phase diagram. This quantity can be computed numerically for large $N$~\cite{latorre2005entanglement} and compared with analytical calculations in the large-$N$ limit based on semiclassical fluctuations~\cite{vidal2004entanglement}. This analytical procedure can be deduced as particular case of the more general discussion on entanglement dynamics in Sec.~\ref{sec_415} below; for this reason, we do not discuss this here.)

\summary{The collective spin low-energy spectrum is described by bosonic excitations, obtained by a Holstein-Primakoff expansion around the classical ground state.  }

\subsubsection{``Spin-wave'' excitations}

The analysis \mpar{Complete many-body spectrum} above concerns collective spin quantum fluctuations and excitations within a fixed sector with collective spin length $S$ --- and we are ultimately interested in the ground state sector with maximal $S=Ns$. Different families of spin excitations lower the collective spin length to $S=Ns-n_{\rm sw}$, with $n_{\rm sw}=0,1,2,\dots$. (For reasons that will become clear below, we will refer to the quantum number $n_{\rm sw}$ as the \emph{total occupation of spin-wave modes with non-vanishing momenta}.)
Their spectrum can also be straightforwardly obtained from semiclassical arguments:
Recalling the definition $\rho=S/(Ns)$ above, we have
\beq
\label{eq_epsilonNsw}
\rho =
1- \frac{n_{\text{sw}}}{Ns}.
\eeq
Substituting into Eqs. \eqref{eq_expansionH>} and \eqref{eq_expansionH<} and consistently neglecting terms of higher order in $1/N$, we obtain the complete spectrum of low-lying excitations above the ground state to leading order in $n/N$ and $n_{\text{sw}}/N$: 
\beq
\label{eq_spectrum}
\begin{split}
H_{\alpha=0} &\approx - N h +   \frac{\omega_>-\omega_{>}^{(0)}}{2s} + \frac{1}{s} \big( \omega_> \, \hat n + h  \, \hat n_{\text{sw}}  \big), \\
H_{\alpha=0} &\approx - \frac N 2 \bigg(\frac{h^2}{(1+\gamma)J_0}   +  J_0(1+\gamma) \bigg) 
+  \frac{\omega_<-\omega_{<}^{(0)}}{2s}
+ \frac{1}{s} \big(  \omega_< \, \hat n + (1+\gamma)J_0 \,  \hat n_{\text{sw}} \big),
\end{split}
\eeq
valid for $h>(1+\gamma)J_0$ and $h<(1+\gamma)J_0$, respectively. (Here $\omega$'s are taken at $\rho=1$.)
In Fig.~\ref{fig2c} we additionally reported the ``spin-wave'' excitation gap $\omega_{\rm sw}=h$ or $J_0(1+\gamma)$ in the two phases. Note that the Hilbert space sector dimension grows exponentially with $n_{\rm sw}$ [cf. the exact expression in Eq.~\eqref{eq_sectorsdim}]; however, because of permutational invariance, these energy levels are exactly degenerate. 
\\


As discussed so far, the properties of  infinite-range spin Hamiltonians can be efficiently computed either analytically (via a large-$N$ asymptotic expansion) or numerically (via exact diagonalization of the single-spin problem for $S \le  N/2 \approx 10^5$).
In closing this Subsection it is worth to briefly mention that the Hamiltonian~\eqref{eq_Halpha0} is equivalent to the Lipkin-Meshkov-Glick model of nuclear physics~\cite{lipkin1965validity,meshkov1965validity,glick1965validity}, which is actually Bethe-ansatz solvable~\cite{ortiz2005exactly}; however, this solution is not practically useful for large $N$, and semiclassical or numerical techniques give much easier access to the relevant information.
 
 \summary{
``Spin-wave'' excitations --- lowering the collective spin length --- remain gapped and dispersionless throughout the phase diagram for $\alpha=0$.
 }

\subsection{Finite-range interactions ($\alpha>0$)}
\label{low_en_theory}

The tendency \mpar{Parameter $\alpha$ as a perturbation} of long-range interactions to form collective spin alignment and to preserve it even in  excited states becomes increasingly prominent as $\alpha$ is decreased. To quantify this aspect, it is convenient to view a long-range interacting system with finite exponent $\alpha$ as a ``perturbation'' of the infinite-range interacting system with all-to-all interactions ($\alpha=0$). 

\subsubsection{Perturbation to mean-field}

This viewpoint can be made explicit by rewriting the Hamiltonian in momentum space. To this aim, we Fourier transform the spin operators $\hat s^{\mu}_{\mathbf r} $ for $\mu=x,y,z$:
\beq
\label{eq_fourierspin}
 \tilde S^{\mu}_{\mathbf k} = \sum_{\mathbf{r}} e^{i\mathbf k \cdot \mathbf r} \hat s^{\mu}_{\mathbf r} \, ,
\eeq
with 
\beq
\mathbf k \equiv \mathbf k_{\boldsymbol{\ell}}=2\pi \boldsymbol{\ell} / L, \quad \boldsymbol{\ell}=(\ell_1,\dots,\ell_d), \quad  \ell_a=0,\pm1,\pm2,\dots,\pm \floor{L/2}
\eeq
(for $L$ even $\ell_a={\pm L/2}$ coincide).
We also define $\tilde S^{\pm}_{\mathbf k }=\tilde S^{x}_{\mathbf k } \pm i\tilde S^{y}_{\mathbf k }$.
Note that 
\beq 
\tilde {\vec{S}}_{\mathbf k=\mathbf 0} \equiv \hat {\vec{S}} = \sum_{\mathbf r} \hat {\vec{s}}_{\mathbf r}
\eeq
is the system's collective spin.
It is straightforward to separate the variable-range quantum XY Hamiltonian~\eqref{eq_H} into the  $\alpha$-independent collective part --- given by the $\mathbf k=\mathbf 0$ terms --- and the ``perturbation'' controlled by $\alpha$:
\beq
\label{eq_lrxyfourier}
\hat H_\alpha = \hat H_{\alpha=0} + \hat V_\alpha
\eeq
with\footnote{Note that in this expression the various ${\mathbf k}$-modes are \emph{not} dynamically decoupled, since $\big[\tilde S^\mu_{\mathbf k },\tilde S^\nu_{\mathbf q }\big] = i \epsilon^{\mu\nu\lambda} \tilde S^\lambda_{{\mathbf k + \mathbf q}}$.}
\beq
\label{eq_Valpha}
\hat V_\alpha = - \frac{J_0}{4s^2 N}  \sum_{\mathbf k\neq \mathbf 0} f_{\mathbf k}(\alpha) \Big[ \big(\tilde S^+_{\mathbf k}   \tilde S^-_{-\mathbf k} + \tilde S^-_{\mathbf k}  \tilde S^+_{-\mathbf k}\big)
+\gamma \big(\tilde S^+_{\mathbf k}  \tilde S^+_{-\mathbf k} + \tilde S^-_{\mathbf k}  \tilde S^-_{-\mathbf k}\big) \Big] \, .
\eeq
In Eq.~\eqref{eq_Valpha} we defined the function
\beq
\label{eq_fkalpha}
f_{\mathbf k}(\alpha) =  \sum_{\mathbf r \neq \mathbf 0} \frac {\cos(\mathbf k \cdot \mathbf r)}{ \lvert\lvert \mathbf r \rvert\rvert^\alpha}  \Bigg/  \sum_{\mathbf{r}\neq\mathbf{0}} \frac{1}{\lvert\lvert\mathbf{r}\rvert\rvert^{\alpha}}
\eeq
which depends implicitly on the dimensionality $d$ of the lattice.
By construction, $f_{\mathbf k=\mathbf 0}(\alpha)=1$.

\begin{figure*}[]
\centering
\includegraphics[width=\textwidth]{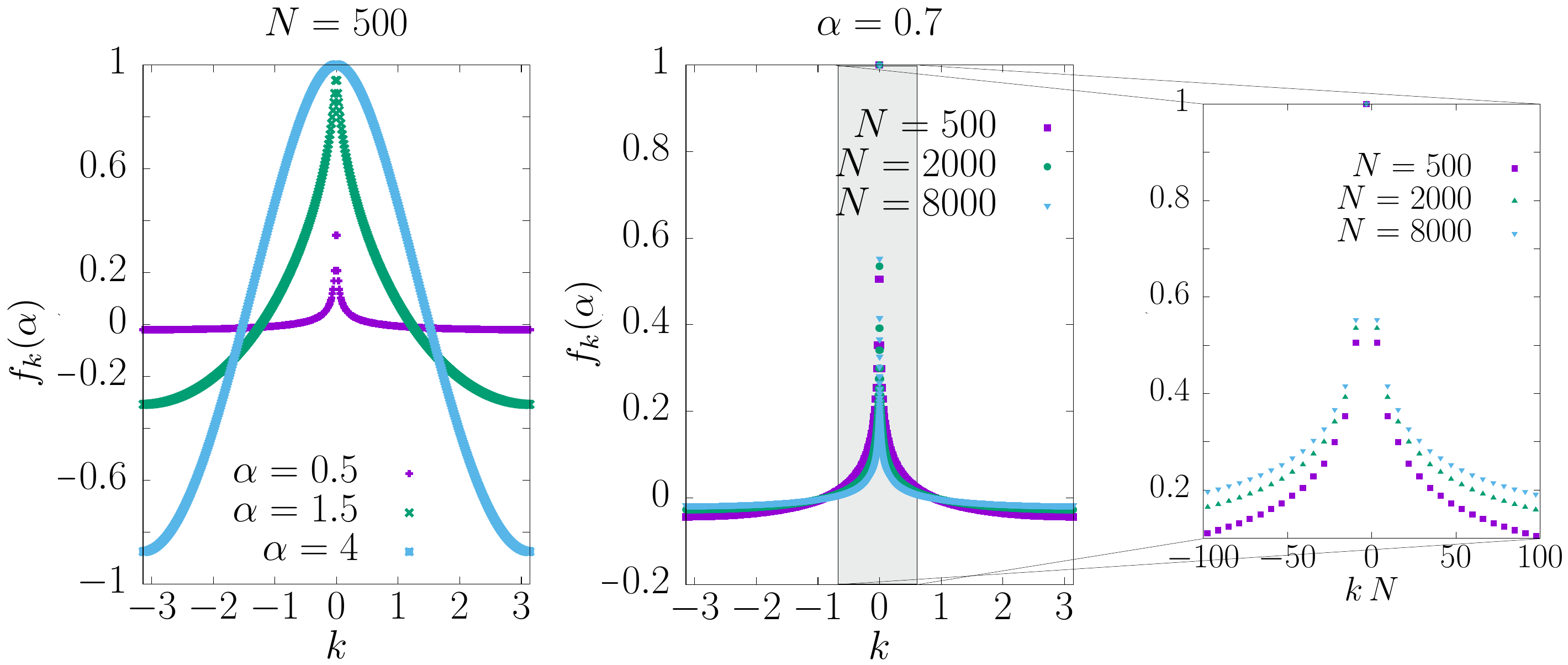}
\caption{Plots of the function ${f}_{\alpha,k}$ \eqref{eq_fkalpha} for $d=1$.
(Left panel):  ${f}_{k}(\alpha)$ is shown for several values of $\alpha$, for $N=L=500$.
The function squeezes towards $k=0$ for $0\le \alpha \le 1$. For $1<\alpha<2$, ${f}_{k}(\alpha)$ becomes a finite function with a cusp behavior for small $k$, while for $\alpha\gg2$ it is a cosine-like function.
(Central panel): ${f}_{k}(\alpha)$ is shown for $\alpha=0.7$ and increasing values of $N$.
Qualitatively similar behavior occurs for $0\le \alpha \le 1$.
Squeezing towards a delta function as $N\to\infty$ occurs with a speed $N^{-(1-\alpha)}$ for $\alpha<1$ and $1/\ln N$ for $\alpha=1$.
(Right panel): a ``zoom'' of the plot in the bottom left panel is shown, for larger values of $N$.
The {rescaled} function in the vicinity of $k=0$  converges to a finite limiting curve as $N\to\infty$.
This \emph{discrete} structure approaches a continuum as $\alpha \nearrow 1$.}
\label{Fig1}
\end{figure*}

When $\alpha \to 0$ \mpar{Properties of $f_{\mathbf k}(\alpha)$ } the couplings $f_{\mathbf k\neq \mathbf 0}(\alpha)$ turn off [Eq.~\eqref{eq_falpha0}], and $\hat H_{\alpha}$ reduces to a Hamiltonian  describing a single collective degree of freedom. 
The effect of spatially modulated interactions $\alpha\neq0$ is then to couple the collective spin to all finite-wavelength modes describing spatially non-trivial spin fluctuations, resulting in complex interacting many-body dynamics. 
The form of this coupling is dictated by the function $f_{\mathbf k}(\alpha)$.
While the specific choice of the lattice may influence the detailed form of $f_{\mathbf k}(\alpha)$, the physics of long-range interacting systems is only affected by the asymptotic behaviour at small ${\mathbf k}$:  
In \ref{app:boundsF} we derive 
\begin{eqnarray}
\label{eq_estimateftildealpha}
{f}_{\mathbf{k}_{\boldsymbol{\ell}}\ne\mathbf{0}} (\alpha) \equiv {f}_{\boldsymbol{\ell}\ne\mathbf{0}} (\alpha) &\underset{ |\boldsymbol{\ell}|\to\infty}{\thicksim}&   \frac {A(\alpha)} {|\boldsymbol{\ell} |^{d-\alpha} }+ \frac { B(\alpha)} { |\boldsymbol{\ell} |^{(d+1)/2} }\,  \qquad\qquad \text{for } 0<\alpha<d  \, ;\\
\label{eq_estimateftildealphabigger}
{f}_{\mathbf{k}\ne\mathbf{0}} (\alpha) & \underset{\mathbf k \to \mathbf 0}{\thicksim}& 1 -  \bar A(\alpha) |\mathbf{k}|^{\alpha-d} - \bar B(\alpha)  |\mathbf{k}|^2 \quad   \text{for } \alpha>d  \, .
\end{eqnarray}
The sharp changes in behavior are summarized in Fig.~\ref{Fig1}, where we plot ${f}_{k} (\alpha)$ for a range of values of $\alpha$ and $d=1$. 
Its shape shrinks from $f_{k}(\alpha\to\infty)=\cos k$ to 
\beq
\label{eq_falpha0}
f_{k}(\alpha\to0)=\delta_{k,0}\, ,
\eeq
becoming increasingly singular at $k=0$ as $\alpha$ is decreased: 

\begin{subequations}
\label{eq_fkalphaasympt}
	\begin{align}
	\label{eq_discrete}
f_{k_\ell}(\alpha)  \equiv f_{\ell}(\alpha)  & \sim c(\alpha) |\ell|^{-(1-\alpha)}  &\quad \text{for } \alpha<1 ;\\
f_{k}(\alpha) & \sim 1-c(\alpha) |k|^{\alpha-1}  &\quad \text{for } 1<\alpha<3 ;\\
f_{k}(\alpha) & \sim 1-c(\alpha) k^{2}  &\quad \text{for } \alpha>3 .		
	\end{align}
\end{subequations}

For long-range interactions $0<\alpha< 1$, the values of $f_{k}(\alpha)$ progressively squeeze onto the vertical axis as  $L\to\infty$; upon zooming near $k=0$ one finds a sequence of \emph{discrete} finite values, see the right panel of Fig.~\ref{Fig1}~\cite{lerose2020origin,defenu2021metastability}. 
This phenomenon can be physically interpreted as follows: interactions decay so slowly with the spatial distance that the system behaves as a permutationally invariant system over finite length scales, hence observables are unable to resolve finite wavelengths.
Only modes with extensive wavelengths $k_\ell \propto 1/L$ may impact the physical properties. As $\alpha$ is increased to values larger than $d$, all modes $k\neq0$ get eventually activated. 
Despite its simplicity, \mpar{Discrete spectrum for $0<\alpha<d$} the result in Eq.\,\eqref{eq_estimateftildealpha} has significant physical implications: As we will show below, {the low-energy spectrum of a quantum system with long-range interactions remains discrete in the thermodynamic limit}. In this precise sense we may say that \emph{long-range interacting systems with $0<\alpha<d$ interpolate between few-body and many-body physics}.

At the same time, \mpar{Divergent velocity for $\alpha<d+1$} Eq.~\eqref{eq_estimateftildealphabigger} showcases another fundamental properties of long-range interacting systems: the singularity at small momenta gives rise to a divergent velocity of propagation of quantum information across the system for $\alpha<d+1$, violating the famous Lieb-Robinson light-cone bound of short-range interacting systems~\cite{lieb1972finite}. This property is actually completely general, as it does not rely on any low-energy description. We will further discuss its consequences in Sec.~\ref{corr_weak_lr}.

\summary{Finite-range interactions can be seen as a perturbation to the fully-connected Hamiltonian. Physically, this perturbation couples the collective spin to spin fluctuation modes at non-vanishing momentum. Long-range  interactions preferentially generate coupling to long-wavelength modes only.}

\subsubsection{Quantum paramagnetic phase}
\label{sec_241}

We are \mpar{Bosonization of paramagnetic spin fluctuations} now ready to compute the low-energy spectrum and properties of the variable-range quantum XY model~\eqref{eq_H}.
For large $h$ and arbitrary $\alpha$ the paramagnetic ground state is the spin-coherent eigenstate of $S^z$, 
\beq
\label{eq_fullup}
| {\rm GS} _{h=\infty} \rangle = | \boldsymbol{\Uparrow} \, \rangle  \equiv \bigotimes_{\mathbf r} | \up_\mathbf{r} \rangle.
\eeq
For finite $h/J_0$ the ground state is a ($\alpha$-dependent) distortion of this state dressed by spin-lowering excitations. 
A convenient approach to describe spin fluctuations in $| {\rm GS}_{h<\infty} \rangle$ is by mapping them to bosonic modes using the Holstein-Primakoff transformation~\cite{wannier1987statistical}.

Unlike our discussion in Sec.~\ref{sec_232}, we now have to keep track of spatially-resolved fluctuating spin modes, which we can conveniently do by working at the level of individual microscopic spins.
Recalling the standard definitions $\hat s^{\pm}_{\mathbf r} = \hat s^{x}_{\mathbf r} \pm i \hat s^{y}_{\mathbf r} $, we can bosonize individual spin fluctuations around the positive $\mathbf z$ axis by setting
\beq
\label{eq_HPlocal}
 \hat s^-_{\mathbf r} 
 \mapsto
      \hat b^\dagger_{\mathbf r} \sqrt{2s- \hat b^\dagger_{\mathbf r} \hat b_{\mathbf r} }, \quad
   \hat s^+_{\mathbf r} 
   \mapsto   \sqrt{2s- \hat b^\dagger_{\mathbf r}  \hat b_{\mathbf r} } \;  \hat b_{\mathbf r}, \quad
     \hat s^z_{\mathbf r} 
     \mapsto s  -   \hat b^\dagger_{\mathbf r} \hat b_{\mathbf r},
\eeq
where $\hat b_{\mathbf r},\hat b^\dagger_{\mathbf r}$ are canonical bosonic annihilation and creation operators.
Performing the substitutions~\eqref{eq_HPlocal} in Eq.~\eqref{eq_H} we obtain an exact representation of the variable-range XY quantum spin model as a non-linear bosonic Hamiltonian, where the state $| \boldsymbol{\Uparrow} \, \rangle$ in Eq.~\eqref{eq_fullup} corresponds to the Fock space vacuum $|\emptyset\rangle$.

The mapping~\eqref{eq_HPlocal} should be understood as an \emph{embedding} of the two-dimensional Hilbert space of a  spin-$1/2$ in the infinite-dimensional Hilbert space of a bosonic mode. The states $|\up\,\rangle$ and $|\down\,\rangle$ are mapped onto $|0\rangle\equiv |\emptyset\rangle$ and $|1\rangle \equiv b^\dagger|\emptyset\rangle$. The operators on the right-hand sides of Eqs.~\eqref{eq_HPlocal} act non-trivially on the full bosonic space; however, they are block-diagonal, as their matrix elements between the physical spin subspace and its orthogonal complement are vanishing; their action on the physical spin subspace coincides with the operators on the left-hand sides. 

It is convenient \mpar{Bosonic representation of XY quantum spin model} to write the bosonic Hamiltonian directly in momentum space. To this aim, we define the Fourier-transformed bosonic modes\footnote{Note that we take a unitary Fourier transformation on the bosonic modes, while the convention for spins in Eq.~\eqref{eq_fourierspin} was such that $\tilde S^{x,y,z}_{\mathbf k=\mathbf 0} =  \hat S^{x,y,z}$ (collective spin projections).} 
\beq
\label{eq_fourier}
   \tilde b^\dagger_{\mathbf k} =  \frac 1 {\sqrt{N}}\sum_{\mathbf r} e^{i\mathbf k \cdot \mathbf r} \hat b^\dagger_{\mathbf r} \, .
\eeq
We now formally expand the Holstein-Primakoff mapping~\eqref{eq_HPlocal} in $1/s$ and Fourier-transform term by term:
 \beq
 \label{eq_HPfourier}
 \left\{
\begin{split}
 \tilde S^-_{\mathbf k} &\approx
      (2Ns)^{1/2} \, \tilde b^\dagger_{\mathbf k} - \frac 1 {2 (2Ns)^{1/2}} \sum_{\mathbf q_1,\mathbf q_2} \tilde b^\dagger_{\mathbf q_1} \tilde b^\dagger_{\mathbf q_2} \tilde b_{\mathbf q_1+\mathbf q_2-\mathbf k},
    \\
   \tilde S^+_{\mathbf k} &\approx   (2Ns)^{1/2} \, \tilde b_{-\mathbf k} - \frac 1 {2 (2Ns)^{1/2}} \sum_{\mathbf q_1,\mathbf q_2} \tilde b^\dagger_{\mathbf q_1+\mathbf q_2+\mathbf k} \tilde b_{\mathbf q_1} \tilde b_{\mathbf q_2},
    \\
     \tilde S^z_{\mathbf k} &= Ns \; \delta_{\mathbf k,0}  -   \sum_{\mathbf q} \tilde b^\dagger_{\mathbf q+\mathbf k} \tilde b_{\mathbf q}.
\end{split}
\right.
 \eeq
 It is worth to stress here the connection with the previously introduced expansion. First of all, we immediately recognize that the bosonic mode with $\mathbf k=\mathbf 0$ coincides with the previously introduced collective bosonic mode in Eq.~\eqref{eq_hpcollectiveferro}. 
 Furthermore, by expanding $\hat S^2$ using Eqs.~\eqref{eq_HPfourier}, one can check that $\hat n_{\mathbf k = \mathbf 0}= \tilde b^\dagger_{\mathbf k= \mathbf 0} \tilde b_{\mathbf k= \mathbf0}$ cancels to leading order\,\cite{wannier1987statistical}:
 \beq
 \label{eq_nsw}
  \hat n_{\rm sw} \equiv Ns - S = \sum_{\mathbf k \neq \mathbf 0} \tilde b^\dagger_{\mathbf k} \tilde b_{\mathbf k}.
 \eeq
 This equation asserts that the total occupation of non-zero momentum spin-wave modes represents the collective spin depletion, explaining the notations in Eq.~\eqref{eq_epsilonNsw}.
 
Making the substitutions~\eqref{eq_HPfourier} into Eq.~\eqref{eq_lrxyfourier} we obtain an expression of the form
\beq
\label{eq_bosonizedHpm}
\hat H_\alpha =  \frac 1 s \bigg[ (Ns)^1 \mathcal{E}_{0}  +  (Ns)^0 \hat H_2 + (Ns)^{-1} \hat H_4 + \dots \, \bigg] ,
\eeq
where:
\beq
\mathcal{E}_{0} = \mathcal{H}(\mathbf z)= -h
\eeq
is the classical (mean-field) energy density of the paramagnetic state;
\beq
\label{eq_gsH2}
\hat H_2 =  \sum_{\mathbf k} h \, \tilde b^\dagger_{\mathbf k} \tilde b_{\mathbf k} -  \sum_{\mathbf k}  {J_0}   \, f_{\mathbf k}(\alpha)
\left( \frac{ \tilde b_{\mathbf k} \tilde b^\dagger_{\mathbf k} +  \tilde b^\dagger_{-\mathbf k}  \tilde b_{-\mathbf k} }2 + \gamma \frac{  \tilde b_{\mathbf k} \tilde b_{-\mathbf k} +  \tilde b^\dagger_{-\mathbf k}  \tilde b^\dagger_{\mathbf k} }2  \right)
\eeq
 describes semiclassical (Gaussian) spin fluctuations;
\begin{multline}
\label{eq_gsH4}
\hat H_4 = \frac{ J_0} 2 \sum_{\mathbf k,\mathbf q_1,\mathbf q_2}  f_{\mathbf k}(\alpha) \; \times \\
\Big[
\Big(
\tilde b^\dagger_{-\mathbf k+\mathbf q_1+\mathbf q_2} \tilde b_{\mathbf q_1}  \tilde b_{\mathbf q_2}  \tilde b^\dagger_{\mathbf k} 
+ \tilde b^\dagger_{\mathbf q_1} \tilde b^\dagger_{\mathbf q_2}  \tilde b_{\mathbf k+\mathbf q_1+\mathbf q_2}  \tilde b_{-\mathbf k}  
+  \tilde b_{\mathbf k}  \tilde b^\dagger_{\mathbf q_1}  \tilde b^\dagger_{\mathbf q_2}  \tilde b_{-\mathbf k+\mathbf q_1+\mathbf q_2}
+ \tilde b^\dagger_{-\mathbf k} \tilde b^\dagger_{\mathbf k+\mathbf q_1+\mathbf q_2}    \tilde b_{\mathbf q_1} \tilde b_{\mathbf q_2}   
\Big)
\\ + \gamma 
\Big(
\tilde b^\dagger_{-\mathbf k+\mathbf q_1+\mathbf q_2} \tilde b_{\mathbf q_1}  \tilde b_{\mathbf q_2}  \tilde b_{-\mathbf k} 
+ \tilde b_{\mathbf k} \tilde b^\dagger_{\mathbf k+\mathbf q_1+\mathbf q_2}  \tilde b_{\mathbf q_1} \tilde b_{\mathbf q_2}     
+  \tilde b^\dagger_{-\mathbf k}  \tilde b^\dagger_{\mathbf q_1}  \tilde b^\dagger_{\mathbf q_2}  \tilde b_{-\mathbf k+\mathbf q_1+\mathbf q_2}
+  \tilde b^\dagger_{\mathbf q_1} \tilde b^\dagger_{\mathbf q_2}  \tilde b_{\mathbf k+\mathbf q_1+\mathbf q_2}   \tilde b^\dagger_{\mathbf k}  
\Big)
\Big]
\end{multline}
represents the 2-body non-linear interactions between spin fluctuations. One can similarly derive $(Ns)^{-2}\hat H_6$ etc.

While \mpar{Linear spin-wave theory} the full exact bosonic representation is cumbersome, its usefulness rests on the approximability of highly polarized spin states with bosonic states. 
To this aim we introduce the number of bosons 
\beq
\hat n_{\rm tot}  \equiv \hat n_0 + \hat n_{\rm sw} = Ns - \hat S^z 
\eeq 
and we approximate well-polarized states with $n_{\rm tot}\ll Ns$ with dilute Fock states with $n_{\rm tot}$ boson.
 In such corner of the Hilbert space, the bosonic modes turn out to provide an accurate description of spin states and operators. 
Intuitively, by inspecting Eqs.~\eqref{eq_HPfourier}, one recognizes that the action of the non-linear terms (second on the right-hand side) on a dilute Fock state is suppressed by a density factor $n_{\rm tot}/N$ compared to the action of the leading terms.
Thus, up to an error of order $\mathcal{O}[(n_{\rm tot}/N)^2]$, we may identify $\tilde S^-_{\mathbf k} \propto  \tilde b^\dagger_{\mathbf k}$, $\tilde S^+_{\mathbf k} \propto \tilde b_{-\mathbf k}$.
We now show that the ground state of long-range interacting spin models lives exactly in this corner of the spin space.

An approximate solution of the bosonic Hamiltonian~\eqref{eq_bosonizedHpm} can be found by neglecting the terms with $\hat H_4$ and higher order --- an approximation usually termed \textit{linear spin-wave (LSW) theory}.
The quality of the result heavily depends on the parameters and in particular on $\alpha$.
Our purpose is to show that the LSW description of low-energy properties becomes \emph{exact} for $\alpha<d$, and quantify its accuracy for $\alpha>d$.

The quadratic spin-wave Hamiltonian can be diagonalized via a standard Bogolubov transformation,
$\tilde b_{\mathbf k} = \cosh \theta_{\mathbf k }\beta_{\mathbf k} + \sinh \theta_{\mathbf k} \beta^\dagger_{-\mathbf k}$, with 
\beq
\tanh (2\theta_{\mathbf k})\equiv \frac {\gamma J_0 f_{\mathbf k}(\alpha)}{h - J_0 f_{\mathbf k}(\alpha)} \, .
\eeq
The result is 
\beq
Ns \, \mathcal{E}_{0} + \hat H_2 = Ns \, \mathcal{E}_{2} + \sum_{\mathbf k} \omega_{\mathbf k,>}(\alpha) \,\hat \beta^\dagger_{\mathbf k} \hat \beta_{\mathbf k} \ ,
\label{sw_h}
\eeq
where we identify the excitation spectrum 
\beq
\label{equilibrium_exc_spectra}
\omega_{\mathbf k,>}(\alpha) =  \sqrt{\big[ h-J_0 f_{\mathbf k}(\alpha)\big]^2- \gamma^2 \big[ J_0 f_{\mathbf k}(\alpha)\big]^2 } = \sqrt{\big[ h-J_0 f_{\mathbf k}(\alpha) (1-\gamma) \big]\big[ h-J_0 f_{\mathbf k}(\alpha) (1+\gamma) \big]},
\eeq
and the ground-state energy
\beq
N s \,  \mathcal{E}_{2} =  N s \, \mathcal{E}_{0} +   \frac 1 2 \sum_{\mathbf k}  \big[\omega_{\mathbf k,>}(\alpha) - \omega_{>}^{(0)}\big]
\eeq
where $\omega_{>}^{(0)}=h$ [cf. Eq.~\eqref{eq_MFsmallosc>}].

Within LSW theory, the ground-state wavefunction is given by
\beq
\label{eq_bogolubovgs}
\begin{split}
| {\rm GS}_2 \rangle &= \prod_{\mathbf k} \exp\bigg[\frac {\theta_{\mathbf k}}{2} \big(\tilde b_{\mathbf k} \tilde b_{-\mathbf k} - \tilde b^\dagger_{-\mathbf k} \tilde b^\dagger_{\mathbf k} \big) \bigg] \,  |\emptyset\rangle 
\propto \prod_{\mathbf k} \exp\bigg( -\frac{\epsilon_{\mathbf k}}{4\gamma  J_0 f_{\mathbf k}(\alpha)}  \tilde b^\dagger_{-\mathbf k} \tilde b^\dagger_{\mathbf k}  \bigg) \,  |\emptyset\rangle \,
 \end{split}
\eeq
where
\beq
\epsilon_{\mathbf k} \equiv 2h - 2J_0 f_{\mathbf k}(\alpha) -\omega_{\mathbf k,>}(\alpha) \ge 0 \, .
\eeq

The meaningfulness of the LSW solution is determined by $\omega_{\mathbf k,>}$ being real. This requires $h\ge h_{\rm cr}$, where
\beq
h_{\rm cr} \equiv J_0(1+\gamma).
\eeq
The minimum of $\omega_{\mathbf k,>}$ is attained as $\mathbf k\to\mathbf 0$.
To expand around this limit we write $f_{\mathbf k}(\alpha)\equiv 1-\sigma_{\mathbf k}(\alpha)$. 
Calculation gives
\beq 
\label{para_disp}
\omega_{\mathbf k,>}(\alpha)  \underset{\mathbf k\to\mathbf 0}{\thicksim} 2h \sqrt{a + b \sigma_{\mathbf k}(\alpha)}, 
\eeq
with dimensionless coefficients $a$ and $b$.\footnote{Explicitly, $a=1-2J_0/h + (1-\gamma^2)(J_0/h)^2$ and $b=2(J_0/h)[1-(1-\gamma^2)(J_0/h)]$.}
For $h>h_{\rm cr}$ one has $a>0$ and thus 
$\omega_{\mathbf k,>}  \underset{\mathbf k\to\mathbf 0}{\thicksim} 2h \sqrt{a} + h (b / \sqrt{a})\sigma_{\mathbf k}$. For short-range interactions $\alpha\ge d+2$ the gapped dispersion relation is parabolic, $\sigma_{\mathbf k} \sim |\mathbf k|^2 $; for longer range $d<\alpha<d+2$ it behaves as $\sigma_{\mathbf k} \sim |\mathbf k|^{\alpha-d}$; for $\alpha < d$ the spectrum becomes discrete, $\sigma_{\mathbf k_{\boldsymbol{\ell}}} \equiv \sigma_{\boldsymbol{\ell}}$.\footnote{It is interesting to note that the LSW description of the ground state is meaningful even for large $\alpha$, where LSW theory completely fails to capture the possible topological nature of excitations.}
At the critical point $h=h_{\rm cr}$, one has $a=0$ and hence $\omega_{\mathbf k,>} \underset{\mathbf k\to\mathbf 0}{\thicksim} 2h  \sqrt{b \sigma_{\mathbf k}}$, signaling closure of the spectral gap at $\mathbf k = \mathbf 0$. However, for $\alpha<d$, the spectrum of spin-wave excitations with $\mathbf k \ne \mathbf 0$ is discrete.

To assess \mpar{Self-consistency of LSW theory} the accuracy of LSW theory we evaluate the depletion of spin polarization, i.e.
\beq
\langle \hat n_{\rm tot} \rangle =
 N s - \langle \hat S^z  \rangle  
     = \sum_{\mathbf k} \langle  {\rm GS} |
     \tilde b^\dagger_{\mathbf k} \tilde b_{\mathbf k} |  {\rm GS}
     \rangle.
\eeq
Approximating the ground-state $| {\rm GS} \rangle$ by the LSW theory ground-state $| {\rm GS}_2 \rangle$ in Eq.~\eqref{eq_bogolubovgs} we obtain the explicit expression
\beq
\label{eq_gsdepletion}
\langle \hat n_{\rm tot} \rangle =  \frac 1 2 \sum_{\mathbf k}
\frac {\epsilon_{\mathbf k}} {\omega_{\mathbf k,>}}
\, .
\eeq
This quantity depends on $h$, $\alpha$ and  $\gamma$; in particular, it is suppressed as $h\to\infty$  or $\alpha\to0$ or $\gamma\to0$.

\begin{figure*}[]
\centering
\includegraphics[width=0.4\textwidth]{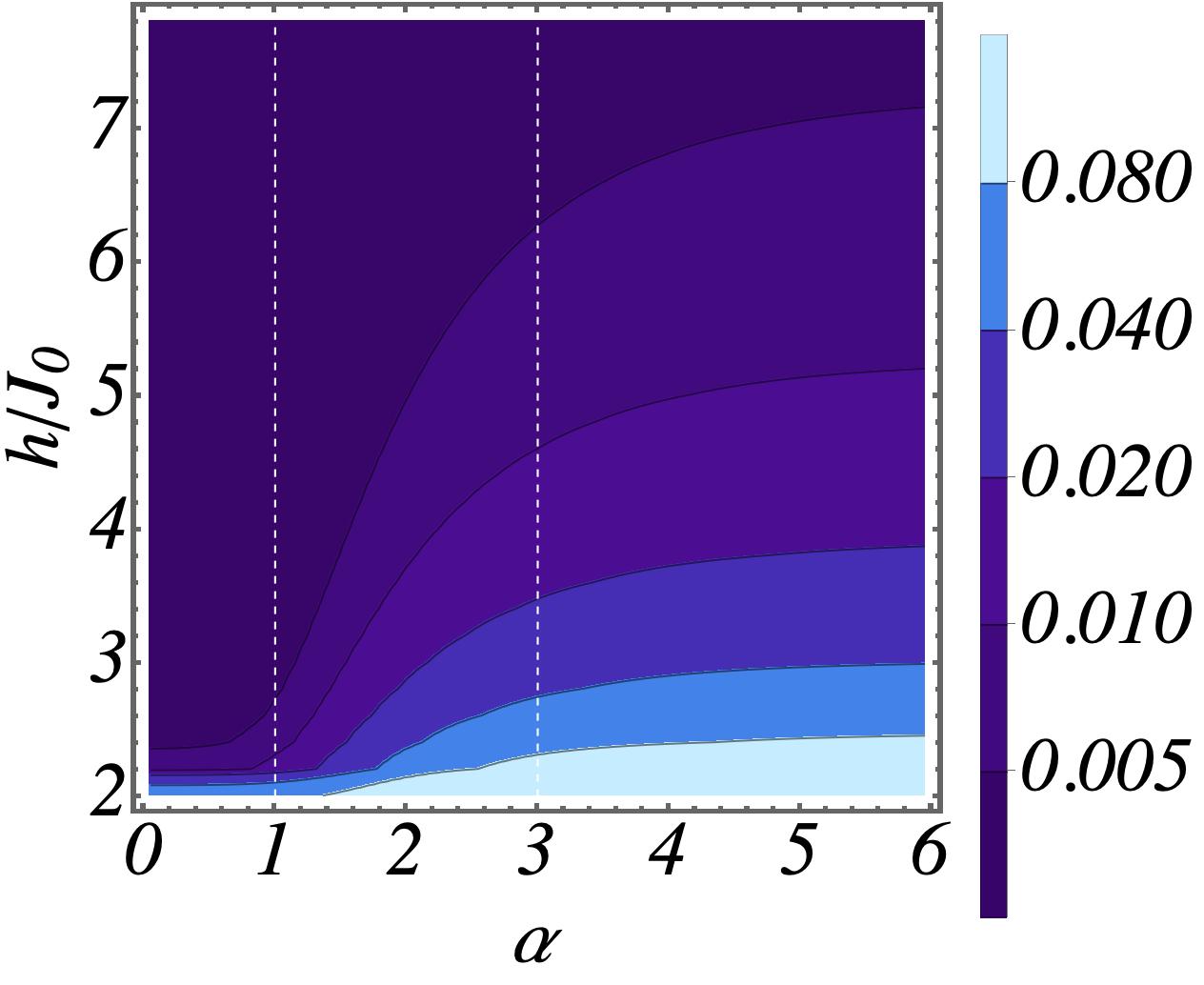}
\caption{
Ground-state spin depletion density $\langle \hat n_{\rm tot} \rangle/N$ in the quantum paramagnetic phase, cf. Eq.~\eqref{eq_gsdepletion}, for $\gamma=1$ and $d=1$ (variable-range quantum Ising chain).
}
\label{fig_gsdepletion}
\end{figure*}

In Fig.~\ref{fig_gsdepletion} we plot the depletion per spin $\langle \hat n_{\rm tot} \rangle/N$ given by Eq.~\eqref{eq_gsdepletion} at fixed $\gamma=1$ (quantum Ising model) and for $d=1$. As is evident, the effect of spin fluctuations is enhanced as the interactions become short-ranged, i.e. $\alpha\to\infty$, or as the critical point $h_{\rm cr}=2J_0$ is  approached.
All the qualitative aspects of this plot can be understood analytically. Specifically, for $h>h_{\rm cr}$, we have
\beq
\frac {\langle \hat n_{\rm tot} \rangle}{Ns} 
= \frac 1 {16} \bigg( \frac{h_{\rm cr}}{h} \bigg)^2 \frac 1 {Ns} \sum_{\mathbf k} f^2_{\mathbf k}(\alpha)  + \mathcal{O}  \Bigg( \bigg( \frac{h_{\rm cr}}{h} \bigg)^3 \frac 1 {Ns} \sum_{\mathbf k} f^3_{\mathbf k}(\alpha) \Bigg).
\eeq
The behavior of  the right-hand side as $N\to\infty$ depends qualitatively on $\alpha$: For $\alpha>d$ the limit is a finite number, 
\beq
\frac 1 N \sum_{\mathbf k} f^2_{\mathbf k}(\alpha) \sim \int_{-\pi}^\pi\dots\int_{-\pi}^\pi \frac{d k_1 \dots dk_d}{(2\pi)^d} f^2_{\mathbf k}(\alpha)
\eeq
 (cf. left panel of Fig.~\ref{Fig1}).
As $\alpha\searrow d$ the function $f_{\mathbf k}(\alpha)$ squeezes on the vertical axis (cf. Fig.~\ref{Fig1}), suppressing the value of the integral. This means that the spin depletion becomes subextensive for $\alpha<d$:  using $f_{\boldsymbol{\ell}}(\alpha) \sim |\boldsymbol{\ell}|^{-(1-\alpha)}$ [cf. Eq.~\eqref{eq_fkalphaasympt}], one finds
\beq
\label{eq_subextensivegsdepletion}
\langle \hat n_{\rm tot} \rangle \sim \sum_{|\boldsymbol{\ell}|<L/2} |f_{\boldsymbol{\ell}}(\alpha)|^2 \sim
\left\{
\begin{split}
\mathcal{O}(1) & \quad \text{for } 0<\alpha<d/2 ,\\
\log L & \quad \text{for } \alpha=d/2 ,\\
L^{2\alpha-d} & \quad \text{for } d/2<\alpha<d. \\
\end{split}
\right.
\eeq
On the other hand, for $h=h_{\rm cr}$, we have
\beq
\frac {\langle \hat n_{\rm tot} \rangle}{Ns}  = \frac 1 {4Ns} \sum_{\mathbf k} \Bigg( \frac{1+ \sigma_{\mathbf k}(\alpha)}{\sqrt{ \sigma_{\mathbf k}(\alpha)}}  -1\Bigg) .
\eeq
Here the behavior of  the right-hand side as $L\to\infty$ depends even more strongly on $\alpha$, in particular for one-dimensional systems $d=1$: For $\alpha>3$ the sum is divergent, $ \frac 1 N \sum_{ k} \frac 1 {\sqrt{\sigma_{ k}(\alpha)}} \sim \int_{-\pi}^\pi \frac{d k}{2\pi}  \frac 1 {| k|} = \infty $. This divergence witnesses the inadequacy of LSW theory to describe  critical behavior of one-dimensional systems with short-range interactions.
Contrarily, for $1<\alpha<3$ one has $\sigma_k(\alpha)\sim |k|^{\alpha-1}$, and the integral is convergent. As $\alpha\searrow1$ the depletion per spin is suppressed, making LSW theory increasingly accurate. Finally, for $\alpha<1$, one finds the same subextensive scaling as in Eq.~\eqref{eq_subextensivegsdepletion}. Note, however, that the collective spin mode $k=0$ yields an additional divergent (but still subextensive) contribution $\langle \hat n_0 \rangle \sim N^{1/3}$ to $\langle \hat n_{\rm tot} \rangle$ at the critical point, which can be shown by semiclassical analysis (see~\ref{app_seiclassicalspectrum}); such a contribution is thus dominant for $0<\alpha<\frac 2 3 d$ and subleading for $\alpha>\frac 2 3 d$. 

The bottom line of this Section is that the Holstein-Primakoff description of spin fluctuations is \emph{exact} in the thermodynamic limit for $0<\alpha<d$, and otherwise an increasingly accurate approximation as $\alpha$ is decreased towards $d$. Importantly, this result is true \emph{regardless} of the value of $s$, down to $s=1/2$. Accuracy for low $s$ may be surprising at first sight, considering Eq.~\eqref{eq_HPlocal}. Its origin can be traced back to the observation that the truncated Holstein-Primakoff mapping gives exact matrix elements within the subspace with at most one boson on each site, for arbitrary $s$. Thus, what really controls the quality of the approximation is the ground-state spin-wave \emph{density}: For $0<\alpha<d$ the probability of finding two or more bosons in a given site in $|{\rm GS}_2\rangle$ is vanishingly small in the thermodynamic limit, and it is finite but parametrically small for $\alpha\gtrsim d$.

\summary{The paramagnetic ground state can be determined via linear spin-wave theory, which becomes exact as the strong long-range regime is approached. }


\subsubsection{Quantum ferromagnetic phase}
\label{sec_242}

To derive  \mpar{Bosonic representation of ferromagnetic spin fluctuations} the low-energy spectrum in the quantum ferromagnetic phase for $\alpha>0$, we promote the frame rotation in Eq.~\eqref{eq_rotatedspins} from the level of the collective spin to the level of individual spins:
\begin{equation}
\label{eq_rotatedspinslocal}
\hat s^x_\mathbf{r} = \cos \theta \, \hat s^X_\mathbf{r} + \sin \theta\,  \hat s^Z_\mathbf{r}, \qquad \hat s^y_\mathbf{r} =  \hat s^Y_\mathbf{r}, \qquad \hat s^z_\mathbf{r} = -\sin \theta \, \hat s^X_\mathbf{r} + \cos \theta \, \hat s^Z_\mathbf{r} \, .
\end{equation}
Hence we perform a Holstein-Primakoff expansion of individual spins with quantization axis $\mathbf Z$ and Fourier-transform, 
 \beq
 \label{eq_HPfourierferro}
 \left\{
\begin{split}
 \tilde S^X_{\mathbf k} &\approx
      (Ns)^{1/2} \, \frac{\tilde b^\dagger_{\mathbf k} + \tilde b_{-\mathbf k}}{\sqrt 2}\, ,
    \\
   \tilde S^Y_{\mathbf k} &\approx   (Ns)^{1/2} \, \frac{\tilde b^\dagger_{\mathbf k} - \tilde b_{-\mathbf k}}{\sqrt 2 i} \, ,
    \\
     \tilde S^Z_{\mathbf k} &= Ns \; \delta_{\mathbf k,0}  -   \sum_{\mathbf q} \tilde b^\dagger_{\mathbf q+\mathbf k} \tilde b_{\mathbf q} \, ,
\end{split}
\right.
 \eeq
and substitute into the Hamiltonian~\eqref{eq_lrxyfourier}.
As in Eq.~\eqref{eq_bosonizedHpm} we obtain a formal series in inverse powers of $Ns$, including the classical energy $(Ns)^1 \mathcal{H}(\mathbf Z )$, the quadratic bosonic Hamiltonian
\begin{multline}
\label{eq_H2ferro}
\hat H_2 = \left( h\cos\theta +J_0 (1+\gamma) \sin^2\theta \right) \sum_{\mathbf k}  \tilde b^\dagger_{\mathbf k} \tilde b_{\mathbf k} \\
- J_0 
 \sum_{\mathbf k} f_{\mathbf k} (\alpha)
\bigg[ \left(  \frac{1+\gamma}2 \cos^2\theta  + \frac{1-\gamma}2 \right)
\frac{  \tilde b_{\mathbf k} \tilde b^\dagger_{\mathbf k} +  \tilde b^\dagger_{-\mathbf k}  \tilde b_{-\mathbf k}  } 2 \\
+
\left(  \frac{1+\gamma}2 \cos^2\theta  - \frac{1-\gamma}2 \right)
 \frac{  \tilde b_{\mathbf k} \tilde b_{-\mathbf k} +  \tilde b^\dagger_{-\mathbf k}  \tilde b^\dagger_{\mathbf k}  } 2 \bigg] \, ,
\end{multline}
as well as quartic and higher-order interactions involving an even number of bosons.
However, unlike in Eq.~\eqref{eq_bosonizedHpm}, we also get an additional term $(Ns)^{1/2} \hat H_1$ linear in $\hat q_{\mathbf k = \mathbf 0}=( \hat b_{\mathbf k = \mathbf 0}+\hat b^\dagger_{\mathbf k = \mathbf 0})/\sqrt{2}$ --- cf. the first line of Eq.~\eqref{eq_Hferrocollectiveboson} --- as well as other odd terms $(Ns)^{-1/2} \hat H_3$ and so on.

The rotation angle $\theta^*$ must be determined by imposing that the expectation value of $\hat q_{\mathbf k = \mathbf 0}$ vanishes.
To lowest order this gives the mean-field solution in Eq.~\eqref{eq_thetastar}.
Paralleling the derivation in the previous Section we can then solve the LSW Hamiltonian $\hat H_2 (\theta=\theta^*)$, which yields the spectrum
\beq
\label{eq_swspectrumferro}
\omega_{\mathbf k,<}(\alpha) = \sqrt{
\left[ J_0^2 (1+\gamma)^2  - h^2 f_{\mathbf k}(\alpha) \right] \left[ 1-  f_{\mathbf k}(\alpha)  \frac{1-\gamma}{1+\gamma}\right]  
}
\eeq
as well as the zero-point energy shift $\frac 1 2 \sum_{\mathbf k} (\omega_{\mathbf k,<}(\alpha) - \omega_{<}^{(0)})$, where
$
\omega_{<}^{(0)} = J_0(1+\gamma)
$ [cf. Eq.~\eqref{eq_MFsmallosc<}].

The analysis \mpar{Self-consistency of LSW theory and beyond} of spectral properties and of the spin depletion in the quantum paramagnetic phase can be repeated for the quantum ferromagnetic phase, with qualitatively similar conclusions.
The mean-field description of local observables is \emph{exact} for $0<\alpha<d$ in the thermodynamic limit. For $\alpha>d$ finite corrections to the mean-field results arise.
Such corrections can be evaluated within the bosonic formalism\,\cite{lerose2019impact,lerose2019prethermal}. 
In particular, the downward shift of the quantum critical point $h_{\rm cr,\alpha}=h_{\rm cr,\alpha=0} -\delta h_{\rm cr,\alpha}$ due to quantum fluctuations amounts to\,\cite{lerose2019impact,lerose2019prethermal}
\beq
\label{eq_shiftphaseboundary}
\frac {\delta h_{\rm cr,\alpha}}{ h_{\rm cr,\alpha=0}} \; = \; \frac \gamma s \, \frac{2\ + 3\gamma}{4(1+\gamma)^2} \, \int_{-\pi}^\pi\dots \int_{-\pi}^\pi \frac{dk_1\dots dk_d}{(2\pi)^d} f_{\mathbf k}^2(\alpha) \, .
\eeq
The right-hand side is in fact vanishing for $0<\alpha<d$ and grows finite for $\alpha>d$.
Note that effects of quantum fluctuations are suppressed as $s\to\infty$.

For completeness, let us mention that for $\alpha<d$, the ground state shares the same basic properties of the fully-connected limit~\cite{frerot2017entanglement}. 
Long-range interactions however can induce unexpected entanglement properties. 
For instance, for $d<\alpha<d+1$, the ground state entanglement entropy the  long-range Dyson Hierarchical model obeys an area law at criticality \cite{pappalardi2019entanglement}, due to its special Tree Tensor Network structure \cite{monthus2021properties}. On the other hand, numerical studies for antiferromagnetic long-range systems have shown violations of area-law scaling also in the gapped phase  \cite{koffel2012entanglement, vodola2014kitaev, gabrielli2010quasistationary, roy2019effect}.

\summary{The ferromagnetic ground state can be determined via linear spin-wave theory in a rotated frame. This approach is exact in the strong long-range regime and it determines corrections to the location of quantum critical point.}


\subsection{Structure of the spectrum beyond linear spin-wave theory}
\label{sec_beyondlsw}

In this final Subsection we comment on the structure of the many-body low-energy spectrum beyond LSW theory.
To grasp such effects we will make use of degenerate perturbation theory --- i.e., of the Schrieffer-Wolff transformation --- around points in the two phases where $\hat H_\alpha$ becomes diagonal.

Quite generally,\mpar{Beyond LSW in paramagnetic phase} 
spin waves provide a rather complete description of low energy properties in the quantum paramagnetic phase, even beyond LSW theory.
This is best understood in the regime of large external field $h$, where the ground state is $|{\rm GS}\rangle=| \boldsymbol{\Uparrow}\rangle$ and excited states can be described as a set of individual spin lowering excitations. The degeneracy of blocks with multiple spin excitations is split by the interactions.
The effective block Hamiltonian for large $h$ is obtained by projecting out interaction terms that do not conserve $S^z$, i.e. the part of $\hat H_\alpha$ proportional to $\gamma$.\footnote{E.g., it can be checked that $\gamma$ drops out from the LSW spectrum
$\omega_{\mathbf k,>}(\alpha)$
in Eq.~\eqref{equilibrium_exc_spectra} to lowest order in $1/h$. } 
The Holstein-Primakoff transformation maps this effective XX quantum spin model to a model of lattice bosons with variable-range hopping $\propto 1/(\Delta r)^\alpha$. Non-linearity of the mapping is associated with the suppression of matrix elements for hopping processes from/to multiply occupied sites.\footnote{In particular, transition amplitudes to states with more than $2s$ bosons at any site are strictly vanishing.}
Such unconventional multi-boson interactions [cf. the second line of Eq.~\eqref{eq_gsH4}] produce a scattering phase shift for quantized spin-wave excitations, which can be determined by solving the few-body problem
\footnote{We note that for local systems in one dimension, the exact phase-shifts fully determine the many-body eigenstates via Bethe-Ansatz  \cite{franchini2017introduction}. }.
A qualitatively similar scenario is expected for finite but large enough $h/J_0$, upon rotating the bare Holstein-Primakoff bosons $b_{\mathbf k},b^\dagger_{\mathbf k}$ to the dressed spin-wave basis $\beta_{\mathbf k},\beta^\dagger_{\mathbf k}$ in $\hat H_\alpha$.

On the other hand,\mpar{Beyond LSW in ferromagnetic phase}
 the phenomenology is drastically different in the quantum ferromagnetic phase, due to the strong binding tendency of local spin excitations.
This is most easily understood starting from the ``classical'' Ising model, i.e. Eq.~\eqref{eq_H} with $\gamma=1$, $h=0$:
\beq
\hat H_\alpha = -   \sum_{\mathbf{r},\mathbf{r'}} J_{\lvert\lvert\mathbf{r}-\mathbf{r'}\rvert\rvert}(\alpha) \, \hat\sigma^x_{\mathbf{r}} \hat\sigma^x_{\mathbf{r'}} \, .
 \eeq
 This Hamiltonian is diagonal in the $ x$-basis. It has two degenerate ground states $|{\rm GS}_+\rangle=| \boldsymbol{\Leftarrow}\rangle$ and $|{\rm GS}_- \rangle=| \boldsymbol{\Rightarrow}\rangle$ with energy $E_{\rm GS}=-J_0 N$, and excited states can be described as a set of spin-changing excitations with respect to either ground state. Here the unperturbed energy levels depend on the details of the spin configuration: The lowest elementary excitations are individual isolated spin excitations, 
  \beq
 E \Big( s^x_{\mathbf r}=s-1, \, s^x_{\mathbf R \neq\mathbf r}=s \Big) = E_{\rm GS}  + \frac {J_0} s \, ;
 \eeq
 the subspace with two spin excitations has a configuration-dependent energy,
 \beq
E \Big( s^x_{\mathbf r}=s-1, \, s^x_{\mathbf r'}=s-1, \, s^x_{\mathbf R \neq \mathbf r,\mathbf r'}=s \Big) = E_{\rm GS}  + 2\frac {J_0} s - \frac{J_{\mathbf{r},\mathbf{r'}}(\alpha)}{s^2} \, .
 \eeq
 \beq
E \Big( s^x_{\mathbf r}=s-2, \, s^x_{\mathbf R \neq \mathbf r}=s \Big) = E_{\rm GS}  + 2\frac {J_0} s  \qquad \text{(for $s>1/2$ only)} \, .
 \eeq
 This implies an attractive potential between spin excitations,\footnote{Note that spin excitations on the same site (relevant for $s>1/2$ only) do not feel any attraction or repulsion, unless the Hamiltonian features self-interaction terms.}
 \beq
 V_{\mathbf{r},\mathbf{r'}}(\alpha) =  - \frac{J_{\mathbf{r},\mathbf{r'}}(\alpha)}{s^2} = -   \frac{J_{0}}{2 s^2 \mathcal{N}_{\alpha,L}} \, \frac{1}{\lvert\lvert\mathbf{r}-\mathbf{r'}\rvert\rvert^\alpha} \, .
 \eeq
 Similarly one can compute the unperturbed energy of more complex spin configurations with three or more spin excitations.
 
 Such excited states acquire a non-trivial dispersion relation upon turning on $h\neq0$ or $\gamma\neq1$. Using lowest-order degenerate perturbation theory, it is straightforward to check that processes generated by $1-\gamma\neq0$ induce a variable-range hopping of individual spin excitations, 
 whereas processes generated by $h\neq0$ do not induce any resonant transitions to lowest order.
 We thus retrieve a dispersion relation $\sim \frac {J_0} {s} [ 1-  f_{\mathbf k}(\alpha) (1-\gamma)/2 ]$ for individual spin excitations, in agreement with Eq.~\eqref{eq_swspectrumferro} from LSW theory.
 The number $N_b$ of stable spin-wave bound states depends on the relative magnitude of ``classical potential well depths'', controlled by $\alpha$, and ``quantum hopping amplitudes'', controlled by $1-\gamma$ and $h$. 
 This number grows unbounded upon reducing the quantum fluctuations. Estimating $N_b$ as well as the lifetime of unstable bound states depending on the interaction range and in one or higher dimensions is in general a challenging problem which, to the best of our knowledge, has not been discussed extensively; see however Refs.~\cite{liu2019confined, kranzl2022observation}.
 


All the observations above \mpar{Short-range limit} carry over to short-range interacting systems $\alpha=\infty$, \emph{provided} the system dimensionality is large enough ($d\ge 2$).

In one dimension \mpar{1d: Domain walls and confinement} LSW is still a meaningful description of the paramagnetic spectrum (asymptotically exact for large external field). In the quantum ferromagnetic phase, however, LSW theory completely misses the relevant degrees of freedom, i.e. topological domain-wall-like excitations. In the simplest case $s=1/2$ these fractionalized spin excitations can be described as fermions, as the exact solution of the XY quantum spin chain~\cite{lieb1961two} makes manifest. 
The qualitative effect of longer-range interactions is then to create an effective attractive potential $v_{\Delta r}(\alpha)$ between domain walls at a distance $\Delta r$ (not to be confused with the attractive potential $V_{\Delta r}(\alpha)$ between individual spin excitations introduced above). Taking for simplicity the classical Ising limit $\gamma=1$, $h=0$ as a reference, one can straightforwardly compute the excess energy of a spin configuration with two domain walls separated by $\Delta r$ sites: Assuming $\alpha>1$,\footnote{For $0<\alpha<1$ we have $v_{\Delta r}(\alpha) =2J_0 \Delta r$ in the thermodynamic limit, in agreement with naive LSW theory. In this case, however, the spatial configuration of the $\Delta r$ flipped spins becomes immaterial. Thus, it is not meaningful to speak about ``domain-wall confinement''.}
\beq
v_{\Delta r}(\alpha) = 
\frac{2J_0}{\zeta(\alpha)} \bigg( \sum_{r=1}^{\Delta r} r  \frac1 {r^\alpha} + \Delta r \sum_{r = {\Delta r}+1}^\infty \frac 1 {r^\alpha} \bigg)\, .
\eeq
This potential grows from $v_{\Delta r=1}(\alpha)=2J_0$ to $v_{\Delta r=\infty}(\alpha)=2J_0 \zeta(\alpha-1)/\zeta(\alpha)$ for $\alpha>2$, or to $\infty$ for $\alpha\le 2$. The asymptotic behavior of $v_{\Delta r}(\alpha)$ at large distance $\Delta r$ is
\beq
\label{eq_potentialLR}
v_{\Delta r}(\alpha) \; \underset{\Delta r\to\infty}{\thicksim}   \; 
\frac{2J_0}{\zeta(\alpha)} \times 
\left\{
\begin{array}{ll}
\displaystyle \frac{\Delta r^{2-\alpha}}{(2-\alpha)(\alpha-1)} & \text{for } 1<\alpha <2, \vspace*{1mm} \\
\log \Delta r & \text{for } \alpha =2, \\
\displaystyle \zeta(\alpha-1)-\frac{\Delta r^{-(\alpha-2)}}{(\alpha-2)(\alpha-1)} & \text{for } \alpha > 2. 
\end{array}
\right.
\eeq
For $\alpha>2$ finitely many bound states coexist with unbound deconfined domain walls. As anticipated above, the cost of having a deconfined domain wall blows up as $\alpha\searrow2$, which witnesses the stabilization of long-range order by long-range interactions in 1d. Upon decreasing $\alpha$ the lowest excitation in the spectrum --- the tightest bound state between two domain walls --- is increasingly well described by LSW theory.

We finally note that while long-range interacting quantum spin chains do not naturally map to local lattice gauge theories\,\cite{lerose2020quasilocalized}, except in special cases\,\cite{muschik2017wilson}, the spatial confinement, the spatial confinement of domain walls bears qualitative resemblance with quark confinement in high energy physics\,\cite{mccoy1978two}. This bridge led to insights on the anomalous non-equilibrium dynamics of these systems\,\cite{liu2019confined,lerose2019quasilocalized,verdel2020real}.
Furthermore, although domain-wall deconfinement prevents finite-temperature ordering for $\alpha>2$, it has been shown that the existence of low-lying bound states 
is associated with a drastic suppression of the dynamical melting rate of the order parameter after shallow quantum quenches~\cite{collura2022discrete}. 

\summary{The low-energy spectrum in the quantum ferromagnetic phase hosts a rich structure of spin-wave bound states for $d\ge2$, or for $d=1$ provided $\alpha$ is low enough. In $d=1$ deconfined domain-wall-like excitations appear for $\alpha>2$ along with confined spin-wave-like bound states. }

\section{Low-energy dynamics}
\label{sec_3}

The previous Section shows how the main impact of long-range interactions on low-energy equilibrium properties of the Hamiltonian in Eq.\,\eqref{eq_H} can be traced back to the quasi-particle spectrum, in turn determined by the function $f_{\mathbf k}(\alpha)$, see Fig.\,\ref{Fig1}. Interestingly, the same is true for several types of non-equilibrium phenomena at low energies. In the present Section, we focus on dynamics following weak perturbations of the ground state, which is captured by the quadratic quasi-particle Hamiltonian (such as Eq.\,\eqref{eq_gsH2}).
This allows us to capture the dynamics of quantum correlations at low energies\,\cite{cevolani2016spreading,cevolani2017universal,frerot2018multispeed,storch2015interplay}, see Sec.\,\ref{sec_31}, the appearance of long-live metastable state\,\cite{kastner2011diverging,defenu2021metastability}, i.e. the quasi-stationary states (QSSs), see Sec.\,\ref{sec_metastability}, the universal defect formation upon slowly traversing criticality\,\cite{acevedo2014new,hwang2015quantum,bachmann2017dynamical,defenu2018dynamical,defenu2021quantum}, see Sec.\,\ref{kzm_gen}, and the appearance of dynamical quantum phase transitions in the Loschmidt echo (DQPTs)\,\cite{weidinger2017floquet,syed2021dynamical}, see Sec.\,\ref{sec_34}. Since Sec.\,\ref{sec_31} and\,\ref{sec_metastability} focus on super-critical quenches, the dynamics occurs in the near equilibrium regime, where the spin-wave expansion around the equilibrium state remains applicable. On the other hand, universal defect scaling and DQPTs are observed for quenches across the critical point, making the applicability of the low-energy theory a priori questionable. Nevertheless, we are going to show how the salient features of those critical quenches actually arise from a low density of excitations above the ground state. 

\subsection{Spreading of correlations}
\label{sec_31}

 \mpar{Lieb-Robinson bound}
In systems governed by local Hamiltonians, out-of-equilibrium quantum correlations are known to spread within a ``light cone'': 
The propagation of information in non-relativistic quantum lattice systems with bounded local Hilbert space obeys a speed limit given by the \emph{Lieb-Robinson theorem} \cite{lieb1967exact}. This states that the support of an operator $\hat A_{\bold r}$ initially localized in a finite region around site $\bold r$ and evolving in the Heisenberg representation with a local Hamiltonian $\hat H$, spreads in space with a finite (model-dependent) velocity $v_{\text{LR}}$. Formally, for any locally interacting lattice system there exist positive constants $\xi, \mu$ and $v_{\text{LR}}$ such that the commutator between two locally supported operators $\hat A_{\bold r}$ and $\hat B_{\bold r'}$ separated by a distance $ \Delta r= |\bold r'-\bold r|$ obeys
\beq
\label{eq_LRB}
||[\hat A_{\bold r}(t), \hat B_{\bold r'}(0)] || \leq \xi ||\hat A_{\bold r}||\;|| \hat B_{\bold r'}|| \;e^{-\mu \, \max \left (0, \Delta  r - v_{\text{LR}} t  \right ) } \ ,
\eeq
where $||\cdot||$ is the operator norm.
Namely, the weight of the time-evolved operator outside the ``light-cone'' region $t \geq \Delta r\, / v_{\text{LR}} $ is exponentially suppressed with $\Delta r- v_{\text{LR}} t \to \infty$. In other words, it takes at least a time  proportional to the distance $t\propto  \Delta r$ to send information at a distance $\Delta r$. Such light-cone propagation of information is by now theoretically well understood in short-range interacting systems, and it goes hand-in-hand with a linear dynamical increase of bipartite entanglement out of equilibrium~\cite{calabrese2004entanglement, luchli2008spreading, kim2013ballistic, nahum2017quantum}. Experimental observation of linear light-cone propagation\,\cite{cheneau2012no,fukuhara2013microscopic} has been accompanied by abundant numerical confirmations\,\cite{manmana2007strongly, barmettler2012propagation, carleo2014light, kormos2017confinment}.

In presence of long-range interactions, the standard behavior of locally interacting systems changes substantially: the bounds on the group velocity may not hold anymore, and the spreading of correlations and information scrambling, defined as the loosing of local information into many-body quantum entanglement\,\cite{landsman2019verified}, may be drastically boosted.
\begin{figure}[t]
\fontsize{12}{10}\selectfont
\centering
\includegraphics[width = 1\columnwidth]{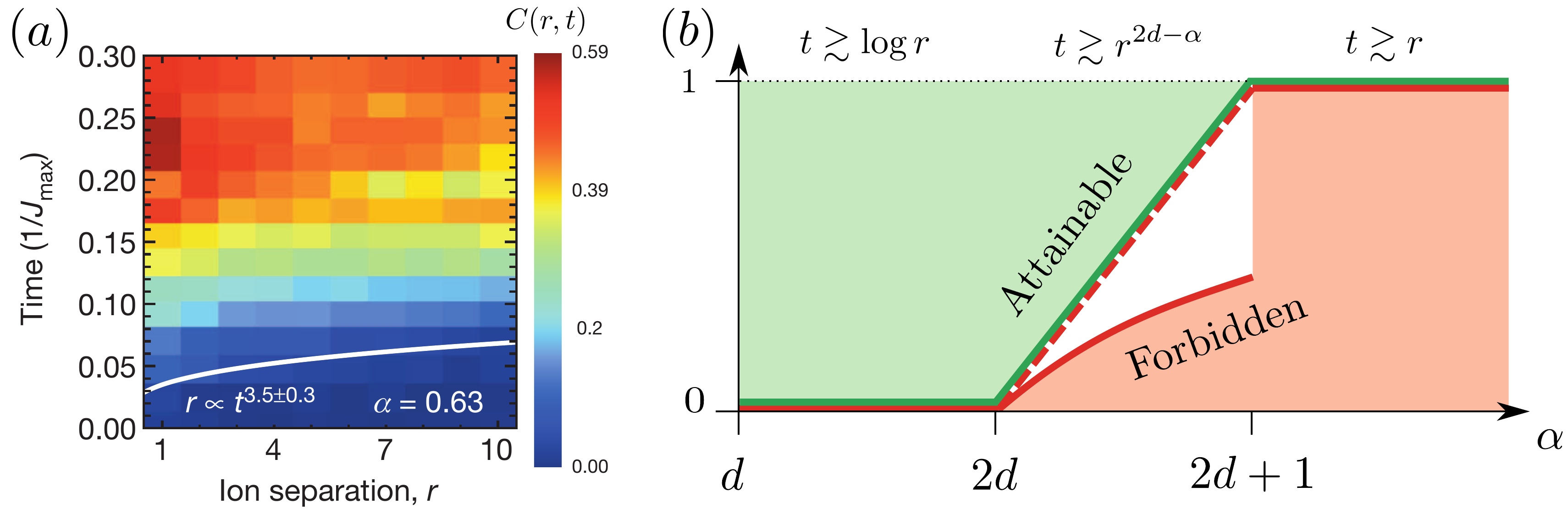}
\caption{ Spatial spreading of correlations in systems with power-law interactions. (a) Connected correlation function in a long-range trapped ion platform following a global quench with $\alpha \approx 0.64$. Image adapted from Ref.~\cite{richerme2014nonlocal}.  (b) Violation of the Lieb-Robinson bound in Eq.~\eqref{eq_LRB} for long-range interacting systems with power-law interactions for $\alpha>d$. Image adapted from Ref.~\cite{tran2021lieb}.}
\label{fig:4_6}
\end{figure}
The study of the impact of algebraically decaying interactions on correlation spreading has been addressed as a function of the different values of the power-law exponent $\alpha$.
Part of the current understanding is based on assessing the behavior of the spatial spreading of connected correlations, e.g. 
\begin{equation}
	\label{eq_Gspreading}
	G_{\alpha\beta}(r, t) = \langle \hat\sigma^\alpha_{i+r}(t) \hat\sigma^\beta_{i}(0)\rangle -\langle \hat\sigma^\alpha_{i+r}(t)\rangle \langle \hat\sigma^\beta_{i}(0)\rangle\ ,
\end{equation}
in paradigmatic quantum spin chains or tight-binding models; the expectation value is taken over some initial state $\ket{\psi_0}$ and $\alpha, \beta=x,y,z$. 

Generalized bounds have been derived for long-range systems\,\cite{hastings2006spectral, foss-feig2015nearly}, see Ref.\cite{chen2023speed} for a recent comprehensive review. The related experiments and  numerical investigations have, however, led to conflicting pictures\,\cite{hauke2013spread,eisert2013breakdown, jurcevic2014quasiparticle,richerme2014nonlocal,cevolani2015protected,cevolani2016spreading,buyskikh2016entanglement}.
For instance, experiments on ion chains\,\cite{richerme2014nonlocal} and numerical simulations within truncated Wigner approximation\,\cite{schachenmayer2015dynamics}
for the one-dimensional long-range XY  model
point towards bounded, super-ballistic, propagation for all values of $\alpha$. In contrast, experiments on the long-range transverse Ising model reported ballistic propagation of correlation maxima with, however, observable leaks that increase when $\alpha$ decreases\,\cite{jurcevic2014quasiparticle}.
Moreover, time-dependent density matrix renormalization group (t-DMRG) and variational Monte-Carlo (t-VMC) numerical simulations indicate
the existence of three distinct regimes, namely instantaneous, sub-ballistic, and ballistic, for increasing values of the exponent $\alpha$, see Ref.\,\cite{schachenmayer2013entanglement,hauke2013spread,eisert2013breakdown,cevolani2015protected,cevolani2016spreading,buyskikh2016entanglement}. \\
In the following, we will see how these difficulties can be overcome in the restricted setting of near-equilibrium dynamics, by studying correlation spreading within linear spin-wave theory.

\summary{The Lieb-Robinson bound forbids super-ballistic spreading of quantum correlations in locally interacting systems. Long-range interactions allow to circumvent this constraint.  }

\subsubsection{Weak long-range regime ($\alpha>d$)}
\label{corr_weak_lr}
Let us first consider the case of the Ising  Hamiltonian, i.e. Eq.\,\eqref{eq_H} with $\gamma=1$, but restrict our study to the spin-wave representation in Eq.\,\eqref{sw_h}. In this Section, we aim at characterizing the universal scaling of correlations following Ref.\,\cite{cevolani2017universal}. Let us simplify the spin-wave dispersion relation in Eq.\,\eqref{equilibrium_exc_spectra} by considering its low-momentum asymptotic expression,
\begin{align}
\label{gapped_dispersion}
\omega_{\mathbf k}  \mapsto \omega^{\rm low}_{\mathbf k} = \Delta + c  { k}^\zeta,
\end{align}
where the gap
$\Delta =  \sqrt{h \left( h + 2J_{0} f_{0} (\alpha) \right)}$ is finite,
$c = \sqrt{\frac{h}{h+2J_{0} f_{0} (\alpha)}}J_{0} \frac{\partial f_{0} (\alpha)}{\partial { k}}$,
and
\begin{align}
\zeta=\begin{cases}
\alpha-d &\qquad\mathrm{if}\quad d<\alpha\leq d+2,\\
2 &\qquad\mathrm{if}\quad \alpha>d+2.
\end{cases}
\end{align}
As long as $\alpha>d$ the 
quasi-particle energy remains finite, while the group velocity diverges for $d<\alpha<d+1$. For any $\alpha>d+2$ one has $\zeta=2$ and the conventional picture of nearest neighbour interactions is recovered.
\mpar{Quenches within the paramagnetic phase}
The system is prepared in its ground state and the coupling is suddenly quenched from $J_{0}^{i}\to J_{0}^{f}\equiv J_0$ at the initial time $t=0$. When considering longitudinal spin correlations, i.e. Eq.\,\eqref{eq_Gspreading} with $\alpha=\beta=x$ one can employ the formula
\begin{align}
\label{eq:gen-corr-tev}
G_{xx}\left(r,t\right)  = g(r) -
\int_{\mathcal{B}}\frac{d^d {\mathbf k}}{\left(2\pi\right)^{d}}\mathcal{F}\left({ k}\right)\frac{e^{i\left(k\cdot r+2\omega_{{\mathbf k}}t\right)}  + e^{i\left(k\cdot r-2\omega_{{\mathbf k}}t\right)}}{2},
\end{align}
where the integral spans the first Brillouin zone $\mathcal{B}$. In the following we are going to ignore the time-independent function $g(r)$ and focus on the time evolution of the correlations, which can be readily obtained
\begin{align}
\label{eq:LRTI.Fk}
 \mathcal{F}(k)=\frac{2h\left(J_0^{\rm i}-J_0^{\rm f}\right)f_k \left(\alpha\right)}{\left[h+2J_0^{\rm f}f_k \left(\alpha\right)\right]\sqrt{h \left[ h + 2J_0^{\rm i} f_k \left(\alpha\right) \right]}}.
\end{align}

\begin{figure}[t]
\centering
\subfloat[ Ising Model]{\includegraphics[width=.48\linewidth]{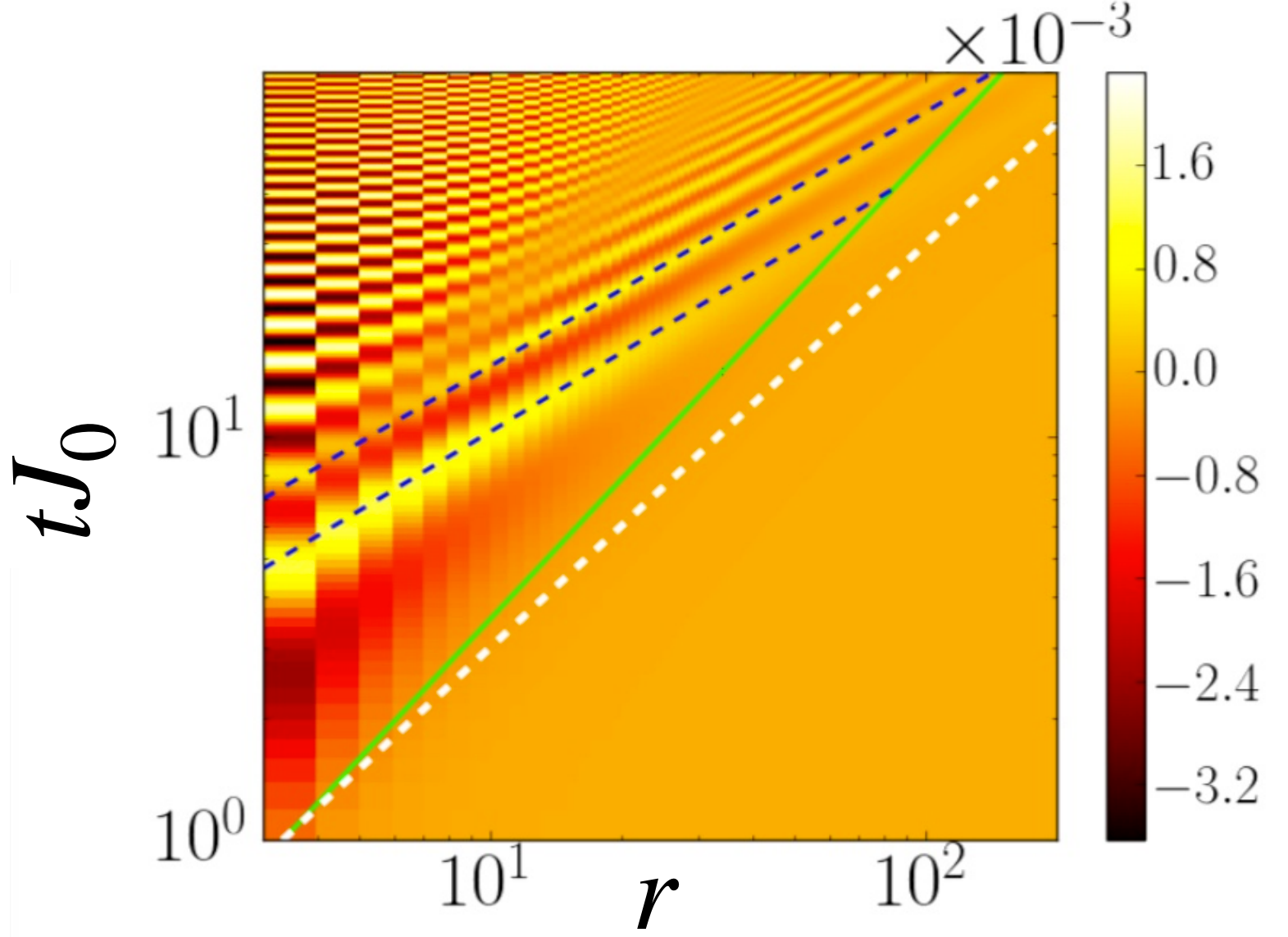}\label{Fig_SecIII_1a}}
\subfloat[XY Model ]{
\includegraphics[width=.48\linewidth]{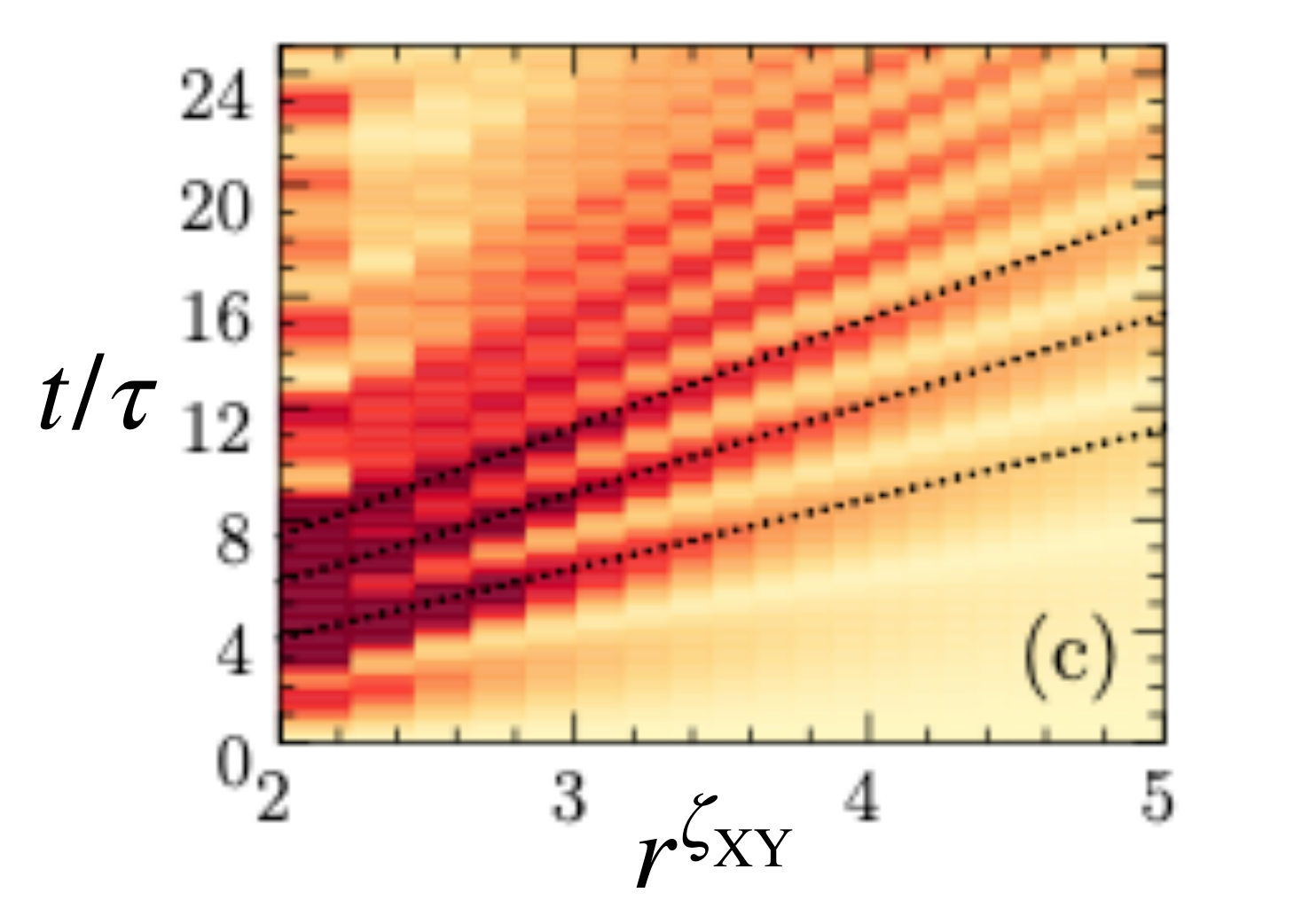}\label{Fig_SecIII_1b}}
\caption{\textbf{Spreading of connected spin-spin correlation function.}  Panel \textbf{(a)} displays $G_{xx}(r,t)$ for the quantum Ising chain with $\alpha=1.7$ for a sudden quench for the quench in the paramagnetic phase from $ h/J_{0}^{\rm i}=50$ to $ h/J_{0}=1$.
The green line is the correlation front which scales sub-ballistically (white dashed line represents ballistic propagation).
The dashed black line are the analytic scaling $r^{\zeta}$ obtained in Eq.\,\eqref{eq:LRnogapmax}. Panel \textbf{(b)} reports $G_{zz}( r,t)$ for the quantum $XY$ model with $\alpha=3$ in $d=2$ for a quench starting from the fully polarized state along $x$ and evolved with the Hamiltonian in Eq.\,\eqref{eq_H} with $\gamma=h=0$ (i.e. a quench from $\gamma=1$ to $\gamma=0$).
The dashed black lines show the scaling of the maxima which is linear in the axis $r^{\zeta_{XY}}$. Panel \textbf{(a)} is adapted from Ref.\,\cite{cevolani2017universal} and panel \textbf{(b)} from Ref.\,\cite{frerot2018multispeed}.}
\label{Fig2}
\end{figure} 

The amplitude of the quench is directly proportional to the difference between the initial $J_{0}^{\rm i}$ and final $J_{0}^{\rm f}$ couplings. Both these values are chosen to maintain the system within the paramagnetic phase $h>h_{\rm cr}$.
The time-dependent correlation function $G_{xx}(r,t)$ of the long-range Ising model obtained by Eq.\,\eqref{eq:gen-corr-tev} is displayed in Fig.\,\eqref{Fig_SecIII_1a} for $\alpha=1.7$. The front of the correlation is highlighted by a green line. Its scaling is not linear but algebraic as expected for long-range interactions\,\cite{hastings2006spectral,foss-feig2015nearly}. 
Nevertheless, the front propagation does not saturate the conventional super-ballistic bounds\,\cite{tran2022hierarchy}, rather it displays a sub-ballistic scaling, i.e. $t \sim r^{\beta_{\rm front}}$ with $\beta_{\rm front}>1$, which is represented as a solid green line.
Inside the correlation front the scaling changes and for the Ising model the correlation maxima (light yellow areas) propagate ballistically with $t \sim r$.

It is interesting to use the stationary phase approximation in order to evaluate Eq.\,\eqref{eq:gen-corr-tev} in the large size and long-time limit. Indeed, for $t,r\to \infty$ the integral in Eq.\,\eqref{eq:gen-corr-tev} is dominated by the configurations with
\begin{align}
\label{stat_phase}
\nabla_{k}\omega_{k}=r/t
\end{align}
where the group velocity diverges as $k^{\zeta-1}$ in the $k \to 0$ limit. Thus, quasi-particles with momentum $k_{\rm sp}=\left(2|c|\zeta t/r\right)^{1/\left(1-\zeta\right)}$ fulfil Eq.\,\eqref{stat_phase} at any given point $(t,r)$ in space-time. The leading contribution to the correlation front propagation comes from the low-energy divergence of quasi-particle group velocity. In order to evaluate the leading contribution to the correlation function we assume that the amplitude function obeys $\lim_{k\to 0}\mathcal{F}(k) \sim k^{\eta}$, leading to
\begin{equation}
\label{eq:statphaseLR}
G_{xx}(r,t) \propto \frac{t^{\gamma}}{r^{\chi}}\cos\left[A_\zeta\left(\frac{t}{r^{\zeta}}\right)^{\frac{1}{1-\zeta}}-2\Delta t + \frac{\pi}{4}\right],
\end{equation}
with
    $\gamma=\frac{\eta+d/2}{1-\zeta}$,
    $\chi=\frac{\eta+d(2-\zeta)/2}{1-\zeta}$, and
   $A_{z}=2|c| (1-\zeta) (2|c|\zeta)^{\frac{\zeta}{1-\zeta}}$.
It follows from Eq.\,\eqref{eq:statphaseLR} that the correlation front  obeys the relation $t^{\gamma}\approx r^{\chi}$ and
\begin{equation}
\label{eq:LRfront}
t  \propto r^{\beta_{\rm front}},
\qquad \beta_{\rm front}=\chi/\gamma.
\end{equation}
Interestingly, the propagation of the wave-front does not depend only on the universal scaling exponent $\zeta$ but also on the specific correlation function under consideration,  since the exponent $\eta$, which describes the low energy scaling of $\mathcal{F}\left({ k}\right)$ in Eq.\,\eqref{eq:statphaseLR},  enters in the determination of the ratio $\chi/\gamma$. As the local limit is approached, the quasi-particle velocity ceases to diverge $\zeta\to 1$ and linear spreading of the wave-front is recovered so that Eq.\,\eqref{eq:statphaseLR} reproduces the Lieb-Robinson expectation\,\cite{lieb1972finite}.
The relation $\chi=\gamma+d/2$ yields $\beta_{\rm front}>1$ and imposes sub-ballistic wave-front propagation.

The quench protocol under consideration stays within the paramagnetic phase, which is characterized by a gapped dispersion relation, see Eq.\,\eqref{gapped_dispersion}. This, in turns, leads to $\lim_{k\to 0}\mathcal{F}(k) \sim O(1)$ and the scaling exponent of the correlation function vanishes, i.e. $\eta=0$. Thus, only the exponent $\zeta$ determines the front propagation scaling $\beta_{\rm front}=2+d-\alpha$. The theoretical prediction for $\alpha=1.7$ and $d=1$ produces $\beta_{\rm front} = 1.3$, which is in perfect agreement with the one observed in the numerical computation displayed in Fig.\,\ref{Fig_SecIII_1a}.
The formula $\beta_{\rm front}=2+d-\alpha$ also matches the exact result obtained in Ref.\,\cite{cevolani2015protected} for $d=1$ and $\alpha=3/2$, also confirmed by t-VMC calculations.

 Within the causal region delimited by the wave-front, the local maxima are determined by the maxima of the cosine function in Eq.\,\eqref{eq:statphaseLR}. Thus, the correlation maxima occur at the time $t_\textrm{max}$, whose value does not depend on the shape of $\mathcal{F}(k)$, but only on the value of the $\zeta$ exponent, yielding 
\begin{equation}
\label{eq:LRnogapmax}
t_\textrm{max} \propto r^{\zeta},
\end{equation}
at least for a gapless dispersion relation. According to this analysis, the maxima of the correlations, located at the time $t_{\max}$ spread super-ballistically for weak long-range interactions. This has to be contrasted with the sub-ballistic scaling obtained in Eq.\,\eqref{eq:LRfront} for the front propagation.

The result in Eq.\,\eqref{eq:LRnogapmax} is consistent with the experimental observation on trapped ions\,\cite{richerme2014nonlocal} as well as with the truncated Wigner approximation analysis\,\cite{schachenmayer2015dynamics,schachenmayer2015many}. However, for the long-range Ising model the dynamical protocol under consideration remains within the paramagnetic phase, leading to a finite gap $\Delta\neq 0$. Therefore, the argument of the cosine function in Eq.~(\ref{eq:statphaseLR}) is insensitive to the non-analytic scaling of the dispersion relation in the low-energy limit, becoming constant in the large $t$ and $r$ limit with $t/r\sim const$. Thus,  Eq.\,\eqref{eq:LRnogapmax} has to be substituted with $t_{\rm max} \propto r$ for a gapped dispersion relations.
It follows that the local maxima are always ballistic, $\beta_{\rm max}=1$ for quenches within gapped phases, see Fig.\,\ref{Fig_SecIII_1a}. The ballistic motion of local maxima has also been observed with a trapped ion quantum simulator\,\cite{jurcevic2014quasiparticle}.

Based on the above discussion\mpar{Quenches in the gapless phase}, it can be deduced that the correlations spreading reflect the low-energy properties of the long-range model.
It is evident that the scaling of correlations is universal in long-range systems, in the sense that it reflects the low-energy properties of the model. Then, a very different picture is obtained by studying a quantum quench within a gapless phase. In order to investigate this dependence we prepare the system in the state fully polarized along the direction $x$ and evolve it with the Hamiltonian in Eq.\,\eqref{eq_H} with $\gamma=h=0$, i.e. the long-range XY Hamiltonian, which has $U(1)$ symmetry. Following Ref.\,\cite{frerot2018multispeed} we consider the case $d=2$. The  dispersion relation can be obtained by Eq.\,\eqref{eq_swspectrumferro} by setting $\gamma=h=0$
\begin{align}
\omega_{k} = J_{0} \sqrt{1-f_{{k}}(\alpha)}.
\end{align}
As expected, changing the symmetry of the final Hamiltonian modifies the low-energy dispersion relation, which now scales as $ 
\omega^{\rm low}_k \propto  k^{\zeta_{\rm XY}}$ with 
\begin{align}
\zeta_{\rm XY}=\begin{cases}
(\alpha-d)/2\qquad \mathrm{if} \quad d<\alpha\leq d+2\\
1 \qquad \mathrm{if} \quad \alpha> d+2
\end{cases}
\end{align}
leading to a diverging quasi-particle group velocity in the $k\to 0$ limit for $d<\alpha<d+2$. A straightforward computation for the long-range XY model\,\cite{frerot2018multispeed,cevolani2017universal} produces $\eta=\zeta$ leading to $\beta_{\rm front}=1+d(2+d-\alpha)/(2\alpha)$. On the other hand, Eq.\,\eqref{eq:LRnogapmax} remains unchanged and it yields $\beta_{\rm max}=\zeta_{\rm XY}$, as it is visible in Fig.\,\ref{Fig_SecIII_1b} and verified by DMRG calculations in Ref.\,\cite{frerot2018multispeed}.

\summary{
In the weak long-range regime, the correlations front spreads non-linearly, with exponents that depend on the details of the underlying low-energy dispersion.
}

\subsubsection{Strong long-range regime ($0<\alpha<d$)}
\label{corr_strong_lr}
We now consider the Hamiltonian \eqref{eq_H} in the strong long-range regime $0<\alpha<d$. 
Following Ref.~\cite{cevolani2016spreading}, in this Subsection the interactions are \emph{not} Kac-normalized, i.e. we set $J \equiv J_0$ in Eq.\,\eqref{kac_norm}.
This leads us to discuss the effect of a divergent quasi-particle energy for $k\to 0$ onto the correlation spreading. [Note that this is different from the rest of the Report where we focus on the discrete spectrum in Eq.~\eqref{eq_fkalphaasympt} at low $k$!]
Within this framework, we approximate the low-energy dispersion relation with the expression \mpar{Divergent dispersion}
\begin{equation}
\omega_{k}\approx\frac{e_0}{k^{\gamma}},
\end{equation}
where $e_0=\sqrt{2hJ_{0}}$ and $\gamma=\frac{d-\alpha}{2}$. Including the modified dispersion relation into the Eq.\,\eqref{eq:gen-corr-tev} one gets 
\begin{equation}
\label{slr_correlations}
G_{xx}(r,t) \sim \int_{\omega}d\Omega\int_{\pi/L}^{\pi}dk \, k^{d-1+\gamma}e^{\imath k r\cos\left(\theta\right)}\left[1-\cos\left(2e_{0}tk^{-\gamma}\right)\right],
\end{equation}
where the factor $k^\gamma$ comes from the low energy limit of the amplitude function in Eq.\,\eqref{eq:LRTI.Fk}, i.e. $\mathcal{F}_{}(k) \sim k^{\gamma}$.
Due to the divergent nature of the quasi-particle spectrum, one can introduce a low-energy cutoff $\sim 1/L$ in the momentum integral in Eq.\,\eqref{slr_correlations}. 
After expanding the exponential term Eq.\,\eqref{slr_correlations} in powers of the distance $r$, the integration is performed term by term \,\cite{cevolani2016spreading}. Then, after discarding finite term in the system size one finds
\begin{equation}
G_{xx}(r,t) \sim \lim_{L\rightarrow\infty} \frac{\sin\left(L^{\gamma}\tau\right)}{\tau}\frac{\int d\Omega 
e^{\imath\frac{R}{L}\cos\left(\theta\right)}}{L^{2\gamma+D}},
\end{equation} \label{eq:nonloc}
where is the dimensionless time variable $\tau=2e_{0}t$.

Due to the algebraic divergence of the quasi-particles spectrum the time scale for signal spreading in the system vanishes in the thermodynamic limit. Accordingly, the vanishing of the signal spreading time displays the same scaling exponent $\gamma$ as the divergence of the quasi-particle energy, which for the Ising model reads $\gamma=\frac{d-\alpha}{2}$. This analytic derivation has been also corroborated by numerical analysis of the spin-wave dynamics in Ref.\,\cite{cevolani2016spreading}. It is worth noting that the same scaling has been derived within a generalized Lieb-Robinson bound in long-range fermionic systems\,\cite{storch2015interplay}. 

\summary{In the strong long range regime without Kac normalization, the divergent quasi-particle energy leads to instantaneous correlation spreading in the thermodynamic limit. }

\subsubsection{Other directions}

The 
\mpar{Numerical results on the Ising model}
present Subsection has been devoted to the study of correlation spreading within linear spin-wave theory\,\cite{cevolani2015protected, cevolani2016spreading, cevolani2017universal, frerot2018multispeed}.  This approximation proved capable to capture the salient features of several numerical simulations~\cite{hauke2013spread, eisert2013breakdown, budich2016dynamical, schachenmayer2013entanglement, schachenmayer2015dynamics}. In particular, in the case of the quantum Ising chain [Eq.~\eqref{eq_H} with $d=1$], numerical matrix-product state calculations have shown that the emergence of a short-range-like light-cone behavior for $\alpha>2$\,\cite{schachenmayer2013entanglement} as confirmed by the study in Sec.\,\ref{corr_weak_lr}. 
On the other hand, for $1<\alpha<2$, the model displays clear light cone but with an infinite propagation speed of almost all excitations. This is linked to the divergence of the maximum group velocity, which leads to a scenario of multispeed prethermalization \cite{frerot2018multispeed}, see again Sec.\,\ref{corr_weak_lr}. 
For $\alpha<1$, all studies report a clear nonlocal regime, with instantaneous transmission of correlations between distant sites, in agreement with the study reported in Sec.\,\ref{corr_strong_lr}.

Despite 
\mpar{Extensions}
the successes of linear spin-wave theory, it would be interesting to reconsider these results using the time-dependent framework that we will present in Sec.\,\ref{sec_421}. One may compute $G(r, t)$ by expressing spin operators in a time-dependent frame and expanding them using Holstein-Primakoff bosons. This very analysis has been performed in a related model in Ref.~\cite{lerose2019impact} in connection with dynamical phase transitions; see also the analogous calculation for scrambling dynamics in Eq.~\eqref{eq_cabGauss}.

Finally, 
\mpar{Rigorous results}
a very successful research direction based on {rigorous mathematical investigations} was pursued to generalize the Lieb-Robinson bound for power-law decaying interactions \cite{chen2023speed}. In a seminal work of 2006, Hastings and Koma~\cite{hastings2006spectral} showed that it takes a time $t \gtrsim \log r$ to propagate information at distance $r$ for all $\alpha>d$. However this bound is far from tight, since it does not recover the linear light-cone in Eq.\eqref{eq_LRB} for large $\alpha$. Several efforts in the past years have led to a greatly improved picture \cite{storch2015interplay, nachtergaele2006lieb, nachtergaele2006propagation, nachtergaele2009lieb, premont-schwarz2010lieb_a, premont-schwarz2010lieb_b,  gong2014persistence, foss-feig2015nearly, eldredge2017quantum, else2018improved,  tran2020hierarchy, tran2020locality, tran2021optimal}. Firstly, it was proved the existence of the linear light-cone $t \gtrsim r$ for $\alpha>2d+1$ \cite{chen2019finite,  kuwahara2020strictly}. Secondly, it was shown that this result becomes  $t \gtrsim r^{\min (\alpha-2d , 1)}$ for $\alpha>2d$ \cite{tran2021optimal, tran2021lieb}.
On the other hand, in the strong long-range regime  $0<\alpha < d$, correlations between distant degrees of freedom can propagate instantaneously, since the bounds on the light-cone time-scale can vanish with the system size \cite{storch2015interplay, guo2019signaling, eldredge2017quantum}. So far, the best estimate for interacting systems is $t\gtrsim N^{\alpha/d-1} \log N $, that can be made tighter $t\gtrsim N^{\alpha/d-1/2} $ for free models with $\alpha < d/2$ \cite{guo2019signaling}.
Notably, the violations of Lieb-Robinson bound have been experimentally probed on trapped ions quantum simulators for $0.6 \lesssim \alpha \lesssim 1.2$ in Refs.~\cite{jurcevic2014quasiparticle, richerme2014nonlocal}.

\summary{ In addition to low-energy approximations, the spreading of correlations has been tackled with various approaches ranging from numerical simulations to mathematically rigorous bounds. The current established scenario is rather complete and satisfactory. }

\subsection{Metastability}
\label{sec_metastability}

In the following, we are going to show that metastability in quantum strong long-range systems may be traced back to their discrete quasi-particle spectrum, which hinders the applicability of the kinematical chaos hypothesis\,\cite{lasinio1996chaotic}.

\subsubsection{State of the art}
Long-range interactions are traditionally connected with the appearance with long-lived metastable states in the out-of-equilibrium dynamics. These states, referred to as quasi-stationary states (QSSs),  display long lifetimes, which diverge increasing the system size. QSSs are widespread within the classical long-range physics world\,\cite{dauxois2002hamiltonian, mukamel2005breaking}, but multiple theoretical observation occurred also in the quantum realm\,\cite{kastner2011diverging,bachelard2013universal}. Long-lived pre-thermalization is also expected to occur in cold atom clouds confined into optical resonators\,\cite{schutz2014prethermalization}, where semi-classical analysis of the Fokker-Planck equation directly connects to the Hamiltonian mean-field model\,\cite{schutz2013cooling,schutz2015thermodynamics}, the workhorse of classical long-range physics\,\cite{campa2014physics}. 
 Recent studies have directly linked the absence of equilibration in strong long-range quantum systems to the discreteness of the quasi-particle spectrum, see Eq.\,\eqref{eq_discrete}. This results in a violation of Boltzmann's H-theorem and leads to the emergence of finite Poincaré recurrence times in the thermodynamic limit\,\cite{defenu2021metastability}. This section discusses the appearance of diverging equilibration times for quantum long-range systems in the thermodynamic limit. 
This is consistent with the properties discussed in Section\,\ref{low_en_theory}, which are common to both large long-range systems and finite local ones. Examples include the inability to completely disregard boundary effects over bulk phenomena\,\cite{barre2007ensemble, latella2015thermodynamics}, the existence of concave entropy regions\,\cite{ispolatov2001first}, and the presence of a macroscopic energy gap between the ground state and the first excited state\,\cite{gupta2012one,gupta2012overdamped}.  It is worth noting that our description mostly pertains isolated quantum systems, while multiple theoretical and experimental observations in cavity systems evidenced a substantial role of dissipation\,\cite{schutz2016dissipation, hruby2018metastability}.

The crucial aspect is that the excitation spectrum of non-interacting systems does not become continuous in the thermodynamic limit. The eigenvalues of a long-range coupling matrix have been shown to remain discrete even in the infinite components limit, forming a pure point spectrum\,\cite{last1996quantum} similar to that observed in strongly disordered systems\,\cite{thouless1972anderson,froehlich1983absence,simon1985some,scardicchio2017perturbation}. A discussion on the spectral discreteness of long-range couplings in the thermodynamic limit has been presented in Ref.\,\cite{defenu2021metastability} for a few quadratic models and used to explain the observation of diverging equilibration times in a long-range Ising model, quenched across its quantum critical point\,\cite{kastner2011diverging}. We refer the readers to Sec.~\ref{low_en_theory} and~\ref{app:boundsF}.

\subsubsection{Quasi-stationary states and spectral properties}
The 
\mpar{Quasi-stationary states}
first evidence of QSSs in quantum systems was described in the prototypical example of the long-range quantum Ising chain [see Eq.\,\eqref{eq_H}]. QSS were shown to appear for quenches starting well inside the paramagnetic phase in the $h\to\infty$ limit and ending deep in the ferromagnetic phase at $h = 0$. Here, the system is prepared in the transversally polarized ground state and evolved according to the classical ferromagnetic Hamiltonian in Eq.~\eqref{eq_H} in the absence of the transverse field $h=0$. As a result, the expectation of the global operator $m_{z} = \langle \sum_{i} \sigma^{z}_{i}\rangle/N$ evolves from the initial value $\lim_{t\to 0} m_{z} = 1$ 
to the equilibrium expectation  $\lim_{t\to \infty} m_{z} = 0$, if the system actually equilibrates. These observations have been extended to any choice of the initial and final magnetic fields $h_i$, $h_f$ using the Jordan-Wigner representation of the Ising model.


The appearance of the QSS has been frequently linked to the scaling of equilibration times of critical observables, such as the magnetization~\cite{antoni1995clustering, mukamel2005breaking, campa2009statistical}. However, persistent time fluctuations have also been found in generic thermodynamic observables of classical systems, such as the evolution of internal energy in systems of particles with attractive power-law pair interactions \cite{gabrielli2010quasistationary}. Similar phenomena 
\mpar{Low-energy quenches}
can be observed in our system, by considering just the leading order low-energy theory. In order to simplify the study we restrict our analysis to the paramagnetic quantum Ising chain, whose quasi-particle dispersion is 
\begin{align}\label{sph_ham}
\omega_{k}=\sqrt{h(h-2J_{0}f_{k}(\alpha))} \, ,
\end{align}
cf. Eq.~\eqref{equilibrium_exc_spectra}.
It is worth noting that the present spin-wave approximation corresponds to the time-dependent Hartree-Fock approximation of the Ising and $O(N)$ rotor models. 
Accordingly, several phenomena occurring in the out-of-equilibrium low-energy dynamics of the Ising Hamitlonian can be also observed in the large-$N$ limit of $O(N)$ models\,\cite{berges2007quantum,sotiriadis2010quantum}, including prethermalization\,\cite{chiocchetta2017dynamical, halimeh2021quantum}, defect formation\,\cite{degrandi2010adiabatic}, dynamical phase transitions\,\cite{syed2021dynamical}.
In particular, the dynamics induced by a sudden quench leads to  universal relaxation properties\,\cite{sotiriadis2010quantum,chandran2013equilibration, syed2021dynamical}.

In this regime, equilibration does not occur in the non-additive regime  due to the discrete quasi-particle spectrum $f_{k}(\alpha)$. In order to demonstrate this fact let us consider a sudden magnetic field quench $h^{\rm i}\to h^{\rm f}$ in the Hamiltonian\,\eqref{sw_h}. 
The quench occurs within the normal phase $h>h_{\rm cr}$ so that no magnetization occurs. Nevertheless, a finite spin-wave density will arise due to the sudden quench and will contribute to the evolution of any internal observable of the system. In order to make a direct parallel with the classical case described in Ref.\,\cite{anteneodo1998breakdown} we consider the evolution of the spin-wave kinetic energy
\begin{align}
\label{kinetic_energy}
K(t)=  \sum_{k}\langle\hat{p}_{k}^{2} \rangle/2N = -\frac{\omega_{k}}{4}\left\langle\left(\beta^{\dagger}_{k}-\beta_{k}\right)^2\right\rangle \ ,
\end{align}
where $\beta_{k}$ and $\beta^{\dagger}_{k}$ diagonalize the quadratic Hamiltonian in Eq.~\eqref{eq_gsH2}.
The 
\mpar{Solution of the quadratic dynamics}
calculation is rather straightforward, since the system is assumed to lie in the ground-state before the sudden quench. After the quench, each spin-wave occupies a squeezed state, so that the system lies in the quantum state $\Pi_{k}\hat{S}_{k}(\zeta)|0\rangle$, where $|0\rangle$ is the vacuum and the squeezing operator $\hat{S}_{k}(\zeta)|0\rangle$ reads\,\cite{gerry2005introductory} 
\begin{align}
\label{squezzing_operator}
\hat{S}_{k}(\zeta)=\exp\frac{\left(\zeta^{*}(\hat{\beta}_{k})^{2}-\zeta(\hat{\beta}^{\dagger}_{k})^{2}\right)}{2}.
\end{align}
The squeezing parameter $r$ is defined by rewriting $\zeta$ in polar coordinates $\zeta=re^{i\phi}$. Then, it is rather straightforward to rewrite the squeezing parameter in terms of the effective oscillator length $\xi(t)$\begin{align}
\label{squeezing_parameter}
\tanh{r_{k}}=\sqrt{\frac{\left(\frac{1}{2\xi_{k}(t)^{2}}-\omega_{k}\right)^{2}+\frac{\dot{\xi_{k}}(t)^{2}}{\xi_{k}(t)^{2}}}{\left(\frac{1}{2\xi_{k}(t)^{2}}+\omega_{k}\right)^{2}+\frac{\dot{\xi}_{k}(t)^{2}}{\xi_{k}(t)^{2}}}}.
\end{align}

To 
\mpar{Persistent oscillations}
obtain the spin-wave dynamics is then sufficient to solve the Ermakov equation, which describes the evolution of the effective length\,\cite{dabrowski2016time}
\begin{align}
\label{ermakov_eq}
\ddot{\xi}_{k}(t)+\omega_{k}^{2}(t)\xi_{k}(t)^{2}=\frac{1}{4\xi_{k}(t)^{3}}.
\end{align}
 
\begin{figure}[t]
\centering
\subfloat[ Kinetic Energy]{\label{Fig2a}\includegraphics[width=.48\linewidth]{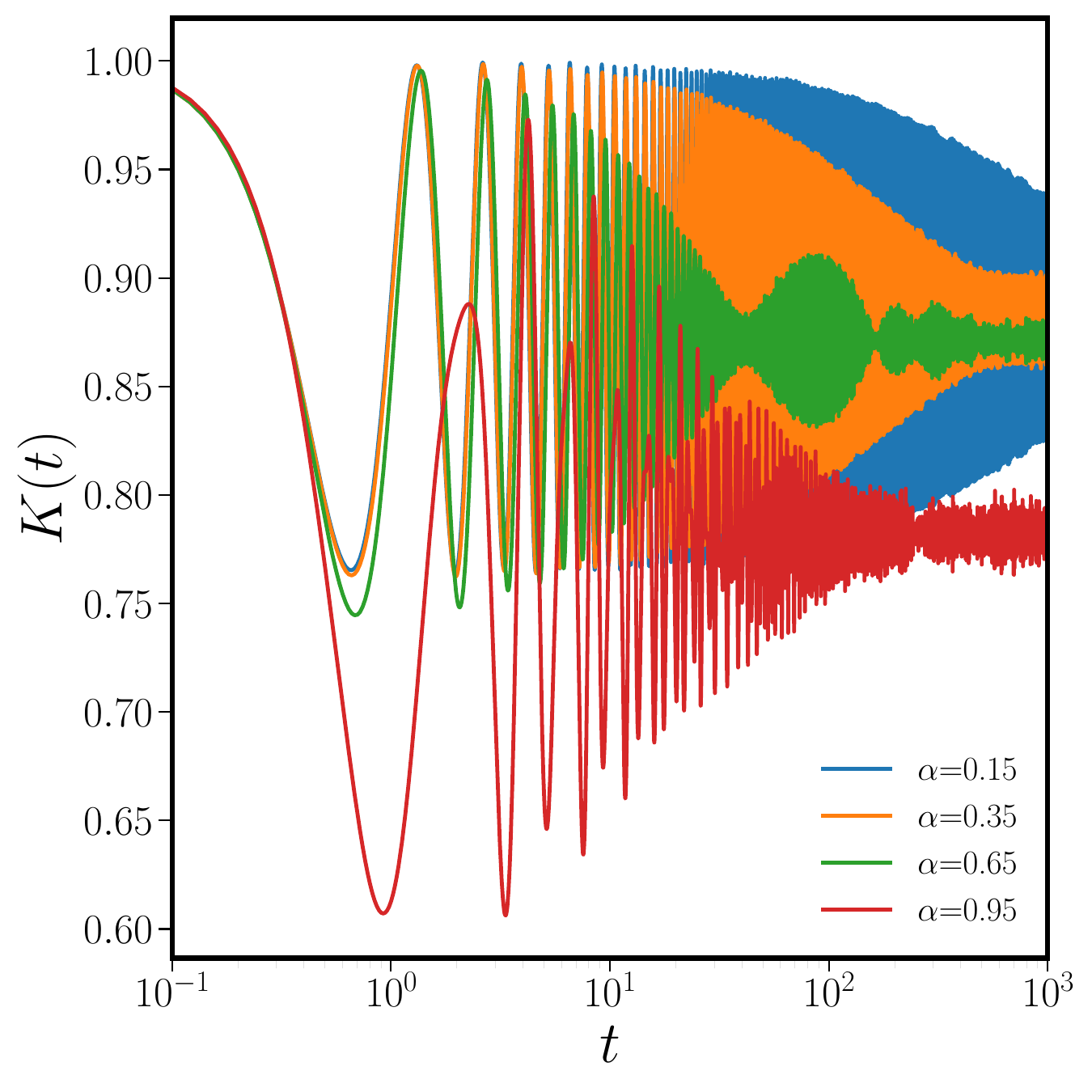}}
\subfloat[Dynamical fluctuations ]{\label{Fig18b}
\includegraphics[width=.48\linewidth]{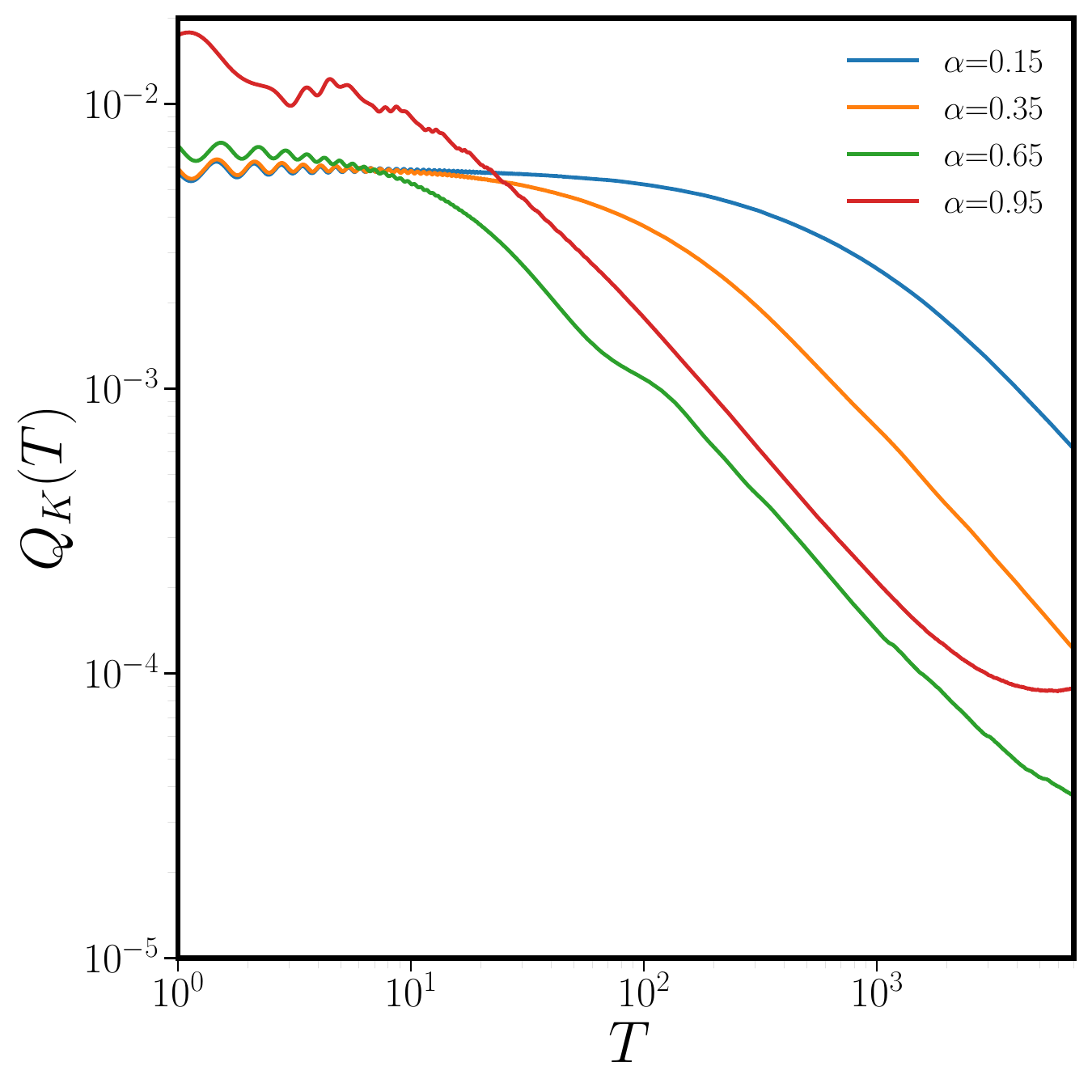}}
\caption{\textbf{Equilibration of long-range spherical model.} Panel (a) displays the dynamical evolution of the kinetic energy following the sudden quench $h_{i}\to h_{f}$. After a steady decay during the initial dynamics $t\lesssim 10^{2}$, dynamical oscillations set to a finite value that remain steady in the long-time regime. The amplitude of dynamical fluctuations for the kinetic energy after a time $T$ is quantified by the quantity $Q_{K}(T)$ see Eq.\,\eqref{eq_eq} displayed in panel (b). }
\label{Fig2}
\end{figure} 

The solution of the sudden quench dynamics is readily obtained by introducing $\omega_{k}(t)=\theta(-t)\,\omega_{k,\rm i}+\theta(t)\,\omega_{k,\rm f}$ in Eq.\,\eqref{ermakov_eq}. 
The resulting dynamical evolution for the spin-wave kinetic energy is displayed in Fig.\,\ref{Fig2} for $\alpha\in[0.15, 0.35, 0.65, 0.95]$. In analogy with the classical case the observable $K(t)$ displays persistent dynamical oscillations, which do not wash out in the thermodynamic limit. The smaller the $\alpha$ the wider the amplitude of those fluctuations.

A 
\mpar{Origin due to the discrete spectrum}
simple explanation of this phenomenon is found in the fully connected limit ($\alpha\to 0$), where the function $f_{k}(\alpha)$ separates between two distinct energy levels in the thermodynamic limit: a non-degenerate ground-state with energy $-J$ and a $N-1$ degenerate excited states with zero energy.

In presence of any given set of boundary conditions, the degeneracy is split  and the system behaves at finite size as a set of harmonic oscillators with discrete energies. As the size increases, the spectrum accumulates at high energy where the eigenvalues $f_{k}(\alpha)$ 
of the coupling matrix become all identical, making the system equivalent to a single quenched harmonic oscillator. 
To characterize equilibration, we introduce the characteristic function of any observable $A$, i.e. $
\chi_{A}(t)=A(t)-\bar{A}$. This quantity captures the dynamical fluctuations around the average value of the observable. Equilibration of the observable $A(t)$ in closed quantum systems occurs when the long-time Cesaro's average of the squared fluctuation vanishes \cite{reimann2008foundation,linden2009quantum,oliveira2018equilibration}:

\begin{align}
\label{eq_eq}
\lim_{T\to\infty}Q_{A}(T)\equiv\lim_{T\to\infty}\langle |\chi_{A}(t)|^{2}\rangle_{T}=\lim_{T\to\infty}\frac{1}{T} \int_{0}^{T}|\chi_{A}(t)|^{2}dt=0\ ,
\end{align}
while metastability shall be associated with a finite value $\lim_{T\to\infty}Q_{A}(T)\neq 0$.
The equilibration of the kinetic energy $K$ in Eq.\eqref{kinetic_energy} -- or other physical observables --  follows a similar argument as the fidelity of a quantum system in the context of the spectrum of long-range systems\,\cite{defenu2021metastability}. \\
In
\emph{weak-long range interacting systems} with translational invariance, the spectrum becomes absolutely continuous in the thermodynamic limit. This implies that $\lim_{t\to\infty}\chi_{K}(t)\to 0$ outside of the Cesaro's average due to the Riemann-Lebesgue lemma. For quantum systems with initial states having no overlap with pure point portions of the spectrum, equilibration as defined in equation \eqref{eq_eq} is ensured by Wiener's theorem \cite{last1996quantum}. 
Considering
 the thermodynamic limit of a \emph{strong long-range interacting} system, where the system size $N$ increases, the eigenmodes $f_{k}(\alpha)$ of the Hamiltonian tend to accumulate at high energy, near $\omega_{k}|_{h_{f}}\sim 2h_f$.
In fact, in 
the case of \emph{flat interactions} ($\alpha=0$), the spectrum consists of a single infinite degenerate eigenstate, resulting in dynamics that precisely correspond to a single harmonic oscillator.

\summary{In the strong long-range regime, the discrete low-energy spectrum yields quasi-stationary states. These manifest in persistent oscillations of time-dependent observables, such as the kinetic energy, with an amplitude that increases as the interaction becomes longer ranged $\alpha \to 0$.  }

\subsubsection{Equilibration in presence of disorder}
\begin{figure}[t]
\centering
\subfloat[Dynamical Fluctuations]{\label{Fig_SecIII_3a}\includegraphics[width=.50\linewidth]{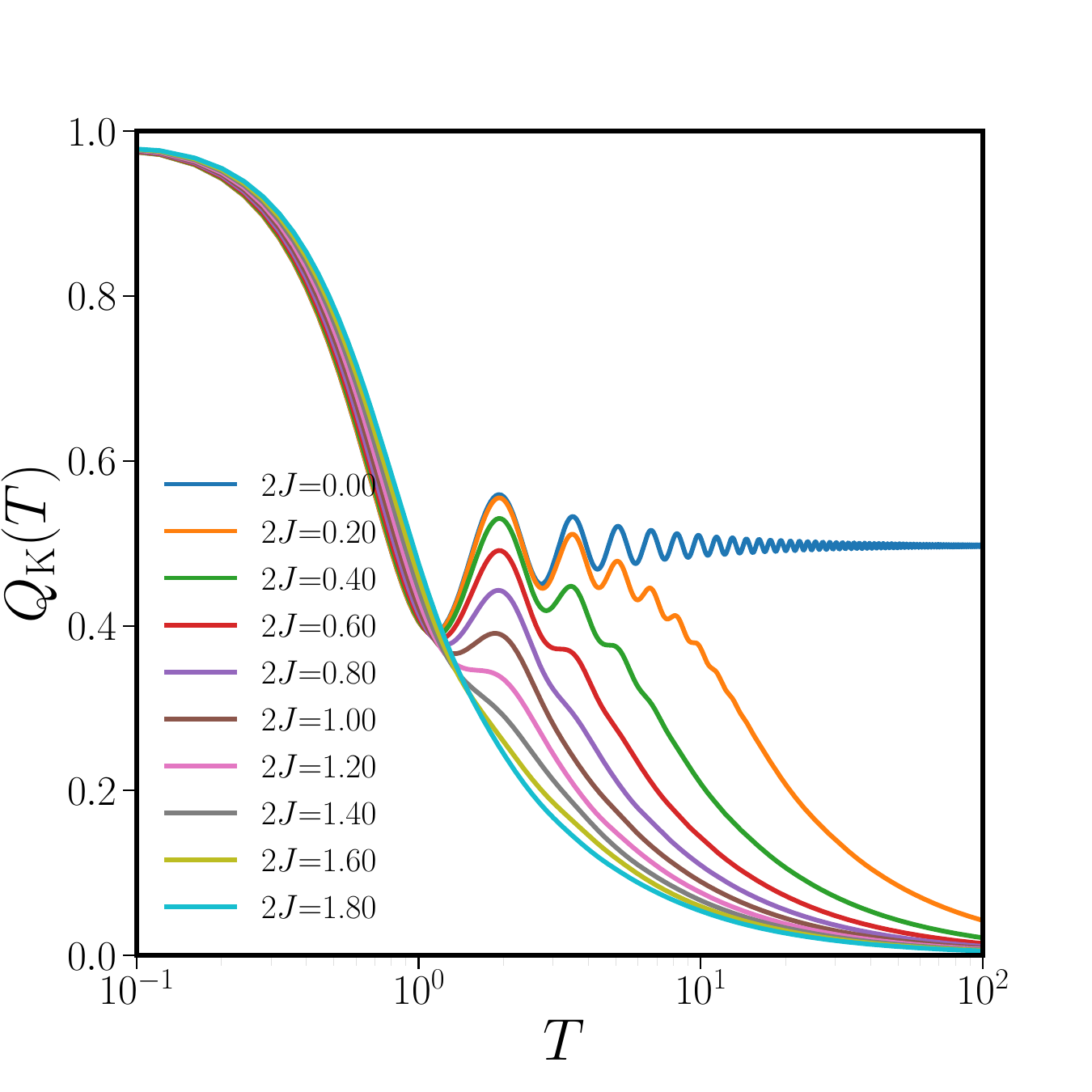}}
\subfloat[Equilibration Time]{\label{Fig3b}
\includegraphics[width=.48\linewidth]{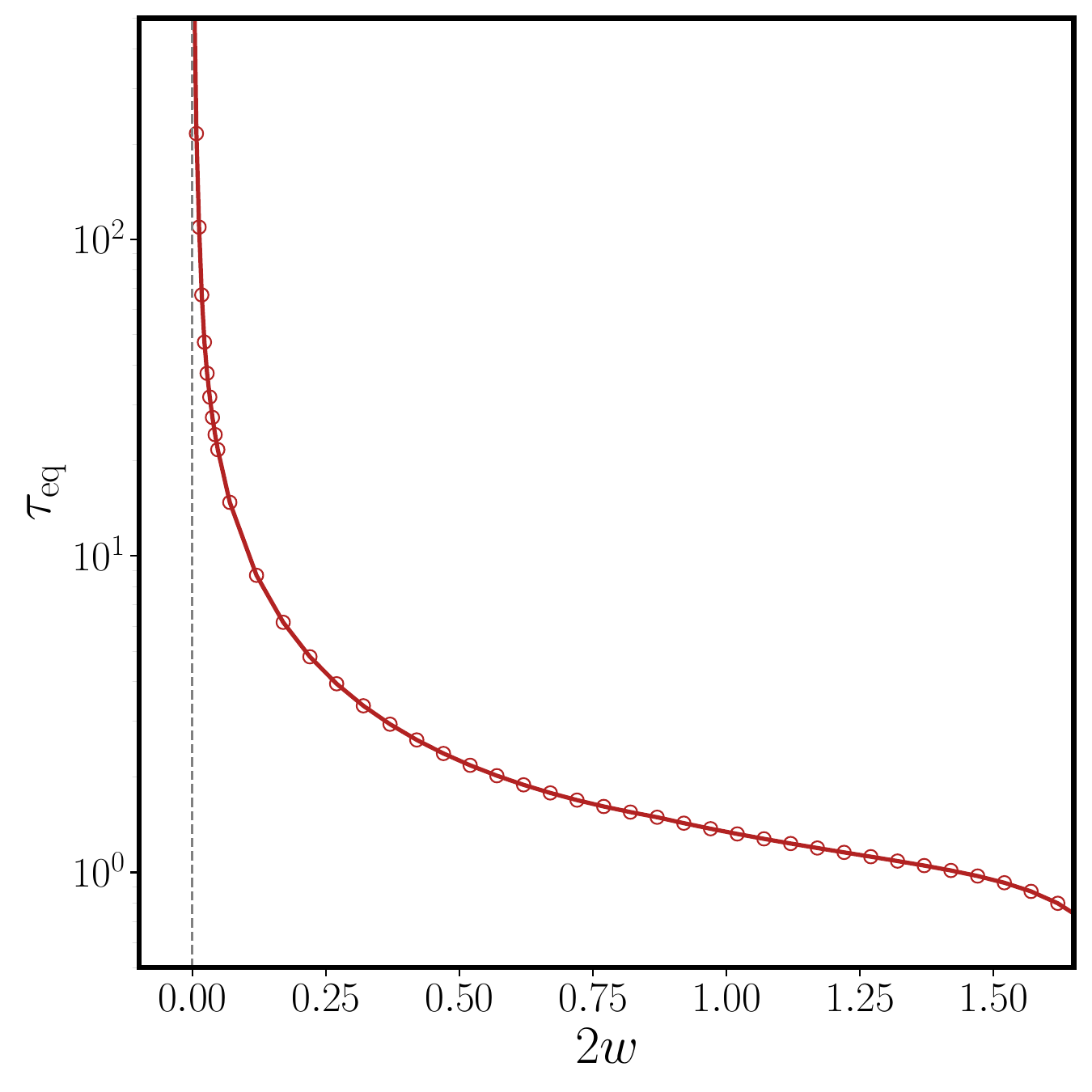}}
\caption{\textbf{Equilibration of the disordered $\boldsymbol{\alpha=0}$ Ising model within spin-wave approximation.} Panel (a): Dynamical fluctuations decay as a function of the disorder strength for the kinetic energy $K$, see Eq.\,\eqref{kinetic_energy}. As the disorder strength is decreased the decay rate also decreases until it vanishes in the zero disorder limit (upper blue curve), where dynamical fluctuations persist at all times and equilibration never occurs. Panel (b): the equilibration time of the kinetic energy observable obtained by fitting the curves in panel (a) via the exponential form in Eq.\,\eqref{eq_time}. The divergence in the clean case ($2w\to 0$) is evident. Figures reproduced from Ref.\,\cite{defenu2021metastability}.}
\label{Fig3}
\end{figure}

We would like to comment on the relation between these results and the metastability that arises in disordered systems. 
From the perspective of spectral discreteness, the metastability observed for quantum systems in the strong long-range regime is fundamentally different. Indeed, disordered couplings tend to lift spectral degeneracy and lead to continuous spectra. A simple example is obtained by perturbing
\mpar{Random fully connected model}
flat interactions ($\alpha=0$) with Gaussian distributed weak couplings $u_{ij}$, namely $J^{\rm dis}_{\boldsymbol{r},\boldsymbol{r}'}=J_{\boldsymbol{r},\boldsymbol{r}'}+u_{ij}$ with
\begin{align}
P(u_{ij})\propto \exp\left(-N\,u_{ij}^{2}/2w^{2}\right)
\end{align}
whose width $2w$ represents the disorder strength.  The disordered couplings lift the infinite ($\sim N-1$) degeneracy of the excited state at zero energy and the spectrum becomes continuous apart from the single non-degenerate ground-state at energy $J_{0}$, where $J_{0}>w$ is the strength of the flat homogeneous interactions\,\cite{kosterlitz1976spherical, edwards1976eigenvalue,dellanna2008critical}. The density of states of the continuous spectrum follows the celebrated Wigner semicircle law\,\cite{metha2004random}. In analogy with the non disordered case, we initialize the dynamics at equilibrium for $h= 2.2 h_{\rm cr}$ at $t\leq0$; then, at $t>0$ the magnetic field suddenly switches at $h_{f}=1.1 h_{\rm cr}$. The continuum nature of the spectrum leads the spin-wave kinetic energy $K(t)$ to exponentially equilibrate, see Fig.\,\ref{Fig_SecIII_3a}.
Indeed, the amplitude of dynamical fluctuations in disordered systems decays exponentially, allowing for the introduction of the equilibration time $\tau_{\rm eq}$:
\begin{align}
\label{eq_time}
Q_{K}(T)\sim e^{-T/\tau_{\rm eq}} \ ,
\end{align}
Numerical analysis shows that the equilibration time $\tau_{\rm eq}$ monotonically decreases with increasing disorder strength $J$, as expected (see Fig.\,\ref{Fig3b}). The exponential decay of dynamical fluctuations and the definition of the equilibration time provide insights into the equilibration dynamics of disordered systems. Interestingly, the results obtained in Fig.\,\ref{Fig3} for quantum systems show remarkable similarities to those obtained for classical spherical models with disordered couplings, as shown in Chapter 4 of Ref. \cite{cirano2006random}. There, the Langevin dynamics of the disordered classical spherical model it is shown not to exhibit metastability, as long as the initial state is not magnetized.

The 
\mpar{Effective dimensionality}
analysis presented here focuses on characterizing the long-time equilibration dynamics of many-body quantum systems, where the thermodynamic limit is generally taken before the long-time limit. However, similar conclusions can be obtained by considering the long-time limit of dynamical fluctuations in finite systems, which yields \cite{reimann2008foundation,linden2009quantum,oliveira2018equilibration}:
\begin{align}
\label{fs_eq}
\lim_{T\to\infty}Q_{K}(T)\propto \frac{1}{\mathcal{N}_{\rm modes}}
\end{align}
Here, $\mathcal{N}_{\rm modes}$ roughly represents the number of modes participating in the dynamics. In the case of finite systems, where the entire spectrum is discrete and only a finite number of modes exist, $\mathcal{N}_{\rm modes}$ captures this finite nature. As the thermodynamic limit is approached and the spectrum becomes continuous, eigenvalues become dense in arbitrarily small energy ranges. In the continuous limit, $\mathcal{N}_{\rm modes}\to\infty$ for dynamics involving initial states in the continuous spectrum.
However, for long-range systems with $\alpha<d$, where $d$ is the dimension of the system, a continuous theory cannot be defined. This is because the only dense region in the spectrum occurs around the energy maximum, where infinitely many degenerate eigenvalues emerge. This violates the assumption of non-degenerate energy gaps underlying equation \eqref{fs_eq}\,\cite{short2011equilibration}.

\summary{
Metastability for strong long-range systems is fundamentally different from the disordered one in that only a finite number of effective degrees of freedom participate in the dynamics.
}

\subsection{Kibble-Zurek mechanism}
\label{kzm_gen}
Within 
\mpar{Landau-Zener problem}
the realm of quantum systems, the Landau-Zener problem provides the earliest, and possibly simplest, example of defect formation during a quasi-static drive\,\cite{landau1965quantum, zener1932non, damski2005simplest}. The problem describes a two-level system slowly driven across an avoided level crossing. Although initialized in the ground state at the initial time $t_{i}< 0$, the system gradually populates the excited state, whose energy separation slowly decreases during the dynamics. The energy gap starts from a minimum at $t=0$ and increases back until the time reaches the endpoint of the dynamics $t_{f}>0$. The dynamical evolution is controlled by the rate parameter $\delta$, i.e. $H(t)\equiv H(\delta\cdot t)$, so that the quasi-static limit is reached as $\delta\to 0$. A straightforward criterion to establish whether a quasi-static transformation remains adiabatic is to ensure that the rate of change of the instantaneous minimal gap $\Delta=|E_{0}-E_{1}|$ remains smaller than the square gap itself
\begin{align}
\label{ad_cond}
\dot{\Delta}(t)\ll \Delta^{2}(t).
\end{align}
The above criterion only involves equilibrium quantities since $E_{\ell}(t)$ represents the spectrum of the instantaneous Hamiltonian at the time $t$. 

While Eq.\,\eqref{ad_cond} has been obtained by heuristic arguments a more rigorous derivation of an adiabatic criterion and a discussion of how it compares with Eq.\,\eqref{ad_cond} can be found in Ref.\,\cite{boixo2010necessary}. Eq.\,\eqref{ad_cond} has been introduced for the Landau-Zener problem, but the argument in Ref.\,\cite{boixo2010necessary} applies to generic quantum systems. In general, as long as Eq.\,\eqref{ad_cond} is satisfied, the excited state population of a quasi-statically driven quantum system decreases with the drive rate and the hypothesis for the quantum adiabatic theorem are satisfied\,\cite{born1928beweis}. However, as the drive approaches a quantum critical point the correlation length of a quantum system diverges and the instantaneous gap vanishes $\Delta\to 0$. As a result, the dynamical scaling of the observables close to the quantum phase transition is reminiscent of the thermodynamic scaling at equilibrium. Yet, in order for such scaling to be displayed, the drive has to be slow enough that the dynamical evolution actually occurs in the vicinity of the equilibrium critical point.

Let us\mpar{Dynamical protocol} focus on the concrete case of the Hamiltonian in Eq.\,\eqref{eq_H} whose internal  control parameter is defined as $\lambda=h(t)-h_{\rm cr}$, such that the ferromagnetic quantum critical point occurs at $\lambda_{c}=0$. For a moment, let us imagine that the system is finite, so that the spectrum remains gapped also at criticality. Then, the hypothesis of the quantum adiabatic theorem\,\cite{born1928beweis} remains fulfilled and any slow enough drive of internal parameters $\lambda(t)\sim\delta t$ only generates adiabatic corrections $\sim\delta^{2}$ to the observables expectations with respect to the equilibrium value, as it can be deduced by simple thermodynamic arguments\,\cite{zwerger2008limited}. However, in the thermodynamic limit, crossing the equilibrium critical point breaks down the conventional adiabatic picture and the residual energy (heat) generated by the drive displays non-analytic behaviour $E_{\rm res}\approx\delta^{\theta}$ with $\theta<2$\,\cite{zurek1996cosmological}. Our task within the present section is the determination of the universal scaling index $\theta$ for quantum long-range systems.

\subsubsection{State of the art}

The 
\mpar{Kibble-Zurek}
Kibble-Zurek mechanism allows to \emph{relate the value of the non-analytic exponent $\theta$ with the equilibrium critical exponents of the model}.  This ingenious theory relies on the adiabatic-impulse approximation, which assumes that the dynamical evolution of a system starting in its  ground-state at $t=-\infty$ adiabatically follows the drive until the freezing time $-\hat{t}$. Beyond the freezing time, the equilibration rate of the system becomes too small with respect to the drive velocity, violating the adiabatic condition in Eq.\,\eqref{ad_cond}. Then, the freezing time satisfies the condition 
\begin{align}
\label{freezing_time}
\dot{\Delta}(\hat{t})/\Delta(\hat{t})^{2}=1.
\end{align} 
Due to the critical scaling of the instantaneous gap $\Delta\propto \lambda^{z\nu}$, Eq.\,\eqref{freezing_time} leads to the freezing time inheriting the equilibrium critical scaling $\hat{t}\propto\delta^{1/(1+z\nu)}$. The state dynamics is assumed to remain frozen for the entire interval $t\in [-\hat{t},\hat{t}]$ until the unfreezing time $\hat{t}$, where adiabaticity is restored (for simplicity we have assumed a symmetric gap).

Once the system has unfrozen the state evolution will resume on the opposite side of the transition, where the Hamiltonian ground-state is supposed to break the Hamiltonian symmetry. Then, the dynamics will induce a transition between the symmetric and a symmetry-broken state. However, this transition will occur at finite correlation length $\hat{\xi}$, since the process can only start at the time $t\geq\hat{t}$
well within the symmetric phase of the model. The dynamics has thus modified the character of the continuous phase transition, making it rather similar to a
first-order one, and the system will likely form topological defects, whose size would be roughly proportional to the (finite) correlation volume $\hat{\xi}^{d}$, as long as the correlation length is well defined, i.e. for $\alpha>d$. Therefore, the total defect density scales according to $N_{\rm exc}\propto \hat{\xi}^{-d}\propto \delta^{d\nu/(1+z\nu)}$\,\cite{dziarmaga2010dynamics}.

Several 
\mpar{Short-range scaling}
verifications of the Kibble-Zurek scaling exist in local systems, both via numerical simulations, exact theoretical studies and experiments\,\cite{delcampo2014universality}. In particular, first studies of defect formation in quantum systems have been pursued on the Hamiltonian in Eq.\,\eqref{eq_H} in the $\alpha\to\infty$ limit, i.e. the nearest neighbour Ising model, where finite size scaling arguments led to the prediction
\begin{align}
\label{fss_kz_arg}
N_{\rm exc}^{\rm fss}\approx \delta^{\frac{1}{2z}}
\end{align}
and the superscript fss stands for finite size scaling. Eq.\,\eqref{fss_kz_arg} produces $N_{\rm exc}\approx \sqrt{\delta}$ in agreement with the Kibble-Zurek prediction $N_{\rm exc} \propto \delta^{d\nu/(1+z\nu)}$ since $z=\nu=1$ in this case\,\cite{zurek2005dynamics}. Soon after this seminal investigation, an exact solution to the universal slow dynamics of the Ising model has been provided by mapping it to a infinite sum of Landau-Zener problems, each representing the dynamics of a single fermionic quasi-particle\,\cite{dziarmaga2005dynamics}. This exact solution provides a different scaling theory for the defect density of the Ising model, which is given by
\begin{align}
\label{kit_ex_kz}
\int N_{\rm exc}(k)dk\approx \delta^{\frac{1}{2z_{\Delta}}}
\end{align}
where we have defined $z_{\Delta}$ from the scaling of the dynamical gap. The result in Eq.\,\eqref{kit_ex_kz} has been also employed to prove validity of the Kibble-Zurek argument in Kitaev chains with long-range pairing terms\,\cite{dutta2017probing} where $z_{\Delta}=z$, as well as in the perfect local case $\alpha=\infty$\,\cite{dziarmaga2005dynamics}.

Apart 
\mpar{Fermi systems}
for the aforementioned results, which explicitly refer to quadratic Fermi systems, the application of adiabatic perturbation theory to slow quenches close to quantum critical points predicts the scaling of the defect density to be in agreement with the classical Kibble-Zurek prediction
\begin{align}
\label{def_sc_kz}
\int N^{\rm KZ}_{\rm exc}(k)\, dk\approx \delta^{\frac{d\nu}{1+z\nu}},
\end{align}
which also leads to the scaling exponent $\theta=z\nu/(1+z\nu)$ for the residual energy\,\cite{polkovnikov2005universal}. Such prediction comes from the assumption that the scaling form of the critical propagator reproduces the equilibrium critical exponents. Since for Fermi systems in $d=1$ one has $z\nu=1$, the perturbative argument yields $d\nu/(z\nu+1)=1/2z$ in agreement with the finite size scaling argument in Eq.\,\eqref{fss_kz_arg}. 
Interestingly, long-range anisotropic interactions with different ranges depending on the type of coupling are known to violate the perturbative assumption\,\cite{defenu2019universal} even in the finite range case\,\cite{divakaran2009defect,deng2009anomalous,dziarmaga2010dynamics}.

In summary, the applicability of the Kibble-Zurek result in quantum systems is supported by two main arguments. The finite size scaling argument reported in Eq.\,\eqref{fss_kz_arg} and the perturbative argument, which reproduces the traditional Kibble-Zurek scaling in Eq.\,\eqref{def_sc_kz}. Both arguments coincide for local quantum many-body systems where the femionic quasi-particle description applies. This is the case of the Ising Hamiltonian in Eq.\,\eqref{eq_H} with $\alpha\gg d+3$ as confirmed by the exact solution obtained at $\alpha=\infty$.

First 
\mpar{ Fully-connected quantum Ising model}
indications that the scaling of the defect density in the $\alpha=0$ Ising model did not follow the Kibble-Zurek prediction appeared in Ref.\,\cite{caneva2008adiabatic}. However, later investigations showed that the residual energy of the the $\alpha=0$ Ising model obeys the Kibble-Zurek mechanism, at least for slow ramps terminating at the critical point, i.e. $t\in [-1/\delta,0]$\,\cite{hwang2015quantum}. This apparent inconsistencies triggered more intensive numerical studies, which unveiled a complicated landscape where the adiabatic crossing of the equilibrium quantum critical point does not display any scaling with the ramp rate $\delta$, but rather featured a novel form of dynamical universality as a function of the scaled variable $\Lambda=N\,\delta $\,\cite{acevedo2014new}.

\begin{figure}[t]
\centering
\subfloat[Slow quench to the quantum critical point]{\label{Fig4a}\includegraphics[width=.48\linewidth]{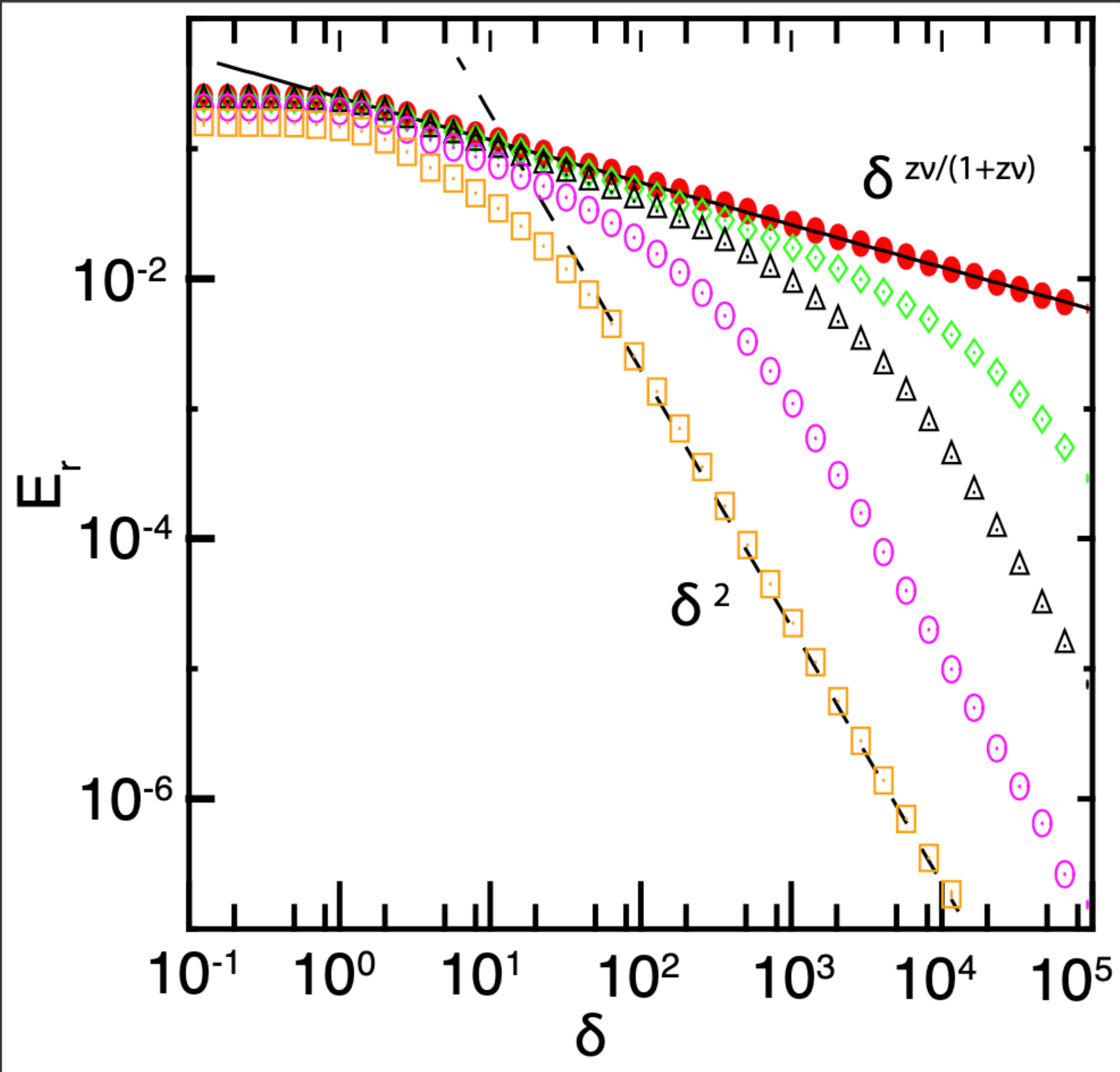}}
\subfloat[Slow quench across the quantum critical point]{\label{Fig4b}
\includegraphics[width=.48\linewidth]{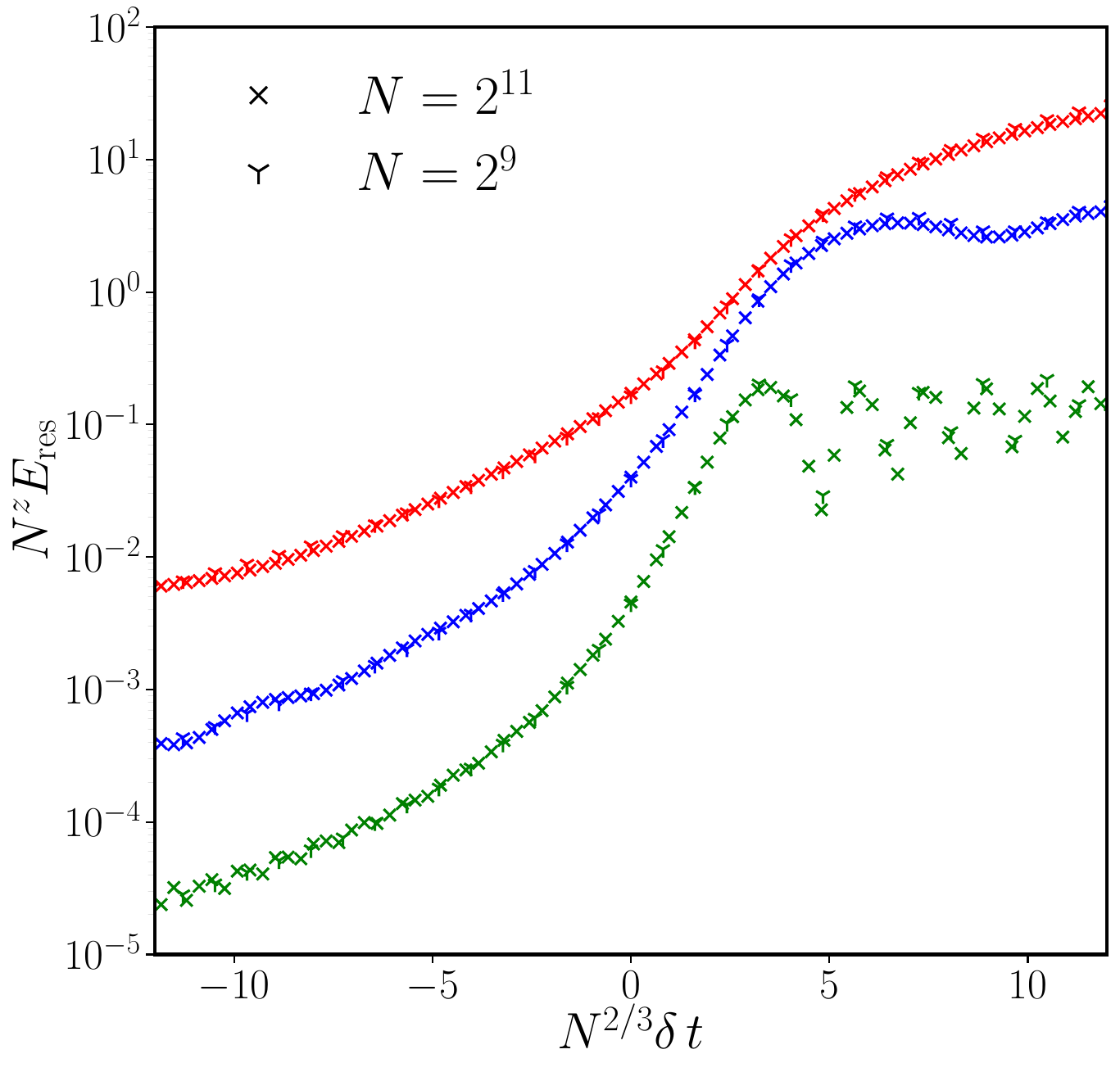}}
\caption{\textbf{Defect formation in the $\alpha=0$ Ising model during a slow quench:} Panel (a): Residual energy as a function of the drive rate $\delta$ for different values of the final gap for a slow dynamics terminating exactly at the critical point. Panel (b): Residual energy after a full ramp across the quantum critical point for two different system sizes and three different values of the universal scaling variable $\Lambda = N\delta = 15, 3.75, 0.94$ from top to bottom. }
\label{Fig4}
\end{figure}

\summary{In locally-interacting systems the Kibble-Zurek mechanism is supported by finite-size scaling and perturbative arguments, which however break down in presence of long-range interactions.}

\subsubsection{Quasi-static dynamics for $\alpha=0$}
The mosaic can be easily recomposed by the study of the slow drive dynamics within the linear spin-wave theory in Eq.\,\eqref{sw_h}. This strategy coincides with the one employed in Sec.\,\ref{sec_metastability}, but with two important differences: first, since we are considering a quasi-static drive, the dynamical evolution remain close to the instantaneous equilibrium state and the only relevant source of deviations from adiabaticity originates from the lowest energy mode. Therefore, we can safely limit ourselves to the case $\alpha=0$, where just a single spin-wave exist. Secondly, and more importantly, we are going to consider a time dependent magnetic field of the form $h(t)=h_{\rm cr}+\delta\,t$ for $t\in [-h_{\rm cr}/\delta, h_{\rm cr}/\delta]$, so that the dynamics initiates in the ferromagnetic state before crossing the critical point. A full treatment of the ferromagnetic state dynamics shall also include the motion of the classical magnetization, whose coupling with the quantum modes is suppressed by a factor $1/N$ in the thermodynamic limit. In the following, we are going to discard the contribution of the classical mode to the dynamics, since a classical variable in a bounded (singular) potential generates a correction which scales at most as $\sim \delta^{2}$\,\cite{landau1982mechanics}, and is, therefore, negligible with respect to the contribution of the quantum mode\,\cite{landau1965quantum}. 

Within the aforementioned assumptions, 
\mpar{Spin-wave evolution}
the quasi-static dynamics of the $\alpha=0$ Ising model reduces to the evolution of a single spin-wave. As for the sudden quench case, see Sec.\,\ref{sec_metastability}, the dynamics initialized in the ground state remains in a squeezed state at all times\,\cite{lewis1967classical, lewis1968class,lewis1969exact}. Then, the dynamics generates any two particle states as follows from Eq.\,\eqref{squezzing_operator}. We focus on a cyclic transformation where the system is initially in the ground state of the equilibrium Hamiltonian, thus, it is convenient to rewrite the single spin-wave state as
\begin{align}
\label{dyn_eigen}
\psi_{0}(x,t)=\left(\frac{1}{2\pi\xi^{2}(t)}\right)^{\frac{1}{4}}e^{-W(t)\frac{x^{2}}{2}}e^{-i\frac{\varphi(t)}{2}}.
\end{align}
with the effective time dependent frequency 
$W(t)=-i\frac{\dot{\xi}(t)}{\xi(t)}+\frac{1}{2\xi^{2}(t)}$,
and the ininfluential phase $\varphi(t)=\int^{t}\frac{dt'}{2\xi^{2}(t')}$. Thus, even in the linear ramp case the entire dynamics is described by the differential Eq.\,\eqref{ermakov_eq}.

In order to determine the excitation density and the ground state fidelity with respect to the instantaneous equilibrium solution of the problem, we define the adiabatic basis $\psi_{n}^{\rm ad}(x,t)$, which is obtained taking the equilibrium spin-wave eigenstates and replacing the constant frequency with the time-dependent one\,\cite{dabrowski2016time}. Accordingly, one can expand the exact time-dependent state in terms of the adiabatic basis $\psi(x,t)=\sum c_{n}(t)\psi^{\rm ad}_{n}(x,t)$, leading to the following results for the excitation density
\begin{align}
\label{exc_den}
N_{\rm exc}(t)=\langle \hat{n}\rangle=\sum_{n\in 2\mathbb{N}}n|c_{n}|^{2}=\frac{\xi(t)^{2}}{2\omega(t)}\left[\left(\frac{1}{2\xi(t)^{2}}-\omega(t)\right)^{2}+\left(\frac{\dot{\xi}(t)}{\xi(t)}\right)^{2}\right],
\end{align} 
and the adiabatic ground-state fidelity
\begin{align}
\label{exp_fidelity}
f(t)=|c_{0}|^{2}=\frac{1}{\xi(t)}\sqrt{\frac{2\omega(t)}{\left(\frac{1}{2\xi(t)^{2}}+\omega(t)\right)^{2}+\left(\frac{\dot{\xi}(t)}{\xi(t)}\right)^{2}}}.
\end{align}
Interestingly, one can relate the former expressions to the squeezing paramenter in Eq.\,\eqref{squeezing_parameter} by the simple relation $\tanh(r)=\sqrt{N_{\rm exc}(t) f(t)^{2}}$.

An 
\mpar{Linear ramp}
analytic solution can be found for a linear ramp across the quantum critical point with the resonant spin-wave having the dynamical frequency 
\begin{align}
\label{freq_scal}
\omega(t)^{2}=4 h(t)(h(t)-2J_{0})\approx 8 \delta |t|
\end{align}
where the last expression on the r.h.s. has been obtained substituting $h(t)=h_{\rm cr}-\delta t$ and expanding for small $\delta t$. The linear scaling of $\omega(t)^{2}$ is the consequence of the gap scaling $z\nu=1/2$ of the equilibrium problem, see Eq.\,\eqref{equilibrium_exc_spectra}. Eq.\,\eqref{freq_scal} represents the perfect crossing of the quantum critical point, since the instantaneous spin-wave frequency perfectly vanishes at $t=0$. In order to effectively incorporate finite size effects, we shall introduce a small deviation from 
perfect degeneracy and rewrite Eq.\,\eqref{freq_scal} as
\begin{align}
\label{freq_scal}
\omega(t)^{2} = 8 \delta |t| + \Delta_{N}^{2},
\end{align}
where $\Delta_{N}$ is the minimal gap of the finite size system. Obviously, $\lim_{N\to\infty}\Delta_{N}\to 0$ and the system attains perfect criticality in the thermodynamic limit. Moreover, due to universality, the minimal gap exhibits power-law scaling of the form  $\Delta_{N}^{2}\approx N^{-1/\nu_{*}}$ with $\nu_{*}=3/2$ as predicted by finite size scaling theory\,\cite{botet1982size} and confirmed by exact studies on the fully-connected quantum Ising model and related flat interacting models\,\cite{dusuel2005continuous,dusuel2005finite,vidal2006finite,ribeiro2008exact}. 

The model in Eq.\,\eqref{freq_scal} describes a cyclic transformation of the single Hamiltonian mode and, in the limit $\delta\to 0$, it can be used to describe a quasi-static cycle in quantum systems with infinitely degenerate spectrum. According to the behaviour of the fidelity and excitation density in the quasi-static limit $\delta\to 0$ the system presents three stages
\begin{enumerate}
\item Perturbative regime ($N<\infty$).
\item Kibble-Zurek regime ($N\to\infty$ and $t\in [-h_{\rm cr}/\delta,0]$).
\item Non-adiabatic regime ($N\to\infty$ and $t\in [-h_{\rm cr}/\delta, h_{\rm cr}/\delta]$).
\end{enumerate}
Regime (1) occurs for a finite minimal gam $\Delta_{N}> 0$: there adiabatic perturbation is applicable and the dynamics remains adiabatic, i.e. $N_{\rm exc}\propto \delta^{2}$\,\cite{degrandi2009adiabatic}. Regime (2) is realised for a thermodynamic system ($\Delta_{N}\to 0$) whose dynamics terminates exactly at the quantum critical point $t=\Delta_{\infty}=0$, where non-analytic corrections of the form $\delta^{\theta}$ appear in the residual energy. As we are gonna see in the following this regime is properly described by the Kibble-Zurek argument. An actual crossing of the quantum critical point only occurs in regime (3) and the system enters the non-adiabatic regime, where the residual energy and the fidelity acquire dynamical correction which do not depend on the drive rate.

The latter result can be easily shown rephrasing Eq.\,\eqref{ermakov_eq} in a rate independent form via the transformations 
\begin{align}
\label{dim_trans}
t = \delta^{-\frac{1}{3}}\tilde{t},\quad \xi= \delta^{-\frac{1}{6}}\tilde{\xi}
\end{align}
which reduce Eq.\,\eqref{ermakov_eq} to the $\delta=1$ case. The expressions in Eqs.\,\eqref{exc_den} and\,\eqref{exp_fidelity} are invariant under the transformations in Eq.\,\eqref{dim_trans} in such a way that the fidelity and excitation density at real times can be obtained by $\tilde{\xi}_{\Delta}(\tilde{t})=\lim_{\delta\to 1}\xi_{\tilde{\Delta}}(t)$. The subscript $\Delta$ has been introduced to explicit the dependence on $\Delta$  of  the solution of Eq.\eqref{ermakov_eq}  In the new variables, the only dependence of the dynamics on the rate $\delta$ remains in $\tilde{\Delta}_{N}=\Delta_{N}/\delta^{1/3}$. While the transformations\,\eqref{dim_trans} have been reported for the case of a linear quench, they can be easily generalized to any non-linear drive $\lambda(t)=(\delta|t|)^{\tau}$, obtaining results analogous to the one described in the present section\,\cite{defenu2021quantum}. Thus, the invariance of Eq.\eqref{ermakov_eq} with respect to the rescaling in Eq.\,\eqref{dim_trans} is enough to demonstrate that the dynamical evolution of the system only depends on the combined variable $\Lambda=\delta N= \tilde{\Delta}_{N}^{-3}$ as first evidenced by the numerical study in Ref.\,\cite{acevedo2014new}.

\summary{
 For a quasi-static drive terminating at the critical point Kibble-Zurek scaling is observed, while for dynamical protocols crossing the critical point the amount of defects is independent of the quench rate.
}

\subsubsection{Adiabaticity breaking}
However, in order to provide estimates for the defect density and fidelity in the quasi-static limit one has to solve Eq.\,\eqref{ermakov_eq} exactly. In the following we are going to drop all the $\sim$ superscripts over the rescaled variables, in order to ease the notation. The crucial condition of adiabatic dynamics is for the system to start in the ground-state at the beginning of the  dynamics, i.e. $\displaystyle{\lim_{t \to -\infty}}\psi(t)=\psi_{0}^{\rm ad}(t)$, leading to the boundary conditions
\begin{align}
\label{bound_cond}
\lim_{t\to-\infty}\xi(t)^{2}=\frac{1}{2\omega(t)}; \quad \lim_{t\to-\infty}\dot{\xi}(t)^{2}=0.
\end{align}
First, 
\mpar{Regime (2)}
it is instructive to consider the case of the quasi-static dynamics terminating at the critical point, i.e. $t\in[-h_{\rm cr}/\delta,0]$. As anticipated, the solution only depends on the combination $\Lambda=N\delta$ and, then, the corrections observed in the dynamics dramatically depend on what limit is taken first, see~\ref{KZM_app}. If one considers the quasi-static limit ($\delta\to 0$) at finite size, the rescaled gap $\Delta_{N}$ diverges and adiabatic corrections arise in all dynamical quantities
\begin{align}
\label{ad_corr}
\lim_{\delta \to 0}N_{\rm exc}(t_{0})= o\left(\delta^{2}\right);\qquad \lim_{\delta \to 0}f(t_{0})=1-o\left(\delta^{2}\right).
\end{align}

More interestingly, if the thermodynamic limit is taken first, the rescaled instantaneous gap vanishes $\lim_{N\to \infty}\Delta_{N}=0$. However, this is not the case for the effective gap $1/\xi(t)^{2}$ nor for its derivative, which attain the finite values
\begin{align}
\label{xi_0}
\lim_{t\to 0^{-}}&\xi^{2}(0)=\frac{\Gamma(p)\Gamma(p+1)}{2\pi p^{2p}},\\
\label{xi_dot_0}
\lim_{t\to 0^{-}}&2\dot{ \xi}(0)\xi(0)=\frac{1}{\sqrt{3}}\ ,
\end{align}
where $p=1/3$.
The finiteness of the results in Eqs.\,\eqref{xi_0} and\,\eqref{xi_dot_0} corresponds to a vanishing fidelity in Eq.\,\eqref{exp_fidelity}. Consequently, the defect density diverges, see Eq.\,\eqref{exc_den}, but the excess energy remains finite
 \begin{align}
 \label{heat}
 \lim_{N\to\infty, \delta\to0}E_{\rm res}(0)\simeq \lim_{t \to 0}\omega(t)N_{\rm exc}(t)\propto\delta^{\frac{1}{3}}.
\end{align}
Eq.\,\eqref{heat} defines regime (2) and is consistent with the outcome of the Kibble-Zurek mechanis\,\cite{degrandi2009adiabatic,dziarmaga2010dynamics} as well as with the numerical result in Ref.\,\cite{hwang2015quantum}. 
\begin{figure*}[t!]
\centering
\subfloat[Half ramp]{\label{Fig_SecIII_5a}\includegraphics[width=.47\textwidth]{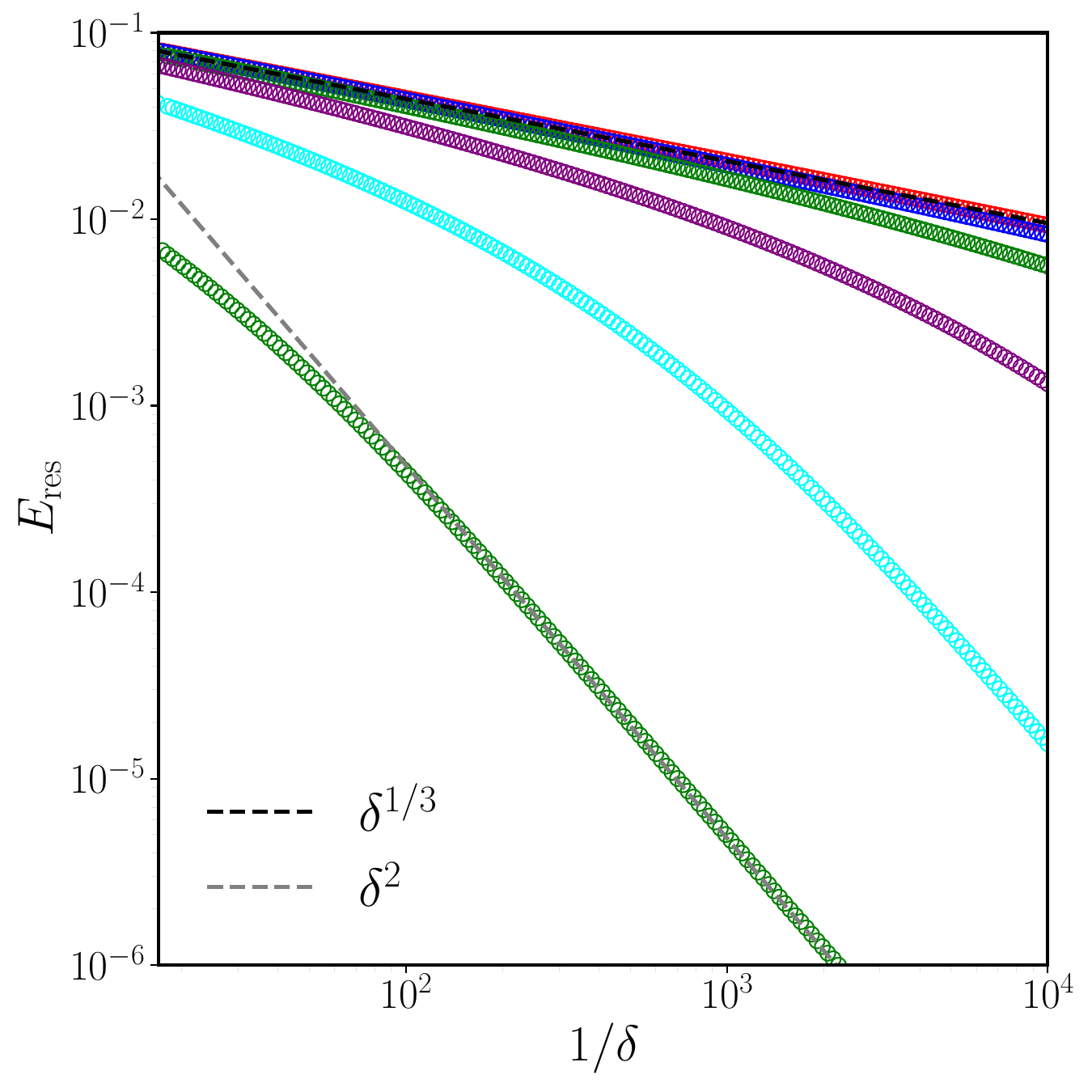}}\,\,
\subfloat[Full ramp]{\label{Fig_SecIII_5b}\includegraphics[width=.49\textwidth]{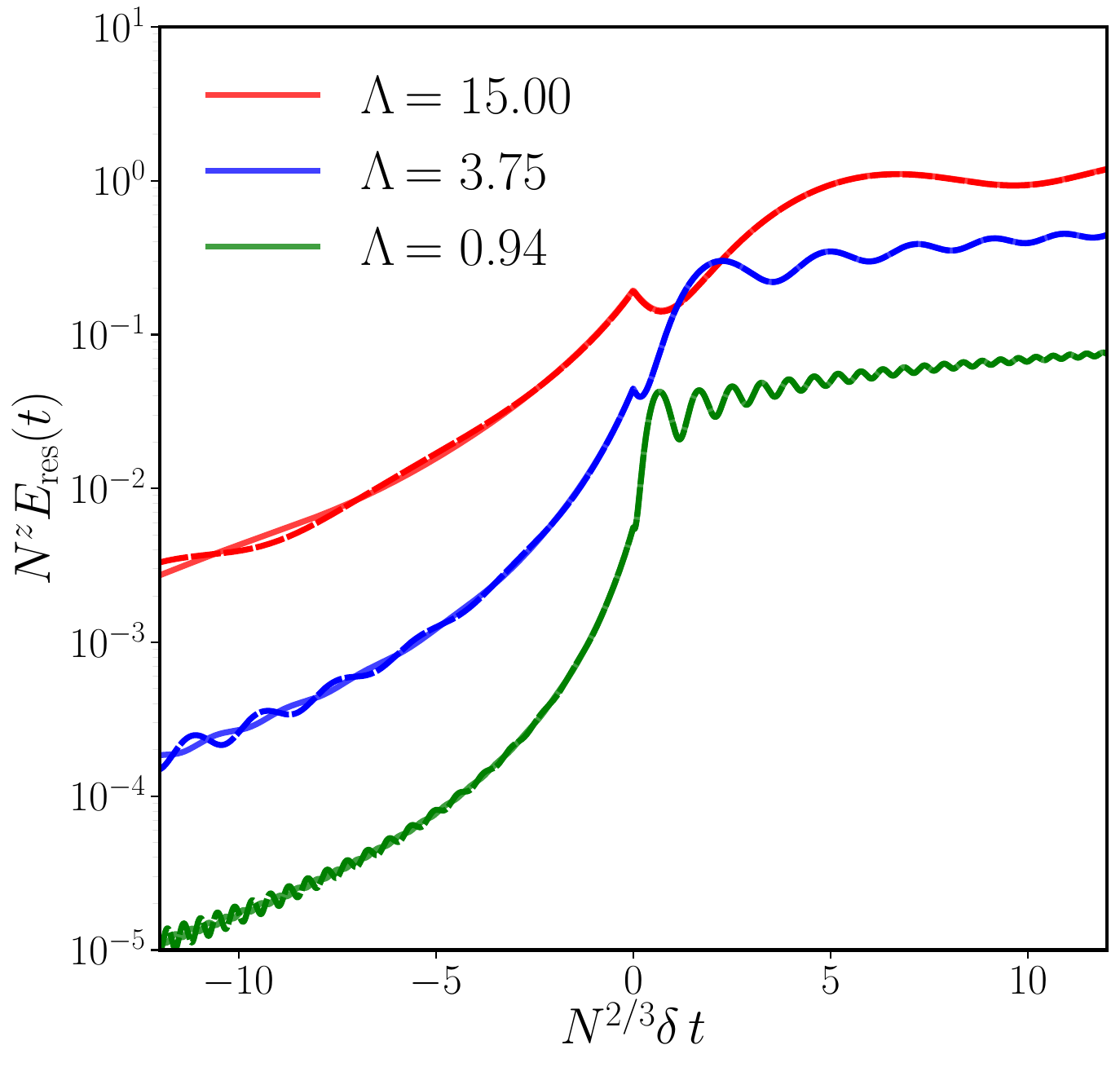}}
\caption{\label{Fig_SecIII_5} 
\textbf{Kibble-Zurek mechanism in the fully-connected model.} Panel \textbf{(a)} shows the residual energy as a function of the ramp speed in the case of a half ramp $t\in[-h_{\rm cr}/\delta,0]$ for $\Lambda=\{ 10^{9}, 3\cdot10^{7}, 10^{6}, 3\cdot10^{4},  10^{3}, 3\cdot10, 1 \}$ from top to bottom. The crossover between the Kibble-Zurek scaling (black dashed line) at large $\Lambda$ and the analytic scaling (gray dashed line) at small $\Lambda$ is evident. Panel\,\textbf{(b)} displays the residual energy after a quasi-static drive obtained by spin-wave theory. The result is obtained within regime (3) and perfectly reproduces the slow-drive universality numerically found in  Ref.\,\cite{acevedo2014new}. Each colour represents a different value of $\Lambda=N\delta$ with $N=2^{9}$ and $N=2^{11}$ (dashed and solid lines in panel b), i.e. the same values displayed for the exact numerical study in Fig.\,\ref{Fig4b}. The curves at different sizes perfectly collapse when drawn as a function of the scaling variables. Moreover, the agreement between the spin-wave theory and the numerical study for the different values of $\Lambda$ is rather remarkable.}
\end{figure*}

Regime (3) 
\mpar{Regime (3)}
is obtained considering directly the thermodynamic limit case $\Delta_{N}=0$ and taking the dynamics in $t=h_{\rm cr}/\delta\approx\infty$ limit, which yields the $\delta$-independent results 
\begin{align}
\label{asy_exc_den}
\lim_{t\to\infty} N_{\rm exc}(t)&=\frac{1}{3}\\
\label{asy_fid}
\lim_{t\to\infty} f(t)&=\frac{\sqrt{3}}{2} \ ,
\end{align}
which characterise the non-adiabatic dynamics as they remain finite in the $\delta\to 0$ limit. The analytical results in Eqs.\,\eqref{asy_exc_den} and\,\eqref{asy_fid} are universal in the traditional of Kibble-Zurek mechanis result. So, they faithfully reproduce the slow drive limit $\delta\to 0$ of any dynamical protocol which crosses the critical point. The universality phenomenon is analysed in details in Ref.\,\cite{defenu2018dynamical}. It can be numerically verified that the analytic solution in~\ref{SM3} accurately describes any drive $\omega'(t_{*})$ such that $|\omega'(\hat{t})-\omega(\hat{t})|^{2}\ll 1$, where $\omega(t)$ is given in Eq.\,\eqref{freq_scal}\,\cite{defenu2021quantum}. It is worth noting that the non-adiabatic regime described by Eq.\,\eqref{asy_fid} and Eq.\,\eqref{asy_exc_den} is profoundly different from the one described in Ref.\,\cite{polkovnikov2008breakdown} for low-dimensional systems. There, the spin-wave description is applied to the case $\alpha\to\infty$. Then, differently from our case one has to integrate over the continuous spin-wave spectrum and non-adiabaticity may arise also for a quench to the critical point due to the flat density of states of 2d systems.

In summary, dynamical corrections to a quasi-static drive  solely depend on the universal variable  $\Lambda=N\delta$. Indeed, the instantaneous minimal gap of a finite system scales as $\Delta_{N}\propto \Lambda^{-1/3}$, it follows that the thermodynamic limit ($N\to\infty$) and the adiabatic one ($\delta\to 0$) do not commute. Accordingly, the observable expectations are universal when displayed as a function of the universal variable $\Lambda$, see Fig.\,\eqref{Fig_SecIII_5b}. This is in perfect agreement with the numerical findings of Ref.\,\cite{acevedo2014new, defenu2018dynamical}.

\summary{Non-adiabatic corrections depend only on the product of the system size and the ramp velocity $\delta$. This leads to a novel form of universality, where no Kibble-Zurek scaling is observed in the thermodynamic limit.}

\subsubsection{Full counting statistics of defects}

Recently, 
\mpar{Defect distribution}
interest has raised around the universality of the higher cumulants of the defect statistics following a quasi-static ramp. In general, the process of defect formation in finite local systems has been argued to follow a Binomial distribution\,\cite{ruiz2020full}, making the process of defect formation across a conventional quantum phase transition akin to the classical process of a coin toss\,\cite{vishveshwara2020defect}. Approaching the thermodynamic limit the probability to 
generate $n$ defects becomes normal and reads
\begin{align}
\label{tfim_full_count}
P_{\rm local}(n)\approx\frac{1}{\sqrt{2\pi (1-p) \langle n\rangle}}\exp\left(\frac{(n-\langle n\rangle)^{2}}{2(1-p) \langle n\rangle}\right)
\ ,
\end{align}
where the average number of defects follows Kibble-Zurek scaling $\langle n\rangle \propto \delta^\frac{d\nu}{1+z\nu}$ and $p$ is the probability for the formation of a single defect. The above theory can be exactly verified in the nearest neighbour transverse field Ising model, whose full counting statistics can be calculated exactly. While for a finite quench rate $\delta$ all moments of the distributions remain finite in the slow drive limit one recovers Eq.\,\eqref{tfim_full_count} with $(1-p)=3/\pi^{2}$ and $\langle n\rangle=\frac{N}{2\pi}\sqrt{\frac{\delta}{2J_{0}}}$. These findings have been also demonstrated on different quantum computing platforms\,\cite{bando2020probing,cui2020kibble}.

The phenomenology of local systems is certainly rich, but does not present any peculiar features due to quantum fluctuations. Indeed, the same theoretical framework can be applied to describe the statistics of the defects generated across a classical and a quantum phase transition\,\cite{ruiz2020full}. As we have already argued in previous section, long-range interaction radically alter this picture as they suppress the defect contribution arising from semi-classical critical dynamics and promote the single quantum model as the leading source of non-adiabatic corrections. Interestingly, using the same methodology employed to derive Eq.\,\eqref{asy_exc_den}, one can obtain the full counting defect statistics. Indeed, the probability to generate $n$ defects in our problem is just given by the $|c_{n}|^{2}$ coefficient in Eq.\,\eqref{exc_den}\,\cite{gherardini2023universal}.
\mpar{Long-range result}
As shown in the appendix, see Eq.\,\eqref{cn_explicit}, the probability to generate $n$ defects reads
\begin{align}
\label{neg_binomial}
P_{\rm LR}(n)\approx {{n+k-1}\choose{n} } \mathrm{sech}(r)^{2k}\tanh(r)^{n},
\end{align}
which is a negative binomial distribution of $k$ successes and $n$ failures. The parameter $\tanh(r)$ defines the squeezing parameter, see Eq.\,\eqref{squeezing_parameter} of the single spin-wave of the system. In the present case, $n$ is the number of defects generated and $k=1/2$.

In light of Eq.\,\eqref{neg_binomial} the parallel between the local and long-range full counting statistics becomes rather striking. The negative binomial distribution described the probability to obtain $n$ failures before a given (non-deterministic) number $k$ of successes occurs. So, the probability for a defect (actually a pair since $n\in 2\mathbb{Z}$) to arise in the quantum long-range system coincides with the probability to observe $n$ failures before the $k$-th success. However, $k=1/2$ in the present problem, see the explicit derivation of Eq.\,\eqref{neg_binomial}, and the equivalence to the classical Binomial process is lost. Negative binomials of fractional $k$ are often dubbed Polya distributions and they do not have any equivalent in classical processes but they naturally emerge in the defect formation of quantum long-range systems due to the quantum nature of the problem. This has to be contrasted with the case of local critical theories where defect formation is a purely classical process\,\cite{vishveshwara2020defect}. The implications of these findings to the quantum thermodynamics of the systems are discussed in Ref.\,\cite{deffner2008nonequilibrium,gherardini2023universal}.

\summary{Long-range interactions suppress the creation of defect, leading to
 a negative binomial defects distribution, without a classical counterpart. 
}

\subsection{Dynamical quantum phase transitions - Loschmidt echo}
\label{sec_34}

Up to this point, our discussion focused on the most traditional examples of dynamical critical phenomena, but, recently, experimental advancements in quantum simulations with cold atoms\,\cite{gring2012relaxation, langen2013local, hild2014far, bordia2016coupling, bordia2017periodically, bordia2017probing,muniz2020exploring} and trapped ions\,\cite{monroe2021programmable} raised the interest on novel form of dynamical criticality\,\cite{polkovnikov2011colloquium}. This is the case of dynamical phase transitions. 
On one side, the name referred to the study of the out-of-equilibrium behavior of order parameters\,\cite{yuzbashyan2006relaxation, barmettler2009relaxation, eckstein2009thermalization, sciolla2010quantum, mitra2012time, marino2022dynamical}, which we refer to \emph{dynamical phase transition (DPT) in the order parameter} and that will be discussed in details in Section \ref{sec_412}. On the other, a novel form of dynamical criticality was discussed, where \emph{nonanalytic cusps in the Loschmidt echo} rate function appear after a quantum quench\,\cite{heyl2013dynamical, heyl2014dynamical, heyl2018dynamical}. We refer to the latter here as \emph{dynamical quantum phase transitions (DQPT) in the Loschmidt echo}.

During 
\mpar{Loschmidt amplitude and echo}
a quench, the system initially prepared in an initial state $\ket{\Psi_{0}}$ is evolved through a time independent final Hamiltonian $H$, i.e. $\ket{\Psi(t)}=\exp(-\ir Ht) \ket{\Psi_{0}}$. The Loschmidt amplitude describes the amplitude of the system returning in its initial state at time $t$ and reads
\begin{align}
\label{lo_amplitude}
\mathcal{G}(t)=\langle\Psi_{0}|\Psi(t)\rangle=\bra{\Psi_{0}}\exp(-\ir Ht) \ket{\Psi_{0}}
 \end{align} 
 whose expression closely resembles the classical finite-temperature partition function $Z(\beta)=\mathrm{Tr}\exp(-\beta H)$. The Loschmidt echo is simply obtained by squaring the amplitude in Eq.\,\eqref{lo_amplitude} yielding
 \begin{align}
\label{lo_amplitude}
\mathcal{L}(t)=|\mathcal{G}(t)|^{2}.
 \end{align} 
 The Loschmidt amplitude and Loschmidt echo play central roles in the theory of DQPTs and appear in various contexts in quantum many-body theory\,\cite{peres1984stability,gorin2006dynamics,schwinger1951gauge,talkner2007fluctuation,campisi2011colloquium,palmai2015edge}. They exhibit a functional dependence on the system size $N$, and in the limit of large $N$, they can be described by rate functions that capture their scaling behavior\,\cite{heyl2013dynamical,gambassi2012large}. DQPTs are defined as nonanalytic behaviors of the Loschmidt amplitude or Loschmidt echo as a function of time. They can be considered as phase transitions in time, analogous to equilibrium phase transitions associated with nonanalytic structures of the free energy\,\cite{pollmann2010dynamics,heyl2013dynamical}. A DQPT is characterized by a sudden qualitative change in the dynamics of the system, typically accompanied by a kink or nonanalyticity in the rate function of the Loschmidt amplitude or Loschmidt echo. This nonanalytic behavior can vary depending on the system and dimensionality, including power-law singularities, logarithmic singularities, and other forms\,\cite{heyl2013dynamical, vajna2015topological, heyl2015scaling}.

 Theoretical evidences of DQPTs in the Loschmidt echo return rate was found in numerous quantum systems\,\cite{heyl2013dynamical, heyl2014dynamical, halimeh2017dynamical, vajna2014disentangling, vajna2015topological, schmitt2015dynamical, campbell2016criticality, weidinger2017dynamical, lang2018dynamical, abdi2019dynamical} and connected with the behaviour of different local observables\,\cite{halimeh2021local}, including different definitions of the order parameter\,\cite{zunkovic2018dynamical,uhrich2020out}. Given the many successful experimental realizations of this kind of DQPTs, especially in trapped ion systems with long-range interactions\,\cite{jurcevic2017direct,zhang2017observationdpt}, it is not surprising that long-range interacting models were also a privileged tool for the theoretical characterization of DQPTs\,\cite{zunkovic2016dynamical,zunkovic2018dynamical,halimeh2017prethermalization, lang2018concurrence, homrighausen2017anomalous,halimeh2020quasiparticle, vajna2015topological,dutta2017probing,defenu2019dynamical,uhrich2020out}.

 \summary{Dynamical quantum phase transitions (DQPTs) in the Loschmidt echo are defined by non-analytic behavior of this quantity as a function of time. }
 
 \subsubsection{Spherical spin-wave theory}
In the following, we are going to show how the emergence of DQPTs in the long-range Ising model can be captured by the study of the harmonic Hamiltonian in Eq.\,\eqref{sw_h}. However, most DQPTs occur when a parameter of the Hamiltonian is quenched across an underlying equilibrium phase transition and the equilibrium spin-wave theory is not capable to capture both side of the transition within the same formalism. It is worth noting that there are cases where DQPTs can arise independently of conventional phase transitions\,\cite{vajna2014disentangling, canovi2014first, andraschko2014purification, schmitt2015dynamical,zunkovic2016dynamical}. 

\mpar{The spherical model}
In order to capture a sudden quench across the dynamical critical point we are going to consider the spherical model Hamiltonian
\begin{align}
\label{sph_model}
H=\frac{1}{2} \sum_k \hat{p}_k \hat{p}_{-k}+\frac{1}{2 } \sum_k \omega_k^2 \hat{x}_k \hat{x}_{-k}
\end{align}
where the $\hat{x}_k$  and $\hat{p}_k$  are canonically conjugate hermitian operators, such that $[\hat{x}_k, \hat{p}_{k'}]=i\delta_{k,k'}$ and the dispersion relation reads
\begin{align}\label{eq:disp}
	\omega_k = \sqrt{h(\mu + 2J_{0}f_{k}(\alpha))} .
\end{align}
The Hamiltonian in Eq.\,\eqref{sph_model} corresponds to the spin-wave Hamiltonian in Eq.\,\eqref{sw_h} once one introduces the annihilation/creation operators
\begin{align}
\hat{\beta}_{k} & =\sqrt{\frac{\omega_{k}}{2}}\left(\hat{x}_{k}+\frac{i}{\omega_{k}} \hat{p}_{k}\right) \\ \hat{\beta}_{k}^{\dagger},& =\sqrt{\frac{ \omega_{k}}{2}}\left(\hat{x}_{k}-\frac{i}{\omega_{k}} \hat{p}_{k}\right).
\end{align}
In fact, the spectrum in Eq.\,\eqref{eq:disp} roughly corresponds with the spin-wave spectrum  in Eq.\,\eqref{equilibrium_exc_spectra} with $\gamma=1$. However, while Eq.\,\eqref{equilibrium_exc_spectra} only depends on the magnetic field and the coupling, the spherical model's dispersion relation, i.e. Eq.\,\eqref{eq:disp}, contains an additional parameter $\mu$. In the spherical model, the parameter $\mu$ has to be calculated self-consistently by imposing a constraint on the spin-wave potential energy, namely
\begin{align}
\label{eq_constraint}
 \sum_{k} \frac{\left\langle\left(\beta^{\dagger}_{k}+\beta_{k}\right)^2\right\rangle}{2h \omega_{k}}= \frac N4 \ .
 \end{align}

The spherical constraint in Eq.\,\eqref{eq_constraint} may induce a quantum critical point in the quadratic model at a field value $h_{\rm cr}^{\rm sph}$ depending on the features of the function $f_{k}(\alpha)$. Below the critical field strength $h_{\rm cr}^{\rm sph}$ the constraint condition in Eq.\,\eqref{eq_constraint} causes the spin-wave to form a condensate state at $k=0$. In its classical version, the spherical model has been introduced to mimic the finite temperature free-energy of $O(n)$-symmetric spin systems in the $n \rightarrow \infty$ limit\,\cite{stanley1968spherical}. 

\summary{The spherical model corresponds to the low-energy spin-wave Hamiltonian\,\eqref{sw_h} equipped with an additional parameter $\mu$, determined by the self-consistent relation in Eq.\,\eqref{eq_constraint}. The model allows to extend the spin-wave description across the critical point. }

\subsubsection{Step approximation}
\label{step_app}
The DQPT occurs in the spherical model following a sudden quench. Thus, we have to consider the average in Eq.\,\eqref{eq_constraint} over a time dependent state $\ket{\Psi(t)} = \prod_k \ket{\psi_{k, 0}(t)}$, where $ \ket{\psi_{k, 0}(t)}$ is the single-spin wave state given by Eq.\,\eqref{Dyn_Eigen} with $n=0$. The explicit calculation leads to the dynamical constraint equation
\mpar{State evolution after a quench}
\begin{align}\label{eq:sph_t}
\frac{h}{N}\sum_k\frac{\xi_k^2(t)}{2 }=1/4,
\end{align}
where $\xi_k^2(t)$ is the solution of the Ermakov Eq.\,\eqref{ermakov_eq}. Similarly to Sec.\,\ref{sec_31}, we are going to consider a sudden quench of the ferromagnetic coupling $J_{0}^{\rm i}\to J_{0}^{\rm f}$, but now the final coupling value $J_{0}^{\rm f}$ shall drive the system across the quantum phase boundary. In this case, a solution of the differential Eq.\,\eqref{ermakov_eq} together with the dynamical constraint in Eq.\,\eqref{eq:sph_t} is rather complicated and, up to our knowledge, has not been attempted yet. On the other hand, a convenient simplifying assumption consists in assuming that, as the ferromagnetic coupling $J_{0}$ is quenched, the parameter $\mu$ also undergoes a discontinues jump between two constant values $\mu_0$ at $t<0$ and $\mu_f$ for $t\geq 0$. This procedure goes under the name of \emph{step approximation} and has been introduced in Refs.\,\cite{sotiriadis2010quantum,chandran2013equilibration}. In order for this approximation to be sensible, one should choose the final value $\mu_f$ in order to reproduce its expected long-time (equilibrated) value. As long as $\alpha>d$ the system can be safely assumed to equilibrate at long time as shown in Ref.\,\cite{chandran2013equilibration,syed2021dynamical},

Within the framework of the step-approximation, the frequency suddenly change from their initial value $\omega_{k,i}$ to a final value $\omega_{k,f}$, leading to the following sudden 
\mpar{Solution}
quench solution for the effective length of each spin-wave
\begin{align}\label{eq:xi}
\xi_k(t) = \sqrt{\frac{1 + \epsilon_k \sin^2(\omega_{k,f} t)}{2\omega_{k,i}}}
\end{align}
with the quench parameter $\epsilon_k = \left( \frac{\omega_{k,i}}{\omega_{k,f}} \right)^2 - 1$.
In the thermodynamic limit $N \rightarrow \infty$ the sum in Eq.\,\eqref{eq:sph_t} can be turned into an integral. Then, taking into account the explicit solution in Eq.\,\eqref{eq:xi} one obtains
\begin{align}
\label{td_const}
\int\! \frac{dk}{2\pi} \frac{h}{2 \omega_{k,i}} \left[
\frac{\epsilon_k}{2} (1 - \cos 2\omega_{k,f} t) \right] =0  .
\end{align}
This equation cannot be fulfilled at all times due to the oscillatory $\cos 2\omega_{k,f} t$ term. Yet, in the limit $t\to\infty$ the dephasing between the different modes washes away the time dependence in Eq.\,\eqref{td_const}, making the solution in Eq.\,\eqref{eq:xi} exact also for the constraint problem as long as the final value of $\mu$ is chosen in order to satisfy the following expression
\begin{align}\label{eq:eps}
\int \frac{dk}{2\pi} \frac{\epsilon_k}{\omega_{k,i}}= 0.
\end{align}
This implicit equation determines the long-time asymptotic value of $\mu_f$ through the $\mu$ dependence of $\omega_{k,f}$. The consistency of the equilibration assumption and, overall, of the step approximation can be verified by the inspection of the numerical solution of the exact problem, see Ref.\,\cite{syed2021dynamical}.

Eq.\,\eqref{eq:eps} can be used to determine the dynamical critical coupling $J_{0}^{\rm c, dyn}$ at which the dynamical excitation become gapless and the constraint parameter $\mu_f$ in Eq.\,\eqref{eq:eps} approaches its critical value $\mu_c$. Using these definitions, Eq.\,\eqref{eq:eps} can be rewritten as
\begin{align}
\label{eq:gcdyn}
\frac{1}{2} = \sqrt{h} \int \frac{dk}{2\pi} \frac{\sqrt{2\mu_0 + 2J_{0}^{i}f_{k}(\alpha)}}{2\mu_c + 2J_{0}^{\rm c, dyn}f_{k}(\alpha)},
\end{align}
where $\mu_{c}$ is the equilibrium critical value. The existence of a finite value $J_{0}^{\rm c, dyn}$ satisfying Eq.\,\eqref{eq:gcdyn} depends both on the parameters $h, \mu_{0}$ and on the value of $\sigma$. The dynamical phase diagram of the model is reported in Ref.\,\cite{syed2021dynamical}. In the present section, for the sake of simplicity, we are going to assume that $J_{0}^{i}$ lies above its equilibrium critical value $J_{0}^{i}>J_{0}^{\rm c}$, i.e. in the condensate phase, and consider the case $\alpha<d+2$.

Given the quadratic nature of the spherical model, the overlap function can be calculated analytically
\begin{align}
\begin{split}
\mathcal{G}(t) = &\prod_k \Biggl\{ (8\omega_{k,i})^{1/4} e^{-i\varphi_k(t)}  \biggl(
2\omega_{k,i}\xi_k(t) + \frac{1}{\xi_k(t)} -
\ir  2\dot{\xi}_k(t)
\biggr)^{-1/2} \Biggr\} \ ,
\end{split}
\end{align}
where the dyanmical phase $\varphi(t)$ is defined below Eq.\,\eqref{ermakov_eq}. 
The 
\mpar{The Loschmidt echo rate}
Loschmidt echo rate function is obtained by taking the logarithm of the squared overlap, yielding
\begin{align}
\label{rate_eq}
r(t) = -\lim_{N \rightarrow \infty} \frac{1}{N} \log \abs{G(t)}^2 = -\log 2 + \int  \frac{dk}{2\pi} \log\abs{X_k(t)} ,
\end{align}
where
\begin{align}
\label{X_expr}
X_k(t) =  \frac{1}{\sqrt{8\omega_{k,i}}}\biggl(
2\omega_{k,i}\xi_k(t) + \frac{1}{\xi_k(t)} -
\ir  2\dot{\xi}_k(t)
\biggr)
\end{align}

As long as $\omega_{k,i}$ is gapped, the expression in Eq.\,\eqref{X_expr} remains smooth and no cusp appears at finite time for the rate function defined in Eq.\,\eqref{rate_eq}. The non-analytic cusps characterizing DQPTs will only appear for a sudden quench from the broken phase where $\Omega_{k,i}$ is gapless. This result demonstrates how one can observe and characterize DQPTs by just analyzing the quasi-particle spectrum. Therefore,  as already mentioned, we are going to consider a  sudden quench of the coupling $J_{0}^{\rm i}\to J_{0}^{\rm f}$, with the initial coupling within the ferromagnetic phase and the final one above the dynamical critical threshold $J_{0}^{\rm f, dyn}$. Since the dynamics is initiated in the broken phase, a complete treatment of the problem shall also include the evolution of the classical mode representing the condensate fraction of spin-waves as it was done in Ref.\,\cite{weidinger2017dynamical}. 
However, in the present description we have discarded this contribution as it is not necessary to observe the DQPTs.

\begin{figure}[t!]
\includegraphics[width=.99\textwidth]{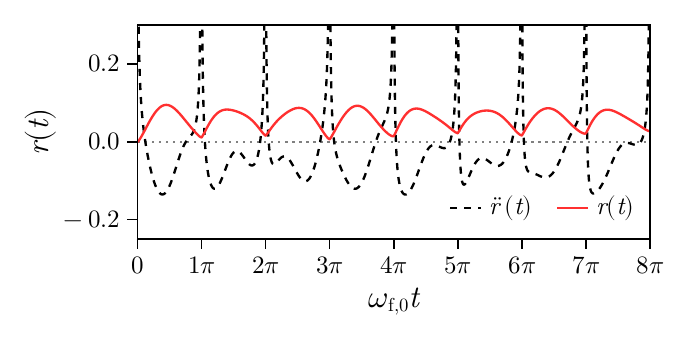} 
\caption{\textbf{DPT in the spherical model.} The Loschmidt echo rate function of the spin-wave theory (in presence of the spherical constraint) after a sudden quench of the ferromagnetic coupling $J_0^{i} = 2 J_0^{c}$ to $J_0^{f} = J_0^{c}/2$. The cusps in the rate function are clearly evident and the second derivative is divergent since the dynamics considered is for $\alpha=2.5$ and $d=1$, see the discussion in the text. Figure adapted from Ref.\,\cite{syed2021dynamical}. \label{Fig_SecIII_6}}
\end{figure}

The signature of DQPT in the Loschmidt echo dynamics is reported in Fig.\,\ref{Fig_SecIII_6}, the Loschmidt echo is shown for a quench from $J_0^{i} = 2 J_0^{c}$ to $J_0^{f} = J_0^{c}/2$ for different values of $\sigma$. The rate function clearly shows non-analyticities at the critical times:
\begin{align}
t^{*}_{m} = \frac{m \pi}{\omega_{k,\mathrm{f}}} \quad m \in \mathbb{N}
\end{align}
which appear due to logarithmic divergences in the integrand in Eq.\,\eqref{rate_eq}. Since the critical time scale is set by the post-quench gap, we do not expect to see nonanalytic cusps in the Loschmidt echo for a quench into the gapless phase, as previously mentioned.

Upon differentiating Eq. (\ref{rate_eq}) with respect to time $n$ times, we encounter terms that are proportional to $1/\omega_{k,i}^n$ when $t = t^*_m$. These terms diverge as $k^{-n (\alpha-d) / 2}$. However, since the integration over $k$ still has to be performed, the $n$-th derivative of the rate function only diverges if $n (\alpha-d) /2 > 1$, or equivalently, $n (\alpha-d) > 2$. Thus, the smaller $\alpha$ the larger the order of the derivative for which the cusps are expected. This analysis holds true throughout the entire region where $d<\alpha<d+2$. On the other hand, it does not apply when $\alpha \geq d+2$ since there is no gapless phase to initiate the calculation.
The discussion presented above offers further evidence that the emergence of nonanalytic cusps is not solely a feature of the step approximation. Instead, it is a consequence of the initial conditions, specifically starting in the gapless phase, as well as the specific form of the function $\xi_{k=0}(t)$ and its time derivatives, which remain unchanged in the exact calculation. Furthermore, the structure of these cusps remains unaltered even in the long-time limit when $\mu(t)$ reaches equilibrium and the step approximation becomes exact.

\summary{Loschmidt echo DQPTs in the spherical model appear for quenches from the broken phase, as results from the exact solution.}

\subsubsection{Strong long-range regime}

The aforementioned analysis cannot be extended to the regime $\alpha<d$ as the system does not equilibrate and the assumptions at the root of the step approximation outlined in Sec.\,\ref{step_app} fail. Moreover, due to the gapped nature of the spectrum the importance of the order-parameter motion of the system at $\alpha<d$ is more prominent and cannot be easily discarded. Several numerical simulations and analytical arguments have been used to show the existence of the DQPT in the Loschmidt echo also for the actual Ising Hamiltonian for different values of $\alpha$\,\cite{halimeh2017dynamical,stauber2017probing,homrighausen2017anomalous,lang2018concurrence}. In particular, extensive numerical studies have been devoted to investigate the connection between the DQPTs occurring in the Loschmidt echo and the DPTs defined via the dynamical scaling of the order parameter\,\cite{zunkovic2018dynamical,halimeh2017prethermalization}, see also Section \ref{sec_412}. Also, the relation between the two different notions and the quasi-particle properties of the model have been largely investigated, but mostly close to the local limit\,\cite{halimeh2020quasiparticle,defenu2019dynamical,uhrich2020out}. In the next section, we are going to introduce the dynamical Holstein-Primakoff transformation, which represents the proper formalism to describe the motion of spin-wave coupled to the classical order parameter and their feedback effect. However, we are only going to use it to describe DPTs in the order parameter leaving aside further comments on singularities of the Loschmidt echo.

\summary{In the strong long-range regime, the leading effect in the Loschmidt echo DQPT comes from the dynamics of the order parameter, and it can be related to other forms of dynamical criticality.  }

\section{Dynamics in highly excited states}
\label{dyn_high_exc}

In this Section we will discuss the treatment of out-of-equilibrium dynamics involving arbitrarily high energy initial states, as in standard quantum quench protocols.

We will begin in Sec.~\ref{sec_classLim} by reviewing mean-field dynamical phenomena for $\alpha=0$.  In Sec.~\ref{sec_alpha0eq} we showed that the fully connected limit of quantum spin systems reduces to the physics of a single collective degree of freedom. In this Subsection, we will discuss how this statement applies to non-equilibrium dynamics as well.

Secondly, in Sec.~\ref{sec_42}, we will discuss how finite-range interactions with $\alpha>0$ affect mean-field dynamical phenomena. Long-range interacting system can be formally viewed as a perturbation of the mean-field limit, as reviewed in Sec.~\ref{low_en_theory}. The perturbation term couples the collective degree of freedom to many spin-fluctuation modes with various wavelengths, resulting in a genuine many-body problem, which can be addressed via the non-equilibrium spin-wave formalism developed in Refs.~\cite{lerose2018chaotic,lerose2019impact}. The coupling strength to mode $\mathbf k$ strongly depends on the interaction range governed by the exponent $\alpha$, as encoded by the function $f_{\mathbf k} (\alpha)$. 
As a result of this tunable decoupling, long-range interactions give rise to non-equilibrium many-body phenomena distinct from the generically expected thermalization: Collective spin ordering is remarkably resilient out of equilibrium, generating long pre-thermal stages of dynamics characterized by long-lived oscillating collective spin polarization. This behaviour was observed in numerical simulations performed with a range of techniques~\cite{zunkovic2018dynamical, buyskikh2016entanglement, schachenmayer2013entanglement} and theoretically understood via the aforementioned approach~\cite{lerose2018chaotic, lerose2019impact}. This analysis shows that the duration of the prethermal stage increases as $\alpha$ is decreased, and diverges with the system size when $\alpha< d$~\cite{mori2018thermalization,lerose2020origin}.

\subsection{Quench dynamics of fully-connected spin systems ($\alpha=0$) }
\label{sec_classLim}

This Section is devoted to the non-equilibrium dynamics of fully connected spin systems. We study the time-evolution starting from ground states  $\ket{\psi_0}$ of a pre-quench Hamiltonian $\hat H(h_0)$ evolving with a different post-quench Hamiltonian $\hat H(h_f)$. For the sake of definiteness, we will mostly consider the Ising model, Eq.~\eqref{eq_H} with $\gamma=1$, and quenches in the transverse field from $h_0$ to $h_f$. As described in Section \ref{sec_232}, ground states of long-range Hamiltonians as \eqref{eq_H} are generically close to spin-\emph{coherent states}.  
Their dynamical behavior is determined by a classical mean-field description emerging in the thermodynamic limit when initialized in fully polarized states, as described in Section \ref{sec_411}. The resulting dynamics of collective observables can give rise to new forms of dynamical criticality, such as dynamical phase transitions, discussed in Section \ref{sec_412}. The semiclassical framework also allows us to describe the growth of quantum fluctuations, which coincides with the flow of linearized shifts around classical trajectories and is thus related to the standard quantifiers of classical chaos, reviewed in Section \ref{sec_413}. When the quantum fluctuations become comparable to the typical length of the phase space, this description breaks down, defining an Ehrenfest time that diverges with $N$ for this class of systems. Remarkably, such semiclassical framework grasps crucial aspects of quantum dynamics, such as the dynamics of scrambling or entanglement as reviewed in Section \ref{sec_414} and \ref{sec_415} respectively.


\subsubsection{Mean-field classical limit}
\label{sec_411}

The dynamics of a system with unbroken full permutational symmetry take place in the totally-symmetric subspace (TSS) of the many-body Hilbert space simultaneously invariant under all permutations. Such dynamics is amenable to an exact representation in terms of few collective degrees of freedom, characterized by an effective Planck constant $\hbar_{\rm eff} \sim 1/N$ suppressed with system size \cite{sciolla2010quantum, sciolla2011dynamical}. We refer to \ref{app_semiCla} for a general discussion.

For systems of interacting quantum spins the limiting semiclassical description may be formulated more directly and intuitively in terms of states with maximal collective spin $S=Ns$ --- the so-called \textit{Dicke manifold}. As discussed in Sec.~\ref{sec_231}, the collective spin approaches a classical limit for large $N$.\footnote{For $s=1/2$, the TSS coincides with the Dicke manifold. For larger $s$ there are more permutationally invariant states with lower $S$, that is outside the Dicke manifold (${\rm dim TSS} \sim N^{2s}$ for large $N$). However, for Hamiltonians without spin self-interactions, one may always consider dynamics within the Dicke manifold.}
For the infinite-range XY Hamiltonian in Eq.~\eqref{eq_Halpha0} the classical limit $\hat H_{\alpha=0}/N \to \mathcal{H}_{\rm cl}$ is given by Eq.~\eqref{eq_Hcl} in Sec.~\ref{sec_231}, where we discussed  equilibrium properties.
In this Section, we will use the same approach to discuss \emph{out-of-equilibrium} properties. 
For definiteness, throughout this Section, we will set $\gamma=1$ (quantum Ising model).

The 
\mpar{Classical dynamics}
non-equilibrium evolution $\braket{ \hat {\vec{S}}(t)}/N$ generated by a sudden change (``quench'') of a Hamiltonian parameter is described by a classical trajectory $\vec{\mathcal{S}}(t)$ on the unit sphere governed by $\mathcal{H}_{\text{cl}}$, i.e.,
\beq
\dot{\vec{\mathcal{S}}} = \big\{ \vec{\mathcal{S}}, \mathcal{H}_{\text{cl}} \big\} \ ,
\eeq
with the canonical Poisson brackets $\{ \mathcal{S}^\mu,\mathcal{S}^\nu \} = \epsilon_{\mu\nu\rho} \mathcal{S}^\rho$, where $\epsilon_{\mu\nu\rho}$ is the totally antisymmetric Levi-Civita tensor. 
Evolution can be recast in terms of the spherical angles $\theta(t),\,\phi(t)$. 
In the case of the Hamiltonian \eqref{eq_Halpha0}, the non-linear precession of the collective spin is described by the classical equations of motion\footnote{For convenience, we rescale time by a factor $s$.}
\begin{align}
\label{eq_classicalMotion}
	\begin{dcases}
		\dot \theta =  2J_0 \sin \theta \cos \phi \sin \phi \, ,\\
		\dot \phi = - h + 2J_0 \cos\theta \cos^2 \phi  \, .
	\end{dcases}
\end{align}
As the Hamiltonian governs a single degree of freedom, the classical limit is trivially integrable and characterized by regular periodic trajectories in phase space. 
Such behaviour corresponds to persistent spin oscillations after a quench, whose period depends on the initial state.
For $|h|<2J_0$, the phase-space also features a separatrix with a diverging classical period, terminating at the saddle point $\theta=0$ and characterized by an exponential instability rate 
\begin{equation}
	\label{lh}
	\lambda = \sqrt{h(2J_0-h)} \ 
\end{equation}
(i.e. the eigenvalue of the stability matrix at the saddle point).

While 
\mpar{Chaotic dynamics in other all-to-all spin models}
such fully-connected spin models generically exhibit periodic orbits, semiclassical chaotic behavior can occur in a number of relevant situations. A standard example comes from introducing time-dependent driving, thus breaking energy conservation: The quantum kicked top~\cite{haake1987classical, haake2010quantum}, corresponding to a step-wise driving protocol applied to the model above, gives a paradigmatic regular-to-chaotic crossover as a function of the driving parameters. Another source of chaoticity comes from coupling the spins to other degrees of freedom, such as a cavity mode, which gives rise to the textbook Dicke model~\cite{dicke1954coherence, deaguiar1992chaos}. Finally, chaotic behaviour can arise from self-interactions of higher spins $s>1/2$, which are generally described by $n=2s>1$ collective degrees of freedom, e.g. Ref.\cite{mori2017classical}, or more general collective models, e.g. Ref.\cite{valencia2023crafting}: In the absence of additional symmetries, self-interactions will break classical integrability. These extended possibilities can be addressed with the method summarized in \ref{app_semiCla}. In the rest of this Section, we will refer to them when discussing the impact of chaos on the quantum dynamics of fully-connected systems. 

\summary{In the thermodynamic limit the quench dynamics of fully connected spin systems are described by classical trajectories of a single collective degree of freedom.
}

\subsubsection{Dynamical phase transitions - Order parameter}
\label{sec_412}

\begin{figure}[t]
\fontsize{12}{10}\selectfont
\centering
\includegraphics[width = 1 \columnwidth]{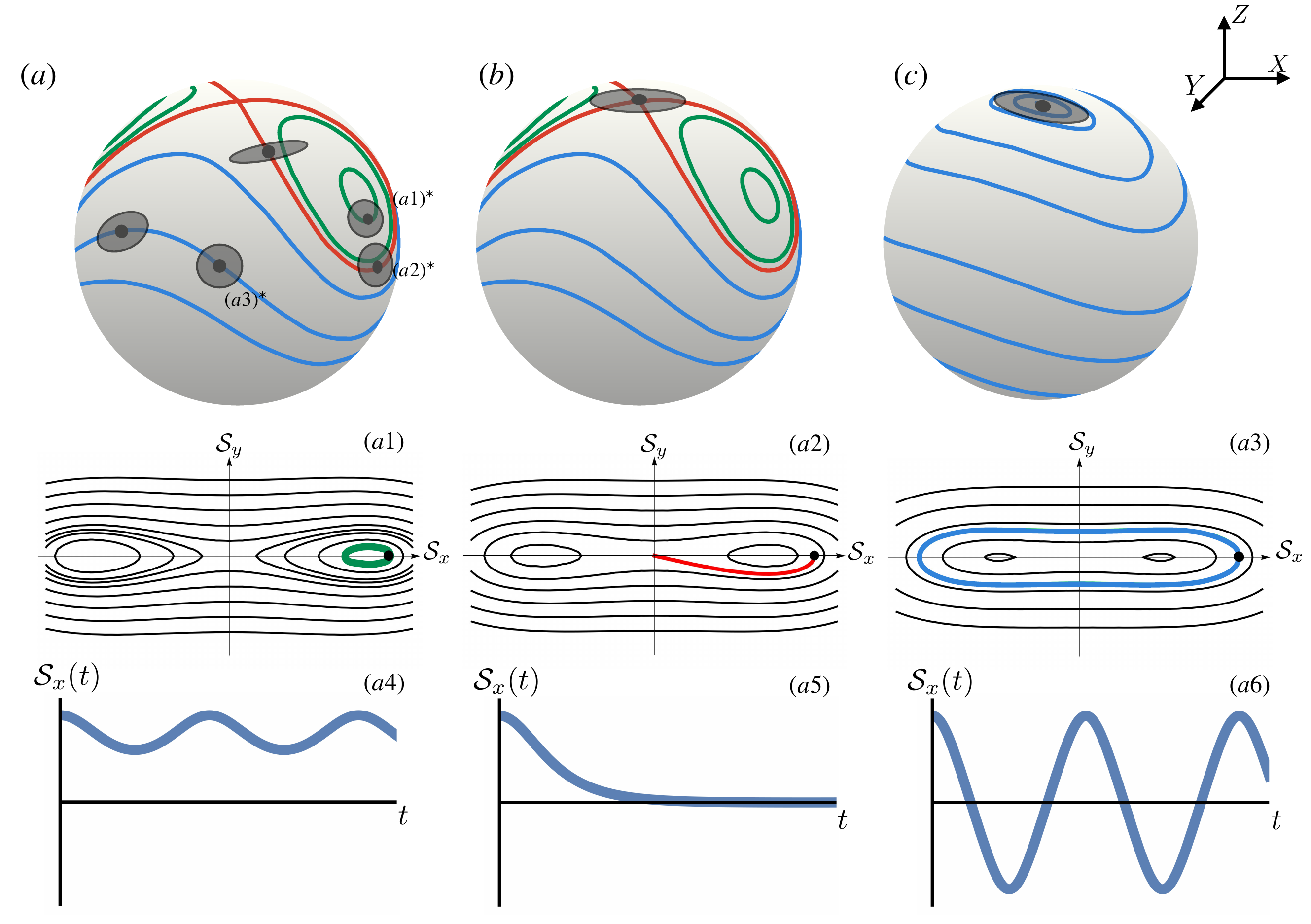}
\caption{Equilibrium configurations and possible instances of non-equilibrium dynamics in the fully-connected quantum Ising model.
(a-c) Pictorial representation on the Bloch sphere of the collective spin for the post-quench Hamiltonian. (a-b) For $h_f<|J|$, the energy possesses two minima characterized by non-vanishing, opposite magnetizations along $x$. (c) For $h_f>|J|$, the system is paramagnetic with a single equilibrium configuration in the direction of the field. Initial fully polarized states at $t=0$ are pictured as a point on the Bloch sphere, surrounded by a small grey circle representing their transverse quantum fluctuations. Labels (a1-3)* represent possible instances of such initial conditions. 
(a1-a3) Semiclassical phase portrait of the ferromagnetic post-quench Hamiltonian, where the initial states move along a nontrivial nonequilibrium trajectory, corresponding to the initial conditions (a1-3)* respectively.
(a4-a5) Associated dynamics of the classical magnetization. 
Labels (a) refer to initial ferromagnetic initial states $h_0<|J|$. Their time-evolution is characterized by ferromagnetic periodic (green) trajectories [see (a1) and (a4)] or paramagnetic (blue) ones [see (a3) and (a6)] with $\overline{S_x(t)}\neq 0$ and $\overline{S_x(t)}=0$, respectively. These are separated by the unstable (red) trajectory occurring at $h_{\rm dcr}$  [see (a2) and (a5)].
Labels (b-c) refer to an initial paramagnetic state $h_0=\infty$ evolved with two different Hamiltonians: (b) quench performed to a ferromagnetic Hamiltonian $h_f<|J|$, the initial state lies on the unstable trajectory, (c) quench performed to a different $h_f>|J|$ paramagnetic configuration. Images adapted from Ref.\cite{lerose2019impact, lerose2020origin}.}
\label{fig:LMG_trajectories}
\end{figure}

The 
\mpar{Magnetic ordering in the asymptotic state}
non-equilibrium evolution described above may or may not result in collective spin ordering at long times, i.e. finite order parameter. An abrupt change of dynamical ordering properties as a function of driving control parameters is referred to as a \textit{dynamical phase transition} (DPT)~\cite{eckstein2008nonthermal, eckstein2009thermalization,  sciolla2010quantum, gambassi2010quantum, schiro2010time, schiro2011quantum, sciolla2011dynamical, schiro2012linear, forster2013quantum, yuzbashyan2015quantum, zunkovic2016dynamical, homrighausen2017anomalous, marino2022dynamical}.  In particular, when a system is quenched from a symmetric state across the equilibrium critical point, dynamical scaling properties associated with aging or coarsening may appear~\cite{chandran2013equilibration, maraga2015aging, chiocchetta2016short, chiocchetta2017dynamical}. Conversely, when a system undergoes a sudden quench from a broken-symmetry state, the resulting out-of-equilibrium dynamics may display two different phases. One can define a non-equilibrium order parameter by time-averaging the corresponding equilibrium order parameter. This quantity may be vanishing or not depending on whether the symmetry is dynamically restored after the quench. The associated \emph{dynamical critical point} is believed to have a universal character. A special interest was placed on systems that fail to rapidly approach thermal equilibrium after the quench, as their dynamical universality may have no equilibrium counterpart~\cite{sciolla2011dynamical, gambassi2010quantum}.

Fully-connected 
\mpar{Fully connected systems}
spin systems provide the simplest instance of genuinely dynamical phase transitions. To illustrate this we consider the infinite-range quantum Ising model [Eq.\eqref{eq_Halpha0} with $\gamma=1$]. The character of non-equilibrium dynamics is encoded in the classical trajectories of the collective spin, which may have paramagnetic or ferromagnetic character. 
Here one studies quenches in the transverse field $h_0\to h_f$ for which DPTs have been extensively studied \cite{zunkovic2016dynamical, sciolla2010quantum, homrighausen2017anomalous, titum2020nonequilibrium}.  The two non-equilibrium phases are distinguished by the time average of the equilibrium order parameter,
\begin{equation}
\label{eq_opDPT}
\overline{\mathcal S^x}	= \lim_{T\to \infty} \frac 1 T  \int_0^T dt \frac{\braket{\hat S^x(t)}}{Ns}
\end{equation}
which serves a non-equilibrium order parameter: It is finite $\overline{\mathcal S^x}\neq 0$ for shallow quenches in the dynamical ferromagnetic phase and it vanishes abruptly $\overline{\mathcal S^x}=0$ at the dynamical critical point $h_{\rm dcr}=(h_0+J_0)/2$, associated with the critical trajectory corresponding to the phase-space separatrix; for deeper quenches $h_f>h_{\rm dcr}$  the system lies in the dynamical paramagnetic phase. 
See Fig.~\ref{fig:LMG_trajectories}(a-b) for an illustration. 
This kind of DPT has been realized experimentally with cold atoms in optical cavities~\cite{muniz2020exploring} or in a superconducting quantum simulator~\cite{xu2020probing}.

The spectral counterparts of these DPTs are given by excited-state quantum phase transitions (ESQPT) \cite{caprio2008excited, bastidas2014quantum, santos2015structure,santos2016excited, perez2017effects, chavez2022spectral, corps2022dynamical}. 
This corresponds to singularities of the density of states at some finite energy density which distinguishes eigenstates with ferromagnetic nature from those with paramagnetic nature: 
See Ref.~\cite{cejnar2021excited} for a recent review. 

The notion of DPT discussed here is in general distinct from that of DQPT discussed in Sec.~\ref{sec_34}, and therefore they may even not occur concomitantly in the same model. However, a connection has been pointed out whenever both phenomena are present\,\cite{zunkovic2018dynamical} (see also Refs.\,\cite{weidinger2017dynamical, lang2018dynamical,lerose2019impact}).
Below, in Section \ref{sec_423}, we will discuss how this out-of-equilibrium phenomenon is affected by decreasing the interaction range.

Let us mention that, due to the  intrinsic semiclassical nature of dynamics in this class of models, observables can be efficiently simulated using phase-space numerical techniques~\cite{curtright2014concise, polkovnikov2010quantum, richter2022semiclassical}, such as the Truncated Wigner Approximation (TWA)~\cite{steel1998dynamics, blakie2008dynamics, polkovnikov2010quantum} or its discrete~\cite{wootters1987wigner, schachenmayer2015many} or clustered~\cite{wurtz2018cluster} versions.
These methods have been intensively used also to explore dynamics with finite $\alpha$ interactions, which we discuss in the next section~\cite{schachenmayer2015many, schachenmayer2015dynamics, pucci2016simulation, pineiro2017nonequilibrium, acevedo2017exploring, mori2019pretermalization, pappalardi2018scrambling, khasseh2020discrete, piccitto2019crossover, sajna2020semiclassical, kelly2021stroboscopic}.

\summary{The asymptotic state after quantum quenches from the ordered phase may or may not display ordering. These dynamical phase transitions are associated with a separatrix in the classical phase-space. 
}
 
\subsubsection{Semiclassical dynamics of quantum fluctuations}
\label{sec_413}

The classical description of dynamics outlined above is exact in the thermodynamic limit $N\to\infty$. In finite systems, however, it has a limited time scale $T_{\text{Ehr}}(N)$ of validity, known as \textit{Ehrenfest time scale}: At long times $t \gtrsim T_{\text{Ehr}}(N)$, quantum fluctuations will dominate the behavior of time-dependent local observables and entanglement quantifiers.
 $T_{\text{Ehr}}$ can be estimated as the time at which the size of quantum fluctuations becomes comparable with a characteristic phase-space scale. This time scale depends on the initial state and on the nature of the underlying classical dynamics. In this Subsection, we discuss the semiclassical dynamics of quantum fluctuations. 
 
To 
\mpar{Time-dependent Holstein-Primakoff}
compute the evolution of quantum spin fluctuations it is convenient to generalize the Holstein-Primakoff approach introduced in Sec.~\ref{low_en_theory}, above, to the non-equilibrium context~\cite{lerose2018chaotic,  lerose2019impact}.
When the system is driven out of equilibrium, the direction of the collective spin configuration [parametrized by $\theta(t)$ and $\phi(t)$] moves along the corresponding classical trajectory on the unit sphere.
We thus let the adapted frame of reference $(\mathbf{X},\mathbf{Y},\mathbf{Z})$ in Eq.~\eqref{eq_rotatedframe} vary in time, in such a way that the $\bold Z$-axis follows the evolution of $\braket{\bold {\hat S}(t)} \propto \mathbf Z(t)$. 
This way, the collective spin components along $\mathbf{X}$ and $\mathbf{Y}$ are associated with quantum fluctuations and will be mapped to canonical bosonic variables.

The time-dependent spin rotation described above is implemented by the time-dependent unitary operator 
\beq 
\label{eq_rotV}
\hat V(\theta(t),\phi(t))= e^{-i\phi(t)\, \hat S^z} \, e^{-i\theta(t)\, \hat S^y}.
\eeq
where the time-dependence of the angles is for the moment unspecified.
The Heisenberg equations for spin components $\hat S^\mu$ with $\mu = X, Y, Z$ in the mobile frame will then read
\begin{equation}
\frac d{dt}\, \hat S^{\mu} = \frac d{dt}\, V \hat S^\mu V^{\dagger} = \frac 1i [\hat S^{\mu}, \tilde { H} ] \, ,
\qquad \text{where} \quad \tilde { H}(t) \equiv \hat V\, \H\, \hat V^{\dagger} +i \hat V \dot{\hat V}^{\dagger} \, .
\end{equation}
The effective time-dependent Hamiltonian
$
\tilde { H}(t) 
$
includes inertial forces arising from the time dependence of $\hat V$. A direct calculation shows
\beq
i \hat V \dot{\hat V}^{\dagger} = - \,  \vec{\omega}(t) \cdot \hat{\vec{ S}} \qquad \text{with} \quad 
\vec{\omega}(t) = \left( -\sin\theta \, \dot{\phi} , \dot {\theta} , \cos\theta \, \dot{\phi}\right).
\eeq

The time-dependent Hamiltonian $\tilde { H}(t)$ is then transformed to a bosonic Hamiltonian via the Holstein-Primakoff transformation, cf. Eq.~\eqref{eq_hpcollectiveferro}. 
This yields an expression of the form\footnote{Here we rescaled time by a factor $s$ for convenience.}
\beq
\label{eq_Hferrocollectivebosondynamical}
\begin{split}
\tilde H(t)  \approx & 
+ (Ns)^1 \; \mathcal{E}\left(\theta(t),\phi(t)\right)  \\
       &
+ ( Ns)^{1/2}\; \Big( \tilde h^{(1)}_Q(t) \hat q + \tilde h^{(1)}_P(t) \hat p \Big)
 \\
     & 
     + (Ns)^0 \left( \tilde h^{(2)}_{QQ}(t) \, \frac{\hat q^2}2 + \tilde h^{(2)}_{PP}(t)\,  \frac{\hat p^2}2 + \tilde h^{(2)}_{QP}(t) \, \frac{\hat q \hat p + \hat p \hat q}2  \right) \\ 
      & +\mathcal{O} \Big( ( Ns)^{-1/2} ).
\end{split}
\eeq
Compared to the ``static'' rotated-frame Hamiltonian (obtained by just rotating the spins and mapping to bosons) [see e.g. Eq.~\eqref{eq_Hferrocollectiveboson}], the additional inertial Hamiltonian modifies the linear terms as
$\tilde h^{(1)}_Q(t)  \equiv h^{(1)}_Q(\theta(t), \phi(t)) + \sin\theta(t) \; \dot {\phi}(t)  $ and 
$\tilde h^{(1)}_P(t)  \equiv h^{(1)}_P(\theta(t), \phi(t)) - \dot {\theta}(t) $,
while quadratic ones are modified as
$\tilde{h}^{(2)}_{QQ, PP
}(t)  \equiv h^{(2)}_{QQ,PP
}\big(\theta(t),\phi(t)\big) - \cos\theta(t) \; \dot{\phi} (t)$ and 
$\tilde{h}^{(2)}_{QP}(t)  \equiv h^{(2)}_{QP}\big(\theta(t),\phi(t)\big) $.

The evolution of $\theta(t)$ and $\phi(t)$ is fixed by the vanishing of the linear terms $\tilde h^{(1)}(t)$, ensuring $\braket{\hat S^X(t)}=\braket{\hat S^Y(t)}=0 $.
 This yields the classical mean-field equations of motion governed by $\mathcal{H}_{\text{cl}}$, i.e. Eq.~\eqref{eq_classicalMotion} for our model.  

On the other hand, the number of collective excitations $\hat n_0=(\hat q^2 + \hat p^2 -1)/2 $ [see e.g. Eq.~\eqref{eq_hpcollectiveferro}] non-trivially evolves in time. Its dynamics are governed by the time-dependent quadratic Hamiltonian parametrized by $\tilde{h}^{(2)}(t)$ above. In order to evaluate them, one computes the Heisenberg equations of motion
\mpar{Dynamics of quantum fluctuations}
\beq
\label{eq_eom}
\begin{dcases}
\dot{\hat q}= +\widetilde{h}^{(2)}_{QP}(t)  \; \hat q + \widetilde{h}^{(2)}_{PP}(t)  \; \hat p\\
\dot{\hat p}=  - \widetilde{h}^{(2)}_{QQ}(t)  \; \hat q - \widetilde{h}^{(2)}_{QP}(t)  \; \hat p
\end{dcases} \ ,
\eeq
with solution 
$ \begin{pmatrix} \hat q(t) \\ \hat p(t) \end{pmatrix}
= U(t)  \begin{pmatrix} \hat q(0) \\ \hat p(0)\end{pmatrix} $, 
where the $2\times2$ propagator $U(t)$ can be formally written as the time-ordered exponential of the matrix defined by the right-hand side of Eq.~\eqref{eq_eom}.  
One can collect the dynamical fluctuations (or ``correlations'') $G^{QQ}(t) \equiv\langle \hat q^2(t) \rangle $, $G^{PP}(t) \equiv\langle \hat p^2(t) \rangle$ and $G^{QP}(t)\equiv\frac{\langle \hat q(t)\hat p(t)+\hat p(t)\hat q(t) \rangle}{2}$
in the $2\times2$ \emph{correlation matrix} 
\beq
\label{eq_Gcorrmatrix}
G(t) =
\begin{pmatrix}
G^{QQ}(t) & G^{QP}(t) \\
G^{QP}(t) & G^{PP}(t) 
\end{pmatrix}
= U(t) \, G(t=0) \, U^T(t) \, .
\eeq
The number of dynamically generated excitations can be expressed as
\beq
\label{eq_ne}
\langle \hat n_0(t) \rangle = \frac{G^{QQ}(t) + G^{PP}(t) - 1}{2} = \frac{1}{2} \Tr \bigg[  G(t) - \frac{\mathbb{1}}{2} \bigg] \ .
\eeq
Note that $\det G(t) \equiv 1/4$, which is an exact property of \emph{pure} Gaussian states preserved by Hamiltonian evolution. For our fully-connected Ising model, the equations of motion for the correlation matrix read 
\begin{align}
\label{eq_motion_feedback_0}
\begin{dcases}
\dot{G}^{QQ} = 2J_0 \cos\theta\sin\phi\cos\phi\, G^{QQ} + 2 J_0  \left(\cos^2\phi-\sin^2\phi \right)\,G^{QP} \\
\dot{G}^{PP} = -2J_0 \cos\theta\sin\phi\cos\phi\, G^{PP} - 2 J_0  \cos^2\phi\sin^2\theta\, G^{PQ}  \\
\dot{G}^{PQ} = -J_0 \cos^2\phi\sin^2\theta\, G^{QQ} + J_0  \left(\cos^2\phi-\sin^2\phi \right)\, G^{PP} \\
\end{dcases} \ .
\end{align}

Crucially, 
\mpar{Relation to the linearized classical flow}
because we obtained these equations by expanding the Hamiltonian in powers of $\heff$ and because classical and quantum evolution generated by quadratic Hamiltonians coincide, \emph{the semiclassical dynamics of quantum fluctuations} --- characterized by the time-dependent correlation matrix $G(t)$ --- \emph{obeys the same equation of motion as the linearized flow of displacements from the classical trajectories.} 
This statement actually applies to arbitrary semiclassical systems with $n$ degrees of freedom, where the correlation matrix $G(t)$ of the quantum fluctuations becomes a $2n \times 2n$ matrix. We refer to Ref.~\cite{lerose2020bridging} or \ref{app_semiCla} for a complete discussion. 
The correlation matrix $G(t)$ is equivalent to the monodromy matrix whose eigenvalues define \emph{the finite-time classical Lyapunov spectrum $\{\lambda_k(t)\}$} \cite{cencini2010chaos}.   

When the classical dynamics is integrable, nearby initial conditions generically separate linearly in time, as it becomes manifest via action-angle variables~\cite{lerose2020origin}. Thus, the temporal growth of the quantum correlations is polynomial, 
$ \langle \hat n_0(t) \rangle \sim t^2 \ $. Isolated unstable trajectories like the separatrix discussed in Sec.~\ref{sec_412} are characterized by exponential sensitivity, and hence $ \langle n_{0}(t) \rangle \sim e^{2\lambda t} $, where $\lambda$ is the largest eigenvalue of the saddle point that controls the instability. 
The asymptotic growth also depends on the initial conditions for systems with a mixed regular-chaotic phase space, e.g. resulting from integrability breaking within a Kolmogorov-Arnold-Moser scenario~\cite{cencini2010chaos}. On the other hand, in systems with fully developed chaos in phase space, the Lyapunov spectrum is uniform and nonvanishing. This implies an asymptotic exponential growth of quantum fluctuations, $ \langle \hat n_{0}(t)  \rangle \sim e^{2 \lambda t} \ $. 
The classification is concluded by the case of stable equilibrium configurations, the linearized dynamics of which are equivalent to that of coupled harmonic oscillators. Accordingly, all the quantities of interest perform bounded (periodic or quasiperiodic) oscillations.
This classification is summarized in the first row of Table~\ref{tab1}.

\begin{table}[t]
\label{tab1}
\centering
\begin{tabular}{lccc}
\toprule
Classical trajectory& Stable & Regular  & Chaotic\\
&&&(Unstable)  \\
\midrule
Collective fluctuations & oscillations & $t^2$ & $e^{2 \lambda t}$\\
Ehrenfest time scale 
& $\mathcal{O}(\sqrt{N})$ &$\mathcal{O}(\sqrt{N})$ & $\mathcal{O}(\ln N)$\\
entanglement entropy
& oscillations & $\ln t$ & $\Lambda_{K}\, t$ \\
square commutator 
& oscillations 
&$t^2$ & $e^{2 \lambda t}$ \\
\bottomrule
\end{tabular}
\caption{
Summary of the dynamical behaviour of entanglement and chaos quantifiers of $N$-particle collective systems in the semiclassical regime.
The growth of the entanglement quantifiers and the square commutator depends on the nature of the limiting classical trajectory in the $2n$-dimensional phase space (stable configuration, regular or chaotic), up to the Ehrenfest time. 
Here, $\lambda \equiv \lambda_1$ is the maximum finite time Lyapunov exponent, and $\Lambda_{K}= \sum_{k =1}^{2K} \lambda_k$ is the sum of the $2K$ largest  Lyapunov exponents, where $K$ is the number of degrees of freedom associated with the considered subsystem.
For $K$=$n/2$, one has the classical Kolmogorov-Sinai entropy rate $\Lambda_{\text{KS}}=\sum_{k \, : \, \lambda_k>0} \lambda_k$. 
}
\label{tab1}
\end{table}

The 
\mpar{Ehrenfest time}
formalism outlined in this Section is quantitatively accurate as long as the number of collective excitations does not grow too large $\langle \hat n_{0} \rangle \ll N$ compared to the system size. As shown in Sec.~\ref{sec_alpha0eq} this assumption is generically valid for ground states, even at the quantum critical points~\cite{dusuel2005continuous, dusuel2004finite}. Out of equilibrium, this condition defines the Ehrenfest time scale, given by 
\beq
\label{eq_TEhr}
\langle \hat n_{0}(T_{\text{Ehr}}) \rangle \sim N \, .
\eeq 
On this time scale the quadratic truncation of the bosonic representation loses accuracy. The non-linear corrections generally lead to saturation of the growth of quantum fluctuations and to revivals on much longer times.
Putting everything together, we have
\begin{equation}
 \label{eq_Tehr}
	\begin{dcases}
	\text{regular trajectories}	 &  \langle \hat n_{0}(t) \rangle \sim t^2  \qquad T_{\text{Ehr}} \sim \heff^{-1/2} \sim \sqrt{N}\\
	\text{unstable (chaotic) trajectories}	&  \langle \hat n_{0}(t) \rangle \sim e^{2\lambda t}  \quad\, T_{\text{Ehr}} \sim \ln \heff^{-1/2} \sim \ln N
	\end{dcases} \ .
\end{equation}

The dynamical growth of collective quantum fluctuations goes hand in hand with the scrambling of quantum information and the dynamics of quantum entanglement. As we will discuss in the next two sections, the approach described here allows us to derive an exact relation between $\langle \hat n_{0}(t)\rangle$, scrambling, and entanglement.

\summary{Collective quantum fluctuations evolve as the linearized flow of displacements around the classical trajectory before the Ehrenfest time scale. The latter is defined as the time for which the amount of quantum fluctuations becomes comparable with the system size. Consequently, this time scale depends on the classical phase-space. }

\subsubsection{Scrambling dynamics}
\label{sec_414}

Scrambling has been recently proposed as a pathway to characterize chaos in many-body dynamics. Generically identified as the delocalization of quantum information, scrambling is commonly quantified by the dynamics of the \emph{square-commutator}
\begin{equation}
	\label{SC}
	c(t) = \langle \left |[\hat A(t), \hat B] \right |^2 \rangle 
\end{equation}
of two observables $\hat A(t)$ and $\hat B$ at different times, where the expectation value is defined as the average over a quantum state $\hat \rho$, i.e., $\langle \cdot \rangle = \Tr( \cdot \hat \rho)$.
Alternatively, scrambling can be studied via the closely related out-of-time order correlators (OTOC) $\langle \hat A(t) \hat B \hat A(t) \hat B\rangle$. The
\mpar{Square-commutator exponential growth}
square commutator was originally introduced by Larkin and Ovchinnikov \cite{larkin1969quasiclassical} to semi-classically describe the exponential sensitivity to initial conditions.\footnote{The heuristics goes as follows: for $\hat A=\hat x$ and $\hat B = \hat p$ in the limit $\hbar \to 0$, upon canonical quantization one has
\begin{equation}
	c(t) \simeq \hbar^2 \{x(t), p(0)\}_{\text{PB}}^2 \simeq \hbar^2 \left | \frac{\partial x(t)}{\partial x(0)} \right |^2 \ .
\end{equation} 
Hence, $c(t)$ encodes the square of the derivatives of the classical trajectory with respect to the initial conditions.}
Thus, whenever the underlying classical limit is chaotic, $c(t)$ is expected to grow exponentially in time as 
\begin{equation}
	c(t) \simeq \hbar_{\text{eff}}^2 \,\,e^{\tilde \lambda t} \ ,
\end{equation}
with a rate $\tilde \lambda$ which may be related to the classical Lyapunov exponent (but it is in principle distinct).  This holds at intermediate times before the Ehrenfest scale $t<T_{\text{Ehr}} \sim \ln \hbar_{\text{eff}}^{-1}$, in this context also referred to as \emph{scrambling time}. Interest in the square-commutator was revived after Kitaev's proposal to use it to characterize many-body dynamics \cite{kitaev2018talk}. In this context, it was shown that for a system at thermal equilibrium the rate $\tilde \lambda$ is upper bounded by quantum effects as $\tilde \lambda \leq \frac{2 \pi T}{\hbar}$, where $T$ is the temperature\,\cite{maldacena2016bound}, as a consequence of the quantum fluctuation-dissipation theorem \cite{tsuji2018bound, pappalardi2022quantum}. This constraint -- now known as ``the bound to chaos'' -- is saturated by models of black holes, including the Sachdev-Ye-Kitaev model (SYK) \cite{sachdev1993gapless, kitaev2018talk}, a system of fully interacting disordered Majorana fermions where $\hbar_{\text{eff}}\sim\ln N$.

In the present case 
of fully-connected systems with a classical limit (Section \ref{sec_411}), scrambling before the Ehrenfest time thus directly probes the sensitivity of the classical trajectories to infinitesimal perturbations.
One can study the square commutator in Eq.(\ref{SC}) by taking the expectation value in pure quasiclassical initial states and by looking at the square commutator between two collective spin projections, namely
\beq
\label{eq_cab}
c_{\alpha \beta}(t) = -  \bigg(\frac 1 {Ns} \bigg)^{2} \bra{\psi_0} \left [ \hat S^{\alpha}(t), \hat S^{\beta}(0)\right ]^2 \ket{\psi_0} \ ,
\eeq
where $\alpha, \beta=x,y,z$ and $\ket{\psi_0}$ is a fully polarized spin-coherent initial state.
Using the expansion of the quantum fluctuations elaborated in Section \ref{sec_413}, we can compute the semiclassical evolution of the out-of-time-order square commutator. By plugging the expansion of the rotated spin operators \eqref{eq_hpcollectiveferro} into the definition \eqref{eq_cab}, one then substitutes the formal solution for the spin fluctuations at time $t$, i.e., $\hat Q(t) = U_{qq}(t)\,\hat Q(0) + U_{qp}(t)\,\hat P(0)$ and ${\hat P(t) = U_{pq}(t)\,\hat Q(0) + U_{pp}(t)\,\hat P(0)}$. The initial fluctuations for coherent states are $\langle \hat Q(0)^2\rangle=\langle \hat P^2(0)\rangle=1/2$ and  $\langle \hat Q(0)\hat P(0)\rangle=\langle \hat P(0)\hat Q(0)\rangle=0$. The resulting out-of-time square commutator in Eq.\eqref{eq_cab} thus reads
\mpar{Semiclassical result}
\begin{align}
\begin{split}
\label{eq_cabGauss}
c_{\alpha\beta}(t) & = 
\Big[  {X}_{\alpha}(t)  \big (\, U_{qq}(t)\,\, 
 {Y}_{\beta} (0)
- U_{qp}(t)\,\, {X}_{\beta}(0)
\, \big) 
+ {Y}_{\alpha}(t) \, \big (\, U_{pq}(t)\, Y_{\beta}(0)  - U_{pp}(t)\, X_{\beta}(0)
\,\big)\Big]^2 
\\ & \quad \quad
+ \mathcal O(\heff) \ .
\end{split}
\end{align}
This expresses a quantitative relation between the square-commutator and the formal evolution $U(t)$ [cf, below Eq.~\eqref{eq_eom}] of the quantum fluctuations, which encodes the of the evolution of linearised displacements and the \emph{finite time} Lyapunov exponent spectrum $\{ \lambda_k(t) \}$, as described in Section \ref{sec_413}. Hence, when the classical limit is integrable, the square-commutator will grow as $c(t)\simeq t^2$, within the Ehrenfest time scale $T_{\text{Ehr}}$. On the other hand in the presence of exponential sensitivity associated with a phase-space separatrix (cf. Sec.~\ref{sec_411} above) or with chaos, the square-commutator $c(t)\simeq e^{2\lambda_1 t}$ grows exponentially before $T_{\text{Ehr}}$ with $\lambda_1=\lambda_1(t)$ the maximal finite-time Lyapunov exponent of the underlying semiclassical trajectory. The different scenarios are summarized in Table~\ref{tab1}.

\begin{figure}[t]
\fontsize{12}{10}\selectfont
\centering
\includegraphics[width = 1\columnwidth]{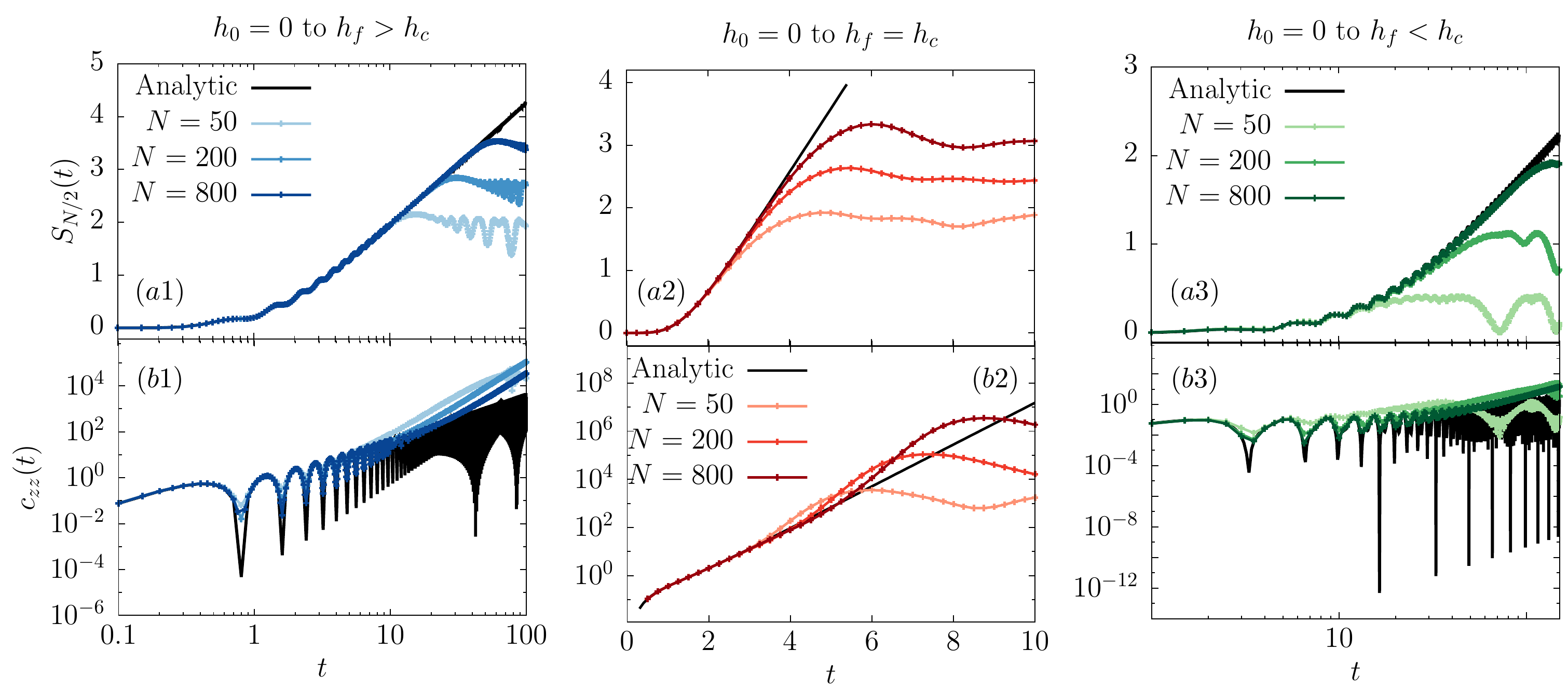}
\caption{Entanglement and scrambling dynamics after a quench of the transverse field in the fully-connected Ising model, from $h_0=0$ to $h_f>0$. The analytical prediction (black lines) is compared with exact numerical results (colours) at finite $N=50,\,200,\, 800$.  
We study the growth of the entanglement entropy \eqref{eq_SA}(top panel) 
and the square commutator \eqref{eq_cab} (bottom panel) 
quenching above, below, and at the dynamical phase transition (DPT) at $h_f=h_{\rm dcr}$, as pictorially shown in Figure \ref{fig:LMG_trajectories} (a).
(a1-b1) Quench above the DPT: $h_f=2J_0>h_{\rm dcr}$. (a2-b2) Quench below the DPT: $h_f=0.2J_0<h_{\rm dcr}$. (a3-b3) Quench at the DPT: $h_f=J_0/2=h_{\rm dcr}$. Time is measured in units of $J_0$. Plots adapted from Ref.\cite{pappalardi2020entanglement}.
}
\label{fig:eeFullyConnected}
\end{figure}

Results 
\mpar{ Comparison with numerics }
for the fully-connected quantum Ising model are shown in Fig.~\ref{fig:eeFullyConnected}, where we consider $c_{zz}(t)$ and compare analytical result (black full line) with numerical exact diagonalization results for finite system sizes. Parameters are specified in the caption. The plot highlights the relation between entanglement entropy (discussed below) and scrambling before the Ehrenfest time. As the considered model has an integrable classical limit, the square-commutator only grows exponentially $c_{zz}(t)\sim e^{2\lambda t}$ for quenches at the dynamical critical point $h_{\rm dcr}$, associated with a classical separatrix with instability rate $\lambda$ given in Eq.~\eqref{lh}.

The exponential 
\mpar{Chaotic fully-connected models}
sensitivity of the square-commutator in the presence of the underlying separatrix has been explored in a number of fully connected mean-field models \cite{pappalardi2018scrambling, quirin2019reversible, pilatowsky2020positive, kidd2021saddle} and recently probed experimentally \cite{li2023improving}.
When the initial state in Eq.~\eqref{SC} is a random permutationally invariant state, the growth rate of the square-commutator corresponds to the average of the finite-time Lyapunov over the whole phase-space. In the presence of an instability, this leads to a modified exponential growth $c{(t)}\sim e^{\lambda t}$ (rather than $ e^{2\lambda t}$) \cite{tianrui2020does}. 
As we discussed above, fully-connected spin systems may exhibit classically chaotic evolution when driven periodically (e.g. quantum kicked top), coupled to other degrees of freedom (e.g. Dicke models), or for larger individual spins $s>1/2$. Analysis as above predicts exponential growth of the square-commutator for underlying classical chaos, as reported in the literature for the quantum kicked top \cite{swingle2016measuring, pappalardi2018scrambling, seshadri2018tripartite, sieberer2019digital, pilatowsky2020positive, lerose2020bridging, sieberer2019digital, yin2021quantum, varikuti2022out, omanakuttan2023scrambling, pappalardi2023quantum}, the Dicke model \cite{chavez2019quantum, lewis2019unifying, buijsman2018nonergodicity, alavirad2019scrambling} and other spin models \cite{yin2021quantum, craps2020lyapunov}. 
Recently, it has been pointed out that scrambling may become \textit{super-exponential} in fully connected models when the average in Eq.\eqref{SC} is done over at infinite temperature state~\cite{qi2023surprises}.

\summary{In fully-connected models, the square-commutator growth is quantitatively determined by the semiclassical phase space before the Ehrenfest time. Hence it encodes finite times Lyapunov exponents.}\\

A similar statement also applies to other quantifiers of quantum information spreading, in particular to entanglement entropy, which we turn to analyze in the next Subsection.

\subsubsection{Entanglement dynamics}
\label{sec_415}
 
It is by now well established that a large body of information about many-body dynamics, their thermalization properties, and the complexity of their numerical simulations, can be inferred from the evolution of bipartite entanglement entropies. 
 For 
 \mpar{Definitions}
 a composite system with Hilbert space ${\mathcal H = \mathcal H_A \otimes \mathcal H_B}$ in a pure state $\ro = \ket{\psi}\bra{\psi}$, the bipartite entanglement between subsystems $A$ and $B$ is encoded in the reduced density matrix ${\ro_A = \Tr_{B}\,{\hat \rho}}$.%
\footnote{The nonvanishing eigenvalues of ${\ro_B = \Tr_{A}\,{\hat \rho}}$ are equal to those of $\ro_A$.}
The system is entangled with respect to the bipartition $(A,B)$ if $\ro_A$ (equivalently, $\ro_B$) is not pure.
The amount of bipartite entanglement can be quantified by the Renyi entropies
\beq
\label{eq_Renyi}
S_A^n = - \frac{1}{1-n} \ln \Tr \, \ro_A^n \ ,
\eeq
parameterized by $n>1$.
The von Neumann entropy is obtained as their limit for $n\to1$, i.e.,  
 \beq
   \label{eq_SA}
   {S_{A} =   - \Tr\big( \ro_A \ln \ro_A \big) } \ .
 \eeq

 As far 
 \mpar{Pictures for local models}
 as local Hamiltonians are concerned, the baseline features of entanglement entropy growth of pure states out-of-equilibrium are well understood. In thermalizing local systems, the entanglement entropy $S_A(t)$ grows linearly in time before saturating to a value proportional to its volume~\cite{luchli2008spreading, kim2013ballistic, mezei2017spread}. This can be viewed as a broad consequence of the light-cone spreading of quantum correlations (see discussion in Sec.\ref{sec_31}). The underlying mechanism is well understood in integrable systems, where it has been explained via a semiclassical picture based on quasi-particle pairs propagation \cite{calabrese2020entanglement}. Analytical insights in chaotic many-body systems came from the study of random unitary circuits~\cite{nahum2017quantum}. On the other hand, the presence of localized integrals of motion (exact or approximate) causes a slowdown of entanglement growth, with a distinguished logarithmic increase for systems exhibiting many-body localization \cite{Abanin2019colloquium}. Neither of these scenarios is adequate for long-range interacting systems, where numerical results exhibited slow logarithmic entanglement growth even in the absence of quenched disorder~\cite{buyskikh2016entanglement, schachenmayer2013entanglement, pappalardi2018scrambling}.

A successful picture to capture entanglement growth in fully-connected systems was only achieved more recently~\cite{lerose2020origin, lerose2020bridging}. The semiclassical growth of quantum fluctuations (described in the section above) allows one to analytically relate the dynamics of the entanglement entropy $S_A(t)$  to the quantifiers of chaos, leading to a general unifying picture. This formalism yields a clean prediction of logarithmic growth in the absence of semiclassical chaos. This constitutes the origin of the slowdown of entanglement growth, which was first observed in numerical simulations of dynamics in long-range interacting quantum spin chains~\cite{schachenmayer2013entanglement, buyskikh2016entanglement,  pappalardi2018scrambling}: We will complete this discussion in Section~\ref{sec_425} below. 
We will illustrate how \emph{$S_A(t)$ asymptotically coincides with the logarithm of the phase space volume spanned by the quantum fluctuations of the subsystem degrees of freedom}, as originally identified in a seminal work by Zurek and Paz~\cite{zurek1995quantum}; see Refs.~\cite{vidal2007entanglement, bianchi2018linear, hackl2018entanglement, lerose2020origin, lerose2020bridging} for more recent literature. 

\begin{figure}[t]
\fontsize{12}{10}\selectfont
\centering
\includegraphics[width = 1\columnwidth]{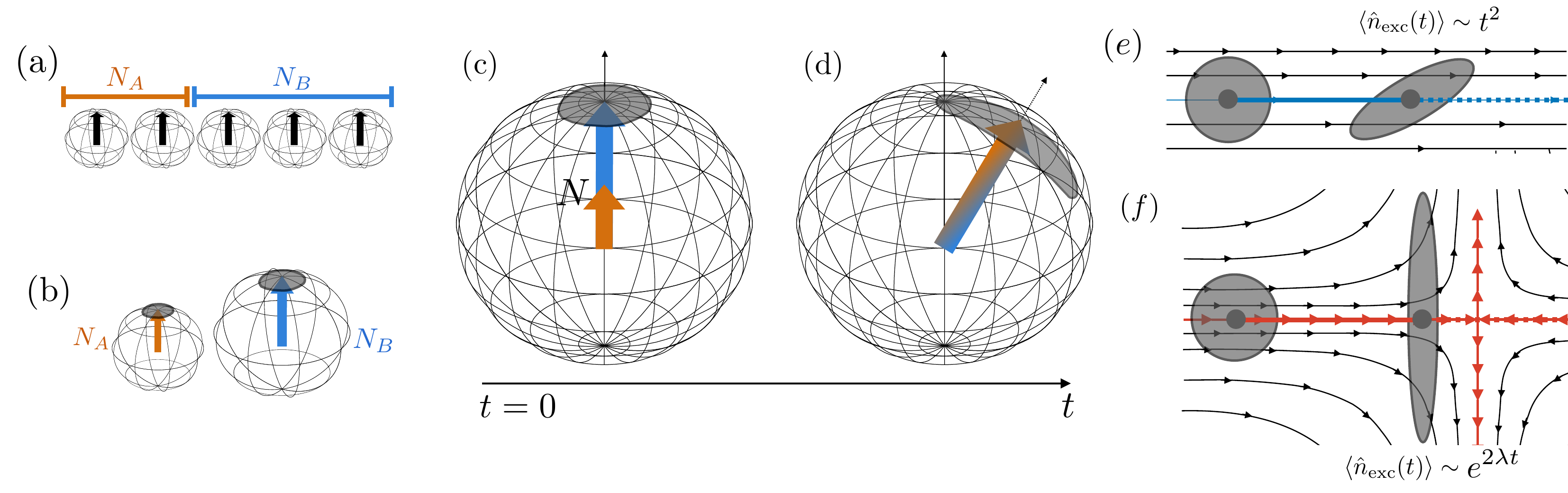}
\caption{ Entanglement dynamics in infinite range spin-chains. (a) The system is partitioned into two blocks of $N_A$ and $N_B$ spins$1/2$, 
initially fully polarized. (b) Collective spins of the two blocks. (c) Collective spin in the factorized initial state, represented on the Bloch sphere. The shaded area represents the quantum uncertainty of transverse components. (d) Nonlinear interactions determine spin squeezing, which makes the two blocks increasingly correlated (entangled). The rate of squeezing is governed by the separation of nearby semiclassical trajectories, and by Eq. (2) it determines the rate of growth of entanglement entropy. Right panels: (e) For generic (noncritical) quenches, nearby trajectories separate linearly in time, leading to a polynomially fast squeezing. (f) For a critical quench, the collective spin lies on the stable manifold of an unstable fixed point in phase space. In this case, nearby trajectories separate exponentially fast in time at a rate $\lambda$ set by the eigenvalue of the linearized flow.
}
\label{fig_4_6}
\end{figure}

In 
\mpar{Entanglement of all-to-all spin models}
the case of fully-connected $N$-particle systems considered here, one considers a bipartition between any two sets of spins, where the only relevant parameter is the number $N_A=f_A N$ of particles in subsystem $A$ (with $N_B = N-N_A = f_B N$).\footnote{Due to permutational symmetry, spatial bipartitions have no meaning.} The collective spin $\hat {\vec S}$ can be correspondingly decomposed as $\hat {\vec S} = \hat {\vec S}_A + \hat {\vec S}_B$ (see Fig.~\ref{fig_4_6}).  Within the semiclassical description, the bipartite system can be represented by bosonic operators $(\hat q_A, \hat p_A)$ and $(\hat q_B, \hat p_B)$, associated with the quantum fluctuations of the two spins $\hat {\vec S}_A $ and $\hat {\vec S}_B$, respectively, via Holstein-Primakoff mapping. 
These quantum fluctuations are characterized by the correlation matrix $G(t)$ defined in Eq.~\eqref{eq_Gcorrmatrix}.
It is convenient to define the subsystem's \emph{reduced correlation matrix} $G_A(t)$ as the $2 \times 2$ matrix of quantum fluctuations built out of the variables of subsystem $A$ alone, i.e.,
\beq
\begin{split}
G_A&=
\begin{pmatrix}
\langle \hat q_A^2 \rangle & \frac{\langle \hat q_A\hat p_A+\hat p_A \hat q_A \rangle}{2} \\
\frac{\langle \hat q_A\hat p_A+\hat p_A \hat q_A \rangle}{2} & \langle \hat p_A^2 \rangle 
\end{pmatrix}
\equiv
\begin{pmatrix}
G^{q_Aq_A} & G^{q_Ap_A} \\
G^{q_Ap_A} & G^{p_Ap_A} 
\end{pmatrix}.
\end{split}
\eeq 
In the semiclassical regime of small $\hbar_{\text{eff}}$, the reduced density matrix $\hat \rho_A(t)$ is asymptotically Gaussian to leading order, and thus fully determined by $G_A(t)$. The entanglement properties can thus be computed via standard techniques~\cite{vidal2007entanglement}, see also Refs.~\cite{bianchi2018linear, hackl2018entanglement}.
The Von Neumann and the second Renyi entropies of a single boson $(\hat q_A, \hat p_A)$ in such a Gaussian state can be expressed in terms of the determinant of $G_A$ as \cite{barthel2006quadratic}
\mpar{Entanglement from quadratic fluctuations}
\begin{subequations}
\begin{align}
\label{eq_EE}
S_A  = 2 \sqrt{ \det G_A } \text{arccoth} & \left(2 \sqrt{ \det G_A } \right) + \frac{1}{2} \log \left( \det G_A - \frac{1}{4}\right) \ ,\\
  \label{eq_S2Gaussian}
  S^{(2)}_A(t) &  =   \frac 1 2   \ln   \; \det \big(2 G_A(t)\big)  \ .
\end{align}
\end{subequations} 
On the other hand, the matrix $G_A$ can be directly related to the correlation matrix $G$ of collective excitations $(\hat q,\hat p)$.\footnote{One can perform a linear canonical transformation to the collective $(\hat q, \hat p)$ and relative $(\delta\hat q, \delta\hat p)$ fluctuation modes:
\begin{equation}
\begin{dcases}
\hat q = +\sqrt{f_A} \; \hat q_A + \sqrt{f_B} \; \hat q_B \\
\delta \hat q = -\sqrt{f_B} \; \hat q_A + \sqrt{f_A} \; \hat q_B 
\end{dcases} 
\begin{dcases}
\hat p = +\sqrt{f_A} \; \hat p_A + \sqrt{f_B} \; \hat p_B \\
\delta  \hat p = -\sqrt{f_B} \; \hat p_A + \sqrt{f_A} \; \hat p_B 
\end{dcases}  \ .    
\end{equation}
Since the Hamiltonian is a function of the collective spin only, the latter bosonic mode is frozen in the vacuum.
}
The explicit computation shows that the determinant can be expressed as
\beq
\label{eq_detASpin}
\det G_A  
= \frac{1}{4} + f_A f_B \; \langle  \hat n_{0} \rangle
\eeq
where $\hat n_{0} = ( \hat q^2 +  \hat p^2  -1 )/ 2 $ represents the number of bosonic excitations of the collective spin [cf. Eq.~\eqref{eq_ne}]. 
While the global evolution preserves the total volume, i.e., $\det \big(2 G(t)\big) \equiv 1$, the information loss generated by projecting the collective quantum fluctuations onto a subsystem yields an increase of entropy. By Eq.~\eqref{eq_S2Gaussian}, this increase may be visualized as an enhancement of the projected volume spanned by the reduced quantum fluctuations within the subsystem phase space, due to the progressive stretching of the phase-space volume spanned by the quantum fluctuations, see Fig.~\ref{fig_4_6}.
The 
\mpar{Entanglement dynamics and classical phase-space}
growth of entanglement entropy out of equilibrium is thus completely determined by the dynamical generation of the collective excitations $\langle  \hat n_{0}(t) \rangle = \frac{1}{2} \Tr \left[  G(t) - \frac{\mathbb{1}}{2} \right]$. As discussed in Section~\ref{sec_413}, $G(t)$ describes the flow of linearized displacements around the classical trajectories. This connection highlights that the entanglement growth in the semiclassical regime is determined by the chaoticity properties of the underlying classical phase space. In fact, the qualitative time-dependence of $\langle  \hat n_{0}(t) \rangle$ depends on the nature of the classical trajectories. See Table~\ref{tab1} for a summary.

The classical dynamics of fully-connected spin-$1/2$ systems are generically integrable (as discussed in Section \ref{sec_411}). Hence the temporal growth of the quantum correlations is at most polynomial, $\langle n_{exc}(t)\rangle \sim t^2$ [see Fig.~\ref{fig:4_6}(e)], leading to 
\begin{equation}
\label{eq_entaSepa}
	S_A(t)  \underset{t\gg1}{\thicksim}  S_A^{(2)}(t) \sim \ln t \ .
\end{equation}

In the (non-generic) case of quenches to dynamical critical points (see discussion in Section \ref{sec_412}), the collective spin moves along an isolated unstable trajectory called separatrix [see Fig. \ref{fig:4_6}(f)]. Out-of-equilibrium generation of collective excitations is thus exponentially fast in such critical quenches, $\langle \hat n_{\text{exc}}(t) \rangle \sim  e^{2\lambda t}$, leading to a linear growth of entanglement entropy with a predicted slope 
\begin{equation}
	S_A(t)   \underset{t\gg1}{\thicksim}  S_A^{(2)}(t) \sim \lambda t \ .
\end{equation}
This phenomenology fully describes the entanglement entropy dynamics of the fully-connected quantum Ising model. This is shown in Fig.~\ref{fig:eeFullyConnected}, where $S_A(t)$ is studied for quenches in the transverse field from $h_0=0$ to different $h_f$ below above and at the dynamical critical point $h_{\rm dcr}$, discussed in Sec.~\ref{sec_412}.

This analysis can be extended to general fully-connected systems whose classical limit has $n>1$ degrees of freedom~\cite{lerose2020bridging} 
and may exhibit chaos. In this case, quantum fluctuations grow as $\langle \hat n_{\text{exc}}(t)\rangle \sim e^{2\lambda t}$, and the growth of the entanglement entropy $S_A(t)$ is generically linear in time with a rate set by the sum of the largest $2n_A$ Lyapunov exponents~\cite{zurek1994decoherence, bianchi2018linear, lerose2020bridging}:
\beq
S_A(t) \underset{t\gg1}{\thicksim} S^{(2)}_A(t) \thicksim \Lambda_{A} t =\bigg(\sum_{k=1}^{2n_A} \lambda_k\bigg) t \ .
\eeq
For $n_A=n/2$, this rate coincides with the classical Kolmogorov-Sinai entropy rate $\Lambda_{KS}=\sum_{\lambda_k \, : \, \lambda_k>0} \lambda_k$ \cite{cencini2010chaos}. 
A linear growth of entanglement entropy thus occurs in chaotic fully-connected spin systems, such as the quantum kicked top~\cite{zarum1999quantum, miller1999signatures, chaudhury1009quantum, ghose2008chaos, piga2019quantum, wang2004entanglement, trail2020entanglement, ghose2004entanglement, kumari2019untangling, lombardi2011entanglement, stamatiou2007quantum, madhok2014information, fiderer2018quantum, pappalardi2018scrambling}, the Dicke model~\cite{furuya1998quantum, wang2004entanglement, lambert2004entanglement, lobez2016entropy, sinha2019chaos, song2012quantum, zhang2015large, bhattacharya2017emergent, mirkhalaf2017entanglement, gietka2019multipartite, lewis2019unifying, lerose2020bridging}, and larger-$s$ fully-connected systems. 

The classification above is concluded by the case of near-equilibrium dynamics around stable equilibrium configurations. In this case, the linearized dynamics are equivalent to that of coupled harmonic oscillators, leading to persistent oscillations in entanglement dynamics. See Table~\ref{tab1} for a summary.

It 
\mpar{Squeezing induced picture}
is worth noting that the stretching of collective quantum fluctuations in phase space, which lies at the origin of bipartite entanglement growth, is very explicitly connected with important witnesses of multipartite quantum entanglement, such as \emph{spin squeezing}~\cite{ma2011quantum} and \emph{the Quantum Fisher Information}~\cite{pezze2018quantum}. 
A popular quantifier of spin-squeezing is defined by the minimal transverse variance of collective quantum spin fluctuations~\cite{kitagawa1003squeezed, wineland1993squeezed}:
\begin{equation}
\label{squeezing}
\xi^2  \equiv \min_{\abs{\mathbf{u}}=1,\mathbf{u}\perp \mathbf{Z}} \frac{\Big\langle \big(\mathbf{u}\cdot \mathbf {\hat S}\big)^2 \Big\rangle}{ N/4} 
. 
\end{equation}
The squeezing parameter $\xi^2$ is equal to one $1$ for coherent (separable) states, while it is smaller for squeezed states $\xi^2<1$.
It has long been known 
\cite{sorensen2001entanglement, pezze2009entanglement}
that collective spin squeezing is a witness of many-body quantum entanglement,
which can be generated by fully connected interactions~\cite{kitagawa1003squeezed, wineland1993squeezed}. Indeed, over the timescale $t \ll T_{\rm Ehr}$, $\xi(t)$, squeezing is explicitly related to $\langle n_{0}(t)\rangle $ in Eq.~\eqref{eq_detASpin} as
\beq
\label{eq:squeezing}
\xi^2(t) = 1+2 \langle  \hat n_{\text{0}}(t) \rangle - 2 \sqrt{ \langle  \hat n_{\text{0}}(t)\rangle( \langle  \hat n_{\text{0}}(t)\rangle +1)} \; .
\eeq
 {Equations (\ref{eq_EE}, \ref{eq_detASpin}) and \eqref{eq:squeezing} express the quantitative link} --- pictorially illustrated in Figure \ref{fig_4_6} ---  between the entanglement entropy $S_A$ and and the spin squeezing parameter $\xi$, in collective spin models in the semiclassical regime in and out of equilibrium.
Following this relation, we will refer to the mechanism of dynamical entanglement entropy growth outlined here as \emph{spin-squeezing picture}~\cite{lerose2020origin}.
For definiteness in this Report, we focus on bipartite entanglement entropies; we refer the readers to Ref.~\cite{lerose2020bridging} for further details on the dynamics of multipartite entanglement.

In conclusion, we recall that the analysis presented above is valid before the Ehrenfest time scale defined in Eq.~\eqref{eq_TEhr}. Over longer times entanglement entropies saturate, $S^\infty_A \propto \log N_A$. 

\summary{In fully connected systems, entanglement dynamics is captured by a semiclassical picture (before Ehrenfest time), which relates it to the squeezing of quantum fluctuations in phase space. This relation generically predicts a logarithmic entanglement growth, in the absence of semiclassical chaos. 
}

\subsection{Quench dynamics of long-range interacting spin systems ($\alpha>0$)}
\label{sec_42}

As soon as the interaction range is finite, the full permutational symmetry of the problem is broken and, in principle, the system could thermalize. The purpose of this Section is to describe the persistent collective character of dynamics in systems with $0<\alpha < d$. 
By formulating a non-equilibrium spin-wave theory (Sec.~\ref{sec_421}), we will be able to develop a physical picture in terms of a semiclassical collective degree of freedom coupled with excitations with finite wavelengths~\cite{lerose2018chaotic,lerose2019impact},.  Analysis of the resulting non-linear coupled equations allows one to demonstrate the freezing of finite-wavelength modes for long times, resulting in lower bounds to thermalization time scales
(Sec.~\ref{sec_422}).  
This scenario affects DPTs (Sec.~\ref{sec_423}), scrambling dynamics (Sec.~\ref{sec_424}), and the characteristic slow growth of entanglement entropy (Sec.~\ref{sec_425}).

\subsubsection{Dynamics of quantum fluctuations with finite interaction range}
\label{sec_421}

To study the impact of finite range interactions on the mean-field dynamics, we resort to a \emph{non-equilibrium spin-wave theory} developed in Refs.~\cite{lerose2018chaotic, lerose2019impact}. 
The goal is to refine the time-dependent formalism of Sec.~\ref{sec_413} to the full momentum-space representation of the finite-range spin Hamiltonian, Eqs.~\eqref{eq_lrxyfourier} and~\eqref{eq_Valpha}. Similarly to the discussion of equilibrium properties in Sec.~\ref{low_en_theory}, one can single out a collective $\alpha$-independent part of the Hamiltonian involving $\mathbf k=\mathbf 0$ spin operators only, and an $\alpha$-dependent perturbation which activates spin fluctuation modes at $\mathbf k\neq \mathbf 0$. 

One can expand the individual spins as bosonic fluctuations around a yet unspecified time-dependent quantization axis ${\mathbf{Z}}$ --- which we will later self-consistently require to coincide with the instantaneous direction of the collective spin $\langle  \vec{\hat{S}}(t) \rangle \propto {\mathbf{Z}}(t)$. 
To achieve this, one performs the time-dependent rotation generated by $\hat V(t)$ in Eq.~\eqref{eq_rotV}. The spin components in this time-dependent frame are governed by the inertial Hamiltonian $\tilde {\H}(t) = \hat V\, \H\, \hat V^{\dagger} +i \hat V \dot{\hat V}^{\dagger}\ ,$ as in Sec.~\ref{sec_413}. One then applies Holstein-Primakoff transformations to the individual rotating spins, as in Eq.~\eqref{eq_HPfourierferro}. 
The resulting transformed Hamiltonian can be organized in the usual form 
\mpar{Time-dependent Holstein Primakoff}
\beq
\label{eq_swGeneric}
\tilde{ H}(t) = \frac 1 s \bigg[ (Ns)^1 \tilde{\mathcal{E}}_{0}(t)  +  (Ns)^{1/2} \tilde H_1(t) +  (Ns)^0 \tilde H_2(t) +  (Ns)^{-1/2} \tilde H_3 (t)+ (Ns)^{-1} \tilde H_4(t) + \dots \bigg]  \ .
\eeq
%
As we already observed in the discussion of equilibrium properties, comparison to the expansion for $\alpha=0$ shows that $\hat q\equiv \tilde{q}_{\mathbf{k}=\mathbf{0}} \ , \, \hat p \equiv \tilde{p}_{\mathbf{k}=\mathbf{0}} \ , \, \hat n_{0} \equiv \hat n_{\mathbf{k}=\mathbf{0}}$, and that the total occupation number of spin-wave excitations $\hat n_{\text{sw}}$ is given by the sum of bosonic occupation numbers of all the other spin-wave modes at finite wavelength, cf. Eq.~\eqref{eq_nsw}
\beq
\label{eq_boso_k}
\hat n_{\text{sw}} = \sum_{\mathbf k\neq \mathbf 0} \hat n_{\mathbf k}  \ , \quad  \text{with}
\quad \hat n_{\mathbf k} \equiv  \frac{\tilde{q}_{\mathbf k} \tilde{q}_{-\mathbf{k}}+\tilde{p}_{\mathbf k} \tilde{p}_{-\mathbf {k}}-1}{2}\ .
\eeq   
The individual occupation numbers quantities $\hat n_{\mathbf k\neq\mathbf 0}$ are exactly conserved by the collective part of the Hamiltonian $\tilde{H}_{\alpha=0}(t)$, which only depends on $\mathbf k\neq \mathbf 0$ bosons through the collective spin length [i.e. through $\hat n_{\text{sw}}$].

The equations of motion of the classical angles $\theta(t)$,  $\phi(t)$ is once again found by imposing the condition that $\braket{\hat S^X(t)}=\braket{\hat S^Y(t)}=0 $~\cite{lerose2018chaotic, lerose2019impact}, namely that the collective bosonic mode describes fluctuations around the instantaneous average spin polarization. This amount to setting the coefficients of $\tilde q_{\mathbf k = \mathbf 0}$ and $\tilde p_{\mathbf k = \mathbf 0}$ equal to zero. 
Taking into account the leading term $\tilde H_1(t)$ only, one retrieves the usual classical mean-field equations of motion~\eqref{eq_classicalMotion}.

However, in the presence of finite-range interactions, the collective spin trajectory gets modified by quantum fluctuations. This effect is the non-equilibrium counterpart of the corrections to the equilibrium spin polarization arising from a finite spin-wave density in the quantum ferromagnetic phase, cf. Sec.~\ref{sec_242}. The corrections arise from the terms in $\tilde H_3(t)$ involving a bosonic operator with $\mathbf k=\mathbf 0$ (such that the remaining two operators have momenta $\pm \mathbf k$). In physical terms, these interactions describe the scattering of a collective spin excitation into a pair of spin waves with opposite finite momenta and vice versa.
Taking into account this ``feedback'' from quantum fluctuations, one obtains a pair of modified equations for the angles $\theta(t), \phi(t)$~\cite{lerose2018chaotic, lerose2019impact}:
\mpar{Equation of motion with the feedback}
\begin{subequations}
\label{eq_motion_angles}
\begin{align}
 \frac{d}{dt}\theta =& +  2J_0 (1-\epsilon)  \sin\theta \cos\phi \sin\phi \nonumber \\
                         & -   2J_0 \bigg( \frac{1}{Ns} \sum_{\mathbf k}  f_{\mathbf k}(\alpha) \;  \left\langle \tilde{p}_{\mathbf k} \tilde{p}_{-\mathbf k} \right\rangle \bigg)  \sin\theta \cos\phi \sin\phi 
                         \\ & 
                         + 2J_0 \Bigg( \frac{1}{Ns} \sum_{\mathbf k}  f_{\mathbf k}(\alpha)   \frac{\left \langle \tilde{q}_{\mathbf k} \tilde{p}_{-\mathbf k} + \tilde{p}_{\mathbf k} \tilde{q}_{-\mathbf k}\right\rangle}{2} \Bigg) \cos\theta  \sin\theta \cos^2\phi,  \nonumber \\             
 \frac{d}{dt}\phi =& -h +  2J_0 (1-\epsilon)   \cos\theta \cos^2\phi  \nonumber \\
                                          & -2J_0\bigg( \frac{1}{Ns} \sum_{\mathbf k\ne0}  f_{\mathbf k}(\alpha) \; \left\langle  \tilde{q}_{\mathbf k} \tilde{q}_{-\mathbf k} \right\rangle \bigg) \cos\theta \cos^2\phi  
                                          \\ & 
                                          + 2J_0 \Bigg( \frac{1}{Ns} \sum_{\mathbf k\ne0}  f_{\mathbf k}(\alpha)  \frac{\left\langle\tilde{q}_{\mathbf k} \tilde{p}_{-\mathbf k} + \tilde{p}_{\mathbf k} \tilde{q}_{-\mathbf k}\right\rangle}{2} \Bigg) \sin\phi \cos\phi  , 	
                                          \nonumber
\end{align}
\end{subequations}
where we introduced the time-dependent \emph{spin-wave density}
\begin{equation}
	\label{eq_swDensity}
\epsilon(t) \equiv \frac{\langle \hat n_{\text{tot}}(t)\rangle }{Ns} = \frac{1}{Ns} \sum_{\mathbf k} \frac{\langle \tilde{q}_{\mathbf k} \tilde{q}_{-\mathbf{k}}\rangle +\langle \tilde{p}_{\mathbf k} \tilde{p}_{-\mathbf {k}}\rangle -1}{2} \, .
\end{equation}
As usual, we observe that the impact of quantum fluctuations on the classical trajectory is suppressed in the classical limit $s\to\infty$, and grows with $\alpha$ at fixed $s$.
Similarly to what we found in equilibrium (Sec.~\ref{low_en_theory}), the properties of $f_{\mathbf k}(\alpha)$ imply that all the quantum backreaction terms in the right-hand sides of the above equations of motion are vanishingly small in the thermodynamic limit for $0<\alpha<d$ and finite for $\alpha>d$ but vanishing as $\alpha \searrow d$. In the latter case, we can make the usual replacement $(1/N)\sum_{\mathbf k} \mapsto \int d^d\mathbf{k}/(2\pi)^d$ in the thermodynamic limit. However, in finite systems, those corrections could be expected to play a role for arbitrary $\alpha$ at sufficiently long times.

In turn, 
\mpar{Spin-waves Hamiltonian}
the evolution of quantum fluctuations is regulated by the spin-wave Hamiltonian $\tilde H_2(t)$ to the same order of approximation as above. It is instructive to report its explicit expression for the considered variable-range quantum Ising model:
\begin{multline}
	\label{eq_hSw2}
	\hat H_{2}(t) = - 2J_0 \sum_{\mathbf k} f_{\mathbf k}(\alpha) \, \times \\ \left ( \cos^2\theta \cos^2 \phi \, \frac{\tilde q_{\mathbf k}\tilde q_{-\mathbf k} }2 + \sin^2\phi\; \frac{\tilde p_{\mathbf k}\tilde p_{-\mathbf k} }2  - \cos \theta \sin \phi\cos\phi\,  \frac{\tilde p_{\mathbf k} \tilde q_{-\mathbf k} + \tilde q_{\mathbf k} \tilde p_{-\mathbf k}}2
	\right ) \\
	+ 2J_0 \cos^2\phi \, \sum_{\mathbf k} \hat n_{\mathbf k} \, .
\end{multline}
This Hamiltonian is equivalent to a set of \emph{externally driven} quantum harmonic oscillators, labelled by the momentum $\mathbf {k}$, where the driving is given by the motion of $\theta(t)$ and $\phi(t)$ and controlled by the couplings $f_{\mathbf k}(\alpha)$.

%
To close 
\mpar{Equations of motion for quantum fluctuations}
the system of equations of motion it is convenient to define the momentum-resolved correlation functions 
\begin{subequations}
\label{eq_swk}
\begin{align}
G^{qq}_{\mathbf k}(t)   \equiv \langle \, \tilde q_{\mathbf k}(t) \tilde q_{-\mathbf k}(t)\, \rangle ,\quad &
G^{pp}_{\mathbf k}(t)   \equiv \langle \, \tilde p_{\mathbf k}(t) \tilde p_{-\mathbf k}(t)\, \rangle , \quad  \\
G^{qp}_{\mathbf k}(t)   \equiv \frac 12 \langle \,  \tilde q_{\mathbf k}(t) \tilde p_{-\mathbf k}(t) &+ \, \tilde p_{\mathbf k}(t) \tilde q_{-\mathbf k}(t)\rangle \ .
\end{align}	
\end{subequations}
Starting from the Heisenberg equations $ \frac{d}{dt} \tilde{q}_{\mathbf k} = i [\tilde H_2(t),\tilde  q_{\mathbf k}] $ and  $ \frac{d}{dt} \tilde{p}_{\mathbf k} = i [\tilde H_2(t),\tilde  p_{\mathbf k}] $ we compute
\begin{equation}
\label{eq_motion_feedback}
\begin{dcases}
\dot G^{qq}_{\mathbf k}  = \,  4 J_0  f_{\mathbf k}(\alpha) \,   \cos\theta \cos\phi\sin\phi \;  G^{qq}_{\mathbf k} 
+4 J_0 \left( \cos^2\phi - f_{\mathbf k}(\alpha)\,  \sin^2\phi \right)\, G^{qp}_{\mathbf k} \, ,\\
\dot G^{pp}_{\mathbf k}  =  -4 J_0 \left( \cos^2\phi -  f_{\mathbf k}(\alpha) \,\, \cos^2\theta \cos^2\phi \right) G^{qp}_{\mathbf k} 
-4 J_0  f_{\mathbf k}(\alpha)\, \,  \cos\theta \cos\phi\sin\phi \,  G^{pp}_{\mathbf k} 	\, ,\\
\dot G^{qp}_{\mathbf k}  =  -2 J_0\left(\cos^2\phi -  f_{\mathbf k}(\alpha)\, \,   \cos^2\theta \cos^2\phi \right) G^{qq}_{\mathbf k} 
+ 2 J_0   \left( \cos^2\phi -  f_{\mathbf k}(\alpha)\,\, \sin^2\phi \right) G^{pp}_{\mathbf k}  \, .
\end{dcases}
\end{equation}
Like Eqs.~\eqref{eq_motion_feedback_0}, these equations are also not independent due to the relation $4(G^{pq}_{\mathbf k})^2 = 4\; G^{pp}_{\mathbf k} \; G^{qq}_{\mathbf k} -1$, 
which is an exact property of Gaussian pure states, and which is then satisfied at all times and for all ${\mathbf k}$'s to the considered level of approximation.

The 
\emph{general physical picture}
  is now clear: 
\begin{itemize}
\item To lowest order the collective spin follows the classical mean-field trajectory;
\item This collective spin motion acts as an external drive for the spin-wave excitations, whereby the couplings $f_{\mathbf k}(\alpha)$ control the \emph{driving amplitude};
\item The dynamically populated non-equilibrium spin-wave ``bath'' may in turn back-react to modify the collective spin dynamics.
\end{itemize}
To quadratic order of approximation in the spin waves, the quantum many-body dynamics of the system is described by the closed set of coupled non-linear evolution equations~\eqref{eq_motion_angles} and~\eqref{eq_motion_feedback}, together with suitable initial conditions (which may be a ground or thermal state of a pre-quench Hamiltonian --- see the discussion on equilibrium states in Sec.~\ref{low_en_theory}). 
This effective decoupling between the dominant zero mode and the finite $k$-suppressed spin waves has recently been exploited in Refs.\cite{roscilde2023entangling, roscilde2023rotor}. There, the quadratic zero mode is replaced by a (fully quantum) rotor, while the $k$ spin waves are kept at the quadratic level. This approach allows to reproduce the dynamics of quantum fluctuations beyond Ehrenfest time.

The formalism derived above is expected to provide an accurate description of the time-evolving many-body wave function whenever the dynamically generated spin-wave {density} $\epsilon(t)$ (see Eq.\eqref{eq_swDensity}) remains small, i.e. $\epsilon(t)  \ll 1 $. This diluteness condition allows us to theoretically describe the many-body dynamics as a self-consistent driven weakly-interacting bosonic gas.
The neglected higher-order terms account for the non-linear scattering among spin waves, which is expected to contribute to the dynamics only over time scales parametrically long in $1/\epsilon$.
Physically, the diluteness condition above corresponds to the requirement that the time-evolving collective spin magnitude $S$ remains close to its maximal value $Ns$, as showcased by Eq.~\eqref{eq_nsw}.
The quality of this approximation is significantly impacted by the interaction range via $f_{\mathbf k}(\alpha)$. 

Below we will use the formalism outlined above to review how the finite-range of interactions impacts the mean-field DPT, scrambling, and entanglement dynamics.

\summary{
Finite-wavelength fluctuations can be modeled as a set of driven bosonic modes, where the drive is given by the collective spin motion. The quantum fluctuations, in turn, affect the collective spin dynamics. The full dynamics is thus described by a set of coupled non-linear evolution equations.  
}

\subsubsection{Prethermal freezing of spin-wave excitations}
\label{sec_422}

The dynamical generation of spin-wave excitations with non-vanishing momenta for $\alpha>0$ is responsible for modifications to the mean-field dynamics. As is manifest in Eqs.~\eqref{eq_motion_angles}, the impact of the quantum backreaction is controlled by the finite-range perturbation via  $ f_{\mathbf{k}\ne\mathbf{0}}(\alpha)$. In turn, the same quantities $ f_{\mathbf{k}\ne\mathbf{0}}(\alpha)$ bound the rate itself of dynamical generation of spin waves: From Eq.~\eqref{eq_hSw2} it is clear that $\big[\hat n_{\mathbf{k}\ne\mathbf{0}}, \tilde{ H}(t) \big] \propto f_{\mathbf{k}\ne\mathbf{0}}(\alpha)$, consistently with the sum of the first two equations in~\eqref{eq_motion_feedback}.

To formulate a more precise bound we can proceed as follows: First, we can eliminate the free spin-wave precession, i.e. the last term in Eq.~\eqref{eq_hSw2}, by switching to the ``interaction picture'': $\tilde q_{\mathbf k} \mapsto \cos \Phi(t) \tilde q_{\mathbf k} + \sin \Phi(t) \tilde p_{\mathbf k}$, $\tilde p_{\mathbf k} \mapsto -\sin \Phi(t) \tilde q_{\mathbf k} + \cos \Phi(t) \tilde p_{\mathbf k}$, where $\Phi(t)=\int_0^t ds \, 2J_0 \cos^2\phi(s)$.\footnote{While the axis $\bold Z$ is fixed by a self-consistency requirement, the orientation of the transverse axes $\bold X$, $\bold Y$ is a ``gauge freedom'' within the formalism. The transformation above amounts to performing a time-dependent rotation of the transverse spin components by angle $\Phi(t)$. With this choice of co-moving frame, the spin-wave Hamiltonian fully vanishes for $\alpha=0$: spin fluctuations look frozen.} Such a ``gauge'' transformation does not change the dynamical population of spin waves. All terms of the right-hand side of the modified Eq.~\eqref{eq_motion_feedback} are now proportional to $ f_{\mathbf{k}\ne\mathbf{0}}(\alpha)$. We can then apply Gronwall's lemma~\cite{gronwall1919note} to this linear system of differential equations to bound the growth of the $G_{\mathbf k}(t)$'s: In particular,
\mpar{Spin waves suppression}
\beq
\label{eq_gronwall}
\langle \hat n_{\mathbf k}(t) \rangle = \frac 1 2 \left( G^{qq}_{\mathbf k} + G^{pp}_{\mathbf k}(t)-1 \right) \le \frac 1 2  \left[ \exp \Big( c  |f_{\mathbf{k}}(\alpha)| J_0  t \Big) -1 \right] \, ,
\eeq
where $c$ is a constant related to the norm of the monodromy matrix on the right-hand side of Eq.~\eqref{eq_motion_feedback}.
Thus, spin-wave excitations at momentum $\mathbf k$ can only be dynamically generated over time scales $J_0 t \gg 1/f_{\mathbf{k}}(\alpha)$.

Of course, the bound in Eq.~\eqref{eq_gronwall} is only useful for $\alpha \lesssim d$.
As discussed in Sec.~\ref{low_en_theory} and in~\ref{app:boundsF}, the couplings $ f_{\mathbf{k}\ne\mathbf{0}}(\alpha)$ are suppressed in the thermodynamic limit $N\to \infty$ when $0<\alpha < d$ and are finite for $\alpha>d$; see Fig.~\ref{Fig1} for an illustration. This straightforwardly provides us with a bound on the rate of growth of the population of bosonic excitations for $0<\alpha<d$ (see~\ref{app:boundsF}): 
\beq
\label{eq_bound}
|f_{\mathbf{k}}(\alpha)|
\le \; \frac{ \text{const} }{(\abs{\mathbf{k}} L)^{\beta}} \ , \quad
\text{with} \quad \beta\equiv\min\big(d-\alpha,(d+1)/2\big)  \, .
\eeq
Therefore, there exists a long time scale 
\beq 
\label{eq_Tsw}
T_{\text{sw}}\sim N^{\beta/d} \ ,
\eeq
during which the dynamical excitation of spin waves with \emph{finite wavelengths} is suppressed.\footnote{Note that this further \textit{a posteriori} justifies  the Holstein-Primakoff approach, as the density of spin waves  remains small over a long time window.}
We can further easily bound the total density of spin-wave excitations \emph{at finite times}:
\beq
\epsilon(t) \le \frac 1 {Ns} \sum_{\mathbf k} \frac 1 2  \left[ \exp \Big( c | f_{\mathbf{k}}(\alpha)| J_0  t \Big) -1 \right] \sim
\frac {J_0 t} N  \sum_{\mathbf k}  |f_{\mathbf{k}}(\alpha)| 
\eeq
For $0<\alpha<d$ we have [recall $\mathbf k \equiv \mathbf k_{\boldsymbol{\ell}} = 2\pi \boldsymbol{\ell}/L$, with $\boldsymbol{\ell}=(\ell_1,\dots,\ell_d)$ integers]
\beq
\epsilon(t)  \le 
\frac {J_0 t} {L^d}  \, \sum_{|\boldsymbol{\ell}|<L/2}  \frac 1{|\boldsymbol{\ell}|^\beta}  \sim \frac{ J_0 t}{L^{\beta}}
\eeq
This bound proves that the spin-wave formalism described here is asymptotically exact in this regime and that the mean-field description of the collective spin polarization dynamics becomes exact in the thermodynamic limit. (We stress that this conclusion requires that $N\to\infty$ must be taken before $t\to\infty$.)

It 
\mpar{Stability analysis of the spin waves}
must be noted that the bound~\eqref{eq_gronwall} above is far from being tight, as it considers exponential instability for \emph{all} the bosonic modes (and with the worst possible rates). In reality, our system can be approximately viewed as a set of quantum harmonic oscillators subject to periodic parametric drive at a frequency given by the classical collective spin precession. Such driven oscillators may or may not meet a resonance; for a given quench, resonances can be detected by performing a stability analysis of the stroboscopic (Floquet) evolution operator for each spin-wave mode~\cite{lerose2020origin}: 
\beq
\begin{pmatrix}
\tilde{q}_{\mathbf{k}}(t_0+T_{\text{cl}}) \\
\tilde{p}_{\mathbf{k}}(t_0+T_{\text{cl}})
\end{pmatrix} 
= U_{\mathbf{k}}(T_{\text{cl}}) \cdot
\begin{pmatrix}
\tilde{q}_{\mathbf{k}}(t_0) \\
\tilde{p}_{\mathbf{k}}(t_0)
\end{pmatrix} 
\eeq
over the period $T_{\text{cl}}$ of the motion of the angles $\theta(t), \phi(t)$. 
The eigenvalues $e^{\pm \lambda_{\mathbf{k}} T_{\text{cl}}}$ of the $2 \times 2$ matrix $U_{\mathbf{k}}(T_{\text{cl}})$ directly give information on the \emph{Floquet quasi-frequency} $\lambda_{\mathbf{k}}$ (see, e.g., Refs.~\cite{landau1965quantum, blanes2009magnus}) of the driven oscillator, which may be purely real  (resonance at mode $\mathbf k$) or purely imaginary $\lambda_{\mathbf{k}}=i\omega_{\mathbf{k}}$ (non-resonance). In the latter case, the mode population performs bounded quasiperiodic oscillations, whereas in the former case, it blows up exponentially.
The real parts of the {Floquet quasi-frequencies} play the role of finite-time Lyapunov exponents for our system.
To quantify the global stability of the system's dynamics for a given quantum quench, one can compute the sum of all \emph{positive} Lyapunov exponents, which gives the Kolmogorov-Sinai entropy rate $\Lambda_{\text{KS}}$. It is convenient to inspect this quantity on varying the fully polarized initial configurations, parametrized by spherical angles: 
\beq
\label{eq_KS_ic}
\Lambda_{\text{KS}}(\theta_0, \phi_0) = \sum_{{\mathbf k}: \lambda_{\mathbf k}>0} 
\text{Re} [\lambda_{\mathbf k}(\theta_0, \phi_0)]   \ .
\eeq
This quantity detects whether resonances are present for the considered quench, and quantifies the actual (initial) instability.
While the stability analysis described here can be done for arbitrary $\alpha$, it is only really meaningful for $\alpha\lesssim d$, as for larger $\alpha$ the density of spin-waves becomes finite in a finite time, and non-linear effects in the full many-body dynamics cannot be neglected.
For $0<\alpha<d$ the spin waves effectively reduce to a \emph{discrete} set of periodically driven quantum oscillators associated with long-wavelength modes with ${\bold k}  \propto 1/L$, cf. the discussion in Sec.~\ref{low_en_theory}.

\begin{figure}[t]
\fontsize{12}{10}\selectfont
\centering
\includegraphics[scale = 0.8]{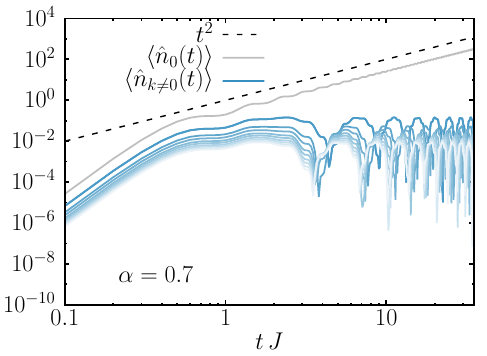}
\includegraphics[scale = 0.4]{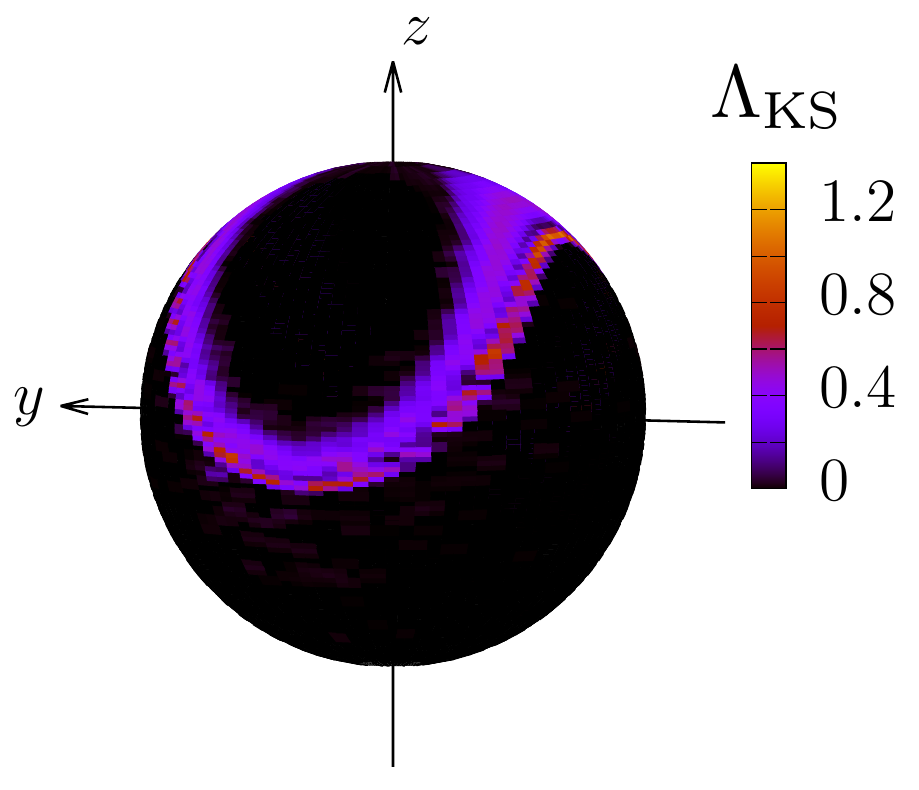}
\caption{Time-dependent $k$-resolved spin-wave population for $\alpha=0.7$ (right panel) after a quench from $h_0=0$ to $h_f=2J$. The blue color gradient for the spin-wave populations in Fourier modes follows the quasimomentum $\abs{k}$ from the darkest ($k=\pm 2\pi/L$) down to smaller-wavelength modes with larger $\abs{k}$ (only the first $20$ modes out of $N=500$ are shown). (Right) Density plot of the Kolmogorov-Sinai entropy rate $\Lambda_{\text{KS}}(\theta_0, \phi_0)$ for different initial conditions $(\theta_0, \phi_0)$ on the Bloch sphere for $\alpha=0.7$, $h=0.5J$. The picture is converged with respect to refining the $k$-space discretization (here $N=100$). Plots adapted from Ref.~\cite{lerose2020origin}.}
\label{fig:populationIsing}
\end{figure}

The stability 
\mpar{Spin waves stability in the Ising chain}
the analysis described above was performed for a long-range quantum Ising chain with $0<\alpha<1$ in Ref.~\cite{lerose2020origin}.
For a large set of initial conditions, spin waves are found to be \emph{stable} (i.e. non-resonant), and consequently, their population remains bounded in time.
While there is no simple rule to predict the existence of resonances, numerical exploration suggests that quenches \emph{near} dynamical criticality typically give rise to resonant excitation of long-wavelength spin waves. %
In other words, the classical separatrix of the mean-field dynamics for $\alpha=0$ typically broadens to a \emph{finite} layer of instability (chaoticity) for $\alpha>0$.
In the left panel of Fig.~\ref{fig:populationIsing} we report the $k$-resolved spin-wave dynamical population for $\alpha=0.7$, obtained by numerically solving  Eq.~\eqref{eq_motion_feedback}. 
The occurrence of resonances is systematically illustrated in the right panel of Fig.~\ref{fig:populationIsing}, which displays the value of $\Lambda_{\text{KS}}(\theta_0, \phi_0)$  as a function of the initial configuration.
These results show that, at least in this case, only quenches near dynamical criticality give rise to resonant excitation of spin waves. 

For quenches that do exhibit resonances at certain non-vanishing momenta, the bound~\eqref{eq_gronwall} captures the qualitative time-dependence of the population of the corresponding modes (the actual rate $\lambda_{\mathbf k}$ is, however, generally lower). Thus, the rapidly growing population of excitations is expected to generate non-linear effects in the full many-body dynamics over comparatively short time scales of order
\beq
T_{\rm Ehr} \sim \log N \, .
\eeq
As anticipated above such resonances involve selected modes with small momenta $\mathbf k \propto 1/L$, as finite-momentum modes are driven very weakly. While all the other modes remain weakly excited (at least) for much longer times $T_{\rm sw}$, the long-wavelength resonant modes form, together with the collective spin, a full-fledged non-linear dynamical system for intermediate time scales $T_{\rm Ehr} \ll t \ll T_{\rm sw}$, which will generally feature semiclassical chaotic behaviour. As the target temporal window lies beyond the Ehrenfest time scale, the self-consistent approximation expounded here is not expected to provide a quantitatively accurate description of dynamics in this regime, and one must resort to numerical simulations to probe this conjectured chaotic behaviour.

A closely related prethermal regime was discussed by Mori in 
\mpar{Quasi-conservation of permutation operators}
Ref.~\cite{mori2019pretermalization} from the point of view of the quasi-conservation of permutation (swap) operators. 
Even though permutational symmetry is dynamically broken for $\alpha>0$, Ref.~\cite{mori2019pretermalization} proves the following bound for long-range interacting quantum Ising chains:
\begin{equation}
	\left | \langle\hat P_{ij}(t) \rangle - \langle \hat P_{ij}(0)\rangle\right | \leq |i-j| \, c_\alpha  \frac {J_0 t}{N^{1-\alpha}}
\end{equation}
where $\hat P_{ij}$ is the swap operator between spins at sites $i, j$ of the chain and $c_\alpha $ is a $N$-independent constant. 
[This result can be easily generalized to arbitrary long-range interacting quantum systems.]
As a result, starting from a permutationally-invariant state with $\langle \hat P_{ij}(0)\rangle=1$ (e.g. fully polarized along a given direction), one deduces that the symmetry under permuting spins \emph{at a finite distance} is approximately conserved, i.e. $\langle\hat P_{ij}(t) \rangle \sim 1$, up to time scales $\sim N^{1-\alpha}$. 
This time scale coincides with $T_{\rm sw}$ defined above upon setting $d=1$. The underlying physical reason is that permutation at finite distances probe finite-wavelength fluctuations, which are consistently frozen over the same time scale. On the other hand, long-wavelength fluctuations are associated with permutations of spins over distances proportional to the length $L$ of the spin chain, for which no long-time quasi-conservation can be guaranteed. Indeed, a severe breakdown of permutational symmetry may occur over a comparatively short time $T_{\rm Ehr} \sim \log N$ over such large length scales [cf. the discussion on resonances above].
For the quench considered in Ref.~\cite{mori2019pretermalization}, it can be shown that spin-wave resonances exist~\cite{lerose2020origin}. This observation points to the onset of a chaotic semiclassical regime over a time scale $T_{\text{Ehr}} \sim \ln N$, in agreement with the detection of semiclassical chaotic behaviour within the numerical simulations of prethermal dynamics in Ref.~\cite{mori2019pretermalization}. 

Finally, it is important to remark that for finite systems, the full self-consistent system of equations~\eqref{eq_motion_angles} and~\eqref{eq_motion_feedback} allows us to investigate non-linear effects arising from dynamical changes in the driving frequency triggered by the quantum backreaction. Long-time numerical computations suggest that the stability diagram based on the resonance picture outlined above is stable to the inclusion of the quantum backreaction.  It is an open problem to characterize conditions for long-time stability in long-range interacting spin systems --- both within and beyond the spin-wave approximation discussed here. An exciting possibility would be to show Kolmogorov-Arnold-Moser-type stability for this class of systems, which would provide a kind of interpolation between scenarios in few-body and many-body physics.

\summary{In the strong long-range regime, there exists a long timescale during which the finite wavelength excitations are suppressed. 
A stability analysis shows that they are stable for generic quenches, while instabilities for finite $\alpha$ can appear in the proximity of quenches to dynamical critical points. 
}

\subsubsection{Impact of finite-range interactions on dynamical phase transitions}
\label{sec_423}

In the last Subsection, we established that for quenches sufficiently far away from dynamical critical points the dynamical generation of spin-wave excitations is non-resonant. This observation leads to a long time window $0<t<T_{\text{sw}}\sim N^{\beta/d}$ (at least) where the spin-wave population remains low, and the collective spin evolution remains close to the unperturbed (mean-field) persistent oscillations. In this scenario, the fate of the $\alpha=0$ dynamical phase transitions (DPT, discussed above in Sec.~\ref{sec_412}) for finite $\alpha$ stands out as a naturally prominent question.

\begin{figure}[t]
\fontsize{12}{10}\selectfont
\centering
\includegraphics[width = 1 \columnwidth]{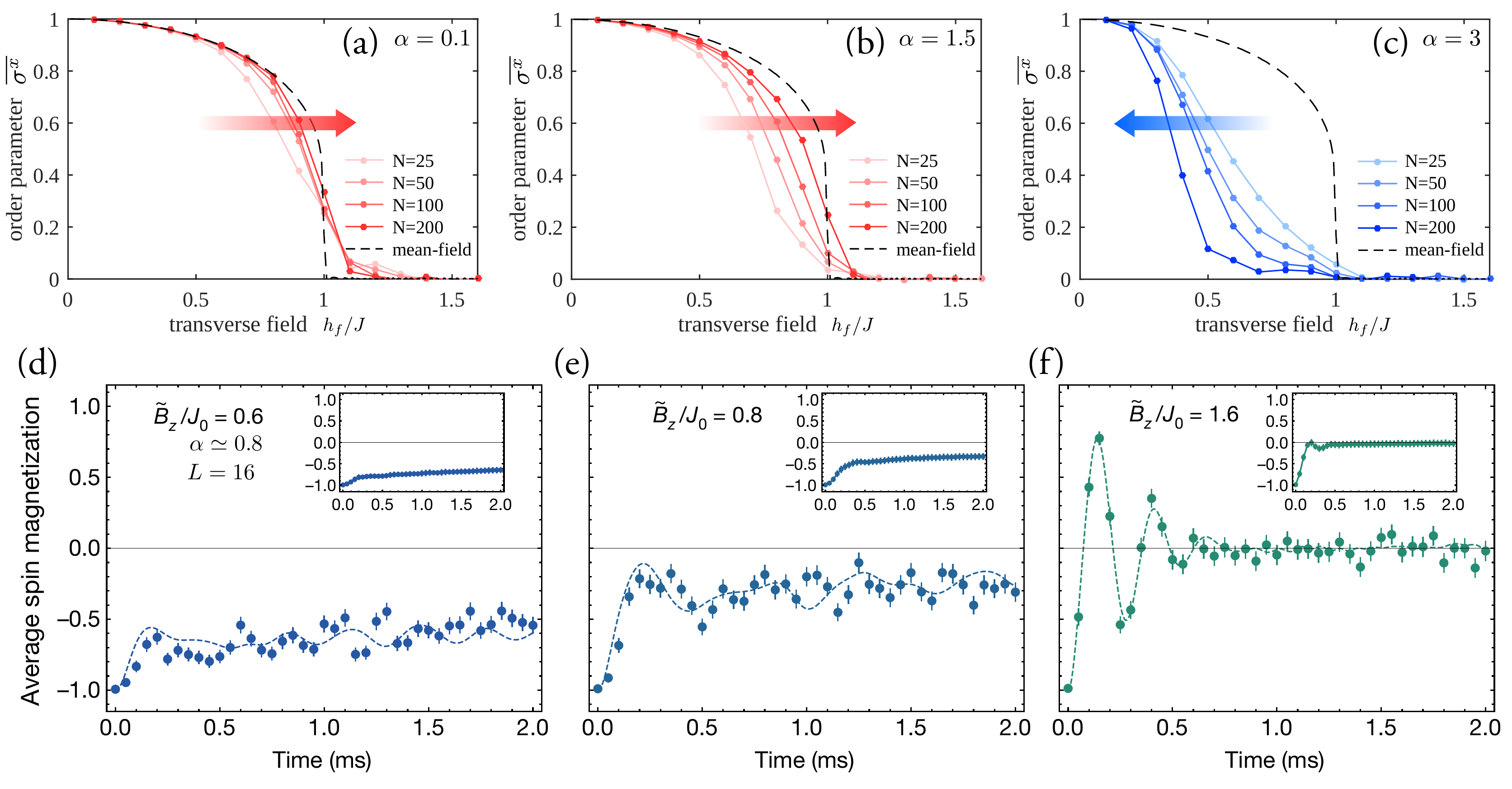}
\caption{ Dynamical phase transition with finite $\alpha$ in the Ising Hamiltonian \eqref{eq_H} with $\gamma=1$. (a-c) Asymptotic value of the order parameter in Eq.~\eqref{eq_opDPT} for different values of the post-quench field $h_f/J$ for quenches from $h_0=-$ with (a) $\alpha=0.1$ (b) $\alpha=1.5$ and (c) $\alpha=3$. Panels adapted from Ref.~\cite{zunkovic2018dynamical}. (d-f) Experimental data of the spin-magnetization $\langle \hat S^x(t)\rangle$ dynamics on a trapped ion quantum simulator. Data with $L=16$ and $\alpha\simeq 0.8$ and transverse field $h_f/J = B_z/J_0=0.6, 0., 1.6$ in (d,e,d) respectively. Panels adapted from Ref.~\cite{zhang2017observationdpt}. }
\label{fig_DPTa}
\end{figure}

This issue has been largely studied as a function of the interaction range parameter $\alpha$ using various numerical approaches~\cite{zunkovic2018dynamical, halimeh2017dynamical, lang2018dynamical, piccitto2019dynamical, piccitto2019crossover, khasseh2020discrete}. For the standard long-range quantum Ising chain, a DPT is found in the thermodynamic limit for $0\leq\alpha < 2$, while it is absent for $\alpha>2$, in qualitative agreement with the equilibrium phase diagram. 
This was shown using matrix-product state dynamical simulations in Ref.~\cite{zunkovic2018dynamical} [see Fig.~\ref{fig_DPTa}(a-c)], where the relation between these DPTs and the singularities in the time-dependence of the Loschmidt echo (DQPTs) has been elucidated, see also Ref.~\cite{halimeh2017dynamical, lang2018dynamical}.  This transition has been studied via the semiclassical truncated Wigner approximation in Ref.~\cite{khasseh2020discrete}, where it was found that the critical exponents for $\alpha \lesssim 0.5$ are the same as the mean-field DPT. 
These DPTs have been experimentally observed with trapped-ion quantum simulators, which realize dynamics described by the long-range quantum Ising chain~\cite{zhang2017observationdpt, smale2019observation, xu2020probing, zhou2023dynamical}.

Finite-range 
\mpar{New dynamical phases}
perturbations can also have a strong impact of the qualitative aspect of the mean-field dynamical phase diagram by inducing new exotic dynamical phases. Refs.~\cite{lerose2018chaotic, lerose2019impact} studied the impact of short-range interactions on top of the fully-connected Ising model, showing the emergence of a \emph{chaotic dynamical phase} within which the asymptotic magnetic ordering is characterized by strong sensitivity to the parameters and initial conditions. It is found that nonequilibrium fluctuations can significantly affect the critical dynamics, resulting in a pseudo-aleatory collective evolution, reminiscent of a particle moving in a multiple-well potential, with a large initial energy, and subject to friction. The nonequilibrium phase diagram universally acquires the basic characteristics of a ``coin toss'' experiment. This result is confirmed by matrix-product-state numerical simulations away from the perturbative regime~\cite{lerose2018chaotic, lerose2019impact}, and a similar scenario was observed in the Dicke model~\cite{lewis2021characterizing}.

\summary{Consistent with the prethermal scenario discussed above, dynamical phases on the two sides of a DPT persist for $0< \alpha \leq 2$, as demonstrated by numerical results. Semiclassical chaos appears in correspondence of the dynamical critical point. }

\subsubsection{Scrambling dynamics with variable-range interactions}
\label{sec_424}
 
Let us briefly discuss how a finite value of $\alpha>0$ impacts the scrambling of quantum information and in particular the OTOC~\eqref{SC} dynamics, introduced in Sec.~\ref{sec_414}. We recall that the square-commutator has been initially put forward due to its exponential growth, i.e.
\[
C(t) = \braket{ |[\hat A(t), \hat B(0)] |^2} \simeq \hbar_{\text{eff}}^2 \,\, e^{\lambda t} \ ,
\]
valid  before the Ehrenfest or scrambling time $t\ll T_{\text{Ehr}}\sim \ln \hbar_{\text{eff}}^{-1}\sim \ln N$ for systems with a semiclassical chaotic limit. This kind of behaviour characterizes as well large-$N$ all-to-all disordered interacting models \cite{kim2021dirac, scaffidi2019chaos, marino2019cavity, davidson2017semiclassical, pappalardi2020quantum, correale2023probing}, despite the absence of an obvious semiclassical limit, including the SYK which saturates the bound to chaos~\cite{maldacena2016bound}.
However, 
\mpar{Fast scrambling}
such exponential growth ---  also known as \emph{fast scrambling}~\cite{sekino2008fast} --- is challenged by finite-range interactions. The square-commutator $C(t)$ was proved to grow at most polynomially in locally interacting systems~\cite{kukuljan2017weak} and in long-range interacting systems with $\alpha>d$~\cite{kuwahara2021absence}. This led to several proposals suggesting fully-connected interactions as an ingredient for fast scrambling \cite{wanisch2023information,  belyansky2020minimal, li2020fast, bentsen2019fast}.

Finite-range interactions lead to a well-defined spatial structure, which allows investigating the space-dependent square commutator
\mpar{Space resolved quare-commutator}
\begin{equation}
	C(r, t) = \langle \left | [\hat A_{\bold x}(t), \hat B_{\bold x_0}(t)] \right |^2 \rangle  \ ,
\end{equation}
where $r=|{\bold x}-{\bold x_0}|$ is the distance between the location of the two considered operators. For locally interacting systems the square commutator $C(r, t)$ becomes appreciable at times $t\sim x/v_B$  \cite{nahum2018operator, keyserlingk2019operator}, where $v_B$, referred to as the \emph{butterfly velocity}~\cite{shenker2014black}, is generally smaller than the Lieb-Robison one $v_B\leq v_{LR}$ [cf. \eqref{eq_LRB}]. By contrast, long-range interactions are found to induce a non-linear light-cone effect, whereby information can spread super-ballistically. This occurrence has been studied numerically in quantum spin chains with variable range interactions~\cite{luitz2019emergent, colmenarez2020lieb, pappalardi2018scrambling, lin2018out} and established via effective hydrodynamics descriptions in disordered models~\cite{chen2019quantum, zhou2020operator, zhou2023hydrodynamic}. All these studies indicate the absence of ballistic spreading of $C(r, t)$ for $\alpha\leq d$. 
For quantum spin systems with sufficiently small $\alpha$, one may use the non-equilibrium spin-wave theory reviewed in Sec.~\ref{sec_421} above to study the dynamics of the space-resolved square commutator $C(r, t)$. In agreement with the onset of semiclassical chaos for near-critical quantum quenches, discussed above in Sec.~\ref{sec_422}, one may expect a concomitant exponential growth of the square commutator in that regime.

\summary{Strong long-range interactions lead to violations of the ballistic growth of the square-commutator and, in addition, have been proposed as an ingredient for fast (exponential) scrambling.}

\subsubsection{Entanglement entropy dynamics: Spin-squeezing vs Quasiparticle picture} 
\label{sec_425}

\begin{figure}[t]
\fontsize{12}{10}\selectfont
\centering
\includegraphics[width = 1\columnwidth]{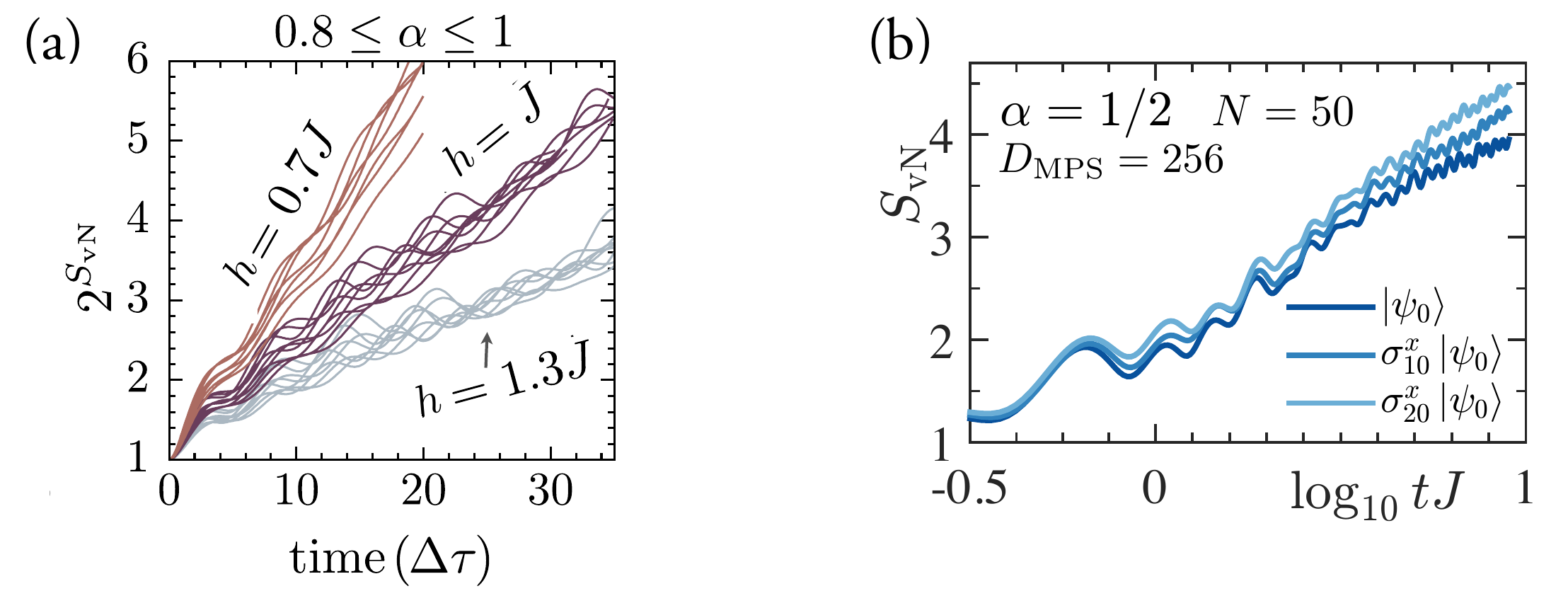}
\caption{Logarithmic growth of the entanglement entropy after a quench with the Ising Hamiltonian \eqref{eq_H} with $\gamma=1$, $d=1$. (a) Exponential of $S_A(t)$  as a function of dimensionless time, for different transverse fields $h=0.7J, 1J, 1.3J$. For each $h$, the results for $L=30,40,50$ and $\alpha=0.8, 0.9, 1$ are plotted. (b) $S_A(t)$ in logarithmic scale for different initial states $|\psi_0\rangle=|\uparrow\uparrow\dots\uparrow\rangle$ and the ones generated by applying single site Pauli operators. Simulation with $L=50$, $\alpha=0.5$ and $h=J$. Image adapted from Refs.~\cite{schachenmayer2013entanglement, buyskikh2016entanglement} for (a), (b) respectively. }
\label{fig:entaA}
\end{figure}

Long-range 
\mpar{Counterintuitive slow growth}
interacting systems exhibit a conceptually different dynamics of the entanglement entropy~\eqref{eq_EE} with respect to locally interacting systems~\cite{luchli2008spreading, kim2013ballistic, mezei2017spread, calabrese2020entanglement, nahum2017quantum}. On one side, their non-local interactions allow quantum correlations between distant degrees of freedom to build up very quickly. As discussed above, this leads to violations of the Lieb-Robinson bound \eqref{eq_LRB} and nonlinear light-cone spreading of quantum correlations, see Sec.~\ref{sec_31} and\,\ref{sec_424}.
 On the other hand, the bipartite entanglement entropy growth after a quench with the Hamiltonian \eqref{eq_H} was found to exhibit a counterintuitive dramatic slowdown as the range of interactions is increased: It becomes logarithmically slow for algebraically-decaying couplings with 
$\alpha$ smaller than the spatial dimensionality $d$~\cite{buyskikh2016entanglement, schachenmayer2013entanglement, pappalardi2018scrambling, lerose2020origin}, see also Fig.~\ref{fig:entaA}. 
Such numerical results can be rationalized using the semiclassical techniques introduced above, which leads to a complete picture of entanglement growth for long-range interacting systems~\cite{lerose2020origin}. Reviewing this framework is the goal of the present Subsection. 
For completeness, we mention in passing that multipartite entanglement associated with algebraically decaying interactions has been studied in depth, e.g. in the form of dynamical spin-squeezing~\cite{foss2016entanglement, perlin2020spin, comparin2022robust, comparin2022multipartite, bornet2023scalable, block2023universal} or via its relation to dynamical susceptibilities in equilibrium~\cite{hauke2016measuring}.

In the fully-connected limit $\alpha=0$, the growth of entanglement is determined by the squeezing of the collective fluctuations stemming from the underlying classical trajectory, see Fig.~\ref{fig:4_6}.  The general framework illustrated in Sec.~\ref{sec_415} predicts logarithmic growth in the absence of semiclassical chaos (Table \ref{tab1}), which is generic in fully-connected spin systems\footnote{without self-interactions if $s>1/2$}.

For finite $\alpha$ the behavior of $S_A(t)$ can be understood at intermediate times by accounting for spin-wave excitations with non-vanishing momentum $\mathbf k$ [cf. Eq.\eqref{eq_swk}] on top of the entanglement dynamics arising from collective spin excitations (or spin squeezing), discussed in Sec.~\ref{sec_415} above. 
As 
\mpar{Spin waves entanglement entropy}
described in Sec.~\ref{sec_421}, the time-evolving state of the spin-wave excitations  is encoded in the correlations in Eq.\eqref{eq_swk}, i.e. $({G}^{qq}_{\mathbf{k}}(t),{G}^{qp}_{\mathbf{k}}(t),{G}^{pp}_{\mathbf{k}}(t))$ defined by
${G}^{\alpha\beta}_{\mathbf{k}}(t) = 
\frac{1}{2} \Big\langle  
\tilde{\alpha}_{\mathbf{k}} (t) \tilde{\beta}_{-\mathbf{k}} (t) + \tilde{\beta}_{\mathbf{k}} (t) \tilde{\alpha}_{-\mathbf{k}} (t)
\Big\rangle$
for $\alpha,\beta=q,p$.
Within the linear spin-wave analysis, the state of a subsystem composed of $M = f_A N < N$ spins contained in a region $A$ of the lattice is a Gaussian bosonic state determined by the instantaneous correlations 
\beq
\bigg\{ G^{\alpha\beta}_{\mathbf r, \mathbf {r'}}(t) =  \big\langle  
\alpha_{\mathbf r} (t) \beta_{\mathbf {r'}} (t) + \beta_{\mathbf r} (t) \alpha_{\mathbf {r'}} (t)
\big\rangle \bigg\}_{\substack{\mathbf r, \mathbf {r'}\in A \\ \alpha,\beta=q,p}}
\eeq
 within $A$, which can be expressed in terms of $\tilde{G}^{\alpha\beta}_{\mathbf{k}}(t)$ via inverse Fourier transform.
This set of correlations uniquely identifies the reduced density matrix $\hat \rho_A (t)$. 
The von Neumann entropy of this Gaussian bosonic state can be computed via standard techniques~\cite{hackl2018entanglement}, namely
\begin{equation}
\label{eq_ee_sw}
S_A = \sum_{i=1}^M\, S(\nu_i) \ , \quad \text{with} \quad 
S(\nu_i) = \frac{\nu_i+1}2\, \ln \frac{\nu_i+1}2 - \frac{\nu_i-1}2\, \ln \frac{\nu_i-1}2 \ ,
\end{equation}
where $\nu_i$ are the symplectic eigenvalues of the correlation matrix.

\begin{figure}[t]
\fontsize{12}{10}\selectfont
\centering
\includegraphics[scale = 0.7]{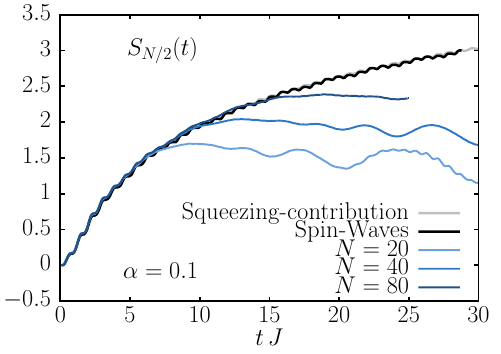}
\includegraphics[scale = 0.7]{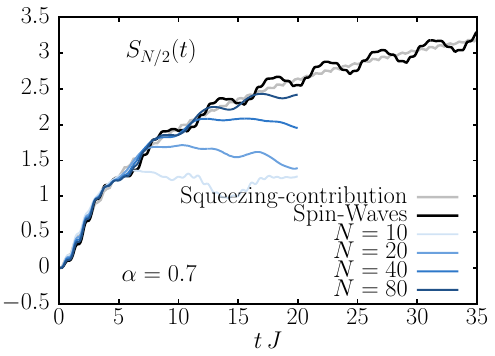}
\caption{Entanglement dynamics after a quench from the ferromagnetic ground state $h_0=0$ with a long-range Ising Hamiltonian \eqref{eq_H} with $\gamma=1$ and $h_f=2$. Comparison between finite-size MPS-TDVP numerical data (light-to-dark blue curves for increasing $N$), the spin-squeezing contribution (grey) and full spin-wave entanglement (black), for $\alpha=0.1$ (left panel) and $0.7$ (right panel), for the quench $h_0=0 \to h_f=2J$, with $N=500$. Figure adapted from Ref. \cite{lerose2020origin}. }
\label{fig:entroAlpha}
\end{figure}

For long-range interactions with $0<\alpha< d$, the growth of $S_A(t)$ turns out to be determined by the stability of the discrete set of long-wavelength excitations, expressed by the Floquet quasi-frequencies $\lambda_{\mathbf{k}}$ with $\abs{\mathbf{k}} \propto 1/L$: see the dedicated discussion in Sec.~\ref{sec_422} and Fig.~\ref{fig:populationIsing}. In particular, one can apply the general semiclassical description entanglement discussed in Sec.~\ref{sec_415} and summarized in Table \ref{tab1}.
If all the modes are stable (i.e., non-resonant), then $S_A(t)\sim \ln t$ exhibits a slow growth dominated by the collective spin fluctuations with $\mathbf k=\mathbf 0$ only. This is indeed the case for typical quenches away from dynamical criticality, as discussed in Sec.~\ref{sec_422}. 
This observation underlies and rationalizes the previous numerical findings of logarithmic growth of the von Neumann entanglement entropy~\cite{buyskikh2016entanglement, schachenmayer2013entanglement, pappalardi2018scrambling}, reported at the beginning of this Subsection. 
On the other hand, if some mode is unstable (i.e., resonant), then ${S_A(t)\sim \Lambda_{\text{KS}} \, t}$ exhibits a fast growth dominated by the instabilities,  with $\Lambda_{\text{KS}}$ in Eq.~\ref{eq_KS_ic}. This is what may happen for quenches in the proximity of dynamical critical points, discussed in Sec.~\ref{sec_412}.

The 
\mpar{Long-range picture}
\emph{physical picture} for the long-range entanglement dynamics is now clear before the Ehrenfest time:
\begin{itemize}
\item The leading contribution comes from the semi-classical squeezing of the collective spin, which grows logarithmically in the absence of classical chaos;
\item In the strong long-range regime, the suppressed long-wavelength spin waves provide a subleading contribution to the entanglement growth.
\end{itemize}
The above analysis shows that slow logarithmic growth of the entanglement entropy should be generally expected in quench dynamics of spin systems with strong long-range interactions starting from a state with large spin polarization\footnote{Subject to the usual caveat of the absence of individual spin self-interactions for $s>1/2$.}.

One 
\mpar{Ising model example}
can solve the spin-wave equation of motion in Eqs~\eqref{eq_motion_feedback} and compute the resulting time-dependent entanglement entropy via Eq.~\eqref{eq_ee_sw}. The results for a typical quench in the long-range quantum Ising chain in a transverse field [cf. Eq.~\eqref{eq_H} with $\gamma=1$, $d=1$] are shown in Fig.~\ref{fig:entroAlpha}, where the exact numerical $S_A(t)$ for finite system size is compared with fully-connected ``spin-squeezing'' contribution and with the result obtained with the inclusion of spin waves. This analysis applies to a wide variety of spin models and quenches, as shown in \ref{sec_mori} where we study the long-range Ising Hamiltonian with transverse and longitudinal field for a quench near the critical point.

We remark that the underlying mechanism crucially relies on the \emph{discreteness} of the set of excitation modes (the long-wavelength spin waves) which result in a bounded, subleading contribution to entanglement growth. 
This property 
\mpar{Other types of perturbations}
is characteristic of strong long-range interactions ($\alpha < d$) and generically does not occur for other types of perturbations.
If, for instance, a \emph{finite-range} perturbation is added on top of a fully-connected model, one can still have stable excitations. However, the presence of the continuous spectrum of excitations results in light-cone spreading of quantum correlations and linear growth of entanglement according to a standard quasiparticle picture~\cite{calabrese2004entanglement}, see e.g. \cite{lerose2019impact} and \ref{app_short} for an example. 
In the weak long-range regime $d<\alpha<d+2$, the growth of entanglement has been related to the nonlinear dispersion relation of quasiparticles \cite{frerot2018multispeed}.

We finally reiterate that the picture of entanglement dynamics for long-range interacting spin systems reviewed here, based on the semiclassical dynamics of quantum spin fluctuations, covers setups not encompassed by other theoretical pictures such as quasi-particles~\cite{calabrese2004entanglement}, spacetime membranes~\cite{nahum2017quantum} or local integrals of motion~\cite{serbyn2013universal}. 

\summary{
For $0\leq \alpha <d$, a semi-classical picture predicts that the entanglement growth is dominated by the collective spin squeezing (logarithmic for generic quenches).}

\

\section{Dynamical phases induced by periodic driving}
\label{sec_5}
In this Section we will expand our analysis to non-autonomous, coherently driven systems. We will show how the previously introduced ideas allow to characterize nonequilibrium phases of spin systems with novel kinds of collective order dynamically stabilized by a periodic drive, which would not be possible in equilibrium~\cite{bukov2015universal}. Here, long-range interactions play the twofold role of protecting long-range order in highly excited states and hindering heating. We explain how this basic mechanism also protects spatiotemporal order such as time-crystalline behavior \cite{russomanno2017floquet, zhang2017observationdtc, kyprianidis2021observation}.\\

 \subsection{Kapitza phases}
 \label{sec_kapitza}
 
 As realized by Kapitza long ago, a rigid pendulum can be stabilized upside down by periodically driving its suspension point with tuned amplitude and frequency. While this dynamical stabilization is feasible in a variety of instances in systems with few degrees of freedom, it is natural to search for generalizations to multi-particle systems. In particular, a fundamental question is whether, by periodically driving a single parameter in a many-body system, one can stabilize an otherwise unstable phase of matter against all possible fluctuations of its microscopic degrees of freedom. 
 
Following Ref.~\cite{lerose2019prethermal}, we report here on such a stabilization in experimentally realizable quantum many-body systems: a periodic modulation of a transverse magnetic field can make ferromagnetic spin systems with long-range interactions stably trapped around unstable paramagnetic configurations as well as in other unconventional dynamical phases with no equilibrium counterparts. 

Specifically, we will study the variable-range quantum Ising chain
\beq
\hat H_{\alpha}(t) = -   \sum_{i,j} J_{i,j}(\alpha)  \, \hat\sigma^x_{i} \hat\sigma^x_{j}
 - h(t)  \sum_{i}  \hat\sigma^z_{i} 
 \label{eq_LRIsingdriven}
\eeq
i.e., the Hamiltonian~\eqref{eq_H} with $d=1$, $\gamma=1$, $s=1/2$ (all such unnecessary restrictions are just chosen for the sake of definiteness and connection with trapped-ion experiments).
Periodic driving is implemented as a cyclic modulation of the magnetic field $h(t)$. Starting from the fully-connected limit $\alpha\to0$ --- akin to the classical Kapitza pendulum --- we employ the non-equilibrium spin-wave theory reviewed above in Sec.~\ref{sec_421} to establish conditions under which dynamical stabilization extends to the quantum many-body domain. 

We conclude by discussing the long (or infinite) lifetime of such quantum many-body Kapitza phases. Elucidating the nature of quantum many-body dynamics in the strong long-range regime --- where no meaningful Lieb-Robinson bounds can be formulated --- these results complement the body of work on Floquet prethermalization in short- and weak long-range interacting spin systems, for which we refer the readers to the original works, see e.g. Refs.~\cite{mori2016rigorous,abanin2017effective,machado2020long}.
 
\subsubsection{Fully-connected limit $\alpha=0$: Non-equilibrium phases by driving}
 \label{sec_mfkapitza}

We first consider the nonequilibrium dynamics of fully-connected spin systems subject to an external periodic drive: 
We start from the infinite-range quantum Ising Hamiltonian
\mpar{ Driven dully-connected Hamiltonian}
 \beq
 \label{eq_lmgdriven}
 \hat H_{\alpha=0}(t) = - \frac{ {J}_0} N \sum_{i,j=1}^N \hat\sigma_i^x \hat\sigma_j^x - h(t) \sum_{i=1}^N \hat\sigma_i^z .
 \eeq
 subject to a monochromatic drive in the transverse field,
  \beq
  \label{eq:protocol}
 h(t) = h_0 + \delta h \cos(\Omega t),
 \eeq
 with amplitude $\delta h$ and frequency $\Omega$.
 
\begin{figure}
\centering
 \includegraphics[width=.48\textwidth]{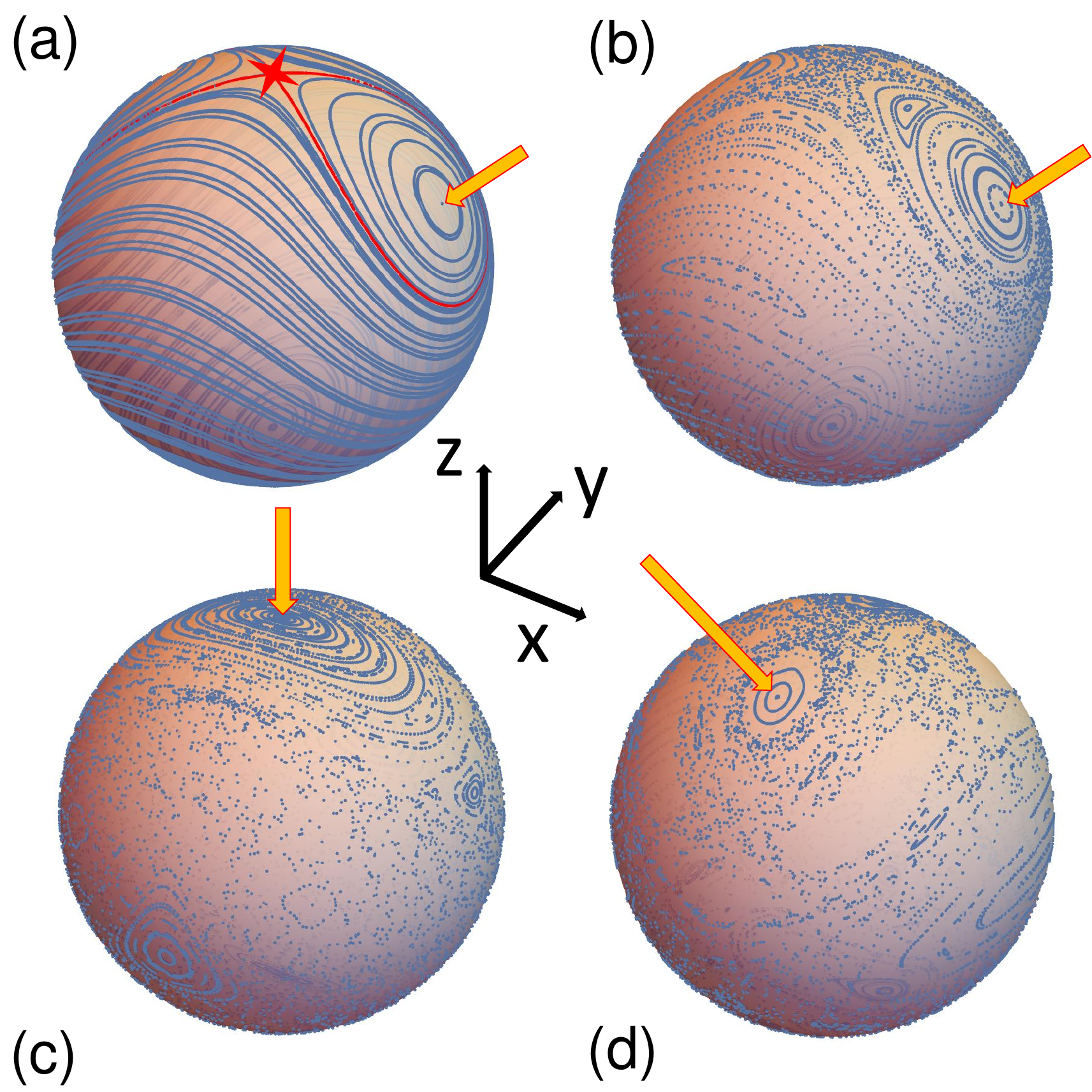} 
\caption{%
  Collective spin dynamics in the infinite-range Ising ferromagnet.
  (a) Classical phase-space trajectories of the static Hamiltonian with $h/{J}_0=1.2$. 
  (b), (c), (d): Stroboscopic trajectories $\{\vec{\mathcal{S}}(t_n)\}$, with $t_n=2\pi n/\Omega$, $n=0,1,2,\dots$ of the collective spin subject to a driving of frequency $\Omega/{J}_0=5$ and amplitudes  $\delta h/{J}_0=0.01$ (b), 3.3 (c), and 5 (d), with $h_0/{J}_0=1.2$. 
Panel (b) shows the presence of a possible ferromagnetic dynamical ordering, corresponding to the evolution occurring within a single ferromagnetic sector $\mathcal{S}^x>0$, with a special synchronized trajectory (appearing as a single point under stroboscopic observations), together with the onset of chaotic behavior around the unstable paramagnetic point~\cite{russomanno2015thermalization}. Panel (c) shows the appearance of a dynamically stabilized phase, akin to the well-known stabilization of the inverted driven Kapitza pendulum~\cite{kapitza1965dynamical,landau1965quantum}. 
Panel (d) shows that for larger driving frequencies, an unconventional dynamical ferromagnetic ordering appears, where the direction of the magnetization is orthogonal to the direction $x$ of the actual ferromagnetic interactions. ``Islands'' with stable stroboscopic trajectories
are indicated by the arrows.
Figure taken from Ref.~\cite{lerose2019prethermal}.
}
\label{fig:1}
\end{figure}

As discussed Sec.~\ref{sec_411}, in the thermodynamic limit $N\to\infty$ the nonequilibrium dynamics are governed by the classical limit $\mathcal{H}_{\text{cl}}(t)$ of the rescaled Hamiltonian $\hat H/S$,
\beq
\mathcal{H}_{\text{cl}}(t) = - {J}_0 \left(\mathcal{S}^x\right)^2 - h(t) \mathcal{S}^z.
\eeq
The quench dynamics in presence of a static field $h(t)\equiv h_0$ has been discussed in Sec.~\ref{sec_412}.
For $0\le {h_0} < 2 {J}_0$ the system supports the ferromagnetic state indicated by the arrow in Fig.~\ref{fig:1}(a), $\vec{\mathcal{S}}(t)$ follows one of the trajectories represented on the Bloch sphere in panel (a), selected by the initial condition $\vec{\mathcal{S}}(0)$. 
Two families of them are characterized by a ferromagnetic-like, symmetry-breaking periodic evolution with opposite signs of the nonvanishing time-averaged order parameter $\overline{\mathcal{S}^x}$. A trajectory (red) passing through the unstable paramagnetic point (red star) separates these two families from the paramagnetic-like orbits with $\overline{\mathcal{S}^x}=0$.
See Sec.~\ref{sec_412} for more details.

Turning on the modulation in Eq.~\eqref{eq:protocol}, representative samples of discrete stroboscopic trajectories $\{\vec{\mathcal{S}}(t_n)\}$, where $t_n=2\pi n/\Omega$, $n=0,1,2,\dots$, of the collective spin are reported in Fig.~\ref{fig:1}(b), (c), and (d).
For small modulation $\delta h$ [see panel (b)], the two ferromagnetic ground states leave room to two periodic trajectories of the collective spin within the corresponding ferromagnetic sectors, \emph{synchronized} with the drive --- hence, appearing as a single point under stroboscopic observations. 
Conversely, initial states in a neighborhood of the unstable paramagnetic point [red star in panel (a)] display chaotic motion as soon as $\delta h \neq 0$  \cite{russomanno2015thermalization, das2006infinite}. 
As $\delta h$ increases, this chaotic region invades an increasingly large portion of the sphere \cite{russomanno2015thermalization}. This behavior can be understood on the basis of classical Kolmogorov-Arnold-Moser theory~\cite{poschel2009lecture,gutzwiller2013chaos}. Related phenomena have been experimentally observed with Bose-Einstein condensates~\cite{tomsovic2017experimental}.

Upon further increasing the modulation [see panel (c)], a region in the parameter space emerges where \emph{dynamical stabilization} 
\mpar{Dynamical stabilization}
of the unstable paramagnetic point occurs, thereby opening up a stability region around it. This phenomenon is analogous to the stabilization of the inverted pendulum discovered by Kapitza~\cite{kapitza1965dynamical,landau1965quantum}.
In addition to this Kapitza-like stabilization, as $\delta h$ increases with $h_0 \approx {J}_0$ 
[see panel (d)], another unconventional regime appears, characterized by dynamical ferromagnetic ordering in the $yz$-plane orthogonal to the direction $x$ of the actual ferromagnetic interactions.

The origin 
\mpar{Analytical understanding}
of the numerical phenomenology described above may be analytically understood by studying the regime of fast-driving limit $\Omega\to\infty$ as a function of the rescaled amplitude 
\beq
 \zeta = \delta h / \Omega \, .
 \eeq 
 In this limit one can easily compute the effective static Hamiltonian governing the stroboscopic evolution, usually termed \emph{Floquet Hamiltonian}: see \ref{app_Magnus}.
When the system is driven rapidly enough at finite driving amplitude, the effective evolution is just governed by the time-averaged Hamiltonian: In physical terms, the system has no time to react to variations of external parameters much faster than its characteristic dynamical time scales. However, if the modulation amplitude $\delta h$ is simultaneously increased with the frequency, keeping a finite ratio
$\zeta \equiv \delta h/\Omega$, 
the effective dynamics may become qualitatively different from those governed by the static Hamiltonian. Such qualitative changes involve a partial resummation of the high-frequency expansion \eqref{eq:Magnus} of the Floquet Hamiltonian~\cite{bukov2015universal}, which is in general an intractable problem. 

Analytic solutions in closed form may be obtained in some cases by performing a convenient time-periodic canonical transformation~\cite{bukov2015universal}. In our case, this strategy is implemented by moving to a time-dependent frame in order to effectively eliminate the oscillating magnetic field:
\beq
\begin{pmatrix}
\hat \sigma^x_{i} \\
\hat \sigma^y_{i} \\
\hat \sigma^z_{i} 
\end{pmatrix}=\begin{pmatrix} \cos\big( 2\zeta \sin(\Omega t) \big) \hat \sigma'^x_{i} + \sin\big( 2\zeta \sin(\Omega t) \big) \hat \sigma'^y_{i} \\
-\sin\big( 2\zeta \sin(\Omega t) \big) \hat \sigma'^x_{i} + \cos\big( 2\zeta \sin(\Omega t) \big) \hat \sigma'^y_{i} \\
\hat  \sigma'^z_{i}
 \end{pmatrix}.
\eeq
The transformation is chosen in such a way that the inertial term in the transformed generator $\tilde{H}(t) $ exactly cancels the driving term. 
Thus $\tilde{H}(t) $ is given by the static part of the Hamiltonian alone [i.e. $h(t) \mapsto h_0$] with $\hat \sigma^x_{i} \hat \sigma^x_{j}$ replaced by
\beq
\begin{split}
&\cos^2\big( 2\zeta\sin(\Omega t) \big) \hat \sigma'^x_{i} \hat  \sigma'^x_{j} + \sin^2\big( 2\zeta \sin(\Omega t) \big) \hat \sigma'^y_{i} \hat \sigma'^y_{j}  \\ 
& + \cos\big( 2\zeta \sin(\Omega t) \big) \sin\big( 2\zeta \sin(\Omega t) \big) \big( \hat \sigma'^x_{i} \hat \sigma'^y_{j} + \hat \sigma'^y_{i} \hat \sigma'^x_{j}\big).
\end{split}
\eeq
Crucially, the modulation $\delta h$ enters $\tilde{H}(t)$ via the finite combination $\zeta$ only, which allows us to perform a standard high-frequency expansion for large $\Omega$. 
\mpar{Effective Hamiltonian}
The effective static Hamiltonian $\hat H_{\text{eff}}$ to lowest order is given by time-averaging: Upon taking the classical limit, this reads  
\beq
\label{eq:effXY_0}
 \mathcal{H}_{\text{eff}} = - {J}_0 \, \left(\frac{1+\gamma(\zeta)}{2}  (\mathcal{S}^x)^2 + \frac{1-\gamma(\zeta)}{2}  (\mathcal{S}^y)^2\right) - h_0 \, \mathcal{S}_z,
 \eeq
i.e., a fully-connected $XY$-model with a ``Floquet-engineered'' anisotropy parameter 
\beq
\label{eq_effanisotropy}
\gamma=\gamma(\zeta)=\mathcal{J}_0(4\zeta),
\eeq
where $\mathcal{J}_0$ is the standard Bessel function of the first kind. 

\begin{figure}
\centering
\includegraphics[width=0.4\textwidth]{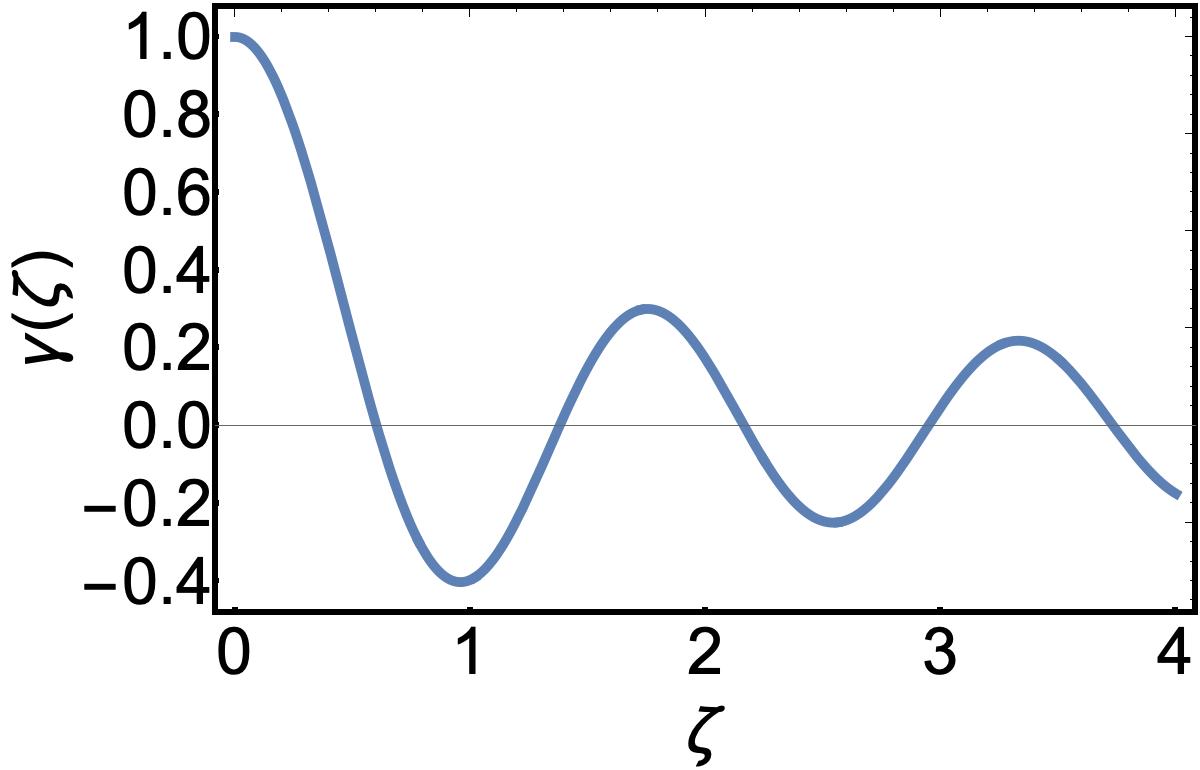}
\caption{
Plot of the anisotropy $\gamma$ in the effective fast-driving Floquet Hamiltonian $\mathcal{H}_{\text{eff}}$, as a function of the rescaled driving amplitude $\zeta$, given by Eq.~\eqref{eq_effanisotropy}.
 }
\label{fig_anisotropy}
\end{figure}

\begin{figure}
\centering
 \includegraphics[width=0.7\textwidth]{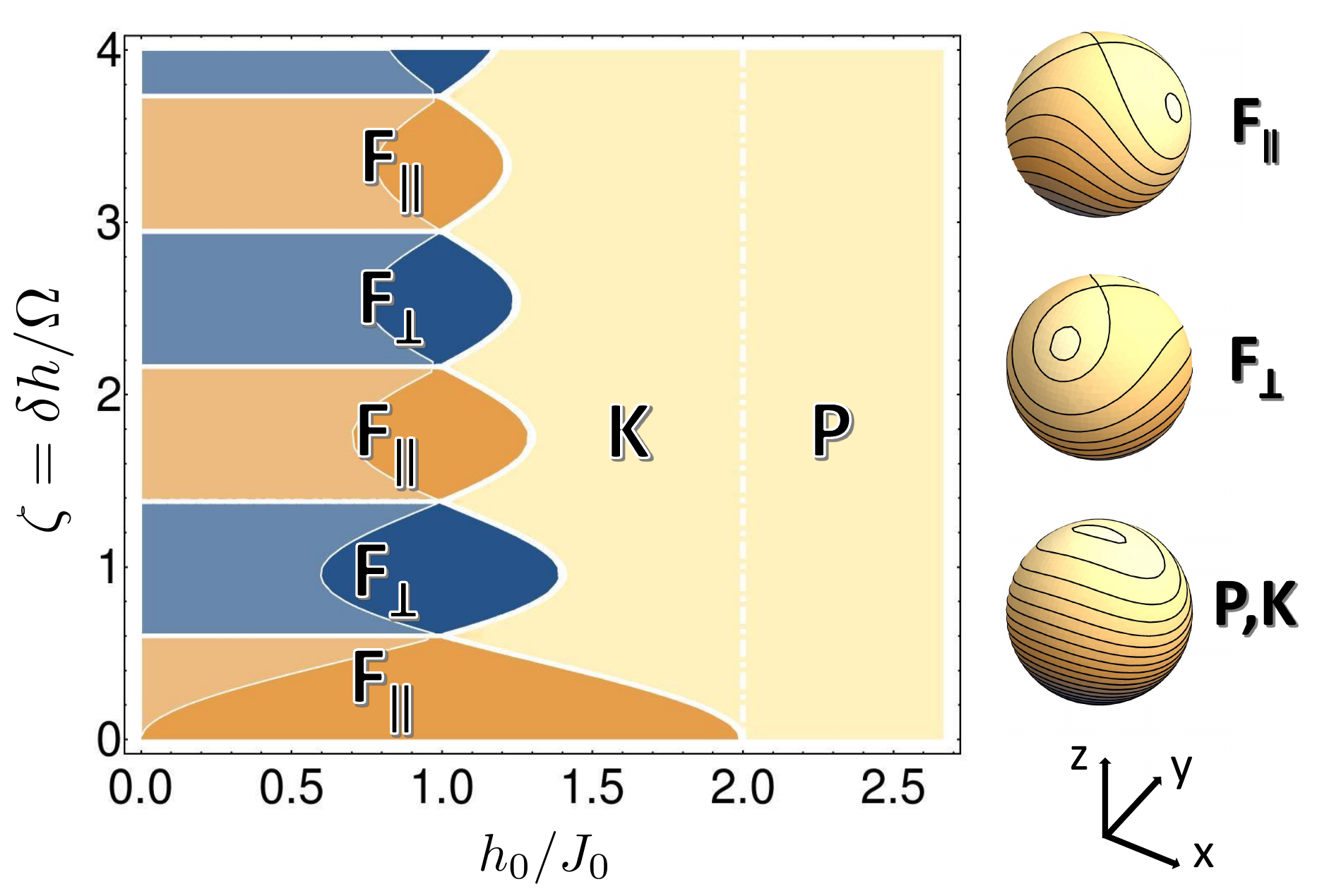} 
\caption{
Left: Fast-driving nonequilibrium phase diagram of the periodically driven infinite-range Ising model defined by Eqs.~\eqref{eq_lmgdriven} and~\eqref{eq:protocol}, taken from Ref.~\cite{lerose2019prethermal}. Upon varying the average magnetic field $h_0$ and the rescaled modulation amplitude $\zeta = \delta h / \Omega$, a dynamical paramagnetic phase $P$, a dynamically stabilized Kapitza paramagnetic phase $K$, a conventional dynamical ferromagnetic phase $F_{\parallel}$ and an unconventional dynamical ferromagnetic phase $F_{\perp}$ with orthogonal magnetization emerge.
The line $\zeta=0$ is the equilibrium phase diagram of the model.
Within the shaded region on the left, a second Kapitza phase coexists with $F_{\parallel,\perp}$. 
(Note that the dashed line separating $K$ and $P$ does not correspond to 
an actual phase transition.)
Right: Schematic phase-space portraits of the effective high-frequency Hamiltonians governing the evolution of the collective spin, highlighting the various phases.
}
\label{fig:2}
\end{figure}

Equation~\eqref{eq:effXY_0} shows that the net effect of the driving is to redistribute the ferromagnetic coupling strength along the  directions $x$ and $y$.
The behavior of the effective anisotropy $\gamma$ as a function of the rescaled driving strength $\zeta$ is shown in Fig.~\ref{fig_anisotropy}.
As $\zeta$ increases from zero, the effective ferromagnetic interaction along $x$ weakens, which makes it possible to dynamically stabilize the paramagnetic configuration. 
The exact boundary $h_0 = h_{\text{cr}}(\zeta) \equiv  {J}_0 \, (1 + \lvert {\cal J}_0(4\zeta) \rvert)$ of the Kapitza phase $K$ is reported in Fig.~\ref{fig:2}. 
Note that this region is continuously connected  with the paramagnetic one $P$ in the phase diagram, see Fig.~\ref{fig:2} --- similarly to the region of dynamical stabilization of the classical Kapitza pendulum, which is continuously connected with the parameter region with a reversed direction of gravity, in which stability is trivial~\cite{landau1965quantum}.

As 
\mpar{Kapitza phases}
$\zeta$ increases further, intervals with a negative anisotropy $\gamma$ appear, favoring ferromagnetic ordering along the direction $y$.
The mechanism is thus elucidated for the occurrence of the unconventional dynamical phases with ferromagnetic ordering in the $yz$-plane, orthogonal to the direction $x$ of the actual ferromagnetic interaction, which builds up whenever $\gamma<0$, $h_0<{J}_0 \, (1-\gamma)$, i.e., within
the regions denoted by $F_{\perp}$ in Fig.~\ref{fig:2}.

The numerical simulations in Fig.~\ref{fig:1} show that these nonequilibrium phases persist at finite driving frequencies, comparable to the characteristic energy scale ${J}_0$ of the system. When the driving frequency $\Omega$ is large but finite,  the effective Floquet Hamiltonian~\eqref{eq:effXY_0} receives perturbative corrections in an expansion in inverse powers of $\Omega$, which cause small quantitative modifications of the boundaries in Fig.~\ref{fig:2}. (For explicit expressions we refer to the original work~\cite{lerose2019prethermal}.)

\begin{figure}[t]
\centering
\includegraphics[width=0.46\textwidth]{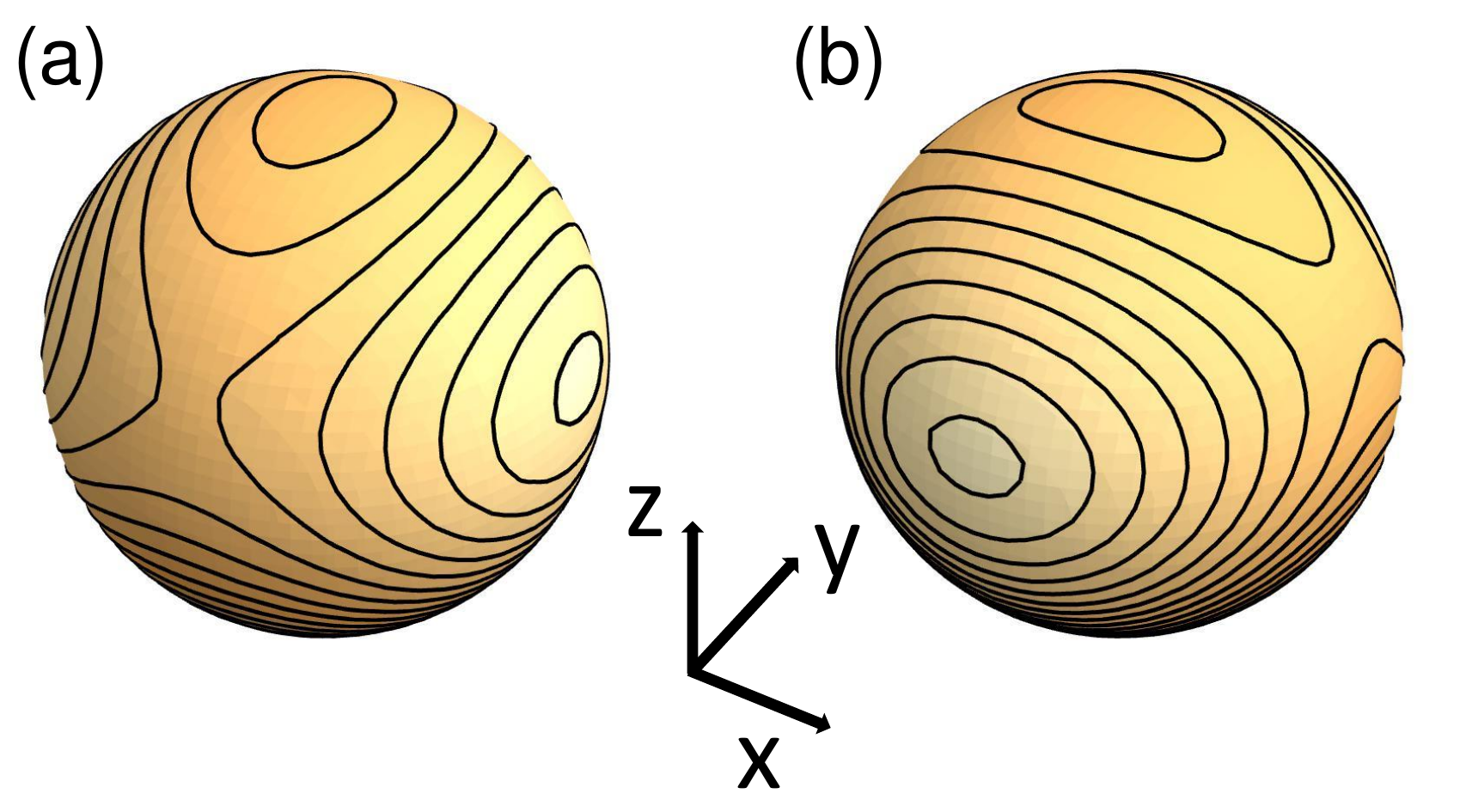}
\caption{%
Schematic phase-space portraits of the effective Hamiltonian $\mathcal{H}_{\text{eff}}$ in Eq.~\eqref{eq:effXY_0} on the sphere, with parameters belonging to the shaded region of the nonequilibrium phase diagram in Fig.~\ref{fig:2}, 
corresponding to the coexistence of a dynamically stabilized Kapitza phase and the ferromagnetic phase $F_{\parallel}$ [(a), shaded blue in Fig. \ref{fig:2}], or $F_{\perp}$ [(b), shaded orange in Fig. \ref{fig:2}]. We emphasize that the paramagnetic configuration is here associated with a \emph{maximum} of $\mathcal{H}_{\text{eff}}$.
Figure taken from Ref.~\cite{lerose2019prethermal}.
}
\label{fig:coexistence}
\end{figure}

A second Kapitza phase coexists with $F_{\parallel,\perp}$ for $h_0 <{J}_0 \, (1-\lvert {\cal J}_0(4\zeta) \rvert)$, i.e., within the shaded region in Fig.~\ref{fig:2}. In this case the effective Hamiltonian \eqref{eq:effXY_0} has a \emph{maximum} at the paramagnetic point in addition to the two ferromagnetic minima in the $xz$- or $yz$-plane, depending on $\gamma$ being positive or negative, respectively. The corresponding phase-space portraits are shown in Fig.~\ref{fig:coexistence}. In particular, in correspondence of the values $\zeta_1, \zeta_2, \dots$ such that $\gamma=0$ (related to the zeros of the Bessel function $\mathcal{J}_0$), the effective Hamiltonian has continuous $O(2)$ symmetry. In this case, stable trajectories exist around the direction of both the ferromagnetic minima and the paramagnetic configuration, which would be unstable in  absence of the drive. 

\summary{Periodic driving the fully-connected model can lead to dynamical stabilization and Kapitza phases. These can be analytically understood in the fast driving limit in terms of an emergent XY model with Floquet-engineered anisotropy parameter.}

\subsubsection{Quantum many-body Kapitza phases for $\alpha>0$}
\label{manybodykapitza}

The dynamically stabilized collective Kapitza phases discussed in Sec.~\ref{sec_mfkapitza} represent a semiclassical realization of the classical Kapitza pendulum with collectively interacting spins. However, it is a priori unclear whether such a Kapitza dynamical stabilization may occur in general quantum many-body systems with finite-range interactions, which give rise to fluctuations at all length scales: While dynamical stabilization of a collective degree of freedom is possible, the presence of many fluctuating degrees of freedom may be expected to give room to heating and destabilize orderly structures.
The existence of dynamically stabilized many-body Kapitza phases was pointed out for a general class of quantum spin systems  with long-range interactions in Ref.~\cite{lerose2019prethermal}; we will here review this phenomenon.

We turn to discuss the full interacting Hamiltonian~\eqref{eq_LRIsingdriven} with $\alpha>0$.
As in Sec.~\ref{sec_mfkapitza}, we shall consider a periodic modulation of the magnetic field,
\beq 
\label{eq:protocol} h(t)=h_0 + \delta h \, \cos(\Omega t).
\eeq
The goal of this Subsection is to demonstrate that (most of) the dynamically stabilized phases persist at least over a parametrically large time scale for $0<\alpha\le 2$, where the many-body dynamics cannot be reduced to those of a single collective degree of freedom.\footnote{These phases are actually more stable in higher-dimensional \cite{britton2012engineered,labuhn2016tunable} and/or higher-spin \cite{senko2015realization} systems (without spin self-interactions), where fluctuations are less effective.}

When $\alpha > 0$, both the collective spin $\vec{\mathcal{S}}$ and the spin excitations with non-vanishing momenta non-trivially participate in non-equilibrium dynamics.
The non-equilibrium spin-wave theory introduced in Refs.~\cite{lerose2018chaotic,lerose2019impact} and reviewed in Sec.~\ref{sec_421} provides a controlled mathodological approach as well as an intuitive physical picture of non-equilibrium dynamics in terms of the coupled evolution of the collective spin and dynamically generated spin waves. This formalism can be straightforwardly extended to systems subject to arbitrary driving protocols, by replacing $h$ with $h(t)$ in Eqs.~\eqref{eq_motion_angles}.
To make the Section more self-contained we report here the expression of the variable-range Hamiltonian $\hat H(t)$ \eqref{eq_LRIsingdriven} expanded to quadratic order in the spin-wave operators:
\mpar{Time-dependent quadratic Hamiltonian}
\begin{multline}
\label{eq:timeindepH}
\hat H(t) =  - N h(t) \bigg( 1- \frac{\hat n_0 + \hat n_{\text{sw}} }{N} \bigg)\cos\theta(t)   \\ - N {J}_0 \bigg[  \bigg( 1- \frac{\hat n_0 + \hat n_{\text{sw}} }{N} \bigg) \sin\theta(t) \cos \phi(t) \bigg]^2  \\
        -  4 J_0 \sum_{k}  f_k(\alpha) \bigg(
       \cos^2\theta(t) \cos^2 \phi(t) \; \frac{\tilde{q}_k \tilde{q}_{-k}}{2} + \sin^2 \phi(t) \; \frac{\tilde{p}_k \tilde{p}_{-k}}{2} \\
       -  \cos\theta(t) \cos \phi(t) \sin \phi(t) \; \frac{\tilde{q}_k \tilde{p}_{-k} + \tilde{p}_k \tilde{q}_{-k}}{2}
       \bigg),
\end{multline}
where we use the same notations as in the rest of the Report.

\begin{figure}[t]
\centering
\includegraphics[width=0.4\textwidth]{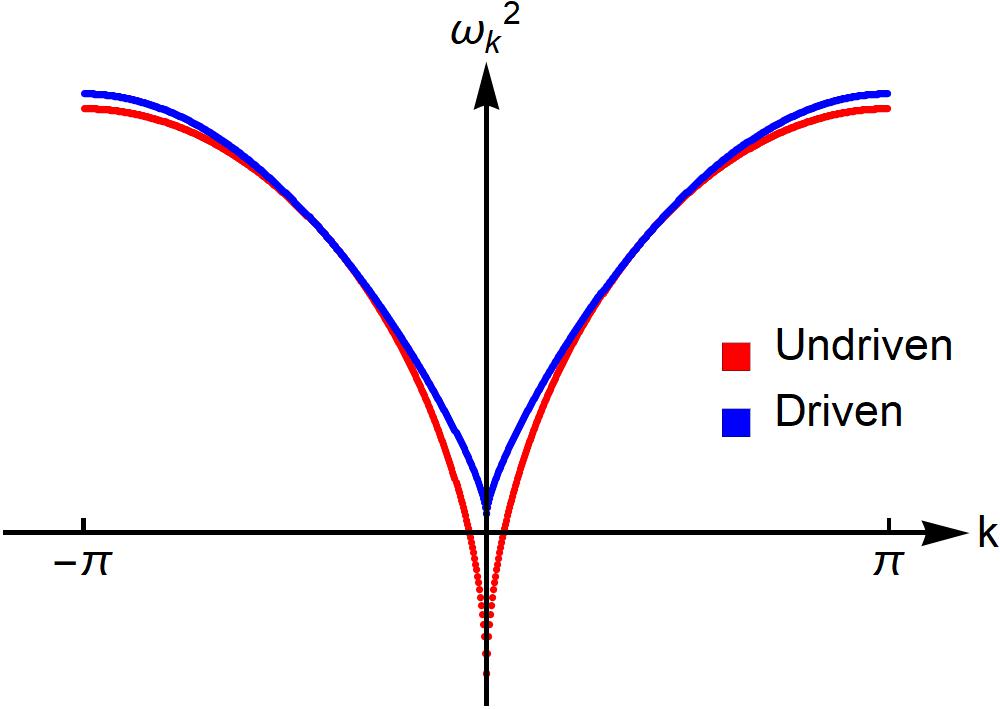} 
\caption{%
Stabilization of \emph{many-body} Kapitza phases. In the presence of suitable periodic driving, the otherwise unstable spectrum of quantum excitations 
around the paramagnetic configuration gets simultaneously dynamically stabilized for all values of $k$.
Here $\alpha=1.5$, $N=400$, and $h_0/{J}_0=1.35$. In absence of driving $\delta h = 0$ the system is in the ferromagnetic phase. 
The red points represent the (squared) frequency spectrum $\omega_k^2 = { h_0\left(h_0-2{J}_0 f_k(\alpha)\right)}$ of the spin-wave excitations, labeled by their wavevector $k$.
An extended interval of long-wavelength modes are unstable (i.e., $\omega_k^2 < 0$ for $k$ near $0$). 
As the driving is turned on with a strength $\delta h$ in a suitable range of values, not only the collective spin mode with $k=0$ discussed in Sec. \ref{sec_mfkapitza}, but also the whole set of modes with $k\ne0$ become \emph{stable} (i.e., $\omega_k^2 > 0$ for all $k$). 
The blue points show the exact effective dispersion relation $\omega_k^2 = \left (h_0-{J}_0 f_k(\alpha)\right)^2$ in the presence of a high-frequency driving $\Omega\to\infty$ with $\zeta=\delta h/\Omega=0.6014$ (corresponding to $\gamma=0$ in the effective Hamiltonian, see the text). 
When ${J}_0 \ll \Omega < \infty$, this effective dispersion relation receives perturbative corrections in inverse powers of $\Omega$, and no qualitative changes occur during the prethermal stage (see Sec.~\ref{sec:heating} and references therein). 
Figure taken from Ref.~\cite{lerose2019prethermal}.
}
\label{fig:manybodystabilization}
\end{figure}

A \emph{many-body} Kapitza phase 
\mpar{Many-body Kapitza phases}
consists of a simultaneous dynamical stabilization of the whole spectrum of quantum excitations around an unstable  configuration. Intuition on this phenomenon can be obtained at the level of linear stability by expanding $\hat H(t)$ to quadratic order in the quantum fluctuations, as in Eq.~\eqref{eq:timeindepH}, around the paramagnetic configuration with $\theta=0$:
\beq
\label{eq:parpoint}
\hat H(t) = \mathcal{E}_{\text{cl}}(t) +
        2 \sum_{k}  \left[ \big( h(t)-2J_0 f_k(\alpha) \big)
        \frac{\tilde{q}_k \tilde{q}_{-k}}{2} + h(t) \frac{\tilde{p}_k \tilde{p}_{-k}}{2}
       \right],
\eeq
where $ \mathcal{E}_{\text{cl}}(t)=-2N h(t)$. In the absence of modulation in the ferromagnetic phase  [i.e., $h(t)=h_0<2{J}_{0}$], an extended interval $[-k^*,k^*]$ around $k=0$ in the spin-wave band is associated with unstable modes,
as their corresponding frequency $\omega_k = \sqrt{ h_0\left(h_0-2J_0{f}_k(\alpha)\right)}$ becomes imaginary for small enough $k$. However, upon introducing the modulation $h(t)$ as in Eq.~\eqref{eq:protocol} with $\delta h \neq 0$, the effective dispersion relation $\omega_{k,\text{eff}}$ is modified. For a suitable choice of the driving parameters, the frequencies $\omega_k$ may become real for all values of $k$. The occurrence of this nontrivial stabilization of an otherwise unstable phase of matter against all possible fluctuations of its degrees of freedom is illustrated in Fig.~\ref{fig:manybodystabilization} and it represents a generalization of the Kapitza pendulum to a genuine many-body system.

\begin{figure}
\centering
 \includegraphics[width=0.5\textwidth]{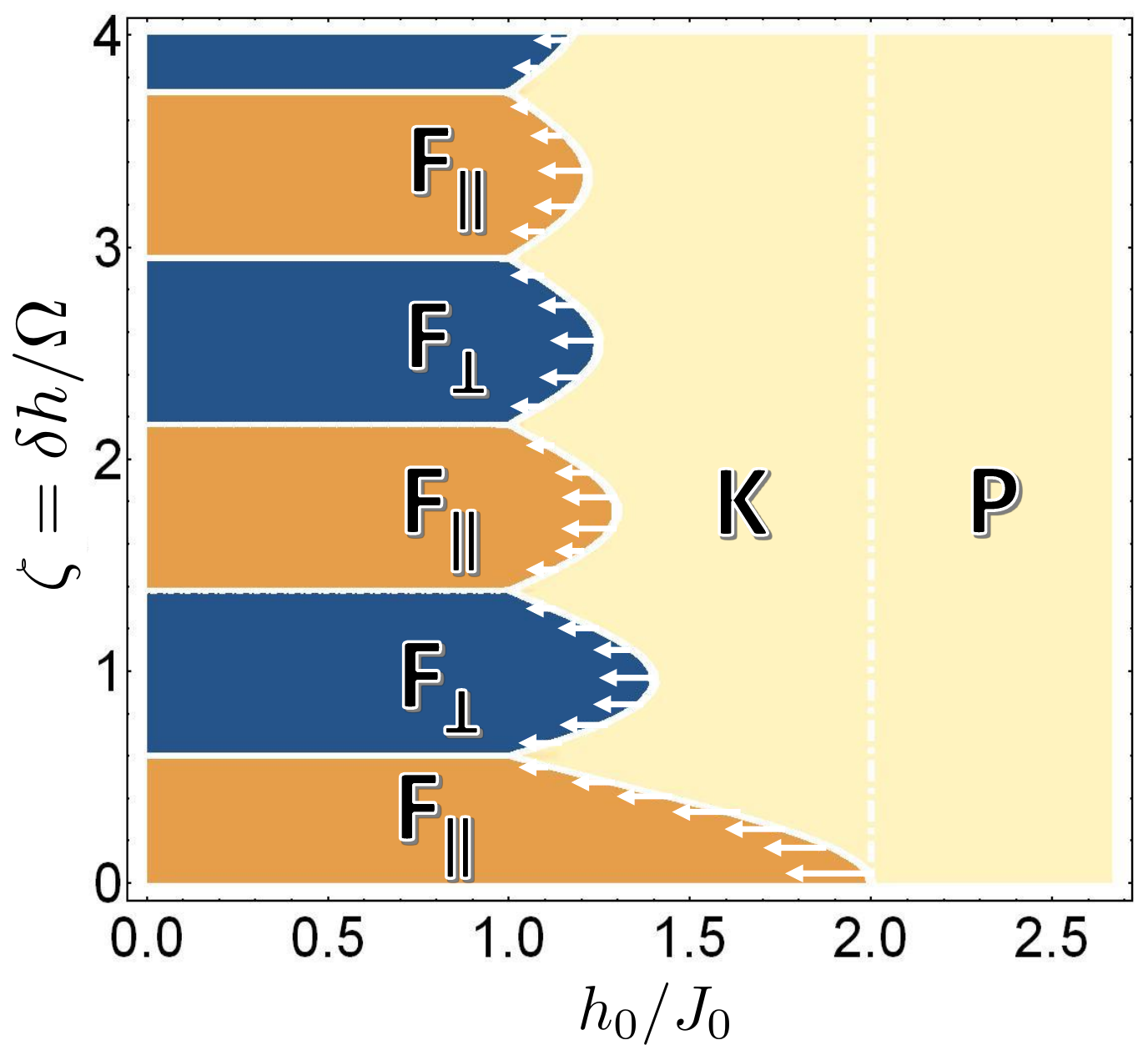} 
\caption{Fast-driving nonequilibrium phase diagram of the periodically driven long-range Ising chain defined by Eq.~\eqref{eq_LRIsingdriven} and \eqref{eq:protocol}, for $\alpha>0$.
Compared to Fig. \ref{fig:2}, the shaded region of coexistence of phase $K$ with $F_{\parallel,\perp}$ has disappeared, and the left boundary of region $K$ moves leftwards upon increasing $\alpha$, as indicated by the white arrows. This displacement is vanishingly small in the thermodynamic limit for $0<\alpha\le1$, and finite for $\alpha>1$. The amount indicated by the arrows corresponds to $\alpha=1.5$ (it is magnified by a factor $2$ for ease of visualization).
Figure taken from Ref.~\cite{lerose2019prethermal}.
}
\label{fig:NEQphasediagramAlpha}
\end{figure}

In order to understand how all the degrees of freedom can get  dynamically and simultaneously stabilized by driving a single modulated global field $h(t)$, we consider the fast-driving limit $\Omega\to\infty$ as a function of the rescaled driving amplitude $\zeta$, which can be studied analytically also for $\alpha\neq 0$.
%
In this regime the stroboscopic evolution of the system at times $t_n = 2\pi n/\Omega$ with $n=0,1,2,\dots$ is governed by an effective static Hamiltonian $\hat H_{\text{eff}}$ obtained via a high-frequency expansion (see \ref{app_Magnus}). 
The computation of $\hat H_{\text{eff}}$, discussed in Sec.~\ref{sec_mfkapitza} for the infinite-range limit, is actually independent of the particular of the interaction range. Consequently, it can be implemented following exactly the same steps, leading to an effective long-range XY spin chain:
\mpar{Effective manuy-body Hamiltonian}
\beq
\label{eq:effXY_alpha}
\hat H_{\text{eff}} 
=
 -  \sum_{i\ne j}^N \frac{J}{\lvert\lvert i-j \rvert\rvert^{\alpha}} 
 \bigg[
\frac{1+\gamma(\zeta)}{2} \hat \sigma^x_{i} \hat \sigma^x_{j} + \frac{1-\gamma(\zeta)}{2} \hat \sigma^y_{i} \hat \sigma^y_{j} 
 \bigg] 
 - h_0 \sum_{i}^N \hat \sigma^z_{i},
\eeq
where the anisotropy parameter $\gamma(\zeta)= \mathcal{J}_0(4\zeta)$ is the same as in Eq. \eqref{eq:effXY_0} and is plotted in Fig.~\ref{fig_anisotropy}.  
Equation \eqref{eq:effXY_alpha} allows us to discuss the modification of the nonequilibrium phase diagram in Fig.~\ref{fig:2} for $\alpha>0$ and large $\Omega/{J}_0 \to\infty$.
The driven dynamics at stroboscopic times is equivalent to the \emph{quench} dynamics governed by the effective static Hamiltonian $\hat H_{\text{eff}}$.
%
As 
\mpar{Dynamical ordering}
we reviewed in Sec.~\ref{sec_422} above concerning dynamical phase transitions, dynamical ordered phases arising from quench dynamics 
exist as long as 
the (post-quench) Hamiltonian supports long-range order at finite energy density above the ground state.
In the present case, such dynamical ordering can be interpreted as dynamically stabilized non-equilibrium ordering and is dictated by the phase structure of $\hat H_{\text{eff}}$:
Initializing the system in a state with a well-defined average polarization close enough to that characterizing an equilibrium state of $\hat H_{\text{eff}}$, its dynamical (stroboscopic) order parameter will be stable in the course of time-evolution.
As we reviewed in Sec.~\ref{sec_eq}, for one-dimensional systems ordering at finite energy density requires $\alpha \le 2$~\cite{dyson1969existence,thouless1969long,dutta2001phase}.
The character of the dynamical magnetic ordering of the system depends upon the driving amplitude $\zeta$: When the effective anisotropy parameter $\gamma(\zeta)$ is small or large enough and negative, there appear dynamically stabilized unconventional ferromagnetic phases with paramagnetic character or with magnetization in the $yz$-plane orthogonal to the direction of actual ferromagnetic interactions, respectively. The latter phase in particular has no equilibrium counterpart in the Ising model.

Upon increasing $\alpha$ up to the value $2$, quantum fluctuations modify the phase boundaries in the nonequilibrium phase diagram in Fig.~\ref{fig:2} as shown by the white arrows in Fig.~\ref{fig:NEQphasediagramAlpha}. The shift of the phase boundary can be quantitatively computed using (equilibrium) spin-wave theory, which is exact for $\alpha\lesssim 1$ and approximate for $1< \alpha\le 2$, see Eq.~\eqref{eq_shiftphaseboundary}.

\begin{figure}[t]
\centering
\includegraphics[width=0.7\textwidth]{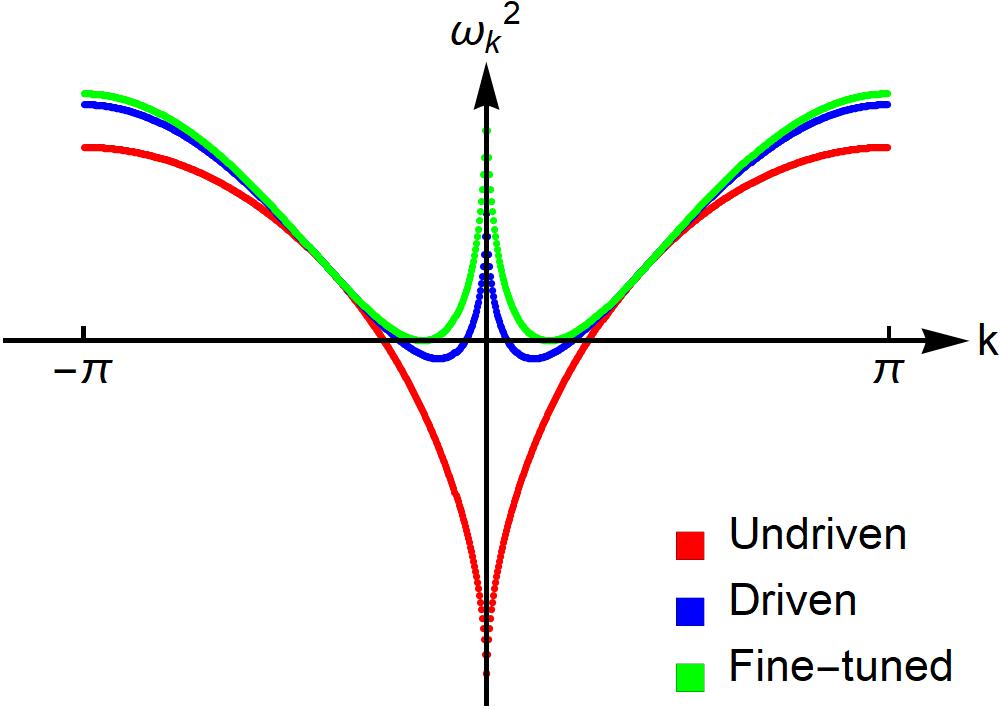} 
\caption{%
Effective spectrum of the quantum spin-wave excitations 
around the unstable paramagnetic configuration for $\alpha=1.5$,  $h_0/{J}_0=0.35$, in the presence of a high-frequency drive with $\delta h/\Omega=0$  (red), $0.4023$ (blue) and $0.6014$ (green), corresponding  to effective anisotropy parameters $\gamma=1$, $0.45$, and $0$, respectively, in the effective Hamiltonian $\hat H_{\text{eff}}$ in Eq. \eqref{eq:effXY_alpha}. 
The blue and green points correspond to parameters within the shaded region in Fig. \ref{fig:2}, in which coexistence of Kapitza and ferromagnetic phases occurs in the infinite-range model. Although the collective $k=0$ mode is dynamically stabilized, for $\alpha>0$  an extended interval in the Brillouin zone appears with modes characterized by imaginary frequencies $\omega_k^2 <0$, as shown, e.g., by the blue points. 
As shown by the green points, this instability disappears only at isolated points $\zeta_1,\zeta_2,\dots$ for which $\gamma=0$ [corresponding to the zeros of the Bessel function, see after Eq. \eqref{eq:effXY_0}], i.e., characterized by an emergent $O(2)$ rotational symmetry. Figure taken from Ref.~\cite{lerose2019prethermal}.
}
\label{fig:coexistenceband}
\end{figure}

Quantum
\mpar{Instability induced by quantum fluctuations}
fluctuations have a further, dramatic effect of the nonequilibrium phase diagram. 
In fact, the second Kapitza phase  which coexists with the ferromagnetic phases at mean-field level,  indicated by the shaded region in Fig.~\ref{fig:2},  turns out to be unstable to many-body fluctuations at finite wavelength. 
Although the driving can stabilize the collective spin,  there appears an extended interval of unstable spin modes in the Brillouin zone with finite wavelength $k\ne0$, which are expected to prevent dynamical stabilization. 
In fact, within a linear stability analysis, the effective spectrum of excitations is given by
\beq
\omega_k^2 = \big[h_0-(1-\gamma(\zeta)) \, J_0 {f}_k(\alpha)\big]  \big[ h_0-(1+\gamma(\zeta)) \, J_0{f}_k(\alpha)\big],
\eeq 
as obtained by expanding Eq.~\eqref{eq:effXY_alpha} in spin-wave operators around the paramagnetic configuration $\theta=0$.
The effective dispersion relation features a finite interval in the Brillouin zone characterized by with imaginary frequencies within the range of parameter values $h_0 < {J}_0 \, \big[ 1-\lvert\gamma(\zeta)\rvert \big]$ under consideration, see Fig.~\ref{fig:coexistenceband}. 
The amplitude of this interval shrinks to zero when the anisotropy $\gamma=\mathcal{J}_0(4\zeta)$ approaches $0$, i.e., when the driving strength $\zeta$ equals one of the zeros $\zeta_n$ with $n=1,2,\dots$ of the Bessel function. 
Away from this discrete set of values, the Kapitza phase coexisting with the ferromagnetic phases turns out to be destabilized by these finite-wavelength fluctuations, at least at the level of linear stability, in spite of the stabilization of the collective $k=0$ mode.

We 
\mpar{Stability}
remark that when $\zeta$ is tuned to an isotropic point $\zeta_n$, the many-body Kapitza phase discussed above becomes stable in the high-frequency limit $\Omega\to\infty$. 
The reason behind such stability may be easily traced back to the stroboscopic conservation of  the collective spin projection $\mathcal{S}^z$ along the field direction, due to the emergent $O(2)$ rotational symmetry. 
Indeed, if the system is initialized in a fully polarized state with a small displacement $\theta_0$ away from the $z$-axis, the collective spin has to remain trapped in a neighborhood of the fully polarized configuration $\theta=0$, because $\mathcal{S}^z(t_n) \approx 1- \theta_0^2/2$ cannot decrease.


Let 
\mpar{$\alpha > 2$}
us finally briefly comment on what happens for $\alpha > 2$. 
For $\alpha=\infty$ the long-range quantum Ising chain~\eqref{eq_LRIsingdriven} reduces to the standard quantum Ising chain with nearest-neighbor interactions (which has been studied in 
Refs.~\cite{russomanno2012periodic,bastidas2012nonequilibrium,benito2014floquet}).
 In this case, the effective high-frequency Hamiltonian \eqref{eq:effXY_alpha} describes the XY quantum spin chain, which is exactly solvable in terms of free fermions~\cite{lieb1961two}. 
 From the exact solution, we see that the quantum critical point $h_{\text{cr}}={J}_0$ is independent of $\gamma$, and thus of the driving strength $\zeta$. 
 Accordingly, it is natural to conjecture that the left boundary of the Kapitza phase  moves leftwards as $\alpha$ exceeds $1$, as shown in Fig.~\ref{fig:NEQphasediagramAlpha}, and eventually approaches the straight vertical line $h_{\text{cr}}(\zeta)={J}_0$ when $\alpha\to\infty$. 
 However, ferromagnetic ordering in not stable at finite energy density for $\alpha>2$. 
 In the nonequilibrium dynamics starting from a polarized state, domain-wall excitations melt the original magnetic ordering. 
In this case the equilibrium (ground-state) phase diagram~\cite{bastidas2012nonequilibrium,benito2014floquet} of the effective Hamiltonian does \emph{not} allow an interpretation in terms of dynamically stabilized many-body Kapitza phases. 

\summary{Kapitza phases can survive many-body long-range interactions with $0\leq \alpha<2$.}

\subsubsection{Prethermalization and heating} 
\label{sec:heating}

We address the footprint of the fast-driving nonequilibrium phase diagram on the finite-frequency dynamics upon reducing $\Omega$ down to a scale comparable with the single-particle energy scale $\tilde{J}_{0}$ of the system.
In 
\mpar{Signatures of heating}
this case, one should expect the system to eventually absorb an ever-increasing amount of energy from the drive \cite{dalessio2014long}. In order to address  this point, we initialize the system in various fully polarized states parameterized by angles $(\theta_0,\phi_0)$ on the Bloch sphere, and study the driven evolution for various values of $\alpha>0$ and driving parameters $h_0,\delta h,\Omega$ by
 numerically integrating the non-equilibrium spin-wave theory evolution equations. 
 In this formalism, heating can be monitored by studying the energy variation rate $\langle \dot H\rangle$, or simply through the depletion $\epsilon(t)$ of the collective spin polarization from its maximal value, cf. Eq.~\eqref{eq_swDensity}: In fact, heating to infinite temperature must be accompanied by deterioration of magnetic ordering, since entropy is maximized by states with low total spin, i.e. large $\epsilon$. 
 
\begin{figure}[t]
\centering
\includegraphics[width=\textwidth]{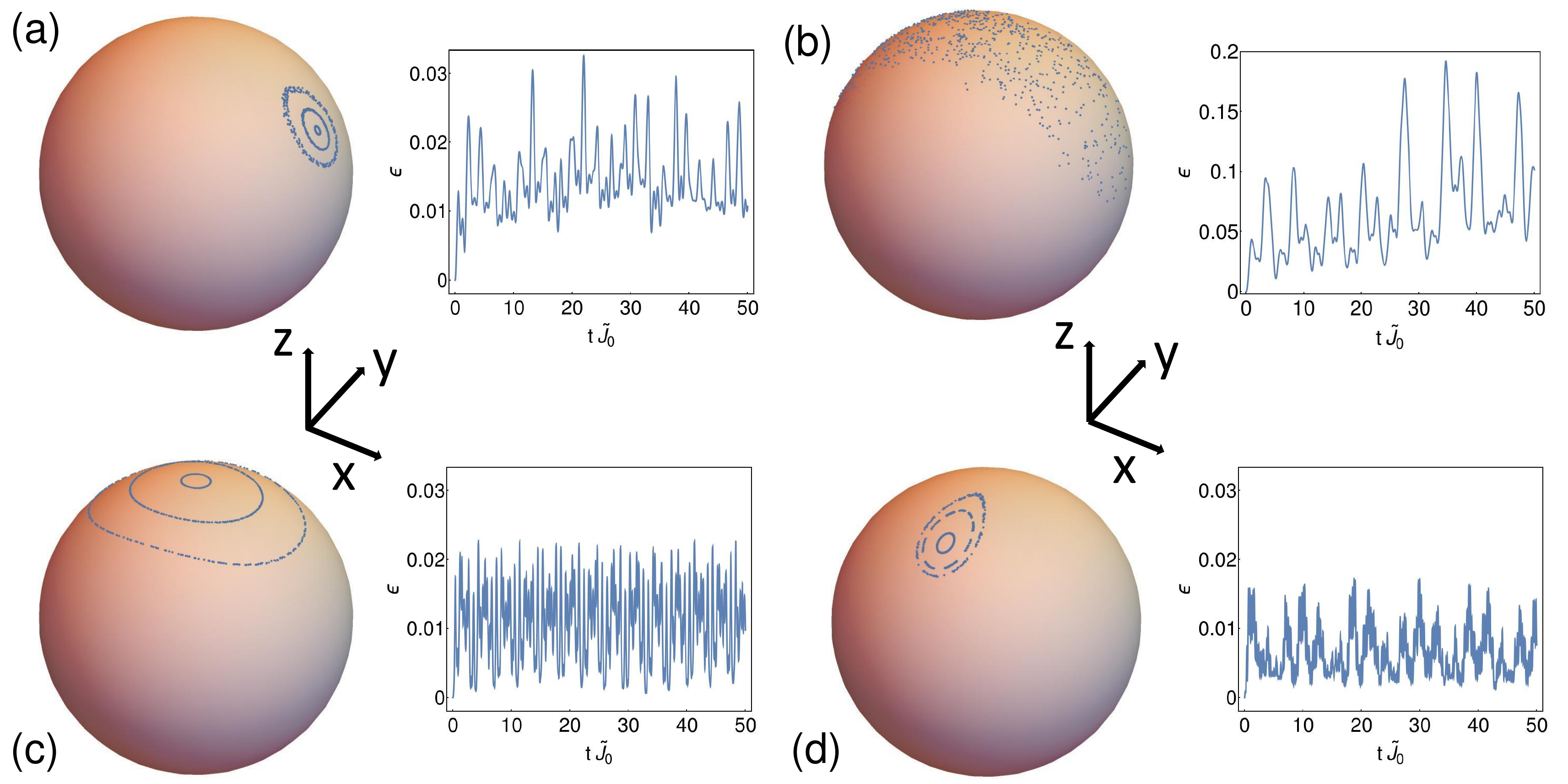} 
\caption{%
Persistence of the dynamically stabilized phases at finite driving frequency.
Left in each panel: 
Stroboscopic time-evolution $\vec{\mathcal{S}}(t_n)$ of the collective spin of the long-range Ising chains in 
Eq.~\eqref{eq_LRIsingdriven} with $\alpha\ne0$, subject to the modulated magnetic field in Eq.~\eqref{eq:protocol}. 
 $\vec{\mathcal{S}}(t_n)$ is obtained by nonequilibrium spin-wave theory and, for simplicity of visualization, is projected onto the unit sphere.
In all simulations, the static field is $h_0/{J}_0=1.2$, as in Fig.~\ref{fig:1}, the driving frequency is $\Omega/{J}_0=8$, and we used $N=100$. The system is initialized in fully polarized states in the $xz$ [panels (a), (b), (c)] and $yz$ [panel (d)] planes. 
Right in each panel: relative departure $\epsilon(t)$ of the total spin from its maximal length $N/2$, due to the dynamical generation of quantum spin-wave excitations, corresponding to the largest trajectory in each panel. 
In particular: 
(a) Dynamical ferromagnetic phase, with 
$\alpha=1$ and $\delta h/{J}_0 = 0.05$. 
(b) Fast heating in the chaotic dynamical regime, with   
 $\alpha=0.8$, $\delta h/{J}_0 = 0.2$. 
(c) Dynamically stabilized Kapitza phase, with 
 $\alpha=1$, $\delta h/{J}_0 = 5.33$. 
(d) Dynamically stabilized ferromagnetic  phase with magnetization in the $yz$-plane orthogonal to the direction $x$ of the actual ferromagnetic interactions, with $\alpha=1$, $\delta h/{J}_0 = 8$. 
Panels (a), (c), and (d) demonstrate that the dynamical phases $F_{\parallel}$, $K$, $F_{\perp}$ (see Fig.~\ref{fig:2}), respectively, continue to exist at finite driving frequency. 
The amount of excitations generated remains small and the total energy remains bounded across many cycles, qualifying these phases as \emph{prethermal}.
In panel (b), instead, heating is witnessed by the growth of $\epsilon(t)$ (notice the different vertical scale in the plot).
The heating rate in this case increases upon increasing $\alpha$. Figure taken from Ref.~\cite{lerose2019prethermal}.}
\label{fig:3ph}
\end{figure}
 
Results are reported in Fig.~\ref{fig:3ph}.
\mpar{Classical phase-space and heating}
Whenever the system is initialized in a stable or dynamically stabilized regime $P$/$K$ or $F_{\parallel,\perp}$, and the frequency $\Omega$ is off-resonant with the spin-wave band, i.e., $\Omega \gg 4 {J}_{0}$, as shown in Fig.~\ref{fig:3ph}(a),(c),(d) the evolution presents a long time interval during which the absorption of energy from the drive, as well as the amount of spin-wave excitations, is bounded. In this regime, heating is suppressed, consistent with a Floquet pre-thermal scenario.
On the other hand, whenever the system is initialized in a chaotic dynamical regime as in Fig.~\ref{fig:3ph}(b), irrespective of the value of $\Omega$ and of  $\alpha>0$, the density $\epsilon(t)$ of dynamically generated spin-wave excitations, as well as the energy $\braket{\hat H(t)}$, increase at a finite rate. 
These observables witness heating, which is expected to be the generic response of a many-body system to external periodic drive in the absence of dissipative mechanisms \cite{shirai2016effective}. 
In the dynamical regimes $F_{\parallel,\perp}$ of panels (a) and (d), the synchronized trajectories of the collective spin $\vec{\mathcal{S}}(t)$ act as an ``internal'' periodic driving at frequency $\Omega$ on the quantum oscillators $(\tilde{q}_k,\tilde{p}_k)$'s through the last interaction terms in the spin-wave Hamiltonian \eqref{eq:timeindepH}. As long as $\Omega$ is off-resonant (see above), the spin waves behave as a periodically driven system of non-interacting bosons, which relaxes to a periodic quasi-stationary state described by a stroboscopic generalized Gibbs ensemble \cite{lazarides2014periodic,russomanno2012periodic}. 
As we discussed 
\mpar{Prethermalization for $0<\alpha<d$}
at length in the context of quench dynamics, for $0<\alpha<d$ the quantum backreaction on the collective spin dynamics is suppressed with system size. The prethermal stage is thus characterized by long-lived oscillations, the duration of which diverges in the thermodynamic limit. 
For $\alpha\gtrsim d$, on the other hand, the description of prethermal dynamics is more complicated due to nonlinear interactions between the collective spin and the bosonic ``bath''; however, relaxation to a quasi-stationary state is typically very slow or absent.
The neglected nonlinear spin-wave interactions are finally expected to ultimately lead to the decay of this {prethermal} quasi-stationary state~
\cite{canovi2014first,
chandran2016integraction,
bukov2015prethermal,
weidinger2017floquet,
citro2015dynamical,
d2013many}. 
The 
\mpar{Slow heating for $\alpha> d$}
fastest heating processes are associated with the absorption of a quantum of energy $\sim \Omega$ from the drive through a high-order resonant transition involving $\Omega/J_0$ elementary local transitions. According to by now standard theoretical arguments~\cite{abanin2015exponentially,abanin2017effective,mori2016rigorous,machado2020long}, the associated heating time scale is expected to scale exponentially as 
$\tau \sim \exp (\, \text{const} \times \Omega/{J}_{0})$.

In conclusion, we note that several numerical studies of quench dynamics in long-range interacting chains with $2<\alpha\ll \infty$ suggested that magnetic ordering survives for surprisingly long times in the prethermal regime~\cite{zunkovic2016dynamical, vanderstraten2018quasiparticles, halimeh2017dynamical, liu2019confined}. This occurrence has been recently analytically understood in terms of a suppressed rate of formation of unbound domain walls~\cite{collura2022discrete}. Furthermore, the expected functional form of the lifetime of such dynamical long-range ordering as a function of the driving protocol and of $\alpha$ has been determined, predicting the possibility of having extremely long-lived order even at finite driving frequency~\cite{collura2022discrete}. When conditions for this phenomenon are met, signatures of dynamically stabilized ordered phases are expected to emerge even for $\alpha>2$.

\summary{Due to the suppression of spin-wave populations in ordered phases, strong long-range interactions prevent heating and rather lead to prethermal regimes. In periodically driven systems, the occurrence of heating for $\alpha>0$ depends on the nature of phase-space trajectories in the classical limit $\alpha=0$: For regular dynamics, the system exhibits a Floquet-prethermal regime with suppressed heating, whereas, for chaotic trajectories, the system displays heating with fast absorption of energy.}

 \subsection{ Discrete time crystals}
 \label{sec_tc}

The concept of spontaneous breaking of (continuous) time-translational invariance in quantum many body systems has been brought to widespread attention in Ref.\,\cite{wilczek2012quantum}. Soon after, these non-equilibrium phases were proven impossible at equilibrium\,\cite{bruno2013impossibility,watanabe2015absence}. Yet, \emph{discrete} time translational invariance, realized in periodically driven systems, can be spontaneously broken\,\cite{sacha2015modeling,else2016floquet,khemani2016phase}. Thus,
the term ``discrete time crystals" (DTC) refers to systems where the discrete time-translation symmetry, encoded in the periodically driven Hamiltonian $\hat H(t) = \hat H(t+T)$, is spontaneously broken. Expectation values of relevant observables exhibit oscillations with a period that is an integer multiple of $T$. Several experimental observation of DTC have been discussed in the last decade\,\cite{bordia2017periodically,zhang2017observationdtc,choi2017observation,rovny2018observation}.
For a general overview of these research efforts, we refer the readers to recent reviews~\cite{Sacha17review,khemani2019brief,zaletel2023colloquium}.

Following Ref.\,\cite{russomanno2017floquet} we say that a DTC phase exists if, for a class of states $\ket{\Psi}$ with short-ranged connected correlations \cite{else2016floquet}, there always exists an observable $\hat{O}$ such that the time-evolved expectation value in the thermodynamic limit $N\to\infty$,
satisfies the following conditions:
\mpar{Definition of discrete time crystal}
\begin{enumerate}
\item \emph{Time-translation symmetry breaking}: $\langle \hat O(t+T)\rangle \neq \langle \hat O(t)\rangle$, even though $\hat H(t) = \hat H(t+T)$, so that long-range correlated Floquet eigenstates of the propagator $\hat U_F = \hat U(t+T,t)$ exist\,\cite{else2016floquet}.
\item \emph{Rigidity}: the periodic oscillations of $\langle \hat O(t)\rangle $, with a period $\tau$, shall persist in a whole finite and connected region of the Hamiltonian parameter space.
\item \emph{Persistence}: the periodic oscillations of $\langle \hat O(t) \rangle$ become stable at long time in the thermodynamic limit $N\to\infty$.
\end{enumerate}
These conditions cannot be fulfilled by a local many-body quantum system due to the presence of external driving, which would lead to relaxation towards an infinite-temperature state, thereby preventing long-lived oscillations. To protect ordering against relaxation, a mechanism is required to control the impact of dynamically generated excitations.

Pre-thermal 
\mpar{DTC in long-range models}
stability can be achieved through long-range interactions, which are known to generate metastable states with lifetimes that grow as the system approaches the thermodynamic limit, see Secs.\,\ref{sec_metastability} and Sec.~\ref{sec_421}. Then, it is natural that the investigation of DTCs in clean systems has been primarily focused on long-range interacting models where the robustness of collective oscillations in presence of periodic drive is guaranteed.  Accordingly, stable DTC phases can only be found for $\alpha < d$\,\cite{russomanno2017floquet, surace2019floquet, munozarias2022floquet, kelly2021stroboscopic}, while for $\alpha > d$, the lifetime of oscillations is expected to be finite in the $N \rightarrow \infty$ limit\,\cite{machado2020long, pizzi2020time,collura2022discrete}. The $\alpha=0$ Ising model is a privileged playground for time-translational symmetry breaking, since it features a kaleidoscope of different DTCs. Indeed, the presence of $p$-order DTC phases (with a period of $pT$, where $p$ is an integer) which were first witnessed in specifically designed $\mathbb{Z}_p$-connected states\,\cite{russomanno2017floquet, surace2019floquet} has recently been detected as well in the standard $\mathbb{Z}_2$ symmetric Ising model~\cite{giergiel2018time, kelly2021stroboscopic}, where $p$ can be fractional and signatures of this behavior persist for finite $\alpha$ and in the classical limit\,\cite{pizzi2021discrete, pizzi2021classical1, pizzi2021classical2}.

\subsubsection{ Mean-field DTC }

In the mean-field limit $\alpha=0$ it is possible to obtain an analytic solution for the periodic dynamics and establish the simplest instance of DTC~\cite{russomanno2017floquet}. 
We consider a step-wise drive of the magnetic field  in Eq.\,\eqref{eq_lmgdriven} of the form
\begin{equation}
\label{drive}
    h(t) = \psi \sum^{\infty}_{n=1} \delta(t-nT) \ ,
\end{equation}
of amplitude $\psi$, and we focus on the evolution of the order parameter $m^a (t) = \frac{1}{N} \sum_{j} \langle\hat{\sigma}_j^a\rangle$ (where $a = x,y,z$), i.e. the components of the magnetization of the system. 

First, we observe that the Floquet operator (see \ref{app_Magnus}) can be expressed as the product of two distinct operators:
\begin{equation} \label{Uf}
\hat U_F = e^{- 2 i \psi \hat{S}_z} e^{i J_0 T \hat{S}_x^2/N} \ ,
\end{equation}
where we used the global spin operators notation defined in Eq.\,\eqref{Stot}.
The term $\exp(-2i\psi \hat{S}_z)$ in Eq.\,\eqref{Uf}, which acts as a rotation around the $z$-axis, represents the effect of the kick term on the observable $\vec{m}$. The other term describes the evolution of $\vec{m}$ induced by the second term on the right-hand side of Eq. \eqref{Uf} over one period $T$. The Heisenberg equations of motion corresponding to this evolution for the operators $\hat{S}_a$ are:
\begin{align} 
\label{HeisS}
\frac{d}{dt}\hat{S}_x &= 0 \,, \\
\frac{d}{dt}\hat{S}_y &= \frac{J_{0}
}{N} \left(\hat{S}_x \hat{S}_z + \hat{S}_z \hat{S}_x \right) \,, \\
\frac{d}{dt}\hat{S}_z &= -\frac{J_{0}
}{N} \left(\hat{S}_x \hat{S}_y + \hat{S}_y \hat{S}_x \right) \,.
\end{align}
As usual, due to the mean field nature of the problem we can neglect the spin-spin correlations in the thermodynamic limit\,\cite{sciolla2013quantum}. Taking averages on both sides of Eq.\,\eqref{HeisS} and using the decoupling relations $\braket{\hat{S}_a\hat{S}_b}\simeq \braket{\hat{S}_a}\braket{\hat{S}_b}$ one obtains the closed set of equations for the magnetization:
\mpar{Classical equations of motion}
\begin{align}
\label{HamM}
\dot{m}_x = 0 \ ,\quad
\dot{m}_y = J_{0}
 m_x m_z \ , \quad
\dot{m}_z = - J_{0}  m_x m_y .
\end{align}
As a consequence, after a time interval $T$, $\vec{m}$ undergoes a clockwise rotation around the $x$-axis by an angle of $J_{0} T m_x(t)$. The $\mathbb{Z}_2$ symmetry of the model is encoded in the dynamical symmetry $\psi \rightarrow \psi + \pi/2$ and $\vec{m}_n \rightarrow R_z (\pi n) \cdot \vec{m}_n$ in Eq. \eqref{HamM}.
Integrating out the equation of motions in Eq.\,\eqref{Uf}, the evolution of the observable $\vec{m}$ is given by
\begin{align}
\label{map2}
 \vec{m}_{n+1} = f(\vec{m}_n) \equiv R_z (2 \psi) \cdot R_x (-J_{0}Tm_{x,n}) \cdot \vec{m}_{n},
\end{align}

Due to the periodic nature of the drive, the map in Eq.\,\eqref{map2} displays an Hamiltonian structure and its action preserves the area of the region on the sphere $|\vec{m}|^2 = 1$ span by the dynamics. Accordingly, one can employ polar coordinates along the $z$-axis, $\vec{m} = (\sin \theta \cos \phi, \cos \theta \cos \phi, \cos \theta)$ in order to express an area element as $dS = d \cos \theta d \phi$. Then, the coordinates $\phi$ and $I = \cos \theta$ serve as natural canonical conjugate variables for our system. Following the discussion in Ref.\,\cite{sciolla2013quantum} the action $I$ can be regarded as the $z$-component of the angular momentum and $\phi$ as the rotation angle around the same axis.

Them, in the small period limit $T\to 0$ the map can be rewritten as
\begin{align}
\label{small_period_map}
I_{n+1} &= I_{n} \, , \\
\phi_{n+1} &= \phi_n + 2 \psi \, ,
\end{align}
with initial conditions $I_0 = 0$ and $\phi_0 = \pi/2$. This corresponds to the Poincaré map obtained by taking stroboscopic section of the integrable dynamics. 
\mpar{Quasi-periodic motion}
In other terms the motion of the order parameter $\vec{m}_n$ at vanishing drive periods is quasi-periodic with a period $\pi/\psi$. Slightly increasing the strength of the kicking period $T$ the map in Eq.\,\eqref{small_period_map} is perturbed and the fate of the system follows the Kolmogorov-Arnold-Moser theorem\,\cite{kolmogoro1954conservation,arnold1963proof,moser1962invariant}. The theorem states that small perturbations in the form of Eq.\,\eqref{drive} only slightly deform the the torus $I = \text{const}$ at least as long as the drive frequency is not resonant. Thus, the motion remains quasi-periodic for drive strength $\psi$ far enough from a rational multiple of $\pi$. However, as soon as a resonance is approached and $\psi \approx \psi_{r} \equiv r \pi$ with $r=q/p$ and $p$ and $q$ are coprime integers,  pairs of elliptic and unstable fixed points emerge in the dynamics due to the Poincare-Birkhoff theorem\,\cite{brown1977proof}.

Then, 
\mpar{$p$-iterated map}
distinct regions in the phase space $(I, \phi)$ can be distinguished depending on the action of the $p$-iterated map $f^p(\vec{m})$, which also correspond to different $\vec{m}_n$ evolutions. Quasi-periodic behaviour persists for initial conditions $(I_0, \phi_0)$ far enough from the fixed points, where a rotation dynamics occurs with $\phi$ periodically spanning the interval $[0,2\pi]$. On the other hand, as the initial conditions $(I_0, \phi_0)$ approach the fixed points, a libration dynamics arises and $\phi$ continuously oscillates around a finite value. As a result, successive map iterations do not substantial alter the magnetization value $\vec{m}_{n+p}\approx \vec{m}_n$ and a DTC phase appears. Finally, the boundary between the DTC and the quasi-periodic regimes are occupied by chaotic regions, which grow and eventually take over the regular ones at large $T$.

\summary{The classical limit of the fully-connected model displays periodic motion in the $T\to0$ limit. Close to resonances, quasi-periodicity is broken but the Poincare-Birkhoff theorem leads to time-crystalline behavior, characterised by time-translational symmetry breaking of the magnetization, stable to small perturbations in the kicking strength. }

\subsubsection{Finite-size and finite-range effects}

The discussion above has been based on the semiclassical analysis which becomes exact in the thermodynamic limit. Yet, it is interesting to perform some numerical simulations at finite $N$ to validate the large-$N$ picture. At each finite size the modulus of the total spin $\hat{S}$ of the system is conserved restricting the dynamics to the subspace of constant $\hat{S}^2 = S (S+1)$, with $S=N/2$. Then, it is relatively straightforward to perform exact diagonalization up to large sizes ($N= 800$)\,\cite{ribeiro2008exact,russomanno2017floquet}. To visualize the eigenstates in this subspace, 
\mpar{Spin coherent representation of the eigenstates}
we introduce the spin coherent states\,\cite{auerbach1994interacting} 
\begin{align}
\label{coherent_states}
    \ket{\Omega (\theta, \phi)}  = e^{-i \vec{n} \cdot  \vec{\hat{S}}} \ket{\boldsymbol{\Uparrow}},  
\end{align}
where $\ket{\boldsymbol{\Uparrow}}$ is the eigenstate corresponding to the maximum projection of the spin along the $z$ direction and $\vec{n} = (\sin \theta \cos \phi, \sin \theta \sin \phi, \cos \theta)$. The overlap between different coherent states remains finite at finite $N$ and reads
\begin{equation}
    \braket{\Omega (\theta, \phi)|\Omega (\theta + \Delta \theta, \phi+ \Delta \phi)} )= \left( \sin \frac{\Delta \theta}{2} e^{-i \Delta \phi} \right)^{2S}, 
\end{equation}
which vanishes in the $N \rightarrow \infty$ limit due to the exponent $S$. However, for any finite $N$ the states in Eq.\,\eqref{coherent_states} form an overcomplete basis for the Hilbert space. Then, one can characterize the dynamics by estimating the projection $|\braket{\Omega (\theta, \phi)|\eta_m}|^2$ for various different Floquet eigenstates $\ket{\eta_m}$.

\begin{figure*}[t!]
\centering
\subfloat[Time crystal]{\label{Fig_SecV_1a}\includegraphics[width=.47\textwidth]{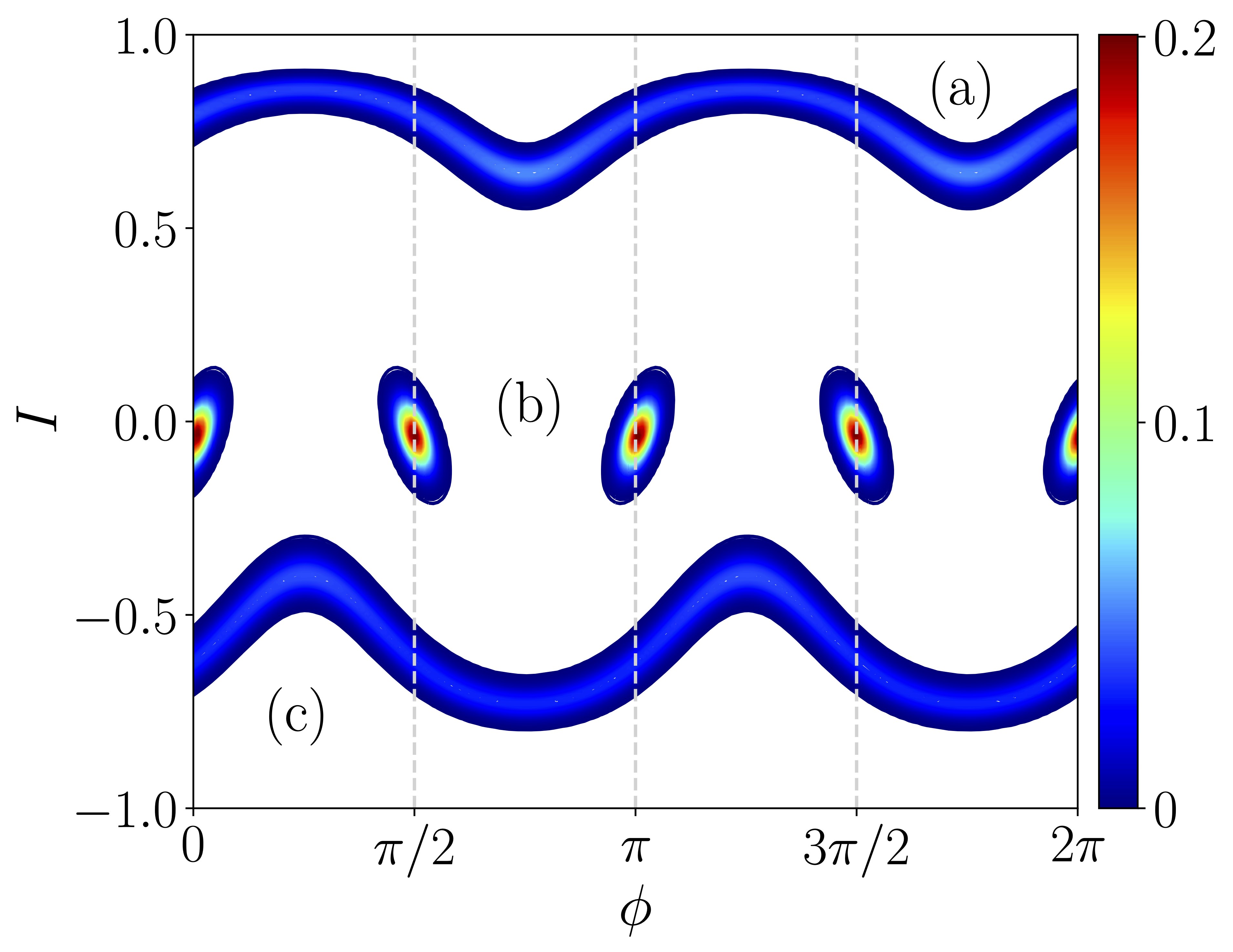}}\,\,
\subfloat[Chaotic phase]{\label{Fig_SecV_1b}\includegraphics[width=.49\textwidth]{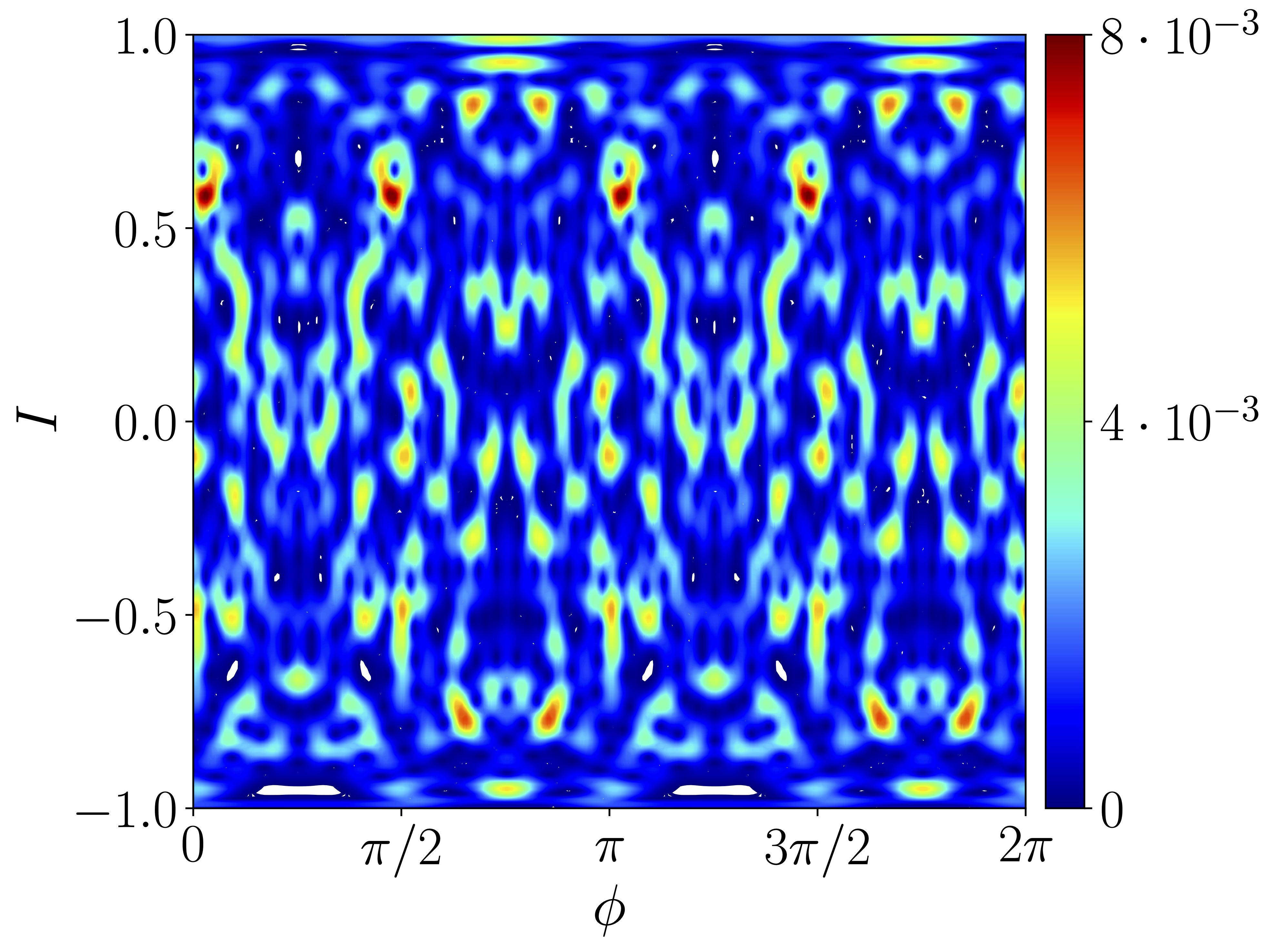}}
\caption{\label{Fig_SecV_1} 
\textbf{Eigenstate structure.} The eigenstate stracture radically changes between the three different phases of the system: while no recognizable pattern is present in the chaotic phase, panel (b), in the quasi-periodic phase the eigenstate is localized in a connected region of the $(I,\phi)$ space (panel (b), curves (a) and (c)), while in the $p=4$ DTC phase it appears localized around four, $\mathbb{Z}_4$ symmetric, points (panel (a): curve (b)).  Image adapted from Ref.\,\cite{giachetti2022highorder}. }
\end{figure*}

The 
\mpar{Eigenstate localization in phase-space}
overlap $|\braket{\Omega (\theta, \phi)|\eta_m}|^2$ for different values of $m$ are shown in Fig.\,\ref{Fig_SecV_1}. The Floquet eigenstates in the $p=4$ DTC phase appear clearly localized around four $\mathbb{Z}_4$ symmetric points, see Fig.\,\ref{Fig_SecV_1a}, while this is no longer the case in the quasi-periodic phase, see Fig.\,\ref{Fig_SecV_1b}. This behavior can be explained semi-classically: close to a resonance, the Floquet evolution can be interpreted as a hopping between $p$ adjacent wells in the classical phase space\,\cite{giergiel2018time}, so that the Floquet eigenstates have the form of tight-binding Bloch wavefunctions. A similar behavior for the $p=2$ case (around $\psi = \pi/2$) has been observed in Ref.\,\cite{russomanno2017floquet}. Let us notice that, given the initial condition chosen in the present study, in the $N \rightarrow \infty$ limit, the only eigenstate which contributes to the dynamics, will be the one with a non-zero overlap with the point $\theta = 0$, $\phi=0$, which in turn can correspond to each of the three phases.

The current picture is not substantially altered by the inclusion of quantum fluctuations due to a finite value of $\alpha$ or by additional local couplings. Indeed, the structure of the low-$T$ DTC regions with $p=2$ is generally resilient to quantum fluctuations. On the other hand, sufficiently high values of $\alpha$ enhance the chaotic phase, leading to the disruption of the DTC phases with $p>2$ for large enough values of the drive period $T$, see Ref.\,\cite{giachetti2022highorder} and Fig.\,\ref{Fig_SecV_2b}.

\summary{
The exact eigenstates for finite system size reflect the time crystalline behaviour of the mean-field result. Their phase-space representation shows localization around symmetric points in the time-crystal phase, while they are delocalized in the chaotic one. 
}

\subsubsection{Order parameter}

 The dynamical phase diagram of different high-order DTC phases exhibits intricate self-similar and fractal structures, where the regular phases are intertwined with the chaotic and quasi-periodic regions.
The characterization of the entire dynamical phase diagram has long remained a difficult challenge. But the introduction of a novel order parameter enabled a comprehensive characterization of DTC phases, irrespective of their order\,\cite{giachetti2022highorder}. 

The 
\mpar{Order parameter}
key to introduce a useful order parameter is to consider different values of the amplitude, $\psi$, $\psi + \delta \psi$, which amounts to consider two nearby initial conditions in the phase space. Then, we introduce the definition
\begin{align}
\label{order_parameter}
 \zeta^2 = \frac{1}{n_{\rm max}} \sum^{n_{\rm max}}_{n=0} \left( m_{x,n} (\psi + \delta \psi) -  m_{x,n} (\psi) \right)^2.
\end{align}
 Both in the DTC phase and in the quasi-periodic one the evolution is not chaotic, so that the two nearby trajectories $m_{x,n} (\psi + \delta \psi)$ and  $m_{x,n} (\psi)$ diverge polynomially in time. Expanding the latter equation for small deviations $\delta \psi$ yields
\begin{equation} \label{zetareg}
    \zeta^2 = \frac{1}{n_{\rm max}} \sum^{n_{\rm max}}_{n=0} \left( m_{x,n} (\psi + \delta \psi) -  m_{x,n} (\psi) \right)^2 \sim \frac{\ell}{n_{\rm max}} \sum^{n_{\rm max}}_{n=0} \delta \psi^2 n^2   \sim \ell (\delta \psi n_{\max})^2 \ ,
\end{equation}
where $\ell$ depends on the average distance between two randomly chosen points of the two nearby trajectories 

The value of $n_{\rm max}$ has to be large enough so that the rightmost term in Eq.\,\eqref{zetareg} remains $O(1)$, i.e. $n_{\rm max} \rightarrow \infty$ as $\delta \psi \rightarrow 0$. However, the value of $\ell$ jumps discontinuously between the libration regime (corresponding to a DTC phase) and the rotation one (corresponding to a quasi-periodic phase). Indeed, close to the fixed point of the iterated map the micro-motion becomes negligible and $\zeta \rightarrow 0$, signalling the emergence of the pure time-crystalline regime. The value $\zeta$ in the two phases is not universal and depends on the value of $ (n_{\rm max}  \delta \psi)\sim O(1)$.
The jump in the value of $\ell$ results in a discontinuity in $\zeta$, which may be observed in the numerical distribution of $\zeta$. Indeed, the order parameter in the DTC is sharply peaked around $\zeta = 0$, but becomes negligible for $\zeta \gtrsim 0.2$. The quasi-periodic phase is signalled by a peak at $\zeta \sim 0.36$, which appears disconnected from the DTC peak at $\zeta = 0$.  The exponential divergence of trajectories in the chaotic phase leads to the memory loss of initial conditions on a time-scale $n_{\rm max} \sim -\log(\delta \psi)$, making the values $m_{x,n} (\psi)$ and $m_{x,n} (\psi + \delta \psi)$ to become two equally distributed random variables with zero mean. Then, due to the central limit theorem, $\zeta^2$ is distributed as a Gaussian in the chaotic phase \begin{equation}
    \langle\zeta^2\rangle = 2 \langle m_{x}^2\rangle 
\end{equation}
and variance $O(n_{\rm max}^{-1})$.  For an isotropic system one can easily derive the peak value for the distribution, since $|\vec{m}|^2 = 1$ one has 
\begin{equation}
    \langle m_{x}^2\rangle = \frac{1}{3} \langle|\vec{m}|^2\rangle = \frac{1}{3} \ ,
\end{equation}
so that $\langle\zeta^2\rangle = 2/3$. 

\begin{figure*}[t!]
\centering
\subfloat[Phase diagram $\alpha=0$]{\label{Fig_SecV_2a}\includegraphics[width=.49\textwidth]{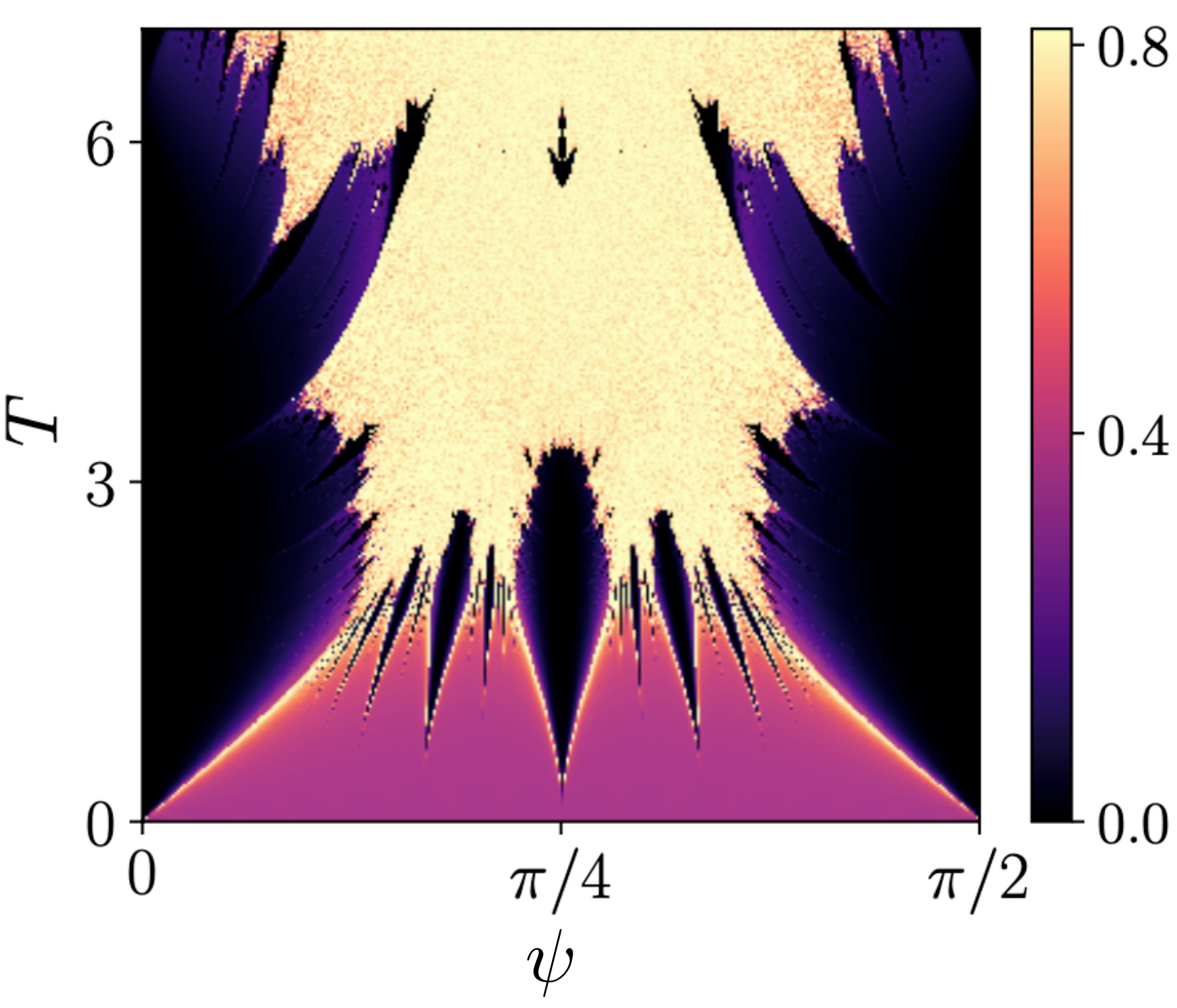}}\,\,
\subfloat[Phase diagram $\alpha=0.5$]{\label{Fig_SecV_2b}\includegraphics[width=.49\textwidth]{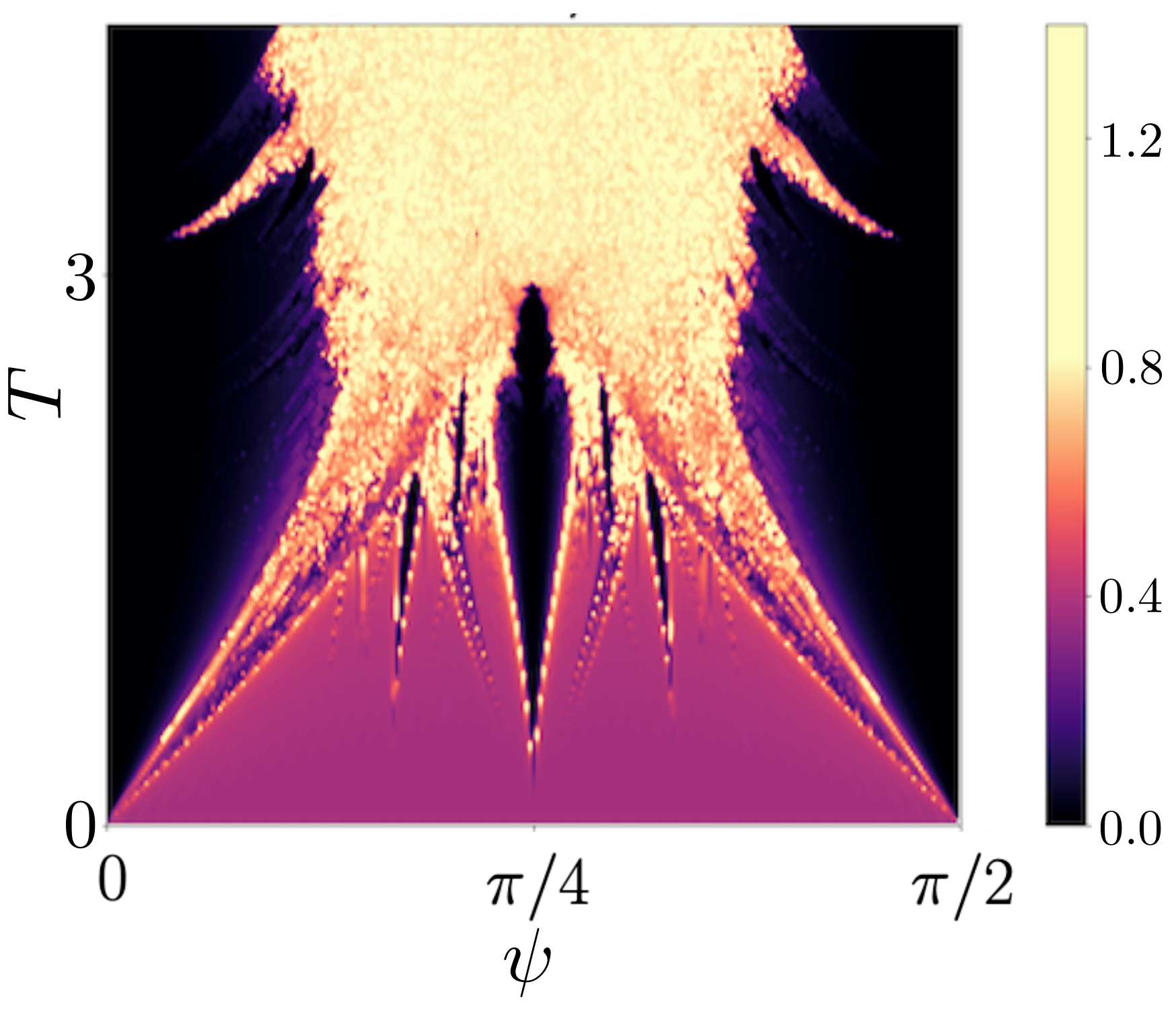}}
\caption{\label{Fig_SecV_2} 
\textbf{Phase diagram.} Panel (a): Color plot of the order parameter $\zeta$ as a function of the amplitude $\psi$ and the period $T$ of the drive, saturated at the value $\zeta=\sqrt{2/3}$, with $n_{\rm max}=300$ and $\delta\psi = 1.6\cdot 10^{-3}$. Panel (b): same as (a) but for $\alpha=0.5$. Image adapted from Ref.\,\cite{giachetti2022highorder}. }
\end{figure*}

Thus, the order parameter $\zeta$ can be used to detect higher-order DTC phases in clean long-range systems, by exploiting the connection between DTC and Poincar\'e-Birkhoff theorem\,\cite{brown1977proof,wisniacki2011poincare}, which rigorously holds in the mean-field $\alpha=0$ limit. Indeed, the phase diagram obtained by the numerical characterization of the order parameter, see Fig.\,\ref{Fig_SecV_2} reproduces and expand the known properties of the DTC phases in the $\alpha=0$ limit.\,\cite{russomanno2017floquet,pizzi2021discrete} 

The 
\mpar{Example in the fully-connected model}
symmetry of the phase diagram around the $\psi = \pi/4$ axis arises from the dynamical $\mathbb{Z}_2$ symmetry, which is a notable feature that would have remained undetectable with a $p$-dependent order parameter. At low values of $T$, the quasi-periodic phase dominates (pink area $\zeta\approx 0.4$), while small islands of the DTC phases emerge around specific values of $\psi$, corresponding to rational multiples of $\pi$ (black areas $\zeta\approx 0$). Initially, the size of these islands increases with increasing $T$, and as they approach each other, chaos begins to emerge along their boundaries (yellow area, $\zeta\approx \sqrt{2/3}$. Ultimately, all islands associated with DTC of order $p>2$ are engulfed by the chaotic phase, with the largest (central) one corresponding to $p=4$ surviving the longest. Interestingly, at certain values of the driving period, we observe a revival of the higher-order DTC phases, particularly pronounced for $p=4$ (small, arrow shaped, black area at high $T$ for $\psi\approx \pi/4$ in Fig.\,\ref{Fig_SecV_2a}).

The boundary between the chaotic and DTC phases is not smooth; instead, it exhibits self-similar patterns that repeat at increasingly smaller scales. The emergence of this fractal scaling in the boundaries of time-crystalline phases draws a direct analogy with similar phenomena observed in traditional critical systems, particularly percolation, self-avoiding random walks, and the Potts model\,\cite{hastings2002exact, duplantier2000conformally}, where a rigorous connection between conformal invariance and stochastic evolution has been established\,\cite{kager2004guide, cardy2005sle}. As previously noted in Ref. \cite{munozarias2022floquet}, the formation of DTC islands can be comprehended within the framework of area-preserving maps\,\cite{mackay1983renormalization}, specifically linked to the existence of Arnold tongues\,\cite{cencini2010chaos, collado2021emergent}. \\

We conclude this Subsection with the remark that, similarly to the Kapitza phases discussed in Sec.~\ref{sec_kapitza}, one can straightforwardly extend the analysis to the variable-range model with $\alpha>0$ using the non-equilibrium spin-wave theory of Refs.~\cite{lerose2018chaotic,lerose2019impact,lerose2019prethermal} reviewed above in Sec.~\ref{sec_421}, see e.g. Fig.~\ref{Fig_SecV_2b}.
For details, we refer the reader to Refs.~\cite{pizzi2021discrete,giachetti2022highorder}.

\summary{Higher-order time-crystals can be detected via an order parameter which accounts for the distance between nearby trajectories. The resulting phase-space displays a fractal pattern.}

\section{Conclusions and perspectives} 
Let us first summarize the salient features that we discussed in this Report; hence, we will mention aspects which have not been covered here; finally, we will point out pending problems that it would be interesting to explore.

\medskip
 {\bf{What we discussed}.\,}
In this Report, we provided a comprehensive pedagogical overview of non-equilibrium phenomena arising in the dynamics of non-random long-range interacting systems. For the sake of definiteness and connection with the theoretical and experimental literature, we took the XY quantum spin model with power-law decaying interactions with exponent $\alpha$. Our primary focus was on the \emph{strong long-range interactions} regime $0<\alpha\lesssim d$, intermediate between the mean-field limit $\alpha=0$ and the quasilocal regime $\alpha \gg d$. It is in this regime that the most surprising and unusual features of out-of-equilibrium dynamics appear. 
Our discussion was divided into three primary setups: Dynamics at low energies; Quantum quenches far away from equilibrium; Periodic driving. 

Sec.\,\ref{sec_eq} - \emph{Equilibrium}: we started by providing a concise but rather exhaustive summary of equilibrium properties exhibited by variable-range ferromagnetic spin Hamiltonians. This encompassed a discussion on the equilibrium phase diagram, the mean-field solution, the expansion in quantum fluctuations, and unusual spectral properties such as discreteness and divergent propagation velocity. 

Sec.\,\ref{sec_3} - \emph{Low-energy dynamics}: we delved into near-equilibrium dynamics in a variety of setups. 
Here, the discrete spectrum of the strong long-range regime induces unusual equilibration dynamics, universal defect formation, and the emergence of non-analytic behavior in the fidelity. 

Sec.\,\ref{dyn_high_exc} - \emph{Quantum quench dynamics}: we introduced a formalism to treat the coupled dynamics of semiclassical collective observables and the dynamics of quantum fluctuations. This allowed us to describe dynamical criticality, quantum information scrambling, and to formulate a squeezing-induced picture for the growth of entanglement.

Sec.\,\,\ref{sec_5} - \emph{Periodic driving}: Finally, we described how strong long-range interactions prevent periodic driving from heating the system or inducing thermalization. Instead, they allow dynamical stabilization of novel phases and time-crystalline behavior.

\medskip
 {\bf What we  \emph{did not} discuss.\,} 
The present memoir covers only a small portion of the varied and lively field of quantum dynamics with long-range interactions. Let us mention (in a non-exhaustive manner) some complementary studies which have not been discussed here.

A large bulk of  literature addresses \emph{mathematically and via quantum-information approaches} the impact of long-range interactions onto the spreading of correlations\,\cite{hastings2006spectral, schachenmayer2013entanglement, hazzard2013far, lashkari2013towards, eisert2013breakdown, sweke2019lieb, storch2015interplay, rajabpour2015quantum, foss-feig2015nearly, matsuta2017improving, chen2019finite, hermes2020dimensionality, else2020improved, guo2020signaling, tran2019localityheating, tran2021optimal, tran2021lieb, tran2022hierarchy}. Interestingly, several of these results pointed out how long-range interactions may not enhance correlation spreading, but remain ``shielded" in dynamical evolution\,\cite{gong2014persistence, cevolani2015protected, foss2015nearly, santos2016cooperative,  luitz2019emergent, kuwahara2020strictly, bhakuni2021suppression}, as partly discussed in Section\,\ref{sec_31}.

While our Report concerns spin systems, a lot of work aims at understanding the dynamics in the presence of \emph{fermionic} long-range Hamiltonians. These studies have been initiated at equilibrium\,\cite{vodola2014kitaev, vodola2015long, maghrebi2016causality, hernandez2017correlation, viyuela2018chiral, pezze2017multipartite, wang2018effective,botzung2021effects} but soon moved into the out-of-equilibrium regime\,\cite{vanregemortel2016information, lepori2017singular, jaschke2017critical, giuliano2018current,defenu2019universal, uhrich2020out, mishra2020disordered, solfanelli2023quantum}. Studies on \emph{bosonic} systems with long-range interactions have a much longer tradition which dates back to the characterization of the thermodynamic and finite-size scaling exponents both in the weak\,\cite{fisher1972critical, sak1973recursion,dutta2001phase,defenu2017criticality} and strong\,\cite{botet1983large,dusuel2004finite,dusuel2005finite} long-range regions. In this perspective, the critical properties of long-range interacting spin-systems have often been obtained through a bosonic theory analogue, especially in the case of spin systems with continuous symmetries\,\cite{maghrebi2017continuous,giachetti2021berezinskii,halimeh2021quantum, marino2022universality}. We have been making use of this analogy, see Sec.\,\ref{sec_34}, where we employed the spherical model to describe dynamical phase transitions\,\cite{weidinger2017dynamical,syed2021dynamical}, but the correspondence extends also to the study of correlation spreading\,\cite{cevolani2016spreading}, prethermalization\,\cite{schutz2013cooling, schutz2014prethermalization, schutz2015thermodynamics} and defect formation\,\cite{chandran2013equilibration, podolsky2014from, silvi2016crossover}.

Our choice to focus on ferromagnetic long-range couplings was mainly motivated by the necessity to evidence those phenomena which are unique to algebraically decaying interactions. On the other hand, \emph{antiferromagnetic} long-range interactions enable phenomena such as structure formation\,\cite{garel1982phase} or frustration\,\cite{chiocchetta2021cavity}, which are also found in systems with finite range interactions. Yet, antiferromagnetic couplings and structural phase formation are common to many experimental platforms, such as trapped ions\,\cite{morigi2004eigenmodes, fishman2008structural, cormick2012structural, silvi2013full, nigmatullin2016formation, birnkammer2022characterizing,kranzl2022observation}, dipolar gases\,\cite{pupillo2008cold, lu2012quantum, cartarius2014structural, natale2019excitation, tanzi2019observation, bottcher2021new} and cold atoms into cavities\,\cite{fernandez-vidal2010quantum,larson2008mott,himbert2019mean}. In these experimental platforms, structure formation phenomena get reshaped by the interplay between long-range interactions, local effects and dissipation, leading to dynamical self-organization\,\cite{fishman2008structural,cormick2012structural,baltrusch2012quantum,silvi2014ab,cartarius2014structural,landig2015measuring,keller2018quenches,landini2018formation,landig2015quantum}, Mott-insulating phases\,\cite{larson2008mott,fernandez-vidal2010quantum,habibian2013bose,dogra2016phase}, supersolids\,\cite{henkel2010three,mottl2012roton,lu2015stable,leonard2017supersolid,tanzi2019supersolid,sohmen2021birth,hertkorn2021density}, dynamical phase transitions\,\cite{brennecke2013real,landig2015measuring,morales2017couplings,leonard2017monitoring,halimeh2021dynamical}, localization and glass physics\,\cite{lesanovsky2013kinetic,habibian2013bose,brunelli2018experimental,sierant2019many, himbert2019mean}, prethermal ordering in presence of driving~\cite{jin2022prethermal}. Some properties on antiferromagnetic long-range interactions may be obtained by their ferromagnetic counterparts, but at a different $\alpha$\,\cite{mendoza2015nature,mendoza2017quantum,diessel2022generalized}. 

A significant part of the research on quantum long-range interactions focuses on \emph{disordered} systems. These studies range from the effect of long-range interactions on many-body localization \cite{deng2009anomalous, monthus2010anderson, pino2014entanglement, 
yao2014many, burin2015localization, Hauke2015many, monthus2016many, wu2016understanding, nandkishore2017many, safavi-naini2019quantum-1, safavi-naini2019quantum-2, roy2019self, sierant2019many, kloss2020spin} (explored experimentally~\cite{smith2016many, brydges2019probing,zu2021emergent}), or features of glassiness in bosonic quantum systems \cite{cugliandolo2022quantum}, e.g. in the quantum Sherrington-Kirkpatrick (SK) model --- and chaotic and holographic properties of fully-connected disordered fermionic systems, à la Sachdev-Ye-Kitaev (SYK)~\cite{ chowdhury2022sachdev}. 
Another class of disordered systems are \emph{random quantum circuits}, which make it possible to study the dynamics through exact or hydrodynamic solutions \cite{fisher2023random}. This is true also in the case of long-range interactions \cite{nahum2021measurement, richter2023transport}, where several rigorous results have been obtained on Brownian all-to-all circuits
\cite{lashkari2013towards, saad2018semiclassical, zhou2019operator, chen2019quantum, sunderhauf2019quantum, piroli2020random, shao-kai2021measurement, jian2021note, stanford2022subleading}.

Many studies have investigated long-range models in \emph{open quantum systems}, where interaction with the surrounding environment induces dissipative effects.
For all-to-all interactions, the man-field approach is correct for systems with collective jump operators\,\cite{davies1973exact,alicki1983nonlinear,mori2013exactness,benatti2016non,kirton2017suppressing,shammah2018open,benatti2018quantum,huybrechts2020validity,wang2021dissipative,piccitto2021symmetries,carollo2021exactness,lundgren2020nature,carollo2018current}. In these settings, quantum correlations  have been studied with methods similar as the ones discussed in Sec.\,\ref{sec_415} in Refs.\,\cite{benatti2017quantum,benatti2018quantum, buonaiuto2021dynamical, boneberg2022quantum} or via Keldysh techniques\,\cite{sieberer2013dynamical, buca2019dissipation, kirton2019introduction}. The impact of finite range interactions has been recently addressed in Refs.\,\cite{passarelli2022dissipative, souza2023sufficient, sulz2023numerical}, while in Refs.\,\cite{zhu2019dicke,seetharam2022correlation,seetharam2022dynamical} it was tackled via a generalization of the non-equilibrium spin-wave approach discussed in Sec.\ref{sec_421}.

\emph{Other studies} in the field include {hydrodynamics} and transport with long-range interactions \cite{kloss2019spin, schuckert2020nonlocal,joshi2022observing,morningstar2023hydrodynamics, ogunnaike2023unifying, gliozzi2023hierarchical}, 
or {hybrid models} stemming for instance from the interplay between short-range and long-range interactions. The interplay between long-range interactions and non-homogeneity has also been studied in quasi-periodic systems\,\cite{deng2019one,roy2021fraction,fraxenet2022localization,wang2023fate}, where the {topological properties} of the system are altered by the long-range couplings\,\cite{liu2020generalized} as it has also been shown in topological superconductors\,\cite{patrick2017topological,jager2020edge,tarantola2023softening,gong2023long}.

To conclude, very recently, a lot of attention in quantum dynamics has been given to \emph{measurement induced phase-transitions}, a new dynamical phenomenon which results from the interplay between unitary (entangling) evolution and (disentangling) measurements\,\cite{fisher2023random}. In this context, Refs.\,\cite{nahum2021measurement, vijay2020measurement, gullans2020dynamical, block2022measurement, sharma2022measurement, minato2022fate, passarelli2023post, sierant2023controlling, sierant2022dissipative, muller2022measurement} have discussed how long-range interactions affect the non-equilibrium phase diagram, which has been probed experimentally in a trapped ion simulator\,\cite{noel2022observation}.

\medskip
 {\bf What shall be done.\, }
Quantum dynamics with long-range interactions is an exciting field of research with challenging open problems that go beyond the ones addressed here. 
We are pleased to conclude with a brief overview of the open questions.

First of all, the methods considered in this Report may be extended to address a series of pending problems about long-range interacting spin systems. 
As mentioned in Section \ref{sec_231}, the spatial \emph{spreading of correlations} could be characterized using the non-equilibrium spin-wave theory~\cite{lerose2018chaotic,lerose2019impact} illustrated in Sec.~\ref{sec_421}. Related approaches could be used  to explore quantum information \emph{scrambling}, where it would be interesting to elucidate how the established exponential growth of the square-commutator~\eqref{SC} at dynamical critical points for $\alpha=0$ is influenced by spatial fluctuations for $\alpha>0$.
A fundamental question concerns \emph{thermalization} in this class of systems. Even though permutational symmetry is broken for $\alpha\neq0$, anomalous dynamics compatible with a prethermal scenario appear, for long -- yet finite -- time scales~\cite{mori2018thermalization,lerose2020origin} (see Sec.~\ref{sec_42}). It is a challenging open problem to understand whether and how some form of non-thermal behavior persists at infinite time. This difficult subject has been addressed in a few numerical explorations of either spectral properties~\cite{fratus2016eigenstate,Schuckert2020Probing,Russomanno2021quantum,Sugimoto2022eigenstate} or finite-time quench dynamics~\cite{halimeh2017prethermalization,zunkovic2018dynamical,piccitto2019dynamical,liu2019confined,collura2022discrete}. However, due to the challenging finite-size effects,  the general question -- especially near the strong long-range regime -- is far from settled. See the recent progress in Ref.\cite{lerose2023theory}. 

Finally, several broader open questions go far beyond what is discussed here. This is the case of long-range \emph{antiferromagnetic} interaction, which characterizes several experimental platforms~\cite{morigi2004eigenmodes, fishman2008structural, cormick2012structural, silvi2013full, nigmatullin2016formation, pupillo2008cold, lu2012quantum, cartarius2014structural, natale2019excitation, tanzi2019observation, bottcher2021new, fernandez-vidal2010quantum, larson2008mott,himbert2019mean} as mentioned above. It is well known that in equilibrium, the competition between anti-ferromagnetism and long-range interactions can result in frustration and phenomena such as spin liquids. It would therefore be important to develop methods to tackle dynamics out of equilibrium.

Concurrently, many remarkable physical phenomena stem from the simultaneous presence of long-range interactions and quenched \emph{disorder}, from glassiness in the SK to fast scrambling in the SYK model. Although these models may be analytically solvable with a mean-field ansatz, their classical limit and the impact of quantum fluctuations remains a challenging problem in general. It would be highly desirable to have a comprehensive framework to understand the quantum dynamics with $\alpha>0$ in this class of models. 
In this regard, we note that the non-equilibrium spin-wave theory of Refs.~\cite{lerose2018chaotic,lerose2019impact} has been extended to disordered spin models as well, see Ref.~\cite{lerose2019thesis}.

To conclude, we remark that in nature long-range interactions always represent an instantaneous approximation for \emph{retarded} interactions mediated by a field (e.g. the electromagnetic field). 
Retardation may give rise to outstanding physical phenomena in certain conditions: For example, finite-frequency long-wavelength modes may non-trivially hybridize with the mediating-field excitations, as happens e.g. for optical phonons and light in ionic crystals ({phonon-polaritons})~\cite{born1966dynamical,lerose2014classical}. Exploration of the full range of dynamical phenomena induced by retarded long-range interactions in AMO platforms stands out as an intriguing direction.

\section*{Acknowledgements}

\noindent We thank F. Carollo, R. Fazio, J. Halimeh, A. Guo, M. Heyl, M. Knap, J. Knolle, J. Marino, L. Piroli, A. Pizzi, P. Poggi, L. Santos, M. Tran and B. Zunkovic for their feedback on the manuscript and suggestions.
N.D. acknowledges funding by the Swiss National Science Foundation (SNSF) under project funding ID:\,200021\_207537, from the Swiss Secretariat for Education, Research and Innovation (SERI) and by the Deutsche Forschungsgemeinschaft (DFG, German Research Foundation) under Germany’s Excellence Strategy EXC2181/1-390900948 (the Heidelberg STRUCTURES Excellence Cluster). 
A.L. acknoledges support by the European Research Council (ERC) under the European Union’s Horizon 2020 research and innovation programme (grant agreement No. 864597) and by the Swiss National Science Foundation.
S.P. acknowledges support by the Deutsche Forschungsgemeinschaft (DFG, German Research Foundation) under Germany’s Excellence Strategy - Cluster of
Excellence Matter and Light for Quantum Computing (ML4Q) EXC 2004/1 -390534769.

\appendix
\section{Semiclassical spectrum}
\label{app_seiclassicalspectrum}

For a system with a single degree of freedom, the spectrum obtained via the semiclassical quantization rule consists
of classical trajectories  with energy $E_n$ such as to enclose an area $S_{\text{cl}}(E_n)$  in  phase space
 equal to an integer multiple of Planck's constant $h$, i.e.,
\beq
\label{eq_semiclquant}
S_{\text{cl}}(E_n) = n h , \quad n \text{ integer},
\eeq
where  $S_{\text{cl}}(E)$ corresponds to the classical action. 
This quantization rule, together with the well-known relation between the action $S_{\text{cl}}$ and the classical period of a trajectory (see, e.g., Ref.~\cite{landau1958course})
\beq
\frac{d S_{\text{cl}}(E) }{dE} = T_{\text{cl}}(E) \equiv \frac{2\pi}{\Omega_{\text{cl}}(E)}
\eeq
yields
the semiclassical level spacing
\beq
\label{eq_semicllevelspacing}
E_{n+1}-E_n \sim \frac{d E_n}{dn} = \hbar \, \Omega_{\text{cl}}(E_n).
\eeq
This equation may be seen as a generalization of the relation valid for the spectrum of a harmonic oscillator to nonlinear dynamics, in which the oscillations are not isochronous and thus the quantum energy spectrum is not equispaced.
The semiclassical density of states $\rho(E)$ is given by the inverse level spacing.

We can use the above semiclassical relation to elucidate the spectral properties of the fully-connected quantum Ising model, cf. Sec.~\ref{sec_alpha0eq}.
For $h>h_{\text{cr}}$, taking the low-energy limit ($n$ finite and $N\to\infty$) of Eq. \eqref{eq_semicllevelspacing}, one recovers the harmonic tower of excitations of Eq. \eqref{eq_expansionH>}, with $\Omega_{\text{cl}}(E_n) \sim_{N\to\infty} \omega_>/s$.
At the quantum critical point $h=h_{\text{cr}}$, the energy gap of the elementary excitations above the ground state vanishes in the thermodynamic limit.
The classical counterpart of this phenomenon is the vanishing of the classical frequency of small oscillations, which occurs because the minimum of $\mathcal{H}_{\text{cl}}$ at criticality is quartic rather than quadratic.
The critical scaling of the energy gap as $N\to\infty$ may be extracted via semiclassical considerations: 
Retaining the quartic terms of order $1/N$ neglected in  Eq. \eqref{eq_expansionH>}, and applying the semiclassical quantization rule  in Eq.~\eqref{eq_semiclquant}, one finds the low-energy asymptotics of quantized energy levels as
\beq
E_n - E_0 \quad \underset{\substack{n \text{ finite} \\ N\to\infty}}{\thicksim} \qquad c \, \frac{n^{4/3}}{N^{1/3}},
\eeq
which shows that the critical gap above the ground state for $h=h_{\text{cr}}$ scales as $N^{-1/3}$ for large $N$.
Along these lines, one also finds $\langle \hat n_0 \rangle \sim N^{1/3}$.
For a more systematic analysis see e.g. Ref.~\cite{vidal2004entanglement}.

\section{Semiclassical description of fully-connected systems}
\label{app_semiCla}
In this Appendix we review the semi-classical descriptions of quantum dynamics, which applies beyond the example of fully-connected spin systems discussed in Sections \ref{sec_411} and \ref{sec_413}.

\subsection{Semiclassical approach}
We focus on quantum systems characterized by a small parameter $\heff$, which controls the impact of the quantum fluctuations.
A system in this class is described by $n$ degrees of freedom, identifyed by $2n$ operators ${\hat{\boldsymbol \xi}=(\hat q_1,\dots,\hat q_n,\hat p_1,\dots,\hat p_n)}$. These satisfy the standard canonical commutation relations ${[\hat q_i , \hat p_j] = i \hbar_{\text{eff}} \, \delta_{ij}}$, o more compactly  $[\hat{\boldsymbol \xi},\hat{\boldsymbol \xi}]=i \hbar_{\text{eff}} \mathbb{J} $, where 
$\mathbb{J}$ is the symplectic unit\footnote{The symplectic matrix  $\mathbb{J}$ is given by the $2n \times 2n$ antisymmetric matrix {\scriptsize $\mathbb{J} = {\begin{pmatrix} \mathbb{0}_n & \mathbb{1}_n \\ -\mathbb{1}_n & \mathbb{0}_n  \end{pmatrix}}$}, which satisfies $\mathbb{J}^2=-\mathbb{1}_{2n}$.}. 
The system is such that allows a re-scaling of the Hamiltonian 
 \beq
 \hat H  = \heff^{-1} \; \mathcal{H}_\cl(\hat{\boldsymbol \xi}) \ ,
 \eeq
that leads to the following the Heisenberg equation of motion\footnote{Subtleties related to the ordering of the operators are not relevant in the following discussion.}
 \beq
 \dot{\hat{\boldsymbol \xi}} = 
 \mathbb{J} \; \partial \mathcal{H}_\cl(\hat{\boldsymbol \xi}) \ .
 \eeq
One could equivalently define a classical system described by $2n$ classical phase-space variables $\boldsymbol \xi_\cl = (q_1,\dots, q_n, p_1,\dots, p_n)$, obeying the canonical Poisson brackets $\{q_i, p_j\} = \delta_{ij}$ and whose dynamics is given by the Hamilton-Jacobi equation of motion $\dot{\boldsymbol \xi}_{\cl} = \{\boldsymbol {\xi}_{\cl}, \mathcal H_{\cl}\} =  \mathbb{J} \; \partial \mathcal{H}_\cl ({\boldsymbol {\xi}_\cl})$.

The full quantum evolution for the expectation value of the operator $\hat {\bxi}(t)$ evaluated on a the generic quantum state $\ket{\psi_0}$ reads
 \beq
 \frac d {dt} 
  \braket{\hat{\boldsymbol \xi}(t)}  = 
 \mathbb{J} \; \braket{\partial \mathcal{H}_\cl (\hat{\boldsymbol \xi}(t) )\, }\ .
 \eeq
This is exactly what stated by the Ehrenfest theorem \cite{wheeler1998remarks}, which describes the exact quantum evolutions of operators at time $t$, without approximations. Even if this relation is reminiscent of the Hamilton's equations for the classical variable $\bxi_\cl$, in principle one has $\braket{\partial\mathcal{H}_\cl (\hat \bxi)} \neq \partial\mathcal{H}_\cl (\braket{\hat \bxi})$.
However, whenever quantum fluctuations are small one can look at the replacements\footnote{Notice that this is always exact the case of quadratic Hamiltonians.}
\beq
\label{eq_QFinstate_qui}
\braket{\partial \mathcal{H}_\cl (\hat \bxi)} \to  \partial \mathcal{H}_\cl (\braket{\hat \bxi}) \ .
\eeq
This substitution is equivalent closing the cumulants at second order, namely to take $\braket{\hat \xi\; \hat \xi'} = \braket{\hat \xi}\braket{\hat \xi'}$.
We consider the case in which the initial state $\ket{\psi_0}$ corresponds to a narrow Gaussian wave-packet, centered around a point with a \emph{small variance} $\Delta^2$ of quantum fluctuations of order $\Delta^2 \sim \mathcal O(\heff)$. 
A large number of relevant initial states lie in this class. For instance, consider coherent states or pure nonentangled ones, such as uncorrelated product states, routinely prepared in cold-atom experiments via standard techniques. Weakly entangled initial states may be treated on equal footing.

Therefore, by virtue of Eq.~\eqref{eq_QFinstate_qui}, the average $\braket{\hat{\boldsymbol \xi}(t)}$ moves along the classical trajectory to the leading order in $\heff$,
 \beq
 \frac d {dt} 
 \braket{\hat{\boldsymbol \xi}(t)}  = 
 \mathbb{J} \; \partial \mathcal{H}_\cl
 \left ( \braket{\hat{\boldsymbol \xi}(t)} \right )
  + \mathcal{O}(\heff)\ ,
 \eeq
that is
\beq
\braket{\hat{\boldsymbol \xi}(t)} 
= \boldsymbol {\xi}_{\cl}(t) + {O}(\heff) \ .
\eeq

According to the standard semiclassical theory \cite{littlejohn1986semiclassical,  polkovnikov2010quantum, richter2022semiclassical},  quantum fluctuations around the classical trajectory $\boldsymbol {\xi}_{\cl}(t) $ will remain approximately Gaussian for a diverging time scale as $\hbar_{\text{eff}}\to0$ during the evolution, the so-called \emph{Ehrenfest time} scale $\Teh=\Teh(\heff)$. At $\Teh$ quantum interference effects become dominant and the semiclassical description breaks down. The Ehrenfest time can be defined as the time scale for which the gaussian approximation breaks down and quantum fluctuations become of the order of one, i.e. $\Delta^2(\Teh)=O(1)$.
This depends on how quantum fluctuations evolve in time that, in turn, is determined by the regularity properties of the classical trajectories, as summarized in Table \ref{tab1}.

\medskip
This semiclassical description is not restricted to phase-space or coherent variables, but it describes the dynamics of several interesting models. In particular, 
Sciolla and Biroli \cite{sciolla2010quantum} formulated a general theory for systems with full permutational invariance in states belonging to the totally-symmetric sector.

\subsection{Classical limit of permutationally invariant systems}
\label{sec_permutaModels}
We recall how the permutational symmetries 
allow for exactly mapping collective quantum models to systems of few degrees of freedom characterized by a vanishingly small effective Planck constant in the thermodynamic limit \cite{sciolla2010quantum}.

We consider a Hamiltonian $\hat H$ characterizing a uniform all-to-all interaction of $N$ elementary constituents, such as spins or particles. The symmetry under permutations of the degrees of freedom makes the mean-field treatment of the quantum dynamics exact for large $N$.
To show how the semiclassical description emerges, 
we consider an ensemble of $N$ identical $q$-level quantum systems.
A basis of the many-body Hilbert space can be constructed as the tensor product of identical single-unit bases $\{ \ket{\alpha} \}$ with $\alpha=1,\dots,q$. Binary permutation operators are unitary transformations that exchange a pair of units in the system. Their action is defined by
\beq
\hat P_{ij} \ket{\alpha_1, \dots , \alpha_i, \dots \alpha_j, \dots, \alpha_N} =
\ket{\alpha_1, \dots , \alpha_j, \dots \alpha_i, \dots, \alpha_N} \ ,
\eeq
for all pairs $i>j$. 
A system has full permutational invariance if its Hamiltonian $\H$ commutes with all permutation operators.
The totally-symmetric subspace (TSS) of the many-body Hilbert space is simultaneously invariant under all permutations \footnote{Unless permutational symmetry is spontaneously broken or fragmentation phenomena take place \cite{vidal2004entanglement}. }.
A basis of the TSS 
can be obtained by symmetrizing the many-body configurations $\ket{\alpha_1,\dots,\alpha_N}$ with respect to all permutations.
It 
can be labelled by the numbers $N_1,\dots,N_q$  of units occupying each level with $\sum_{\alpha=1}^q N_\alpha = N$.
The dimension of the TSS,
\beq
\text{dim TSS } = \binom{N+q-1}{q-1} \quad  \underset{N\to\infty}{\thicksim} \quad \frac{N^{q-1}}{(q-1)!}\ ,
\eeq
is only polynomially large in $N$, which allows for the exact numerical analysis of large systems. 
Due to the symmetry of $\hat H$, the time-evolution of totally symmetric initial states never leaves the TSS. 
Typically, such initial states may be simple products of identical single-body states, or ground states, like the ones prepared in experiments.

It was shown by Sciolla and Biroli in Ref.~\cite{sciolla2010quantum} that the dynamics of symmetric observables within the TSS is classical in the thermodynamic limit. 
To show this, observe that possible off-diagonal transitions governed by $\hat H$ are uniquely identified by a set of integers $m_1,\dots,m_q$
\beq
\ket{N_1,\dots,N_q} \to \ket{N_1+m_1,\dots,N_q+m_q}.
\eeq
For convenience, we turn the occupation numbers $N_\alpha$ into fractions $x_\alpha\equiv N_\alpha/N$, with $0\le x_\alpha \le 1$ and $\sum_{\alpha=1}^q x_\alpha = 1$,
and denote basis states by $\ket{\mathbf x}$, where $\mathbf x=(x_1,\dots,x_q)$.
Hence, we write the matrix elements of $\hat H$ as\footnote{For simplicity, we assume time-reversal invariance, which results in real matrix elements $T_{\mathbf m}(\mathbf x) \in \mathbb{R}$.}
\beq
\label{eq_matrixelementsemiclassical}
H_{\mathbf x, \mathbf x'} \equiv \braket{\mathbf x| \hat H | \mathbf x'} = V(\mathbf x) \, \delta_{\mathbf x, \mathbf x'} - \sum_{ \mathbf m \ne \mathbf 0 } T_{\mathbf m}(\mathbf x) \delta_{\mathbf x, \mathbf x' + \mathbf m/N},
\eeq 
with $\mathbf m = (m_1,\dots,m_q) \in \mathbb{Z}^q$.
 Terms in the Hamiltonian $\hat H$ involving up to $k$ bodies yield ``local'' transitions in the TSS basis, characterized by $\abs{\mathbf m} \equiv \sum_\alpha \abs{m_\alpha} \le 2k$. 
By the extensivity of the Hamiltonian $\hat H$, both  $V(\mathbf x)$ and $T_{\mathbf m}(\mathbf x) $ are extensive,
 \beq
 V(\mathbf x) \sim N \, v(\mathbf x), \qquad T_{\mathbf m}(\mathbf x) \sim N \, t_{\mathbf m}(\mathbf x).
 \eeq
Crucially, the densities $v$ and $t$ are smooth functions of $\mathbf x$, as they generally result from combinatoric factors of the occupation numbers which are insensitive to small changes $N_\alpha \mapsto N_\alpha \pm 1,2,\dots$ to leading order in the thermodynamic limit \cite{sciolla2010quantum}.
This result is based on the smoothness of the matrix elements of $\hat H$ between two TSS states concerning small changes in the occupation numbers $N_\alpha \to N_\alpha \pm 1, \pm 2, \dots$.
These properties allow to rewrite the Schr{\"o}dinger equation in the TSS as
\beq
\label{eq_shroTTS}
\frac 1N \frac{\partial}{\partial t}\psi(\mathbf x, t)   =
\left [
v(\mathbf x) - \sum_{0 \leq |\mathbf m|\leq 2k}t_{\mathbf m}(\mathbf x) \, \cosh \left (\frac{\mathbf m}N \frac{\partial}{\partial \mathbf x}\right )\, 
\right ]\, \psi(\mathbf x, t) \ .
\eeq
Equation \eqref{eq_shroTTS} shows that the dynamics of wave-functions in the TTS is governed by the effective Hamiltonian
\beq
\mathcal H_\cl(\hat {\mathbf q}, \hat {\mathbf p}) \equiv 
v(\hat {\mathbf q}) - \sum_{\mathbf m}t_{\mathbf m}(\hat{\mathbf x}) \, \cosh \left (\mathbf m \cdot \hat{\mathbf p}\right )\ ,
\eeq
expressed in terms of the conjugated canonical operators $\hat \bxi=(\hat {\mathbf q}, \hat {\mathbf p})$
\beq
\frac{N_\alpha}{N} \mapsto \hat q_\alpha, \qquad -i 
\frac{\partial}{\partial  N_\alpha} \mapsto \hat p_\alpha \ ,
\eeq
with an effective Planck constant 
\beq
\hbar_{\text{eff}} \equiv \frac{1}{N} \qquad \text{($\hbar=1$ in our units)} \ ,
\eeq
that approaches zero in the thermodynamic limit.
Thus, the quantum system of the original system of all-to-all interacting $q$-level units is mapped to $n=q-1$ collective degrees of freedom.\footnote{Notice that the exact constraint $\sum_{\a} x_\a=1$ can be solved explicitly, eliminating one degree of freedom.} As outlined in the previous section, its quantum dynamics is equivalent, in the thermodynamic limit, to the one governed by the Hamilton equations generated by $\mathcal{H}_{\cl}$.

\subsection{Beyond global permutational symmetry}
\label{sec_classicalBeyond}

The semiclassical approach reviewed in the previous \ref{sec_permutaModels} and Sec.\,\ref{sec_411}  applies to a much wider class of states and models than discussed therein.

One natural extension 
consists of a \emph{composite system of $M$ collective subsystems}, possibly composed of different kinds of degrees of freedom.
This is possible if interactions couple the various subsystems uniformly in their elementary units, i.e., via collective operators only. Thus, the global system has a semiclassical description.
When each subsystem is large, the global system will be described by $\sum_{m=1}^M  (q_m-1)$ semiclassical collective degrees of freedom, where $q_m$ is the number of levels for the $m$-th degree of freedom.  
For example, the Dicke model, where $N$ spins interact collectively with a cavity mode \cite{dicke1954coherence},
can be viewed as an example of two classical degrees of freedom, one for the collective spin and one for the cavity mode. The same holds for the two-species kicked top \cite{miller1999signatures}.

A second, closely related generalization,
is represented by \emph{non-symmetric states which partially break the full permutational symmetry}.
Such states may be 
obtained by bringing together a number $M\ll N$ of initially separated subsystems.
In this case, the full permutational symmetry breaks down into the product of smaller permutational symmetries acting separately on each subsystem. 
While the full system evolves outside of its totally symmetric subspace (TSS), the restricted symmetry allows a description of the dynamics within the product of the TSSs of the $M$ individual subsystems.
The semiclassical theory can thereby be applied in the thermodynamic limit, and one ends up with a few-body semiclassical system described by $M \times (q-1)$ collective degrees of freedom.
In this case, the  Hamiltonian depends on these variables only via the $q-1$ global collective combinations, leaving all the $(M-1)\times (q-1)$  remaining coordinates frozen in their initial values.
A simple example 
is given by a permutationally invariant system of $N$ spins-$1/2$  initially in a random product state $\ket{\dots \nearrow\nearrow\nearrow \swarrow \nearrow\swarrow\swarrow\nearrow \dots }$ of spins pointing up or down along a given axis. Such a state is far away from the Dicke manifold of maximal total spin length $N/2$. 
Grouping together the spins pointing in the same direction into two subsystems $A$ and $B$, with $N_A$ and $N_B$ spins respectively, the global system may be viewed as two interacting collective spins $\hat{\vec S}_A$, $\hat{\vec S}_B$, of length $N_A/2$ and $N_B/2$ respectively, initially pointing in opposite directions.
In agreement with the above observation, the motion of  the two spins is not independent: the Hamiltonian generates a nonlinear collective precession,
and the angle between 
$\hat{\vec S}_A$ and $\hat{\vec S}_B$  is a constant of motion.


\section{Asymptotic estimates for ${f}_{\mathbf{k}}(\alpha)$}
\label{app:boundsF}



Here we review the properties of the Fourier transform of $J/{\lvert \lvert \mathbf{r} \rvert\rvert^\alpha}$ on a periodic $d$-dimensional lattice of $V=L^d$ sites, which we denote $f_{\mathbf{k}}(\alpha)$:
\beq
\label{eq_fourier}
{f}_{\mathbf{k}}(\alpha) = 
\sum_{\mathbf{r}\neq\mathbf{0}} \frac{e^{-i\mathbf{k}\cdot \mathbf{r}}}{\lvert\lvert\mathbf{r}\rvert\rvert^{\alpha}}
\Bigg/
\sum_{\mathbf{r}\neq\mathbf{0}} \frac{1}{\lvert\lvert\mathbf{r}\rvert\rvert^{\alpha}}
 .
\eeq
The properties derived below only rely on the asymptotic decay of interactions $J_{\mathbf{r},\mathbf{r'}} \sim 1/{\lvert \lvert \mathbf{r} - \mathbf{r'}\rvert\rvert^\alpha}$ --- neither on the details of $J_{\mathbf{r},\mathbf{r'}}$ at short distances nor on the specific lattice.

\subsection{Strong long-range regime ($0<\alpha < d$)}

For $0<\alpha < d$ the leading behavior is captured by approximating sums with integrals in Eq.~\eqref{eq_fourier}. As we are interested in the scaling with $L$ only, we do not keep track of prefactors. 
Following the standard procedure for Fourier transforming a radial function, we switch to spherical coordinates and integrate over all the angles:
\beq
\label{eq_fourierradial}
{f}_{\mathbf{k}\ne\mathbf{0}} (\alpha) \thicksim \frac{1}{L^{d-\alpha}} \int_1^{L} d\rho \, \rho^{d-1-\alpha} \, 
\frac{\mathcal{J}_{d/2-1} (|\mathbf{k}|\rho)}{(|\mathbf{k}| \rho)^{d/2-1}},
\eeq
where  $\mathcal{J}_\nu(x)$ is the standard Bessel function of order $\nu$.

For finite $|\mathbf k |$ the right-hand side always vanishes in the limit $L\to\infty$. 
A finite value of ${f}_{\mathbf{k}\ne\mathbf{0}}$ is only obtained when $|\mathbf k | \propto 1/L$.
Recalling the definition $\mathbf{k}\equiv\mathbf{k}_\mathbf{n}\equiv 2\pi \mathbf{n} /L$,
we make the substitution $\rho=Ls$ and take $L\to\infty$:
\beq
{f}_{\mathbf{k}_\mathbf{n}\ne\mathbf{0}} (\alpha) \equiv {f}_{\mathbf{n}\ne\mathbf{0}} (\alpha) \thicksim  \int_0^{1} ds \, s^{d-1-\alpha} \, 
\frac{\mathcal{J}_{d/2-1} (2\pi |\mathbf n | s)}{(2\pi |\mathbf n | s)^{d/2-1}}.
\eeq
Thus, for $0<\alpha<d$, ${f}_{\mathbf{k}_\mathbf{n}\ne\mathbf{0}}$ is actually a function of the discrete index $ \mathbf n $.
For large $|\mathbf n|$ we obtain the asymptotic estimate
\beq
\label{eq_estimateftildealpha<d}
 {f}_{\mathbf{n}\ne\mathbf{0}} (\alpha) \thicksim   \frac {A(\alpha)} {|\mathbf n |^{d-\alpha} }+ \frac { B(\alpha)} { |\mathbf n |^{(d+1)/2} }\, .
\eeq
The first  [second] term governs the asymptotic decay of the discrete coefficients ${f}_{\mathbf{n}\ne\mathbf{0}} (\alpha)$ for $(d-1)/2 < \alpha < d$ [respectively $0 < \alpha < (d-1)/2$].

\subsection{Weak long-range regime ($\alpha > d$)}

For $\alpha>d$ the function ${f}_{\mathbf{k}}(\alpha)$ attains a finite limit for all $\mathbf{k}$ as $L\to\infty$.
For small $\mathbf{k}$, this function has a singular behavior.
In this ``large-scale'' limit it is again legitimate to replace the sum by the corresponding integral. Proceeding similarly to Eqs.~\eqref{eq_fourierradial}, we find
\beq
{f}_{\mathbf{k}\ne\mathbf{0}} (\alpha) \thicksim  \int_1^{\infty} d\rho \, \rho^{d-1-\alpha} \,
\frac{\mathcal{J}_{d/2-1} (|\mathbf{k}|\rho)}{(|\mathbf{k}| \rho)^{d/2-1}}
\Bigg/
\int_1^{\infty} d\rho \, \rho^{d-1-\alpha} C_d
\eeq
where $C_d=2^{-(d/2-1)}/\Gamma(d/2)$.
Here the short-distance part gives a regular contribution $\mathcal{O}(|\mathbf k |^2)$ and the long-distance part gives a singular contribution $\mathcal{O}(|\mathbf k |^{\alpha-d})$:
\beq
\label{eq_estimateftildealpha>d}
{f}_{\mathbf{k}\ne\mathbf{0}} (\alpha) \thicksim 1 -  \bar A(\alpha) |\mathbf{k}|^{\alpha-d} - \bar B(\alpha)  |\mathbf{k}|^2.
\eeq
The first  [second] term governs the asymptotic low-momentum behavior of  ${f}_{\mathbf{k}\ne\mathbf{0}} (\alpha)$ for $d < \alpha < d+2$ [respectively $ \alpha > d+2$].

\section{Exact solution of quasi-static drive for a single mode}
\label{KZM_app}
\subsection{Fidelity and defect density}
\label{AppA}
The dynamics of a single spin-wave corresponds to the one of a single Harmonic oscillator and can be solved exactly \cite{lewis1967classical,lewis1968class,lewis1969exact} for any time dependent frequency. Any dynamical state $\psi(x,t)$ in the representation of the coordinate $x$ can be expressed as
\begin{align}
\label{dyn_exp}
\psi(x,t)=\sum \alpha_{n}\psi_{n}(x,t),
\end{align}
where $\alpha_{n}$ are time independent constants and the dynamical eigenstates are given by
\begin{align}
\label{Dyn_Eigen}
\psi_{n}(x,t)=\frac{1}{\sqrt{2^{n}n!}}\left(\frac{1}{2\pi\xi^{2}(t)}\right)^{\frac{1}{4}}e^{-W(t)\frac{x^{2}}{2}}
H_{n}\left(\frac{x}{\sqrt{2}\xi(t)}\right)e^{-i\left(n+\frac{1}{2}\right)\lambda(t)},
\end{align}
the expression for the effective frequency $W(t)$ and the (in-influential) phase $\Phi(t)$ are given in the main text.
If the initial state is a pure state of the basis \eqref{Dyn_Eigen}, specifically the ground state in our case, then one has $\alpha_{n}=0,\quad \forall n>0$ and we recover the single squeezed state generated by the operator in Eq.\,\eqref{squezzing_operator}. This state describe the dynamics at all times, and thus in the exact dynamical basis \eqref{Dyn_Eigen} no excited states will be generated. However, at each finite time $t>-h_{\rm cr}/\delta$ the squeezed state $\psi_{0}(x,t)$ will have a finite overlap with all states in the adiabatic basis $\psi_{n}^{\rm ad}(x,t)$, which is defined as 
\begin{align}
\label{Eq_Eigen}
\psi_{n}^{\rm ad}(x,t)=\frac{1}{\sqrt{2^{n}n!}}\left(\frac{\Omega(t)}{\pi}\right)^{\frac{1}{4}}e^{-\Omega(t)\frac{x^{2}}{2}}H_{n}\left(x\sqrt{\Omega(t)}\right).
\end{align}
It is convenient to write the expression of the defect density as~\cite{dabrowski2016time}
\begin{align}
\label{Exc_N}
n_\text{exc}(t)=\sum_{n\in 2\mathbb{N}}n|c_{n}(t)|^{2},
\end{align} 
where the coefficients
\begin{align}
\label{trans_amp}
c_{n}(t)=\int_{-\infty}^{+\infty}dx\,\psi_{n}^{*}(x,t)\psi_{0}(x,t)
\end{align}
are the transition amplitudes between the dynamical state and the instantaneous equilibrium basis.
It is rather straightforward to get an exact expression for these coefficients
\begin{align}
c_{n}(t)=&\int_{-\infty}^{+\infty}dx\psi_{n}^{\rm ad *}(x,t)\psi_{0}(x,t)
=\frac{1}{\sqrt{2^{n}n!\pi}}\left(\frac{\Omega(t)}{2\xi^{2}(t)}\right)^{\frac{1}{4}}\int_{-\infty}^{+\infty}dx e^{-(\Omega(t)+\tilde{\Omega}(t))\frac{x^{2}}{2}}H_{n}\left(\sqrt{\Omega(t)}x\right).
\end{align}
Performing a change of variable the integral can be cast into the form
\begin{align*}
\int_{-\infty}^{+\infty}dx e^{-(\Omega(t)+\tilde{\Omega}(t))x^{2}}H_{n}\left(\sqrt{\tilde{\omega(t)}}x\right)=(\Omega(t))^{-\frac{1}{2}}\int_{-\infty}^{+\infty} e^{-\left(\frac{\tilde{\Omega}(t)}{\Omega(t)}+1\right)\frac{s^{2}}{2}}H_{n}\left(s\right)ds.
\end{align*}
Next we employ the generating function for Hermite polynomials in the integral,
\begin{multline}
\int_{-\infty}^{+\infty} e^{-\left(\frac{\tilde{\Omega}(t)}{\Omega(t)}+1\right)\frac{s^{2}}{2}}H_{n}\left(s\right)ds=\lim_{t\to0}\frac{d^{n}}{dt^{n}}\int_{-\infty}^{+\infty} e^{-\left(\frac{\tilde{\Omega}(t)}{\Omega(t)}+1\right)\frac{s^{2}}{2}}e^{2st-t^{2}}ds=\sqrt{\frac{2\pi}{\left(\frac{\tilde{\Omega}(t)}{\Omega(t)}+1\right)}}\lim_{t\to0}\frac{d^{n}}{dt^{n}}e^{-t^{2}\frac{\left(\tilde{\Omega}(t)-\Omega(t)\right)}{\left(\Omega(t)+\tilde{\Omega}(t)\right)}}\\
=\begin{cases}
\displaystyle\sqrt{\frac{2\pi}{\left(\frac{\tilde{\Omega}(t)}{\Omega(t)}+1\right)}}\frac{n!}{\frac{n}{2}!}\left(\frac{\tilde{\Omega}(t)-\Omega(t)}{\tilde{\Omega}(t)+\Omega(t)}\right)^{n/2} & \text{for $n\in 2\mathbb{Z}$},\\
0 & \text{for $n\in 2\mathbb{Z}+1$}.
\end{cases}
\label{cn_explicit}
\end{multline}
Thus the probability of having $n$ excitations in the evolved state at the time $t$ is given by
\begin{align}
|c_{n0}(t)|^{2}=\frac{(n-1)!!}{n!!}\frac{\sqrt{2\Omega(t)}}{\xi(t)\left|\tilde{\Omega}(t)+\Omega(t)\right|}\left|\frac{\tilde{\Omega}(t)-\Omega(t)}{\tilde{\Omega}(t)+\Omega(t)}\right|^{n},
\end{align}
which leads to Eqs.\,\eqref{exc_den} and\,\eqref{exp_fidelity} in the main text.
\subsection{Slow quench to the critical point}
\label{SM2}
Here we consider the half ramp with $t\in[-h_{\rm cr}/\delta, 0]$. After the rescaling, this problem coincides to solving the Ermakov-Milne equation 
\begin{align}
\label{ermakov_eq_app}
\ddot{\xi}(t)+\Omega(t)^{2}\xi(t)=\frac{1}{4\xi^{3}(t)},
\end{align}
with the rescaled frequency 
\begin{align}
\label{eq_rescaling}
\Omega(t)^{2}=t+(N\delta)^{-2/3}
\end{align}
with the simplified extended time interval $t\in [-\infty,0]$. 
The solution of Eq.~\eqref{ermakov_eq_app} can be constructed from that of the associated classical harmonic oscillator
\begin{align}
\label{cl_h_osc}
\ddot{x}(t)+\Omega(t)^{2}x(t)=0.
\end{align}
This equation admits the two independent solutions
\begin{align}
x_{1}(t)=\mathrm{Ai}\left(-\Omega^{2}(t)\right),\qquad
x_{2}(t)=\mathrm{Bi}\left(-\Omega^{2}(t)\right)
\end{align}
in terms of the Airy functions $\mathrm{Ai}$ and $\mathrm{Bi}$. The two functions $x_{1}(t)$ and $x_{2}(t)$ have the constant and finite Wronskian
\begin{align}
\mathrm{Wr}(x_{1},x_{2})=\frac{1}{\pi}.
\end{align}
It is convenient to rewrite the solutions of Eq.~\eqref{ermakov_eq_app} as a pair of complex conjugate solutions $w$ and $w^*$ with
\begin{align}
\label{sol_def}
w=a x_{1}(t)+b x_{2}(t),
\end{align}
where $a\in\mathbb{C}$ and $b\in\mathbb{R}$ are constants. Since Eq.~\eqref{cl_h_osc} is homogeneous one can rescale the two solution by a constant factor and subsequently, without loss of generality, impose $b=1$. The function
\begin{align}
\label{xi_sol}
\xi(t)=\sqrt{ww^*}
\end{align}
is a solution of the Ermakov-Milne equation \eqref{ermakov_eq} if
\begin{align}
\label{wronsk_condition}
\mathrm{Wr}(w,w^{*})=2i\mathrm{Im}(a)\mathrm{Wr}(x_{1},x_{2})=i,
\end{align}
which uniquely fixes the imaginary part of $a$. To completely define the solution, it is required to find the appropriate value of $\mathrm{Re}(b)$ which satisfies the boundary conditions
\begin{align}
\label{b_cond}
\lim_{t\to-\infty}\frac{1}{2\xi(t)^{2}}=\Omega(t),\qquad
\lim_{t\to-\infty}\dot{\xi}(t)=0.
\end{align}
These conditions are consistent with the system being in the adiabatic ground state at large $|t|$. In the $t\to\infty$ limit, $\Omega^{2}$ diverges and one must use the asymptotic expansion for the Airy functions
\begin{align}
\lim_{t\to-\infty}x_{1}(t)\sim\frac{\cos\left(\frac{2}{3}\Omega^{3}-\frac{\pi}{4}\right)}{\sqrt{\pi}\Omega^{1/4}},\qquad
\lim_{t\to-\infty}x_{2}(t)\sim \frac{\sin\left(\frac{2}{3}\Omega^{3}-\frac{\pi}{4}\right)}{\sqrt{\pi}\Omega^{1/4}}.
\end{align}
In order to satsify \eqref{b_cond}, the oscillatory terms in the expression for $\xi$ must cancel for large $s$, implying
\begin{align}
\mathrm{Re}(a)=0,\qquad
\mathrm{Im}(a)=b.
\end{align}
Moreover one has to impose the condition 
\begin{align}
\label{wronsk_condition}
\mathrm{Wr}(w,w^{*})=2i\mathrm{Im}(a)b\mathrm{Wr}(x_{1},x_{2})=i,
\end{align}
which fully determines the coefficients in Eq.~\eqref{sol_def},
\begin{align}
\mathrm{Im}(a)=b=\sqrt{\frac{\pi}{2}}.
\end{align}
The resulting expression for the scale factor is
\begin{align}
\xi(t)^{2}=\frac{\pi}{2}\mathrm{Ai}\left(-\Omega(t)^{2}\right)^{2}+\frac{\pi}{2}\mathrm{Bi}\left(-\Omega(t)^{2}\right)^{2},
\end{align}
and the number of defects is given by Eq.\,\eqref{exc_den}. The number of defects at the final point of the ramp (which is the critical point) is obtained by evaluating Eq.~\eqref{exc_den} at $t=0$. At this instant the rescaled frequency is given by its finite-size correction $\Omega(0)=\Lambda^{-1/3}=(N\delta)^{-1/3}$, and the scale factor reads
\begin{align}
\label{xi_0}
\xi(0)^{2}=\frac{\pi}{2}\mathrm{Ai}\left(-\Lambda^{-2/3}\right)^{2}+\frac{\pi}{2}\mathrm{Bi}\left(-\Lambda^{-2/3}\right)^{2}.
\end{align}
Let us consider the thermodynamic limit $\Lambda\to \infty$ first. In this case the argument of the Airy functions goes to zero and the terms in the square brackets of Eq.~\eqref{exc_den} read
\begin{align}
\frac{1}{4\xi(0)^{4}}=\frac{3^{8/3}\Gamma(2/3)^{4}}{16\pi^{2}},\qquad
\left(\frac{\dot{\xi}(0)}{\xi(0)}\right)=\frac{3^{2/3}\Gamma(2/3)^{2}}{\Gamma(1/3)^{2}},
\end{align}
leading to
\begin{align}
n_{\mathrm{exc}}(0)=\frac{\pi\,\Lambda^{1/3}}{3^{2/3}\Gamma(1/3)^{2}},
\end{align}
where we restricted to the leading term in the $\Lambda\to\infty$ limit. Therefore the result for the number of excitations diverges in the thermodynamic limit with a power $N^{1/3}$. However the residual heat is finite since it is obtained by multiplying the divergent defect density with the vanishing oscillator frequency $E_{\rm res}(0)=\Delta(0)n_{\mathrm{exc}}(0)$, leading to
\begin{align}
E_{\rm res}=\frac{\pi\,\delta^{1/3}}{3^{2/3}\Gamma(1/3)^{2}},
\end{align}
which agrees with the KZ scaling of Ref.~\cite{hwang2015quantum}. For a finite-size system $N<\infty$, the slow ramp limit $\delta\to 0$ coincides with the $\Lambda\to 0$ limit of Eq.~\eqref{exc_den} evaluated at $t=0$. The leading term in this case is generated by the velocity correction to the effective frequency
\begin{align}
\lim_{\Lambda\to 0}\frac{\dot{\xi}(0)}{\xi(0)}=-\frac{5}{24}\Lambda^{2/3},
\end{align}
which, substituted into Eq.~\eqref{exc_den} evaluated at $t=0$, gives
\begin{align}
n_{\mathrm{exc}}(0)=\frac{25}{2304}\Lambda^{2}\propto \delta^{2},
\end{align}
which leads to the expected adiabatic correction for the residual energy $E_{\rm res}\propto\delta^2$ in a finite-size system\,\cite{zwerger2008limited}.
\subsection{Full ramp}
\label{SM3}
Along previous sections we have depicted the analytic solution of a semi-infinite ramp with frequency $\omega(t)^{2}=|t|$ starting at $t=-\infty$ and terminating at $t=0$. Now, we are gonna extend such treatment to the entire interval $t\in(-\infty,+\infty)$. It is worth noting that in the case of a full ramp $t\in(-\infty,+\infty)$, we do not consider the case of a finite $\Delta_{N}$, since its calculation do not present any relevant difference from the half-ramp case. Taking the thermodynamic limit first $N\to\infty$, we consider a general solution in the form of Eq.\,\eqref{sol_def} satisfying the boundary conditions
\begin{align}
\lim_{t\to 0^{+}}&\xi^{2}(t)=\frac{\Gamma(p)\Gamma(p+1)}{2\pi p^{2p}},\\
\lim_{t\to 0^{+}}&2\dot{\xi}(t)\xi(t)=\cot(p\pi)
\end{align}
in order to ensure continuity with the solution with the $t<0$ case discussed in Sec.\,\ref{SM2}. Interestingly, this result is accomplished by the coefficients choice
\begin{align}
a&=a_{+}=\sqrt{\frac{3\pi}{2}}\\
\mathrm{Re}(b)&=b_{1}=0\\
\mathrm{Im}(b)&=b_{2}=b_{+}=\sqrt{\frac{\pi}{6}}
\end{align}
which automatically satisfies the Wronskian condition in Eq.\,\eqref{wronsk_condition}.

The defect density in the large time limit $t\approx +\infty$ can be obtained by the asymptotic behaviour of the scale $\xi(t)$, which reads
\begin{align}
&\lim_{t\to\infty}\xi(t)^{2}\approx\frac{t^{1/2}(1+\cos(\pi/3)^{2}+2\cos(\pi/3)\sin(2\zeta))}{2\sin(\pi/3)^{2}}\\
&\lim_{t\to\infty}\xi(t)\dot{\xi}(t)\approx\frac{\cos(\pi/3)\cos(2\zeta)}{\sin(\pi/3)^{2}}.
\end{align}
where $\zeta=\frac{2}{3}t^{3/2}$. Once these expressions are plugged into Eqs.\,\eqref{exc_den} and\,\eqref{exp_fidelity}, one obtains the two results in Eqs.\,\eqref{asy_fid} and\,\eqref{asy_exc_den},
which prove that the fidelity of a quantum harmonic oscillator driven across its quantum critical point remains finite even in the $\delta\to 0$ limit.

\section{Entanglement dynamics in other long-range interacting models}
\subsection{Quantum Ising chain in a tilted field}
\label{sec_mori}
We further discuss the quenches in long-range quantum Ising chains in a tilted magnetic field, described by the following Hamiltonian
\beq
\label{eq_MoriH}
\hat H = 
 -  \frac{J_0}{ \mathcal{N}_{\alpha,N}}  \sum_{ i < j }^N \frac{\hat \sigma_{i}^x \hat \sigma_{j}^x}{\lvert i-j\rvert^{\alpha}}
 - h_z \sum_{i}^N \hat \sigma_{i}^z - h_x \sum_i^N \hat \sigma_i^x \ ,
\eeq
where now $h_z$ and $h_x$ are respectively the transverse and longitudinal field and $\mathcal N_{\alpha, N}$ is the Ka{c} normalization \eqref{kac_norm}.
This model has been considered by T.Mori in Ref.~\cite{mori2019pretermalization} . There, it is argued that the non-equilibrium dynamics of a long-range quantum Ising chain (with $0<\alpha<1$ and with transverse field $h_z=0.32 J$ and longitudinal field $h_x=0.26 J$) shows signatures of many-body chaos. The dynamics are studied by starting from the paramagnetic state with spins fully polarized along the $z$ axis, i.e., from $h_{z,0}=\infty$.
(Note that $x \leftrightarrow z$ have been exchanged in our conventions.)
 
 We apply here the non-equilibrium spin-wave theory and the theory of entanglement dynamics developed in the present chapter.
Upon adding a longitudinal field, the classical equation of motion of the collective spin [cf. Eq.~\eqref{eq_classicalMotion} of Section \ref{sec_411}] now reads
\begin{align}
\label{eq_motion_angles0Mori}
\begin{dcases}
\dot \theta  =  2 J_0 \sin\theta \cos\phi \sin\phi + 2 h_x \sin \phi \\
\dot \phi = -2h_z  + 2 J_0   \cos\theta \cos^2\phi  + 2 h_x \frac{\cos\theta}{\sin\theta}\cos \phi \ .
    \end{dcases}
\end{align}
and the evolution equations for the spin-wave correlations, ( $\tilde J_k \equiv J_0 \widetilde{f}_{\alpha,k}$ ) are
\begin{equation}
\label{eq_motion_feedback_qui}
\left\{
\begin{split}
\dot{G}_{{k}}^{qq}  = \, & 4 \tilde J_k\,   \cos\theta \cos\phi\sin\phi \, \, \tilde{G}_{{k}}^{qq}
+4\left(J_0 \cos^2\phi+ h_x \frac{\cos\phi}{\sin \theta} -\tilde J_k\,  \sin^2\phi \right)\, \tilde{G}_{{k}}^{qp},\\
\dot {G}_{{k}}^{pp} = & -4\left( J_0 \cos^2\phi + h_x \frac{\cos\phi}{\sin \theta}- \tilde J_k\,\, \cos^2\theta \cos^2\phi \right) \tilde{G}_{{k}}^{qp} 
-4 \tilde J_k\, \,  \cos\theta \cos\phi\sin\phi \,  \tilde{G}_{{k}}^{pp}\\
\dot {G}_{{k}}^{pq} = & -2\left(J_0\cos^2\phi - \tilde J_k\, \,   \cos^2\theta \cos^2\phi \right)\tilde{G}_{{k}}^{qq}
+ 2 \left(J_0 \cos^2\phi + h_x \frac{\cos\phi}{\sin \theta}-\tilde J_k\,\, \sin^2\phi \right) \tilde{G}_{{k}}^{pp}
\end{split} \ .
\right.
\end{equation}

\begin{figure}[t]
\fontsize{12}{10}\selectfont
\vspace{-4mm}
\centering
\includegraphics[scale = 0.75]{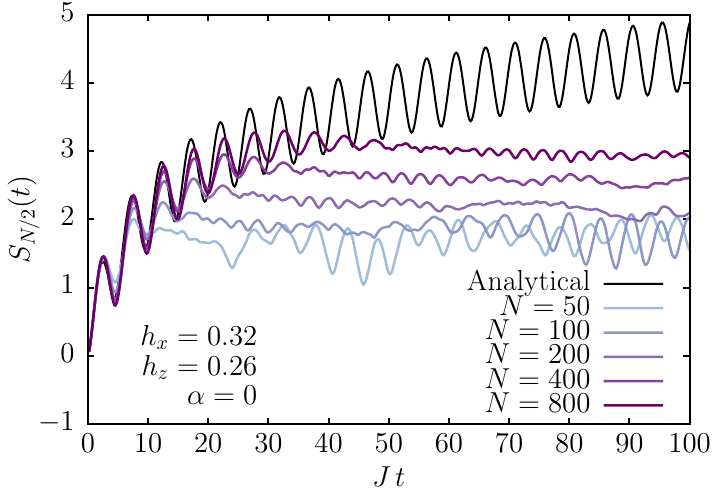}
\caption{Comparison between entanglement entropy growth computed numerically (ED) and analytically (semiclassical formula) for $\alpha=0$, for the quenches in Ref.\cite{mori2019pretermalization} [Eq. \eqref{eq_MoriH} with $h_z=0.32J_0$, $h_x=0.26J_0$, initial state polarized along $z$].
The growth is logarithmic, but finite-size effects are strong due to closeness to a mean-field dynamical critical point.
}
\vspace{-4mm}
\label{fig:5SM}
\end{figure}

\begin{figure}[h]
\fontsize{12}{10}\selectfont
\centering
\includegraphics[scale = 0.58]{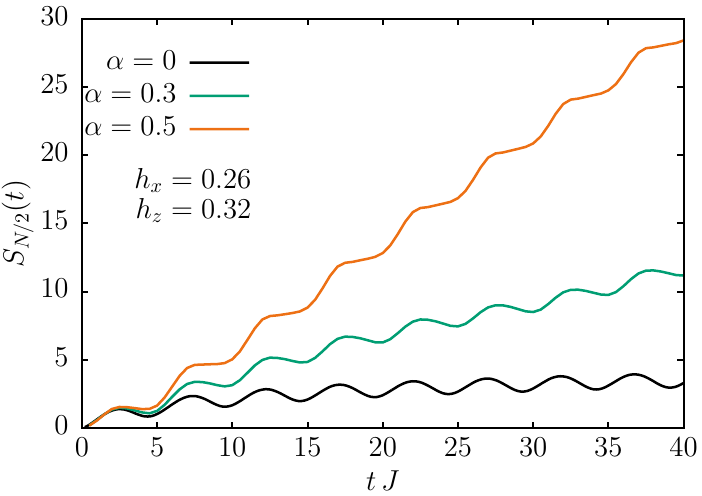}
\includegraphics[scale = 0.25]{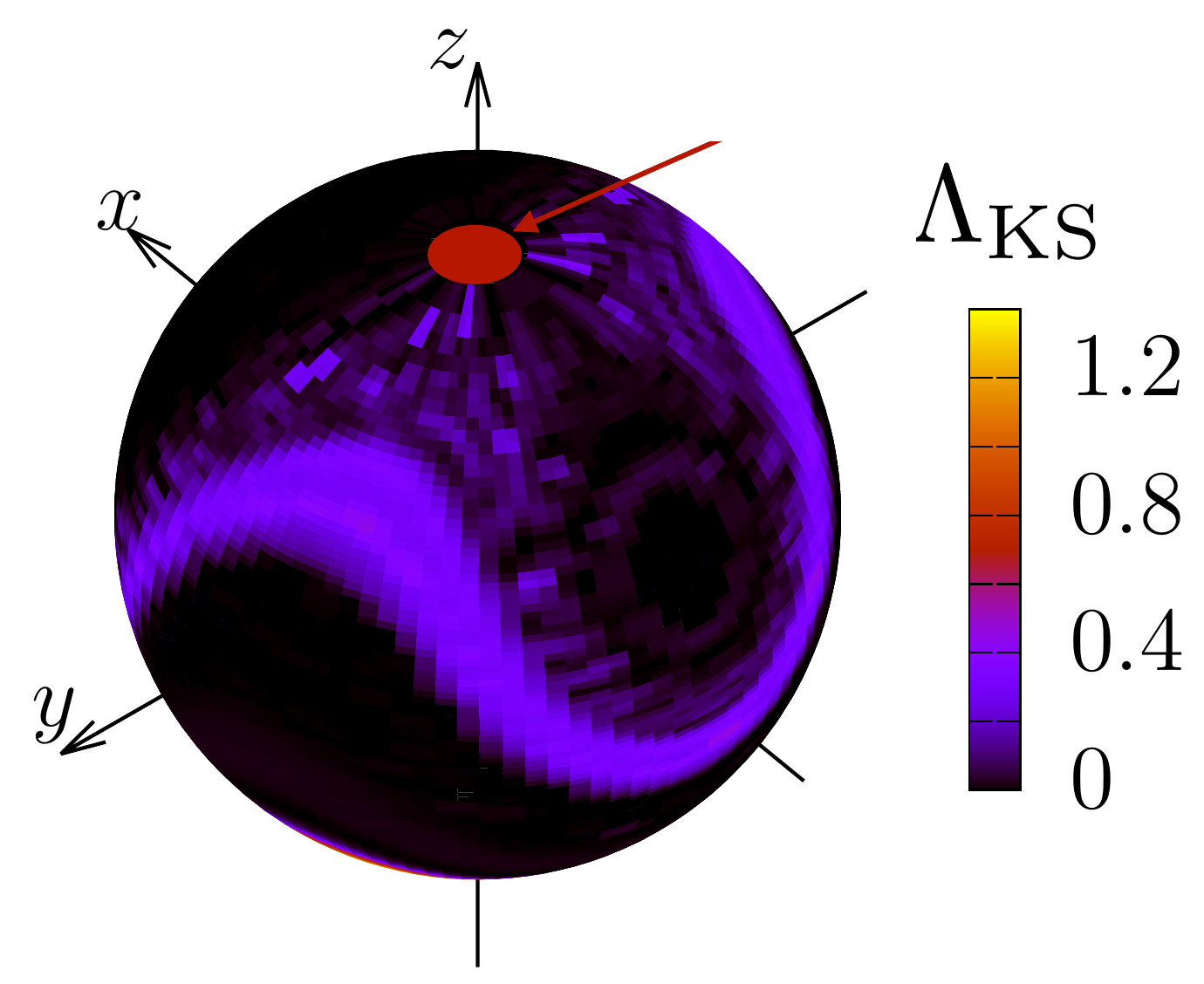}
\caption{Left panel: Comparison between entanglement entropy growth obtained via the full spin-wave computation with $N=500$, for increasing $\alpha=0$, $0.3$ and $0.5$, for the quenches in Ref. \cite{mori2019pretermalization} [Eq. \eqref{eq_MoriH} with $h_z=0.32J_0$, $h_x=0.26J_0$, initial state polarized along $z$].
While the growth is logarithmic in the integrable case $\alpha=0$, the breaking of integrability induced by a finite range triggers a linear growth of $S(t)$, 
due to unstable excitation of long-wavelength spin waves: see the text and Figure \ref{fig:populationMori}. 
Right panel: Spherical plot of the Kolmogorov-Sinai entropy rate $h_{KS}(\theta_0,\phi_0)$ versus the initial spin-polarized configuration, for $\alpha=0.7$. }
\label{fig:6SM}
\end{figure}

 We first study the mean-field case $\alpha=0$, verifying that the growth of entanglement entropy is logarithmic for the considered quench, see Figure \ref{fig:5SM}, as follows from our predictions.
 However, due to the closeness to a nearby dynamical critical point, the short-time dynamics of entanglement is fast, and the universal logarithmic behavior emerges only over longer times. 
 In agreement with our theory, larger system sizes are required to observe the asymptotic behavior, as confirmed by the ED numerical results.
 Because of these strong finite-size effects, we did not attempt for $\alpha>0$ numerical investigations with MPS-TDVP, limited to $N\lesssim 100$, but directly studied the limiting behavior in the thermodynamic limit via a full spin-wave calculation of entanglement dynamics.
 The results are shown in Figure \ref{fig:6SM}, left panel, for increasing values of $\alpha$, and they confirm that the growth of entanglement entropy is linear for $\alpha>0$, as suggested by the results of Ref. \cite{mori2019pretermalization} in view of the interpretation provided by the theory presented here.

\begin{figure}[h!]
\fontsize{12}{10}\selectfont
\centering
\includegraphics[width=0.48\textwidth]{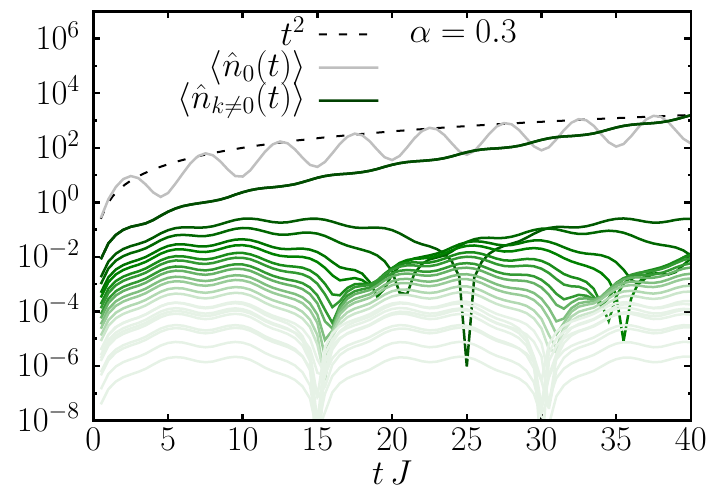}
\includegraphics[width=0.48\textwidth]{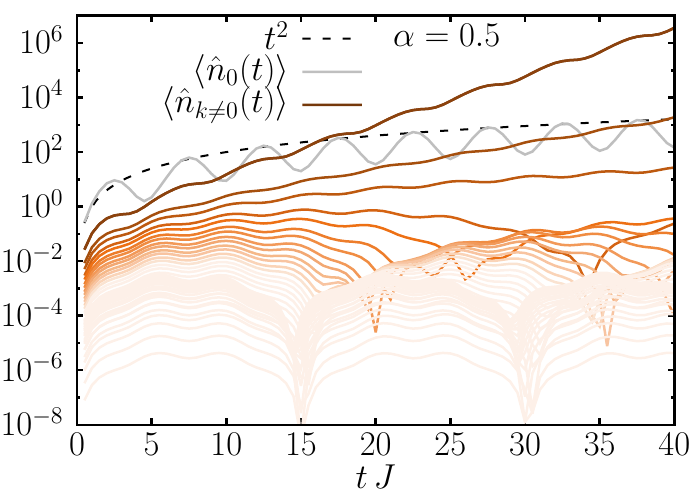}
\caption{Time-dependent $k$-resolved spin-wave population for the quenches in Ref. \cite{mori2019pretermalization} [Eq. \eqref{eq_MoriH} with $h_z=0.32J_0$, $h_x=0.26J_0$, initial state polarized along $z$]. 
Collective quantum fluctuations with $k=0$ grow polynomially, whereas the long-wavelength modes $k=\pm 2\pi/L$ (left, $\alpha=0.3$) and $k=\pm 2\pi/L,4\pi/L,6\pi/L$ (right, $\alpha=0.5$), diverge exponentially fast in time. 
Here we have set $N=500$.}
\label{fig:populationMori}
\end{figure}

To fully corroborate this picture, we presented a similar analysis to that outlined above for the Ising chain in a transverse field.
In Figure \ref{fig:populationMori}, we report the time evolution of the $k$-resolved spin-wave population for the same quench.
The dynamical production of long-wavelength spin-wave excitations is unstable, i.e., exponentially growing.
This occurrence hints at the fact that the quench considered in Ref. \cite{mori2019pretermalization} falls into a layer of instability of the many-body semiclassical dynamics, characterized by a positive Kolmogorov-Sinai entropy rate \eqref{eq_KS_ic} and hence a linear growth of entanglement entropy in time.
This is confirmed by the spherical plot in Figure \ref{fig:6SM}, right panel, of the Kolmogorov-Sinai entropy rate $\Lambda_{\text{KS}}$ as a function of the initial configuration on the Bloch sphere \eqref{eq_KS_ic}.
The considered quench falls inside the instability layer which opens up around the classical separatrix upon increasing $\alpha>0$.
However, we emphasize that a large set of initial configurations show a stable generation of spin waves, and hence slow logarithmic growth of entanglement entropy, even for this Hamiltonian (the black region in the spherical plot).

\subsection{Short-range perturbations to collective spin models }
\label{app_short}

The above analysis shows that slow logarithmic growth of the entanglement entropy can be expected in the quench dynamics of spin-$1/2$ systems with long-range interactions.
The underlying mechanism involves the existence of a \emph{discrete} set of excitation modes (the long-wavelength spin waves) which yield a bounded, subleading contribution to entanglement when non-resonantly driven by the collective spin dynamics.
However, this is an intrinsic property of slowly-decaying interactions, which generically fails for other types of perturbations.
To explicitly show this, we consider \emph{additional} finite-range interactions as perturbations to an integrable system with collective interactions. 
To be specific, we consider a Hamiltonian of the form
\begin{equation}
\label{eq_lrsr}
\hat H_{lr+sr} = \hat H_{\alpha} - \lambda \sum_{i} \hat \sigma_i^x \, \hat \sigma^x_{i+1} \ , 
\end{equation}
where $\hat H_{\alpha}$ is the long-range quantum Ising chain in Eq.~\eqref{eq_H} ($d=1$). 
In Refs. \cite{lerose2018chaotic, lerose2019impact}, it has been shown that the nonequilibrium spin-wave approach adequately describes the dynamics of this Hamiltonian when $\lambda \ll J$.
We show that the two kinds of perturbations, corresponding to raising $\alpha$ or $\lambda$ from $0$, respectively, lead to a radically different scenario of entanglement growth, in accordance with the theory developed above.

\begin{figure}[t]
\fontsize{12}{10}\selectfont
\centering
\includegraphics[width=0.46\textwidth]{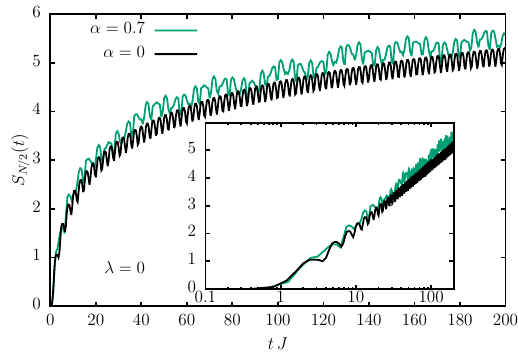}
\includegraphics[width=0.46\textwidth]{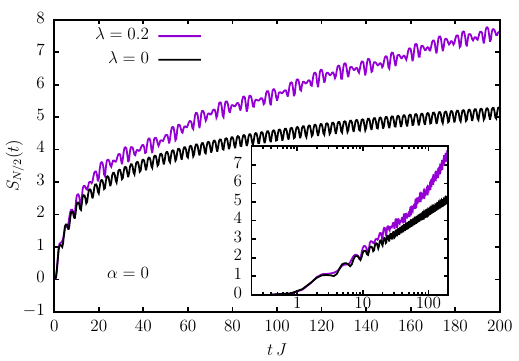}
\caption{Entanglement entropy growth obtained via the full spin-wave analysis, for quenches $h_0=0 \to h_f=0.7 J$ in the Hamiltonian (\ref{eq_lrsr}). We compare the effects of the two kinds of perturbations to the fully-connected quantum Ising model with $\alpha=0$, $\lambda=0$: 
\textit{i)} $\alpha=0.7$, $\lambda=0$, i.e., a slow spatial decay of interactions (left);
\textit{ii)} $\alpha=0$, $\lambda=0.2 J$, i.e., additional weak nearest-neighbor interactions (right). 
In both plots, the black lines represent the behavior of the fully-connected quantum Ising model with $\alpha=0$, $\lambda=0$, for comparison.
It is apparent that the former $\alpha$-perturbation (top) provides only bounded corrections to the logarithmic growth of the permutationally-invariant limit, whereas the latter $\lambda$-perturbation clearly exhibits the onset of a linear-in-time growth (with a small slope) which can be appreciated at long times.
The insets report the same data on a logarithmic time scale, highlighting the different behavior. In this computation  $N=500$. }
\label{fig:SR_LR}
\end{figure}

For the spin-wave analysis of the Hamiltonian $H_{lr+sr}$ in Eq. \eqref{eq_lrsr}, it is actually sufficient to substitute $J_{k\ne0} = J_0 \widetilde{f}_{\alpha,k} +  \lambda \cos k $ in Eqs.~\eqref{eq_motion_feedback}.
In the case $\alpha = 0$, $\lambda \neq 0$, the spin-wave Hamiltonian features two fundamental differences:
firstly, it is equivalent to a system of quantum oscillators with short-range  interactions, hence described by a \emph{continuous} dispersion relation with a finite bandwidth (apart from the singular $k=0$ mode); 
secondly, all excitations with $k \neq 0$ now live on a widely separated energy scale $\lambda \ll J$ with respect to the classical drive.
Therefore, away from fine-tuned resonances, the system typically behaves as a standard model of free bosonic excitations, where the fast, non-resonant drive amounts to modifying their effective dispersion relation. Such a system is expected to exhibit light-cone spreading of quantum correlations and linear growth of entanglement entropy, according to the standard Calabrese-Cardy quasiparticle picture \cite{calabrese2005evolution}, in stark contrast to the perturbation with $\alpha>0$, $\lambda=0$ discussed above. 

To be fair, it should be noted that the $\lambda$-perturbed model features a coexistence of two mechanisms, namely the spin squeezing associated with the singular $k=0$ mode and the traveling quasiparticles associated with the all the remaining $k\neq 0$ modes. Although the second mechanism is clearly dominant [linear over logarithmic $S(t)$], for tiny perturbations $\lambda \ll J$, a long time is required to appreciate this distinction. In practice, for small sizes, short times, weak quenches, and/or weak perturbations, one will always observe a crossover from initial logarithmic growth to an asymptotic linear growth.

We verified the predictions above explicitly: see the comparison between the two perturbations in  Figure \ref{fig:SR_LR}.
We conclude that, as expected based on the present analysis, the nature of the integrability-breaking perturbation is crucial, and the slow growth of entanglement analyzed is a characteristic property of long-range interactions.

\section{Floquet Hamiltonian and high-frequency expansion} 
\label{app_Magnus}
 
Whenever the time-dependent Hamiltonian of a system has a 
period $T$, i.e., $\hat H(t+T)=\hat H(t)$, the resulting time-evolution operator $\hat U(t_2,t_1)$ satisfies
%
\beq
\hat U(t_0+nT,t_0) = 
\big[ \hat U(t_0+T,t_0) \big]^n
\eeq
 for any integer $n$. Accordingly, it is 
 convenient to define an effective static Hamiltonian $\hat H_{\text{eff}}$ \cite{blanes2009magnus, bukov2015universal},
\beq
\hat U_F \equiv \hat U(t_0+T,t_0) = \mathcal{T} e^{ 
-i\int_{t_0}^{t_0+T} d\tau \, \hat H(\tau) 
} \equiv e^ {-i T \hat H_{\text{eff}}},
\eeq
usually referred to as the \textit{Floquet Hamiltonian}.
Its spectrum is defined up to integer multiples of the frequency $2\pi/T$ and it is 
independent of the choice of the reference time $t_0$. 
The state of the system at stroboscopic times $t_n=t_0+nT$ is therefore entirely and unambiguously determined by the  Floquet Hamiltonian $\hat H_{\text{eff}}$.
A series expansion of $\hat H_{\text{eff}}$ in powers of the period $T$,  known as the \textit{Magnus expansion}, 
can be written as 
\beq
\label{eq:Magnus}
\hat H_{\text{eff}} = \sum_{n=0}^{\infty} \hat H_{\text{eff}}^{(n)},
\eeq
with $\hat H_{\text{eff}}^{(n)}$ proportional to $T^n$. Explicitly, the first terms read
\begin{align}
\label{eq:zerothorder}
\hat H_{\text{eff}}^{(0)} &= \int_{t_0}^{t_0+T} \frac{d\tau_1}{T} \, \hat H(\tau_1) ,\\ 
\hat H_{\text{eff}}^{(1)} &= \frac{T}{2}  \int_{t_0}^{t_0+T}  \frac{d\tau_1}{T} \int_{t_0}^{t_0+\tau_1}  \frac{d\tau_2}{T} \, \big[ \hat H(\tau_1),\hat H(\tau_2)\big] ,
\label{eq:firstorder}
\end{align}
with the higher order terms involving a increasing number of nested commutators of $\hat H$ at different times.
This expansion is convergent when $T$ is smaller than the inverse maximal extension of the spectrum of $\hat H(t)$ \cite{blanes2009magnus}. 



\bibliographystyle{elsarticle-num}
\bibliography{draft.bib}

\end{document}